\documentclass[%
 reprint,
superscriptaddress,
preprintnumbers,
 amsmath,amssymb,
 aps,prd,
authoryear,
]{revtex4-2}

\usepackage{graphicx}%
\usepackage{dcolumn}%
\usepackage{bm}%
\usepackage[dvipsnames]{xcolor}%
\usepackage{MathUnicode}
\usepackage{xspace}
\usepackage{booktabs}
\usepackage{natbib}%
\usepackage[colorlinks=true, allcolors=blue]{hyperref}
\usepackage{orcidlink}

\begin{document}
\newcommand{\avg}[1]{\left\langle{#1}\right\rangle}

\newcommand{\vecs}{\ln \boldsymbol{s}}
\newcommand{\vecalpha}{\boldsymbol{\alpha}}

\newcommand{\vecx}{\boldsymbol{x}}

\newcommand{\bsel}{b_\mathrm{sel}}

\newcommand{\beq}{\begin{equation}}
\newcommand{\eeq}{\end{equation}}
\newcommand{\beqa}{\begin{equation}\begin{aligned}}
\newcommand{\eeqa}{\end{aligned}\end{equation}}
\newcommand{\bit}{\begin{itemize}}
\newcommand{\eit}{\end{itemize}}

\newcommand{\hiGpc}{h^{-1} \rm Gpc}
\newcommand{\hiMpc}{h^{-1} \rm Mpc}
\newcommand{\hikpc}{h^{-1} \rm kpc}
\newcommand{\hiMsun}{h^{-1} M_\odot}
\newcommand{\Msun}{M_\odot}
\newcommand{\pimax}{\Pi_{\mathrm{max}}}

\newcommand{\DS}{\Delta\Sigma}
\newcommand{\rp}{r_{\rm p}}

\newcommand{\Sigmasys}{\Sigma_{\rm sys}}
\newcommand{\Sigmanosys}{\Sigma_{\rm no{\text-}sys}}
\newcommand{\Sigmacen}{\Sigma_{\rm cen}^{\rm bary}}
\newcommand{\Sigmamis}{\Sigma_{\rm mis}^{\rm bary}}

\newcommand{\DSsys}{\Delta\Sigma_{\rm sys}}
\newcommand{\DSnosys}{\Delta\Sigma_{\rm no\text{-}sys}}
\newcommand{\DScen}{\Delta\Sigma_{\rm cen}^{\rm bary}}
\newcommand{\DSmis}{\Delta\Sigma_{\rm mis}^{\rm bary}}

\newcommand{\fmis}{f_{\rm mis}}
\newcommand{\fcen}{f_{\rm cen}}
\newcommand{\rmis}{r_{\rm mis}}

\newcommand{\darkemu}{\texttt{DarkEmulator}\xspace}
\newcommand{\redmapper}{{redMaPPer}\xspace} %

\newcommand{\sigmalam}{{\sigma_{\lambda}^{\rm tot}}}
\newcommand{\sigmaDS}{\sigma_{\rm \Delta\Sigma}}
\newcommand{\rholds}{{\rho_{\lambda\Delta\Sigma}}}   %

\newcommand{\ximin}{\xi_{\mathrm{min}}}

\newcommand{\dd}{\,\mathrm{d}}

\newcommand{\mtwoh}{M_{\mathrm{200}c}}

\preprint{FERMILAB-PUB-26-0837-PPD}

\title{No Cluster Left Behind I:\\ Joint Mass Calibration for Dark Energy Survey and South Pole Telescope Clusters}

\author{C.~Zhou\,\orcidlink{0000-0002-5727-2432
}}
\affiliation{Department of Physics, University of California, Santa Cruz, CA 95064, USA}
\affiliation{Santa Cruz Institute for Particle Physics, Santa Cruz, CA 95064, USA}
\email{zhou.conghao@ucsc.edu}

\author{A.~N.~Salcedo\,\orcidlink{0000-0003-1420-527X}}
\affiliation{Steward Observatory, University of Arizona, 933 North Cherry Avenue, Tucson, AZ 85721, USA}
\affiliation{Department of Physics, University of Arizona, 1118 East Fourth Street, Tucson, AZ 85721, USA}

\author{S.~Grandis\,\orcidlink{0000-0002-4577-8217
}}
\affiliation{Universität Innsbruck, Institut für Astro- und Teilchen-physik, Technikerstr. 25/8, 6020 Innsbruck, Austria}

\author{T.~Jeltema\,\orcidlink{0000-0001-6089-0365
}}
\affiliation{Department of Physics, University of California, Santa Cruz, CA 95064, USA}
\affiliation{Santa Cruz Institute for Particle Physics, Santa Cruz, CA 95064, USA}

\author{M.~Costanzi\,\orcidlink{0009-0003-0003-5967}}
\affiliation{Astronomy Unit, Department of Physics, University of Trieste, via Tiepolo 11, I-34131 Trieste, Italy}
\affiliation{INAF-Osservatorio Astronomico di Trieste, via G. B. Tiepolo 11, I-34143 Trieste, Italy}
\affiliation{Institute for Fundamental Physics of the Universe, Via Beirut 2, 34014 Trieste, Italy}

\author{L.~E.~Bleem\,\orcidlink{0000-0001-7665-5079}}
\affiliation{High-Energy Physics Division, Argonne National Laboratory, 9700 South Cass Avenue, Lemont, IL 60439, USA}
\affiliation{Kavli Institute for Cosmological Physics, University of Chicago, 5640 South Ellis Avenue, Chicago, IL 60637, USA}
\affiliation{Department of Astronomy and Astrophysics, University of Chicago, 5640 South Ellis Avenue, Chicago, IL 60637, USA}

\author{A.~Farahi\,\orcidlink{0000-0003-0777-4618
}}
\affiliation{Department of Statistics and Data Science, University of Texas at Austin, Austin, TX 78757, USA}
\affiliation{NSF-Simons AI Institute for Cosmic Origins, University of Texas at Austin, Austin, TX 78757, USA}

\author{M.~Gatti\,\orcidlink{0000-0001-6134-8797
}}
\affiliation{Institut de Ci‘encies de l’Espai, IEEC-CSIC, Campus UAB, Facultat de Ci‘encies, Torre C5 par-2, Barcelona 08193, Spain}

\author{T.~Shin\,\orcidlink{0000-0002-6389-5409}}
\affiliation{McWilliams Center for Cosmology and Astrophysics, Department of Physics, Carnegie Mellon University, Pittsburgh, PA, USA}

\author{F.~Kéruzoré\,\orcidlink{0000-0002-9605-5588}}
\affiliation{High-Energy Physics Division, Argonne National Laboratory, 9700 South Cass Avenue, Lemont, IL 60439, USA}

\author{C.~L.~Reichardt\,\orcidlink{0000-0003-2226-9169}}
\affiliation{School of Physics, University of Melbourne, Parkville, VIC 3010, Australia}

\author{J.~H.~Esteves\,\orcidlink{0000-0003-4373-2386}}
\affiliation{Department of Physics, Harvard University, 17 Oxford St, Cambridge, MA, USA}

\author{M.~Aguena\,\orcidlink{0000-0001-5679-6747}}
\affiliation{INAF-Osservatorio Astronomico di Trieste, via G. B. Tiepolo 11, I-34143 Trieste, Italy}
\affiliation{Laborat\'orio Interinstitucional de e-Astronomia - LIneA, Av. Pastor Martin Luther King Jr, 126 Del Castilho, Nova Am\'erica Offices, Torre 3000/sala 817 CEP: 20765-000, Brazil}

\author{S.~Allam\,\orcidlink{0000-0002-7069-7857}}
\affiliation{Fermi National Accelerator Laboratory, P. O. Box 500, Batavia, IL 60510, USA}

\author{D.~Bacon}
\affiliation{Institute of Cosmology and Gravitation, University of Portsmouth, Portsmouth, PO1 3FX, UK}

\author{J.~Blazek\,\orcidlink{0000-0002-4687-4657}}
\affiliation{Department of Physics, Northeastern University, Boston, MA 02115, USA}

\author{S.~Bocquet\,\orcidlink{0000-0002-4900-805X}}
\affiliation{University Observatory, LMU Faculty of Physics, Scheinerstr. 1, 81679 Munich, Germany}

\author{D.~Brooks\,\orcidlink{0000-0002-8458-5047}}
\affiliation{Department of Physics \& Astronomy, University College London, Gower Street, London, WC1E 6BT, UK}

\author{D.~L.~Burke\,\orcidlink{0000-0003-1866-1950}}
\affiliation{Kavli Institute for Particle Astrophysics \& Cosmology, P. O. Box 2450, Stanford University, Stanford, CA 94305, USA}
\affiliation{SLAC National Accelerator Laboratory, Menlo Park, CA 94025, USA}

\author{R.~Camilleri}
\affiliation{School of Mathematics and Physics, University of Queensland,  Brisbane, QLD 4072, Australia}

\author{J.~Carretero\,\orcidlink{0000-0002-3130-0204}}
\affiliation{Institut de F\'{\i}sica d'Altes Energies (IFAE), The Barcelona Institute of Science and Technology, Campus UAB, 08193 Bellaterra (Barcelona) Spain}

\author{L.~N.~da Costa\,\orcidlink{0000-0002-7731-277X}}
\affiliation{Laborat\'orio Interinstitucional de e-Astronomia - LIneA, Av. Pastor Martin Luther King Jr, 126 Del Castilho, Nova Am\'erica Offices, Torre 3000/sala 817 CEP: 20765-000, Brazil}

\author{T.~M.~Davis\,\orcidlink{0000-0002-4213-8783}}
\affiliation{School of Mathematics and Physics, University of Queensland,  Brisbane, QLD 4072, Australia}

\author{D.~L.~DePoy}
\affiliation{George P. and Cynthia Woods Mitchell Institute for Fundamental Physics and Astronomy, and Department of Physics and Astronomy, Texas A\&M University, College Station, TX 77843,  USA}

\author{S.~Desai\,\orcidlink{0000-0002-0466-3288}}
\affiliation{Department of Physics, IIT Hyderabad, Kandi, Telangana 502285, India}

\author{P.~Doel\,\orcidlink{0000-0002-6397-4457}}
\affiliation{Department of Physics \& Astronomy, University College London, Gower Street, London, WC1E 6BT, UK}

\author{A.~E.~Evrard\,\orcidlink{0000-0002-4876-956X}}
\affiliation{Department of Astronomy, University of Michigan, Ann Arbor, MI 48109, USA}
\affiliation{Department of Physics, University of Michigan, Ann Arbor, MI 48109, USA}

\author{J.~Garc\'ia-Bellido\,\orcidlink{0000-0002-9370-8360}}
\affiliation{Instituto de Fisica Teorica UAM/CSIC, Universidad Autonoma de Madrid, 28049 Madrid, Spain}

\author{D.~Gruen\,\orcidlink{0000-0003-3270-7644}}
\affiliation{University Observatory, LMU Faculty of Physics, Scheinerstr. 1, 81679 Munich, Germany}

\author{G.~Gutierrez\,\orcidlink{0000-0003-0825-0517}}
\affiliation{Fermi National Accelerator Laboratory, P. O. Box 500, Batavia, IL 60510, USA}

\author{D.~L.~Hollowood\,\orcidlink{0000-0002-9369-4157}}
\affiliation{Santa Cruz Institute for Particle Physics, Santa Cruz, CA 95064, USA}

\author{D.~J.~James\,\orcidlink{0000-0001-5160-4486}}
\affiliation{Center for Astrophysics $\vert$ Harvard \& Smithsonian, 60 Garden Street, Cambridge, MA 02138, USA}

\author{K.~Kuehn\,\orcidlink{0000-0003-0120-0808}}
\affiliation{Australian Astronomical Optics, Macquarie University, North Ryde, NSW 2113, Australia}
\affiliation{Lowell Observatory, 1400 Mars Hill Rd, Flagstaff, AZ 86001, USA}

\author{O.~Lahav\,\orcidlink{0000-0002-1134-9035}}
\affiliation{Department of Physics \& Astronomy, University College London, Gower Street, London, WC1E 6BT, UK}

\author{J.~L.~Marshall\,\orcidlink{0000-0003-0710-9474}}
\affiliation{George P. and Cynthia Woods Mitchell Institute for Fundamental Physics and Astronomy, and Department of Physics and Astronomy, Texas A\&M University, College Station, TX 77843,  USA}

\author{J.~Mena-Fern{\'a}ndez\,\orcidlink{0000-0001-9497-7266}}
\affiliation{Aix Marseille Univ, CNRS/IN2P3, CPPM, Marseille, France}
\affiliation{Universit\'e Grenoble Alpes, CNRS, LPSC-IN2P3, 38000 Grenoble, France}

\author{R.~Miquel\,\orcidlink{0000-0002-6610-4836}}
\affiliation{Instituci\'o Catalana de Recerca i Estudis Avan\c{c}ats, E-08010 Barcelona, Spain}
\affiliation{Institut de F\'{\i}sica d'Altes Energies (IFAE), The Barcelona Institute of Science and Technology, Campus UAB, 08193 Bellaterra (Barcelona) Spain}

\author{R.~L.~C.~Ogando\,\orcidlink{0000-0003-2120-1154}}
\affiliation{Observat\'orio Nacional, Rua Gal. Jos\'e Cristino 77, Rio de Janeiro, RJ - 20921-400, Brazil}

\author{A.~A.~Plazas~Malag\'on\,\orcidlink{0000-0002-2598-0514}}
\affiliation{Kavli Institute for Particle Astrophysics \& Cosmology, P. O. Box 2450, Stanford University, Stanford, CA 94305, USA}
\affiliation{SLAC National Accelerator Laboratory, Menlo Park, CA 94025, USA}

\author{A.~Porredon\,\orcidlink{0000-0002-2762-2024}}
\affiliation{Centro de Investigaciones Energ\'eticas, Medioambientales y Tecnol\'ogicas (CIEMAT), Madrid, Spain}
\affiliation{Ruhr University Bochum, Faculty of Physics and Astronomy, Astronomical Institute, German Centre for Cosmological Lensing, 44780 Bochum, Germany}

\author{A.~K.~Romer\,\orcidlink{0000-0002-9328-879X}}
\affiliation{Department of Physics and Astronomy, Pevensey Building, University of Sussex, Brighton, BN1 9QH, UK}

\author{E.~Sanchez\,\orcidlink{0000-0002-9646-8198}}
\affiliation{Centro de Investigaciones Energ\'eticas, Medioambientales y Tecnol\'ogicas (CIEMAT), Madrid, Spain}

\author{D.~Sanchez Cid\,\orcidlink{0000-0003-3054-7907}}
\affiliation{Centro de Investigaciones Energ\'eticas, Medioambientales y Tecnol\'ogicas (CIEMAT), Madrid, Spain}
\affiliation{Physik-Institut, University of Zürich, Winterthurerstrasse 190, CH-8057 Zürich, Switzerland}

\author{I.~Sevilla-Noarbe\,\orcidlink{0000-0002-1831-1953}}
\affiliation{Centro de Investigaciones Energ\'eticas, Medioambientales y Tecnol\'ogicas (CIEMAT), Madrid, Spain}

\author{E.~Suchyta\,\orcidlink{0000-0002-7047-9358}}
\affiliation{Computer Science and Mathematics Division, Oak Ridge National Laboratory, Oak Ridge, TN 37831}

\author{M.~E.~C.~Swanson\,\orcidlink{0000-0002-1488-8552}}
\affiliation{National Center for Supercomputing Applications, University of Illinois, 605 E. Springfield Ave., Urbana, IL 61801, USA}

\author{V.~Vikram}
\affiliation{Central University of Kerala, Kasaragod, Kerala 671325, India}

\author{N.~Weaverdyck\,\orcidlink{0000-0001-9382-5199}}
\affiliation{Berkeley Center for Cosmological Physics, Department of Physics, University of California, Berkeley, CA 94720, US}
\affiliation{Lawrence Berkeley National Laboratory, 1 Cyclotron Road, Berkeley, CA 94720, USA}

\author{J.~Weller\,\orcidlink{0000-0002-8282-2010}}
\affiliation{Max Planck Institute for Extraterrestrial Physics, Giessenbachstrasse, 85748 Garching, Germany}
\affiliation{Universit\"ats-Sternwarte, Fakult\"at f\"ur Physik, Ludwig-Maximilians Universit\"at M\"unchen, Scheinerstr. 1, 81679 M\"unchen, Germany}

\collaboration{DES and SPT Collaborations}

\noaffiliation

\date{\today}%

\begin{abstract}
Multiwavelength observations of galaxy clusters are among the most promising
probes of the growth of structure and dark energy. Optically identified
clusters have a high number density and can thus probe the halo mass function
well into the group regime, but their weak-lensing mass calibration suffers
from a selection bias induced by line-of-sight projection effects. Clusters
detected via their intracluster medium (ICM) --- through X-ray emission or the
thermal Sunyaev--Zel'dovich effect (SZe) --- are less affected by such
selection biases, but current survey sensitivity restricts ICM-selected
samples to high-mass systems.
Previous multiwavelength analyses tend to limit to clusters with millimeter detections and thus to the high-mass population. 
In this paper, we propose a flexible modeling framework in which all clusters
from multiwavelength surveys can be combined and modeled, even if they are
detected in only one survey.
As a pilot study, we jointly model richness and lensing of clusters from the Dark Energy Survey and the SZe signal from the South Pole Telescope.
This approach yields precise cluster mass calibration while self-consistently
constraining optical selection biases.
We have found a $\sim20$\% bias in the large-scale lensing signal of clusters selected by richness, 
and a bias ranging from $\sim-20\%$ at low richness to zero at high richness for clusters selected by both richness and SZe. 
Compared to current approaches, this framework is highly scalable and readily extensible, paving the way for
the full utilization of multiwavelength cluster data in next-generation
surveys.
\end{abstract}

\maketitle

\section{Introduction}

The number counts of galaxy clusters across cosmic time are sensitive probes of the growth of structure and cosmic acceleration [see, e.g.,~\cite{allenCosmologicalParametersObservations2011, weinbergObservationalProbesCosmic2013, hutererGrowthCosmicStructure2015} for reviews]. Galaxy clusters are detectable across the electromagnetic spectrum: X-rays reveal the ionized intra-cluster medium (ICM) \cite{kelloggClustersGalaxiesWide1973, mitchellAriel5Observations1976, sarazinXrayEmissionClusters1986, rosatiEvolutionXrayClusters2002, vikhlininChandraSampleNearby2006, bulbulSRGEROSITAAllSky2024}; optical bands capture the stellar light from member galaxies and weak gravitational lensing \cite{zwickyMassesNebulaeClusters1937, lyndsLuminousArcsClusters1989, gonzalezIntraclusterLightNearby2005, zibettiIntergalacticStars0252005, wittmanFirstResultsShearselected2006, kneibClusterLenses2011, oguriCombiningClusterObservables2011, oguriClusterFindingAlgorithm2014, rykoffRedMaPPerAlgorithmSDSS2014, rykoffRedMaPPerGalaxyCluster2016, bellagambaAMICOOptimizedDetection2018, wenCatalogueClustersGalaxies2018, zhangDarkEnergySurvey2019, umetsuClustergalaxyWeakLensing2020, aguenaWaZPGalaxyCluster2021, klugePhotometricDissectionIntracluster2021, liReachingEdgeProbing2022, montesFaintLightGroups2022, zhouIntrinsicAlignmentRed2023, fuLoVoCCSIIWeak2024, zhangDarkEnergySurvey2024, benoistFinalWaZPGalaxy2025, chenWeakLensingShearSelectedGalaxy2025, englertIntraclusterLightAbell2025}; microwaves trace the thermal Sunyaev-Zeldovich effect (SZe) and cosmic microwave background (CMB) lensing \cite{sunyaevSmallscaleFluctuationsRelic1970, sunyaevObservationsRelicRadiation1972, zaldarriagaReconstructingProjectedMatter1999, carlstromCosmologySunyaevZeldovichEffect2002, lewisWeakGravitationalLensing2006, baxterMeasurementGravitationalLensing2015, bleemGalaxyClustersDiscovered2015, madhavacherilEvidenceLensingCosmic2015, planckcollaborationPlanck2015Results2016, bleemSPTpolExtendedCluster2020, hiltonAtacamaCosmologyTelescope2021, bleemGalaxyClustersDiscovered2024, madhavacherilAtacamaCosmologyTelescope2024, quAtacamaCosmologyTelescope2024}; and radio observations detect synchrotron radiation from relativistic electrons \cite{govoniMagneticFieldsClusters2004, ensslinMagneticTurbulenceCool2006, ferettiClustersGalaxiesObservational2012, vanweerenDiffuseRadioEmission2019, pandey-pommierComprehensiveSKASurvey2026}. The process of mapping these multiwavelength observables to the underlying cluster halo mass is called mass calibration. 

Current cluster surveys at optical, X-ray, and SZe wavelengths cover increasingly large overlapping areas, creating an opportunity to improve cluster mass calibration through the joint exploitation of multiwavelength observables \citep{cohnMultiwavelengthClusterMass2020}. This approach has already been successfully applied to ICM-selected cluster samples, where X-ray and/or SZe information are combined with optical data in a unified framework \cite{bocquetSPTClustersHST2024a, bocquetSPTClustersHST2024b, bocquetMultiprobeCosmologyAbundance2025, grandisSRGEROSITAAllSky2024, ghirardiniSRGEROSITAAllSky2024}. For optically selected samples, however, joint analyses have generally been restricted to clusters with complete multiwavelength coverage, thereby discarding a significant fraction of systems with observations only in optical and reducing the statistical power of the available data \citep[e.g.][]{costanziCosmologicalConstraintsY12021}. Building on \citet{grandisExploringContaminationDESY12021}, \citet{grandisSelectionFunctionClusters2025}, and \citet{zhouForecastingConstraintsOptical2024}, to fully leverage the power of multiwavelength cluster surveys, we introduce an inference framework that integrates clusters detected only in a subset of bands, ensuring no cluster is left behind.

In this paper, we focus on the joint modeling of optical richness $\lambda$, SZe signal-to-noise ratio (SNR) $\xi$, and the weak gravitational lensing of clusters binned by $\lambda$ and $\xi$.
Weak gravitational lensing is the coherent distortion of background galaxy images caused by the foreground matter \cite{bartelmannWeakGravitationalLensing2001, johnstonCrosscorrelationWeakLensing2007, mandelbaumPrecisionClusterMass2010, rozoStackedWeakLensing2011, aguenaCLMMLSSTDESCCluster2021, wuCosmologyGalaxyCluster2021}. Gravitational lensing probes the projected total mass in clusters without assumptions about hydrostatic equilibrium or dynamical state which makes it ideal to calibrate the relation between cluster mass and other cluster observables \cite{mandelbaumPrecisionClusterMass2010, rozoStackedWeakLensing2011,mantzWeighingGiantsGalaxy2016, melchiorWeaklensingMassCalibration2017, simetWeakLensingMeasurement2017, mcclintockDarkEnergySurvey2019, miyatakeWeaklensingMassCalibration2019, mulroyLoCuSSScalingRelations2019, huangOuterStellarMass2022, chiuEROSITAFinalEquatorialDepth2022, ansarinejadMassCalibrationYear32024, grandisSRGEROSITAAllSky2024, kwiecienImprovingGalaxyCluster2025, chiuSRGEROSITAAllSky2025}. 

The cluster richness, $\lambda$, is a proxy for the number of cluster member galaxies in optical and near-infrared surveys, computed by summing the membership probabilities of individual candidate galaxies \cite{oguriClusterFindingAlgorithm2014, rykoffRedMaPPerAlgorithmSDSS2014, rykoffRedMaPPerGalaxyCluster2016, aguenaWaZPGalaxyCluster2021}. It can be measured for low-mass clusters found in optical surveys, which can enable us to probe the low-mass end of the halo mass function \cite{weinbergObservationalProbesCosmic2013, wuCosmologyGalaxyCluster2021}. Richnesses estimated from photometric data are affected by line-of-sight projections, which correlate with the weak lensing signal and induce a bias in the lensing measurements of $\lambda$-selected clusters — a phenomenon known as the optical selection effect \cite{sunayamaImpactProjectionEffects2020, wuOpticalSelectionBias2022, leeOpticalGalaxyCluster2024, zhouForecastingConstraintsOptical2024, ndeImpactProjectioninducedOptical2026}. 
The optical selection effect, if left uncalibrated, could bias the observable-mass relation and the constraints on cosmology \citep{abbottDarkEnergySurvey2020}. 

The SZe SNR, $\xi$, is the amplitude of the thermal Sunyaev-Zeldovich effect signal extracted via a matched spatial filter relative to the local map noise \cite{vanderlindeGalaxyClustersSelected2010, bleemGalaxyClustersDiscovered2015, bleemSPTpolExtendedCluster2020, bleemGalaxyClustersDiscovered2024}. 
$\xi$ scales tightly with cluster mass, can be measured for high-redshift clusters, and is practically free from the large-scale structure induced projection effects that plague richness and lensing measurements \cite{bleemGalaxyClustersDiscovered2015, bleemSPTpolExtendedCluster2020, costanziCosmologicalConstraintsY12021, bleemGalaxyClustersDiscovered2024}. However, at the current survey sensitivity, the SZe measurements are mostly restricted to high mass clusters, which puts a limit on the statistical constraining power and the potential of breaking the degeneracy between $\Omega_{\mathrm m}$ and $\sigma_8$. 

Previous joint analyses of cluster richness, SZe SNR, and weak lensing have typically been restricted to clusters with significant SZe detections \cite{bocquetClusterCosmologyConstraints2019, costanziCosmologicalConstraintsY12021, bocquetSPTClustersHST2024a, bocquetSPTClustersHST2024b}. 
Compared to an analysis that only uses $\lambda$ and cluster lensing \cite{costanziMethodsClusterCosmology2019}, this approach mitigates the optical selection effect since the cluster sample is mostly high-mass clusters that suffer less from projection effects. However, this approach misses the statistical power and the potential to probe low-mass clusters from the vast population of clusters detected in optical bands but undetected by SZe. Furthermore, combining the stacked lensing, $\xi$, and $\lambda$ of all clusters, regardless of their SZe detection status, can effectively calibrate optical selection bias by tightly constraining the scatter between lensing and $\lambda$ at a given halo mass \citep{zhouForecastingConstraintsOptical2024}. This critical property remains unconstrained if the analysis is limited to SZe-detected clusters only.

In this paper, we present a joint modeling of weak lensing, $\lambda$ from the Dark Energy Survey (DES) Year 3 (Y3) and $\xi$ from the South Pole Telescope (SPT), in a framework where cluster observables can be flexibly combined and modeled. We show that we can obtain a good fit to the data . We demonstrate that our model places tight constraints on the correlated scatter between $\lambda$ and lensing at fixed halo mass, which shows the power of the inference framework in which no cluster is left behind. 

The paper is organized as follows. In Sec.~\ref{sec:data}, we present the DES and SPT data used in this work. %
In Sec.~\ref{sec:cluster_density} and Sec.~\ref{sec:lensing_likelihood}, we describe the multiwavelength modeling framework. In Sec~\ref{sec:inference}, we show our implementation, sampling, and posterior distribution. In Sec.~\ref{sec:results}, we show our constraints on optical selection bias, mass-observable relations, and mass distribution of $\lambda$ and $\xi$ clusters. In Sec.~\ref{sec:discussion}, we discuss the scalability and extendability of our model, its complementarity to other approaches to cluster cosmology, and forecast the application of the framework on future data sets.

Throughout the paper, we assume a flat $\Lambda$CDM cosmology with $\Omega_m = 0.3$ and $h=0.7$. All masses refer to $M_{200c}$, the mass enclosed in a sphere in which the density is 200 times the critical density of the universe. Unless noted, all distances are in comoving units. $\rp$ denotes projected distance. $r$ denotes three-dimensional distance.

\section{Data}
\label{sec:data}

For optical detections, we use the full volume-limited DES Year 3 \redmapper cluster catalog \citep[][]{abbottDarkEnergySurvey2018, abbottDarkEnergySurvey2025}, with 
a richness threshold $\lambda_{\rm min} = 20$, 
a redshift range $0.2 < z < 0.85$, 
and a sky coverage of 4,946 deg$^2$. The choice of richness threshold is consistent with previous cosmological analyses \citep{abbottDarkEnergySurvey2020, costanziCosmologicalConstraintsY12021, abbottDarkEnergySurvey2025}. 72 percent of this area overlaps with SPT-SZ and SPTpol. 

For SZe detections, we use the SPT cluster sample from SPT-SZ, SPTpol-ECS, and SPTpol-500d \cite{bleemGalaxyClustersDiscovered2015, bleemSPTpolExtendedCluster2020, bleemGalaxyClustersDiscovered2024}. For SPT fields overlapping with DES, we use the positional association with the DES \redmapper clusters of \citet{grandisSelectionFunctionClusters2025}.
Each \redmapper cluster is matched to the nearest SPT cluster candidate within $0.6\,R_\lambda$, where $R_\lambda = (\lambda/100)^{0.2}\,h^{-1}$Mpc, with a minimum search radius of 2 arcmin.
SPT cluster candidates with a secure redshift $z>0.9$ are excluded beforehand, and no redshift information enters the match.
Because the match is purely positional, we screen the 488 associations and remove those that fail a consistency check: 16 whose SPT redshift differs from the \redmapper redshift by more than 0.1, 24 whose SPT position lies more than 0.5 Mpc from the \redmapper center (30 fail at least one of the two).
These 30 clusters are removed from the sample. 
The SPT-only clusters outside the DES footprint are taken directly from the SPT-SZ, SPTpol-ECS, and SPTpol-500d catalogs \cite{bleemGalaxyClustersDiscovered2015, bleemSPTpolExtendedCluster2020, bleemGalaxyClustersDiscovered2024}: confirmed clusters with $\xi \ge 5$ and $0.2 \le z < 0.85$.

\begin{figure}
  \centering
  \includegraphics[width=\columnwidth]{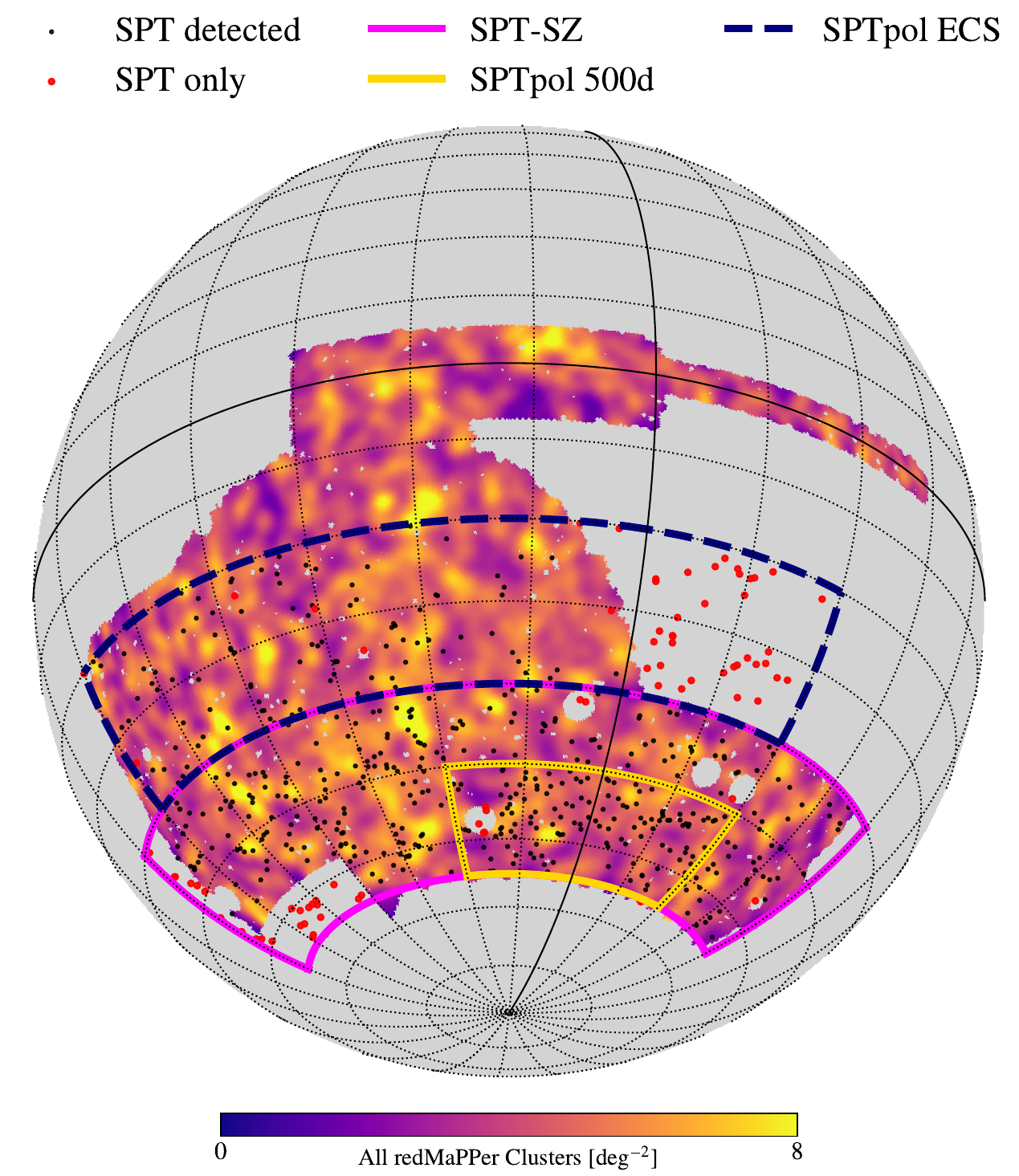}
  \caption{The DES and SPT data used in this study. 
  The color map shows the number density of the DES \redmapper clusters. 
  The black points show the \redmapper clusters detected by SPT, combining 
  SPT-SZ (magenta),
  SPTpol-ECS (dashed box),
  SPTpol-500d (yellow).
  The red points show the SPT-only clusters outside the DES footprint [Eq.~(\ref{eq:pdf_spt_only})]. The two SPTpol-ECS fields at 10--14\,h, which contain 38 of the SPT-only clusters, lie outside the view and are not shown.
  The DES-SPT overlap covers 3567 deg$^2$, the DES-only region is 1379 deg$^2$, and the SPT-only region is 1407 deg$^2$. 
  }
  \label{fig:overlap}
\end{figure}

\begin{figure*}
  \centering \includegraphics[width=\textwidth]{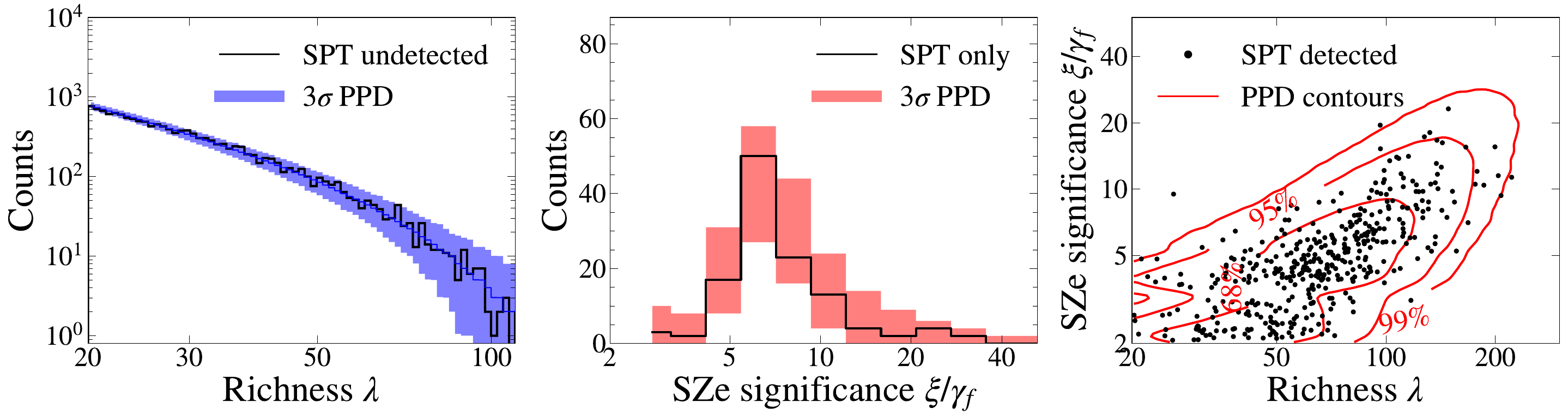}
  \caption{Posterior predictive distribution (PPD) of the cluster observables compared with observations, all redshifts combined.
  The PPD is produced by generating mock observables with Monte Carlo at sampled posterior steps.
  All histograms are differential counts per bin. Because the counts likelihood is not used in the fit, the PPD bands show the fitted \emph{shape} only: each posterior sample is renormalized to the observed total of the panel.
  \textbf{Left}: Marginal distribution of $\lambda$ for SPT-undetected clusters (in an SPT field but below $\ximin$). The black histogram is the data, and the blue band is the 3-$\sigma$ PPD.
  \textbf{Middle}: Marginal distribution of $\xi/\gamma_f$ for the 120 SPT-only clusters outside the DES footprint (SPT-SZ, SPTpol-ECS, and SPTpol-500d; no richness or lensing), starting at the lowest detection threshold $\ximin/\gamma_f$ in the sample, where $\gamma_f$ is the field-depth factor of Sec.~\ref{sec:scaling_relations} that puts clusters from SPT fields of different depth on a common scale The red band is the 3-$\sigma$ PPD of the SZe-only density term [Eq.~(\ref{eq:pdf_spt_only})].
  \textbf{Right}: Joint distribution of $\lambda$ and $\xi/\gamma_f$ for SPT-detected clusters. The black points are the observed values. The red contours enclose 68\%, 95\%, and 99\% of the PPD observable density. The Bayesian $p$-value of 0.53 shows no evidence of a misfit.
  }
  \label{fig:density_ppd}
\end{figure*}

The DES clusters with (without) a positional match to the SPT clusters are labeled ``SPT-detected'' (``SPT-undetected'').
The clusters only in the DES or SPT footprints are labeled ``DES-only'' or ``SPT-only''. 
The overlap between DES and various SPT fields is shown in Fig.~\ref{fig:overlap}, and our cluster sample is summarized in Table~\ref{tab:overlap}. The $\lambda$ and $\xi$ data used in this paper are shown in Figure~\ref{fig:density_ppd}.

\begin{table*}[]
\caption{Summary of our cluster samples, selected from DES Y3 \redmapper catalog with $\lambda\ge20$ and $0.2<z<0.85$ and  SPT-SZ, SPTpol-ECS, and SPTpol-500d with $\xi \ge 5$ and $0.2<z<0.85$.
The SPT-only column lists the combined area of the SPT-SZ, SPTpol-ECS, and SPTpol-500d footprints outside the DES footprint. The SPT surveys (SPT-SZ and SPT-pol) cover 72\% of the DES area but only detect 3.4\% of the \redmapper clusters in their footprint. 
We note that the DES-Y3 source catalog only covers 4,143 deg$^2$ because of different masking for lensing sources. 
\label{tab:overlap}
}
\smallskip
\renewcommand{\arraystretch}{1.5}   %
\setlength{\tabcolsep}{10pt}         %
\begin{ruledtabular}
\begin{tabular}{cccccc}
 & \multicolumn{2}{c}{DES-SPT overlap} & DES-only & SPT-only & Total \\ \colrule
Area {[}deg$^2${]} & \multicolumn{2}{c}{3,567} & 1,379 & 1,407 & 6,353 \\ \colrule
Cluster counts     & \multicolumn{1}{c}{\begin{tabular}[c]{@{}c@{}}SPT-detected \\ 488\end{tabular}}  & \begin{tabular}[c]{@{}c@{}}SPT-undetected\\ 13,926\end{tabular} & 5,452 & 120 & 19,986 \\ 
\end{tabular}
\end{ruledtabular}
\end{table*}

\begin{figure*}
  \centering
  \includegraphics[width=\textwidth]{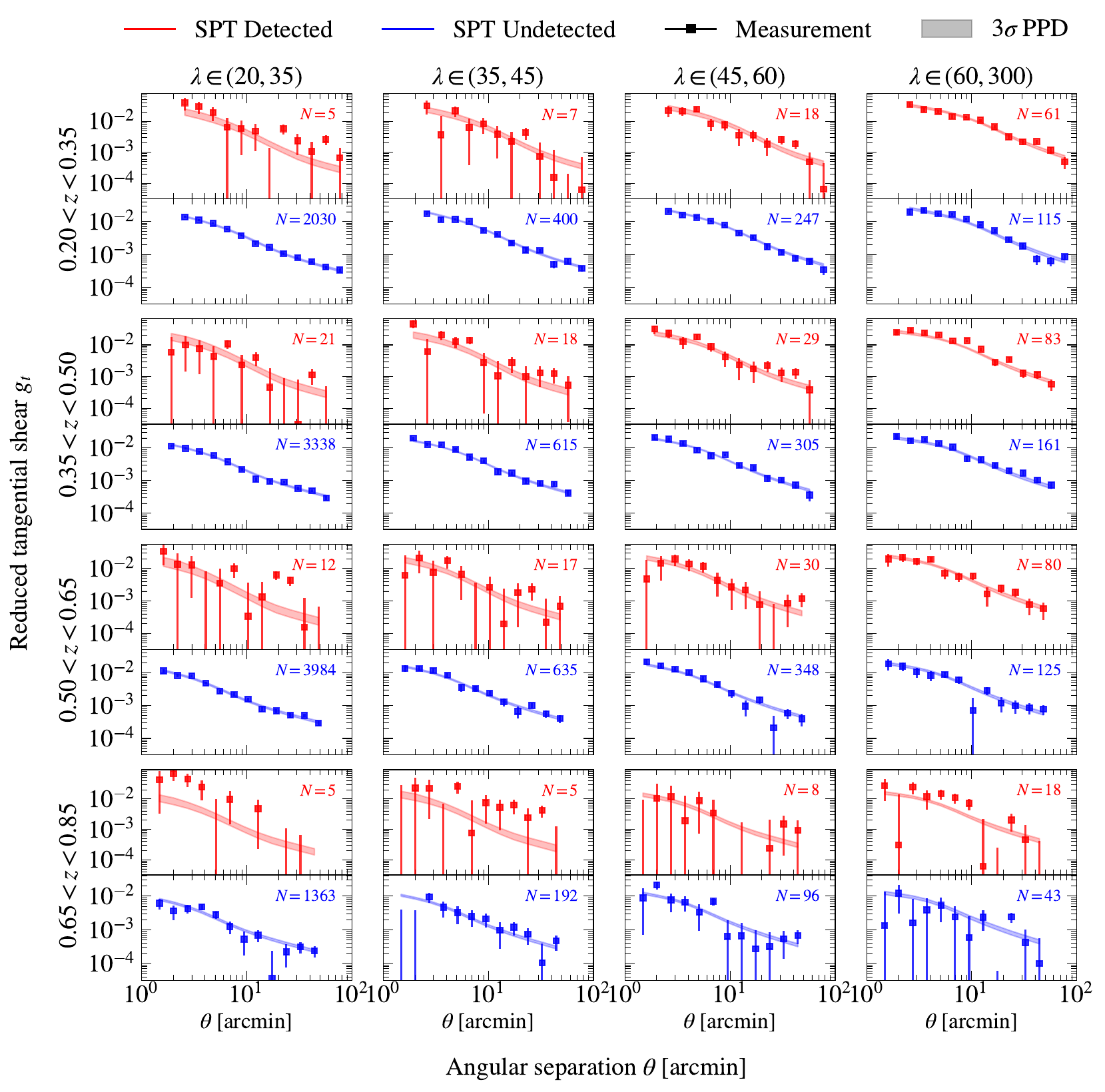}
  \caption{Stacked lensing signal for SPT-detected (red, upper panels) and SPT-undetected clusters (blue, lower panels), in redshift bins (rows) and richness bins (columns).
  Solid squares with error bars are lensing from observations, and the bands are the PPD. The number of clusters in each bin is shown in the top right corner of each panel. Our model fits the data well with a reduced $\chi^2$ of 1.17 calculated at the posterior-predictive median.
  }
  \label{fig:lensing}
\end{figure*}

To measure the lensing signal, we follow the procedure in \cite{grandisSRGEROSITAAllSky2024}, which is detailed in Appendix~\ref{app:lensing_measurement}.
We use the DES~Y3 shear catalog \cite{gattiDarkEnergySurvey2021}, which is built from the $r,i,z$-bands using the \textsc{metacalibration} algorithm \cite{huffMetacalibrationDirectSelfCalibration2017, sheldonPracticalWeaklensingShear2017}, covering 4,143 deg$^2$. 
This area is slightly smaller than that of the cluster catalog due to masking.
The total source density is approximately 5--6~arcmin$^{-2}$.
We use sources in tomographic redshift bins constructed by Self-Organizing Maps (SOMPz) \cite{mylesDarkEnergySurvey2021}.
We measure the reduced tangential shear $g_t(\theta)$ in 15 radial bins: an inner bin out to $0.2\,h^{-1}$Mpc and 14 logarithmically spaced bins between 0.2 and $15\,h^{-1}$Mpc, in physical units computed with the fiducial cosmology at the redshift of the cluster. The fit uses the 12 bins beyond $0.37\,h^{-1}$Mpc. 
The stacked lensing signals are shown in Fig.~\ref{fig:lensing} (for DES-SPT overlap) and Fig.~\ref{fig:lensing_no_spt} (for DES-only). 
We also use \textsc{dnf} \cite{devicenteDNFGalaxyPhotometric2016} and \textsc{bpz} \cite{benitezBayesianPhotometricRedshift2000} source redshifts to measure the so-called boost factor, the correction of lensing dilution caused by cluster member galaxies; see  Appendix~\ref{appendix:contamination}.


\section{Cluster observable densities}
\label{sec:cluster_density}

In cosmological analyses, the cluster observable likelihood $\mathcal{L}$ can be, in principal, written as \cite{barlowExtendedMaximumLikelihood1990, mantzObservedGrowthMassive2010, bocquetSPTClustersHST2024a}

\begin{align}
\mathcal{L} &=\frac{N_{\text {exp }}^{N_{\text {obs }}} e^{-N_{\text {exp }}}}{N_{\text {obs }}!}\prod_{i=1}^{N_{\rm obs}}P(\mathcal{O}_i, g_{i} \mid \Omega) \\
& =\frac{N_{\text {exp }}^{N_{\text {obs }}} e^{-N_{\text {exp }}}}{N_{\text {obs }}!}\prod_{i=1}^{N_{\rm obs}} P(\mathcal{O}_i \mid \Omega)\prod_{i=1}^{N_{\rm obs}}P(g_i \mid \mathcal{O}_i; \Omega), \label{eq:full_density}
\end{align}

where $P(Y|X)$ denotes the conditional probability of $Y$ given $X$ and $N_{\text {exp }}$ is the total number of clusters expected in the survey volume under the selection. $N_{\text {obs }}$ is the total number of clusters observed. $\mathcal{O}_i$ is the vector of observables like $\lambda$ and $\xi$ for for cluster $i$, and $g$ is the lensing signal for cluster $i$. $\Omega$ are the unknown model parameters.  

We define the first term as the counts likelihood $\mathcal{L}_{\rm counts}$, 
the second term as the density likelihood $\mathcal{L}_{\rm density}$, 
and third term as the lensing likelihood $\mathcal{L}_{\rm lensing}$. The likelihood is then 
\beq
\mathcal{L} = 
\mathcal{L}_{\rm counts}
\mathcal{L}_{\rm density}
\mathcal{L}_{\rm lensing} .
\eeq
In this analysis, we do not vary the cosmology, and we discard the cluster counts likelihood ($\mathcal{L}_{\rm counts}$), as it is sensitive to cosmological parameters that are held fixed. We use a stacked lensing likelihood instead of per-cluster lensing. In this section we focus on the cluster observable density likelihood, and in the next section we describe in detail the stacked lensing likelihood. 

\subsection{Cluster Observable Density Likelihood}
For a cluster sample with various observables $O_i$ ($\lambda$, $\xi$, or both) and a selection criterion $S_i$ ($\lambda_i \ge \lambda_{\rm min}$ for the DES clusters, $\xi_i \ge \ximin$ for the SPT-only clusters), the cluster density likelihood is given by
\beq
\mathcal{L}_{\rm density}
=
\prod_{i=1}^{N_{\rm obs}}
P(O_i \mid S_i, z_i; \Omega).
\eeq
Here $O_i$ are the available observables for each cluster,

\beq
O_i =
\begin{cases}
\lambda_i, & i\in \mbox{DES-only},\\
(\lambda_i,\xi_i), & i\in \mbox{SPT-detected},\\
(\lambda_i,\xi_i< \ximin), & i\in \mbox{SPT-undetected},\\
\xi_i, & i\in \mbox{SPT-only}.
\end{cases}
\eeq

We use $\Omega$ to denote the vector of model parameters. $\Omega$ and $z$ are always ``given'' and are not involved in the probability integrations. We assume we know the cluster redshift exactly. For notational simplicity, we omit the subscripts $i$ and $z$ in the following equations. 

For each sample, the probability of observing a cluster with observables $O$ is given by 

\begin{align}
P(\lambda \mid \lambda \ge \lambda_{\rm min}; \Omega) 
&= \frac{P(\lambda, \lambda \ge \lambda_{\rm min} \mid \Omega)}
        {P(\lambda \ge \lambda_{\rm min} \mid \Omega)}, 
\label{eq:pdf_des_only}
\\
P(\lambda, \xi \mid \lambda \ge \lambda_{\rm min}; \Omega) 
&= \frac{P(\lambda, \xi, \lambda \ge \lambda_{\rm min} \mid \Omega)}
        {P(\lambda \ge \lambda_{\rm min} \mid \Omega)},
\label{eq:pdf_spt_detected}
\\
P(\lambda, \xi < \ximin \mid \lambda \ge \lambda_{\rm min} ; \Omega) 
&= \frac{P(\lambda, \xi < \ximin, \lambda \ge \lambda_{\rm min} \mid \Omega)}
        {P(\lambda \ge \lambda_{\rm min} \mid \Omega)},
\label{eq:pdf_spt_undetected}
\\
P(\xi \mid \xi \ge \ximin ; \Omega) 
&= \frac{P(\xi, \xi \ge \ximin \mid \Omega)}
        {P(\xi \ge \ximin \mid \Omega)}.
\label{eq:pdf_spt_only}
\end{align}

These equations are for DES-only, SPT-detected, SPT-undetected, and SPT-only, respectively.
The treatment of the SZe signal of DES-detected clusters is a left-censored Type I Tobit model \cite{tobinEstimationRelationshipsLimited1958, amemiyaTobitModelsSurvey1984, grandisSelectionFunctionClusters2025}, which models both the observed signal above the threshold and the unobserved signal below the detection threshold.

\bigskip

The selection probabilities in the denominators are, 
\beqa
P(\lambda \ge \lambda_{\rm min} \mid \Omega)
& = \int \dd M P(M \mid \Omega) P(\lambda \ge \lambda_{\rm min} \mid M ; \Omega)
\eeqa
and 
\beqa
P(\xi \ge \xi_{\rm min} \mid \Omega) 
& = \int \dd M P(M \mid \Omega) P(\xi \ge \xi_{\rm min} \mid M ; \Omega),
\eeqa
where $P(M \mid \Omega)$ is the probability distribution function (PDF) of halo masses, i.e., the halo mass function normalized to 1. We use the $M_{200c}$ halo mass function of \citet{bocquetHaloMassFunction2016} evaluated at the fiducial cosmology. We use a minimum mass of $10^{12}~\hiMsun$ when we normalize $P(M \mid \Omega)$. We have checked that this mass is sufficient to encompass the low-mass tail of our cluster samples.

For the DES-only likelihood [Eq.~(\ref{eq:pdf_des_only})], the numerator is 
\beqa
P(\lambda, \lambda \ge \lambda_{\rm min} \mid \Omega) 
&= P(\lambda \mid \Omega) \; \Theta(\lambda \ge \lambda_{\rm min})\\
&= \int \dd M P(M \mid \Omega) P(\lambda \mid M),
\eeqa
where $\Theta$ is the Heaviside step function. This probability is 0 for a cluster with $\lambda < \lambda_{\rm min}$, but it does not affect our calculation because our sample only includes clusters with $\lambda \ge \lambda_{\rm min}$.

For the SPT-detected likelihood [Eq.~(\ref{eq:pdf_spt_detected})], the numerator is given by

\beqa
&P(\lambda, \xi, \lambda \ge \lambda_{\rm min} \mid \Omega) 
= \int \dd M P(M \mid \Omega) P(\lambda, \xi \mid M; \Omega) \\
& = \int \dd M P(M \mid \Omega) 
P(\lambda \mid M; \Omega) 
P(\xi \mid \lambda, M; \Omega),
\eeqa
where $P(\xi \mid \lambda, M; \Omega)$ is the distribution of $\xi$ at fixed richness and mass. It differs from $P(\xi \mid M; \Omega)$ because we allow the richness and SZe scatter to be correlated at fixed mass (Sec.~\ref{sec:correlated_scatter}). For uncorrelated scatter, it reduces to $P(\xi \mid M; \Omega)$.

For the SPT-undetected likelihood [Eq.~(\ref{eq:pdf_spt_undetected})], the numerator is given by
\beqa
&P(\lambda, \xi < \ximin, \lambda \ge \lambda_{\rm min} \mid \Omega) \\
&= \int \dd M P(M \mid \Omega) P(\lambda \mid M; \Omega) \int_{0}^{\ximin} \dd \xi\, P(\xi \mid \lambda, M ; \Omega) ,
\eeqa
where $P(\xi \mid \lambda, M; \Omega)$ is the same conditional distribution as in the SPT-detected case, so a correlated richness scatter also shifts the probability of a cluster remaining undetected.

Similarly, for the SPT-only likelihood, the numerator is given by 
\beqa
P(\xi, \xi \ge \ximin \mid \Omega)
&= \int \dd M P(M \mid \Omega) P(\xi \mid M; \Omega).
\eeqa

\subsection{Scaling relations}
\label{sec:scaling_relations}

For richness, we assume that $P(\ln\lambda \mid M)$ follows a normal distribution, with a mean 
\begin{align}
\langle\ln \lambda \mid M\rangle= & \ln \left(A_\lambda\right)+B_\lambda \ln \left(\frac{M}{3 \times 10^{14} M_{\odot} h^{-1}}\right) \nonumber \\
& +C_\lambda \ln \left(\frac{1+z}{1+0.45}\right),
\end{align}
and a mass-dependent variance following \citet{grandisSelectionFunctionClusters2025},
\begin{align}
  \bigl[\sigmalam(M)\bigr]^2 &= \sigma_\lambda^2 \left(1 + \frac{M_{\rm break}}{M}\right)^{2 B_\sigma} + \frac{\bar\lambda - 1}{\bar\lambda^2}\\ \qquad \bar\lambda &= \exp\langle\ln\lambda \mid M\rangle,
  \label{eq:sigma_lambda}
\end{align}
where the first term is the intrinsic scatter, constant at $\sigma_\lambda$ for $M \gg M_{\rm break}$ and rising as $M^{-B_\sigma}$ below it, and the second term has the Poisson form that captures the rise of the scatter toward low richness \cite{costanziMethodsClusterCosmology2019}. The parameters $\sigma_\lambda$, $B_\sigma$, and $\log_{10} M_{\rm break}$ are free (Table~\ref{tab:param}). The total scatter $\sigmalam(M)$ is the richness scatter that enters the correlations with the SZe signal and with lensing below. 

For SZe, we first model the unbiased SZe signal-to-noise $\zeta$ \cite{vanderlindeGalaxyClustersSelected2010}. 
We assume that $\ln\zeta$ follows a normal distribution with mean,
\begin{align}
\langle\ln \zeta \mid M\rangle= & \ln \left(\gamma_f A_{\zeta}\right)+B_{\zeta} \ln \left(\frac{M}{3 \times 10^{14} M_{\odot} h^{-1}}\right) \nonumber \\
& +C_{\zeta} \ln \left(\frac{E(z)}{E(0.6)}\right)
\end{align}
and a constant variance $\sigma_\zeta^2$. Here $\gamma_f$ is a factor that accounts for the different SPT field depth \cite{bleemGalaxyClustersDiscovered2015, bleemSPTpolExtendedCluster2020, grandisSelectionFunctionClusters2025}. The observed $\xi$ is related to $\zeta$ via 
\beq
  \xi \mid \zeta \sim \mathcal{N}\left(\sqrt{\zeta^2+3}, 1\right),
\eeq
which accounts for extra degrees of freedom from the matched filter \cite{vanderlindeGalaxyClustersSelected2010}.
This relation cannot hold down to arbitrarily low significance, since it breaks down for $\xi<\sqrt{3}$. Following \citet{bocquetSPTClustersHST2024b}, we apply the maximization bias only above $\zeta_{\rm min}=1$, and use $\xi \mid \zeta \sim \mathcal{N}(\zeta, 1)$ below it. We use Gauss-Hermite integrals to marginalize over $\zeta$ fast, as described in Appendix~\ref{appendix:gauss_hermite}.

\section{Cluster lensing}
\label{sec:lensing_likelihood}

We use a Gaussian likelihood for stacked lensing, 
\begin{align}
  &\ln\mathcal{L}_{\rm lensing} \propto \nonumber \\ &\sum_{j, k,l}-\frac{1}{2}\left(g_{t,jkl}^{\mathrm{model}}(\theta)-g_{jkl}^{\mathrm{data}}(\theta)\right)^T \mathrm{C}_{j kl}^{-1}\left(g_{t,jkl}^{\mathrm{model}}(\theta)-g_{j k l}^{\mathrm{data}}(\theta)\right) , 
\end{align}
where $g_{jkl}(\theta)$ is the stacked reduced shear in $z$, $\lambda$, and $\xi$ bins indexed by $j,k,l$. Here $\mathrm{C}_{jkl}$ is the lensing covariance matrix of the corresponding bin described in Appendix~\ref{app:cov}.

Our model for the stacked reduced shear, $g_{t}^{\mathrm{model}}$, is built in four steps, described in the following subsections:
\begin{enumerate}
  \item For {\em each} cluster with observed $(\lambda, \xi, z)$, we compute the expected surface density $\Sigma(\rp)$ and excess surface density $\DS(\rp)$ as a function of the projected radius $\rp$, by averaging the profiles at fixed halo mass (Sec.~\ref{sec:lensing_mass}) over the multivariate observable--mass distribution, including the selection effect from the correlated scatter [Eq.~(\ref{eq:mean_ind_lensing}) in Sec.~\ref{sec:correlated_scatter}]. We call these the ``no-sys'' profiles, $\Sigmanosys(\rp)$ and $\DSnosys(\rp)$.
  \item We compute correction factors $S_{\rm sys}$ for baryonic effects and miscentering in each $(\lambda, \xi, z)$ bin [Eq.~(\ref{eq:S_factor}) in Sec.~\ref{sec:lensing_sys}] and apply them to obtain the corrected profiles $\Sigmasys(\rp)$ and $\DSsys(\rp)$.
  \item We convert $\Sigmasys(\rp)$ and $\DSsys(\rp)$ into the reduced shear $g_t(\theta)$ at the fiducial cosmology, integrating over the source redshift distribution [Eq.~(\ref{eq:gt}) in Sec.~\ref{sec:gt}].
  \item Finally, we average the individual $g_t(\theta)$ in the $(z, \lambda, \xi)$ bins of the measurement to obtain the stacked $g_{t}^{\mathrm{model}}$.
\end{enumerate}

Below we only show the SPT-detected case (clusters with both $\lambda$ and $\xi$); the other two cases can be derived analogously by modifying Eq.~(\ref{eq:conditional_mass}).

\subsection{Lensing signal at a given mass}
\label{sec:lensing_mass}

For a given halo mass $M$, we assume $\ln \DS$ follows a normal distribution with mean
$\langle\ln \DS \mid M\rangle$ and standard deviation $\sigmaDS$.
We use the mean $\langle\ln \DS \mid M\rangle$ from the \darkemu \cite{nishimichiDarkQuestFast2019}.
The \darkemu interpolates the halo-matter cross-correlation function $\xi_{\rm hm}(r; M, z)$ for 101 cosmological models, for halos above $M_{200m}\ge10^{12}~\hiMsun$, for radial range $ 0.01\le r \le 100~\hiMpc$ and $z<1.48$. We convert our $M_{200c}$ masses to $M_{200m}$ with the mass--concentration relation of \citet{ishiyamaUchuuSimulationsData2021} before querying the emulator. 
It uses $\xi_{\rm hm}$ to compute the surface density profile $\Sigma(\rp)$ as a function of the projected distance $\rp$,
\beq
\Sigma(\rp, M, z) = \bar{\rho}_m \int \xi_{\rm hm }\left(r=\sqrt{\rp^2 + \pi^2}; M, z\right) \dd \pi
\eeq
and the excess surface density profile 
\begin{align}
\DS (\rp) &= \langle \Sigma (< \rp) \rangle - \Sigma (\rp) \nonumber \\
&= \frac{2}{\rp^2} \int_0^{\rp} d\rp' \rp'\Sigma (\rp') - \Sigma(\rp).
\label{eq:Sigma_to_DS}
\end{align}

\subsection{Lensing signal at given observables}
\label{sec:correlated_scatter}

For a given halo mass $M$, we model the joint probability distribution of the three cluster properties, $\lambda$, $\zeta$, and $\Delta\Sigma$, as a multivariate log-normal distribution, following \citet{zhouForecastingConstraintsOptical2024}, 
\begin{align}
&P(\ln\DS, \ln\lambda, \ln \zeta \mid M)
\nonumber\\
&\sim \mathcal{N} \Big(\big\{\avg{\ln\DS\mid M},\avg{\ln\lambda\mid M},\avg{\ln\zeta\mid M}\big\}, \mathsf{C}\Big), 
\label{eq:3var_gaussian}
\end{align}
where the covariance matrix is given by
\beq
\mathsf{C} =
\begin{pmatrix}
\sigmaDS^2 & \rholds\,\sigmaDS\sigmalam & \rholds\,\rho_{\lambda\zeta}\,\sigmaDS\sigma_\zeta \\
\rholds\,\sigmaDS\sigmalam & \sigmalam^2 & \rho_{\lambda\zeta}\,\sigmalam\sigma_\zeta \\
\rholds\,\rho_{\lambda\zeta}\,\sigmaDS\sigma_\zeta & \rho_{\lambda\zeta}\,\sigmalam\sigma_\zeta & \sigma_\zeta^2
\end{pmatrix},
\label{eq:3var_cov}
\eeq

Here $\rho_{\lambda \zeta}$ is the correlation coefficient between $\lambda$ and $\zeta$ at a given halo mass. $\rholds(\rp)$ accounts for the correlated scatter between lensing signal at radius $\rp$ and $\lambda$ at a given halo mass. This correlation includes the intrinsic correlation and projection effect induced correlation between lensing and richness at a given halo mass \citep{farahiCorrelationsDarkMatter2022, zhangImpactPropertyCovariance2024,sunayamaImpactProjectionEffects2020,wuOpticalSelectionBias2022}.  
We define the product of the correlation coefficient between lensing and richness and the lensing scatter to be 
\beq
  \omega(\rp) \equiv \rholds(\rp) ~ \sigmaDS(\rp). 
  \label{eq:omega}
\eeq We assume $\omega(\rp)$ is constant in the 1-halo and 2-halo regimes and transitions smoothly in between,
\beq
  \omega(\rp) = 
  \begin{cases} 
   \omega_1 & \text{if } \rp \in (0, r_{\min}] \\
   \omega_1 + \frac{\omega_2 - \omega_1}{\ln(r_{\max}/r_{\min})} \ln\left(\frac{\rp}{r_{\min}}\right) & \text{if } \rp \in (r_{\min}, r_{\max}] \\
   \omega_2 & \text{if } \rp > r_{\max}
  \end{cases}
\eeq
where we choose $r_{\min} = 1.0 \ h^{-1}\mathrm{Mpc}$ and $r_{\max} = 8.0 \ h^{-1}\mathrm{Mpc}$.
We will fit for $\omega_1$ and $\omega_2$ values. 
Fig.~\ref{fig:varying_omega} shows the impact of $\omega_1$ and $\omega_2$ on the stacked lensing signal (discussed in Sec.~\ref{sec:varying_omega}).

We further assume that the lensing signal is conditionally independent of the SZe observable at fixed mass and richness. The SZe significance therefore enters the lensing prediction only through the mass posterior $P(M \mid \lambda, \xi)$ [Eq.~(\ref{eq:conditional_mass})].
Under these assumptions, conditioning the joint Gaussian on the observed richness gives
\beq
  \begin{aligned}
& \ln \DS \mid M, \lambda, \xi  \\
\sim & \mathcal{N}\Bigl(\langle \ln \DS \mid M\rangle+\frac{\omega(\rp)}{\sigmalam}\bigl(\ln\lambda -\langle\ln\lambda \mid M, z\rangle\bigr), \\
& \qquad \sigmaDS^2 - \omega^2(\rp)\Bigr).
\label{eq:optical_selection_effect}
\end{aligned}
\eeq
Because the data vector is a linear (stacked) average, we require the conditional mean of $\DS$ rather than of $\ln\DS$. For a lognormal variable,
\beq
\langle \DS \mid M, \lambda, \xi \rangle
= \exp\Bigl[\langle \ln\DS \mid M, \lambda, \xi \rangle
+ \tfrac{1}{2}\bigl(\sigmaDS^2 - \omega^2(\rp)\bigr)\Bigr].
\label{eq:lognormal_mean}
\eeq
Defining the mean profile of haloes at fixed mass,
$\langle \DS \mid M \rangle \equiv
\exp\bigl[\langle \ln \DS \mid M \rangle + \sigmaDS^2/2\bigr]$
--- the quantity our emulator-based profiles provide by construction, being ensemble averages over haloes at fixed mass ---
Eqs.~(\ref{eq:optical_selection_effect}) and~(\ref{eq:lognormal_mean}) combine to
\beq
\begin{aligned}
&\langle \DS \mid M, \lambda, \xi \rangle \\
=\; & \langle \DS \mid M \rangle\,
\exp\Bigl[\frac{\omega(\rp)}{\sigmalam}
\bigl(\ln\lambda - \langle \ln\lambda \mid M, z\rangle\bigr)
- \frac{\omega^2(\rp)}{2}\Bigr].
\label{eq:selection_combined}
\end{aligned}
\eeq
Note that $\sigmaDS$ cancels in Eq.~(\ref{eq:selection_combined}): the stacked signal depends on the correlation coefficient and the lensing scatter only through their product $\omega = \rholds\,\sigmaDS$, so the two are not separately identifiable from mean profiles, and $\sigmaDS$ need not be specified anywhere in the likelihood.

From the law of total expectation, for a given $\lambda$, $\xi$, the expected $\DS$ is given by
\begin{align}
\DSnosys(\rp, \lambda, \xi) \equiv 
  \langle \DS | \lambda, \xi\rangle = \nonumber \\
 \int \langle \DS \mid M , \lambda, \xi \rangle P(M \mid \lambda, \xi) \; \dd M ,
  \label{eq:mean_ind_lensing} 
\end{align}
where 
\beqa
\label{eq:conditional_mass}
P(M \mid \lambda, \xi) = \frac{P(\lambda, \xi \mid M) P(M)}{P(\lambda, \xi)},
\eeqa
where each probability has been described in Sec.~\ref{sec:cluster_density}.
For a given cluster $\lambda$ and $\xi$, we calculate this mass PDF. Then we calculate the mean predicted lensing at observed $\lambda$, $\xi$ with the optical selection effect with Eq.~(\ref{eq:mean_ind_lensing}).

We note that this average is over the observable--mass PDFs instead of a stacked average.  We therefore suppress the average symbol from now on.
We denote it as $\DSnosys$ and will add systematics in the next section.

We repeat the same procedure to compute 
$ \Sigmanosys (\rp, \lambda, \xi) \equiv
\langle \Sigma | \lambda, \xi, z\rangle$.

\subsection{Correction factors $S_{\rm sys}$ due to baryonic effects and miscentering}
\label{sec:lensing_sys}

So far we have computed $\DSnosys(\rp, \lambda, \xi)$ and $\Sigmanosys(\rp, \lambda, \xi)$ for {\em each} cluster with given $\lambda$ and $\xi$.  
We now compute the impact of baryonic effects and miscentering. Baryonic feedback redistributes gas and stars within halos and alters the matter profile relative to the dark-matter-only prediction, mainly at small radii. Miscentering arises when the \redmapper center is offset from the true halo center, which flattens the measured profile on scales comparable to the offset. We always use the \redmapper center as the center for computing lensing.
Baryonic effects and miscentering are modeled with the $S_{\rm sys}$ factors in each stacked lensing bin: 
\beqa
S_{\rm sys}^{\Sigma}(\rp)_{jkl} &= 
\frac{ \Sigmasys }{\Sigmanosys},
\\
S_{\rm sys}^{\DS}(\rp)_{jkl} & = 
\frac{ \DSsys }{ \DSnosys },
\label{eq:S_factor}
\eeqa
where $j,k,l$ are indices for $z, \lambda$ and $\xi$ bins.
We note that the impact of miscentering is not separable from that of baryons, and thus $S_{\rm sys}$ cannot be separated into two factors.
Below, we describe how we compute the $S_{\rm sys}$ factors.

{\bf Baryonic effects.}
In this step, we compute the baryonic correction factor on $\Sigma$. We analytically compute the ratio:
\beq
f_b(\rp \mid M, z; \theta_{b}) = \frac{\Sigma_\mathrm{dmb} (\rp \mid M, z; \theta_b) }{ \Sigma_\mathrm{dmo} (\rp \mid M, z)} .
\eeq

The numerator $\Sigma_\mathrm{dmb}$ is the baryonified density profile computed using the model by \cite{giriEmulationBaryonicEffects2021}, which is described in detail in  Appendix~\ref{sec:bcm}. We use the posterior ranges from the X-ray analysis in \citet{grandisDeterminingBaryonImpact2024} as priors.

The denominator, $\Sigma_\mathrm{dmo}$, is the dark-matter-only density computed with the one-halo term from a truncated NFW profile [Eq.~(\ref{eq:nfw})] and the two-halo term from the linear matter correlation function from \cite{eisensteinPowerSpectraCold1999} and halo bias from \cite{tinkerLargescaleBiasDark2010}.
Instead of computing $f_b(\rp)$ for all halo masses, we compute $f_b(\rp)$ at the mean mass in ($\lambda, \xi, z$) bins.
We denote it as $f_{b}(\rp)_{jkl}$, but note that it is constant in each bin. 
At this stage, we obtain the baryonified, well-centered profile,
\beq
\Sigmacen(\rp, \lambda, \xi) = f_b(\rp)_{jkl} \Sigmanosys (\rp, \lambda, \xi). \label{eq:baryon_sigma}
\eeq

{\bf Miscentering.} 
We now compute the full $S_{\rm sys}$ factors combining baryonic and miscentering corrections.   

Let us assume that the density profile is given by
\beq
\Sigmasys(\rp) = 
\fcen \Sigmacen(\rp)
+ (1-\fcen) \Sigmamis(\rp) ,
\eeq
where $\fcen$ is the fraction of well-centered clusters, $\Sigmacen(\rp)$ is profiles of the well-centered cluster from Eq.~(\ref{eq:baryon_sigma}), and $\Sigmamis(\rp)$ is the profiles of miscentered clusters. 
Calculating $\Sigmamis(\rp)$ requires two steps: First, we calculate the azimuthally averaged profile for a given miscentered distance $\rmis$,
\beq
\Sigmamis(\rp | \rmis) = \int_0^{2\pi} \frac{d \theta}{2 \pi} \Sigmacen \left( \sqrt{\rp^2 + \rmis^2 - 2 \rp \rmis \cos \theta}\right).
\label{eq:Sigma_mis}
\eeq
Second, we integrate over the PDF of $\rmis$, $P(\rmis)$,
\beq
\Sigmamis(\rp) = \int d\rmis P(\rmis) \Sigmamis (\rp | \rmis).
\eeq

For $P(\rmis)$, we follow the parametrization of \cite{zhangDarkEnergySurvey2019}
\beq
P(r_{\rm mis})
= R_\lambda^{-1}
P_{\rm mis}(r_{\rm mis}/R_\lambda \mid \tau)
\eeq

\beq
  P_{\rm mis}(x \mid \tau) =\frac{x}{\tau^2} \exp \left(-\frac{x}{\tau}\right),
\eeq

where the parameter $\tau$ describes the characteristic miscentering offset of our \redmapper clusters, and $x$ is the miscentering offset scaled by \redmapper cluster radius \cite{rykoffRedMaPPerAlgorithmSDSS2014}.
\begin{align}
x & =\rmis / R_\lambda \\
R_\lambda & =(\lambda / 100)^{0.2} (1+z) h^{-1} \mathrm{cMpc}.
\end{align}
We evaluate $R_\lambda$ at the mean $\lambda$ in the bin. 
We use the parameter ranges calibrated by \cite{kellyDarkEnergySurvey2024} as priors, which are based on comparing DES Y3 \redmapper centers and centers derived from Chandra and XMM.
We assume both parameters to be redshift- and $\lambda$-independent, which is consistent with the findings in 
\cite{kellyDarkEnergySurvey2024}. 

Similarly, $\DSsys$ is given by
\beq
\DSsys = \fcen \DScen + (1 - \fcen) \DSmis,
\eeq
where $\DScen$ and $\DSmis$ are computed from $\Sigmacen$ and $\Sigmamis$ using Eq.~(\ref{eq:Sigma_to_DS}).

\subsection{From $\Sigma$ and $\DS$ to $g_t$}
\label{sec:gt}

For a given cluster with $ \lambda, \xi$, we apply the baryonic and miscentering correction factors to obtain
\beqa
\Sigmasys (\rp ; \lambda, \xi )
& = S_{\rm sys}^{\Sigma} \Sigma_{\rm no{\text-}sys}(\rp)_{jkl} (\rp ; \lambda, \xi ) \\
\DSsys (\rp ; \lambda, \xi )
& = S_{\rm sys}^{\DS}(\rp)_{jkl} \DSnosys (\rp ; \lambda, \xi ),
\eeqa
where $j,k,l$ are chosen to be the indices of the bin the cluster is in. 

We then compute the expected reduced shear, integrating the source redshift distribution $P(z_s)$ [Eq.~\ref{eq:Pzs}]

\beq
  g_t(\theta ,  \lambda, \xi, z ) = \int \dd z_s P(z_s) \frac{\Sigma_{\mathrm{crit}}^{-1}(z_l, z_s) \DSsys (\rp, \lambda, \xi )}{1 - \Sigma_{\mathrm{crit}}^{-1}(z_l,z_s) \Sigmasys (\rp, \lambda, \xi )} ,
  \label{eq:gt}
\eeq
where $\rp = \chi_l \theta$, 
$\Sigma_{\mathrm{crit}}^{-1}$ is the inverse critical surface density in comoving units
\beq
  \Sigma_{\mathrm{crit}}^{-1} = \frac{4 \pi G}{c^2} (1+z_{\mathrm{l}}) \frac{\chi_{\mathrm{l}}\chi_{\rm ls}}{\chi_{\mathrm{s}}} 
  .
\eeq
Here $\chi_\mathrm{l}$ is the comoving distance from the observer to the lens, $\chi_\mathrm{s}$ is the distance from the observer to the source, $\chi_\mathrm{ls}$ is the distance from the lens to the source, and $z_l$ is the redshift of the lens.

Finally, we compute the stacked lensing signal $g_t^{\mathrm{model}}$ by binning the predicted cluster shear $g_t(\theta, \lambda,\xi,z)$ in the same $z,\lambda$ and $\xi$ bins as the measurements and taking the mean within each bin.

\subsection{Impact of optical selection bias on the lensing signal}
\label{sec:varying_omega}
\begin{figure*}[tp]
  \centering
  \includegraphics[width=\textwidth]{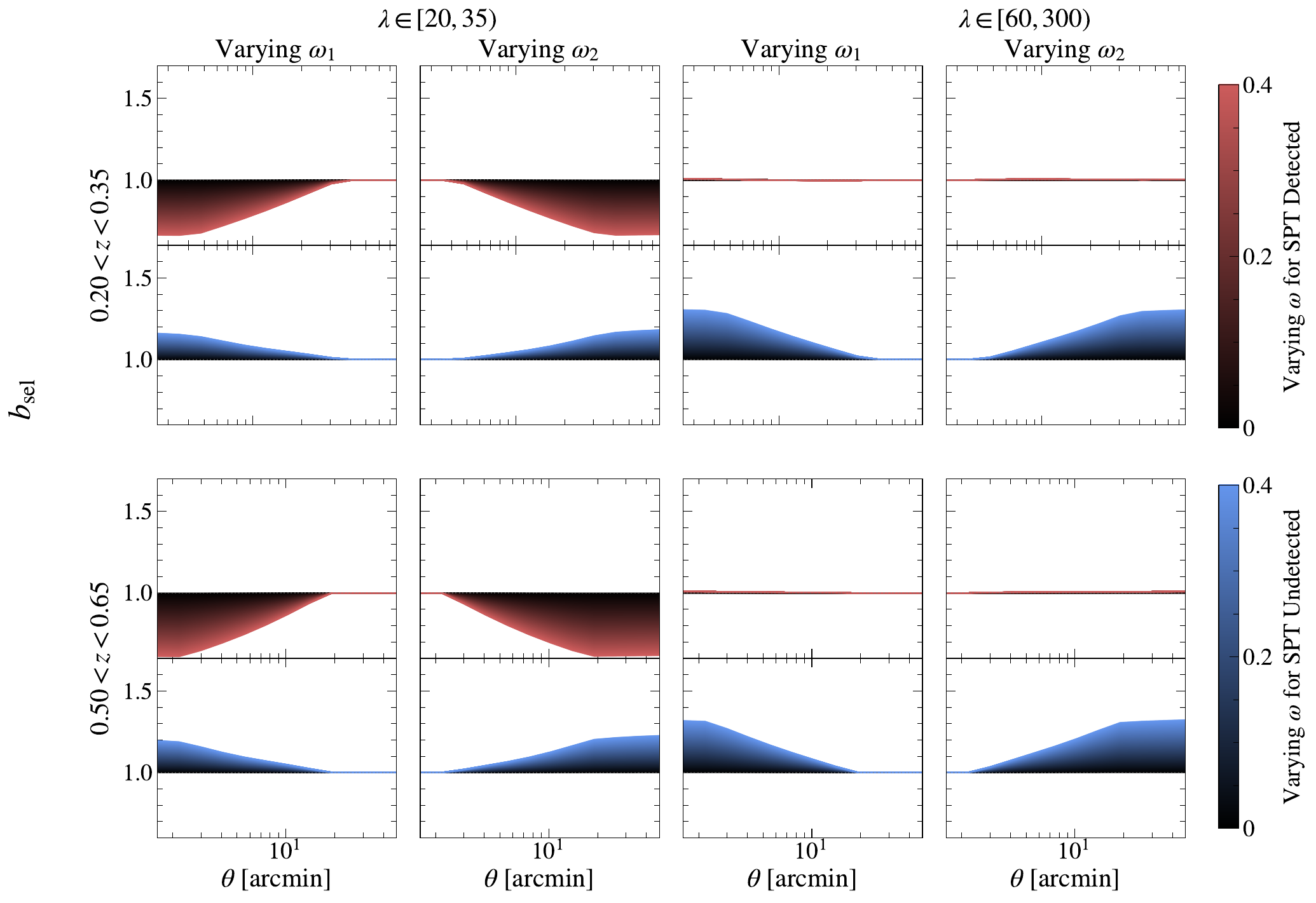}
  \caption{
  Impact of the richness-selection-effect parameters $\omega_1$ (small-scale) and $\omega_2$ (large-scale) on the bias in the lensing signals $\bsel$, for SPT-detected clusters (red) and SPT-undetected clusters (blue).
  For high-richness bins (right-hand columns), the lensing is biased high for SPT-undetected clusters ($\bsel>1$) and unbiased for SPT-detected clusters.
  For low-richness bins (left-hand columns), the lensing is biased low for SPT-detected clusters ($\bsel<1$), which have richness lower than the mean richness-mass relation, and biased high for SPT-undetected clusters.
  The trend persists for lower-redshift bins (upper panels) and higher-redshift bins (lower panels).
  }
  \label{fig:varying_omega}
\end{figure*}

In Eq.~(\ref{eq:omega}), we have introduced $\omega(\rp)$ for describing the correlated scatter between richness and lensing at a given mass.  
In Fig.~\ref{fig:varying_omega}, we show the impact of $\omega_1$ (small-scale) and $\omega_2$ (large-scale) on the lensing signal compared to a sample not affected by selection effects. 
This bias is quantified by the lensing selection bias $\bsel$, the ratio of the stacked lensing model with correlated scatter to the same model without it,
\beq
b_{\rm sel}=\frac
{\left.g_{t,jkl}^{\rm model}(\theta)\right|_{\omega_1,\omega_2}}
{\left.g_{t,jkl}^{\rm model}(\theta)\right|_{\omega_1=\omega_2=0}},
\label{eq:bsel}
\eeq
computed with all other parameters, including the systematics, at their fiducial values.
The curves are colored by the values of $\omega_1$ and $\omega_2$. We vary only one and fix the other at 0.
We vary $\omega_1$ and $\omega_2$ from 0 to 0.4, a range motivated by simulation studies \citep{wuOpticalSelectionBias2022}.  
We show the lowest and the highest $\lambda$ bins on the left and the right columns, and the $0.20 < z < 0.35$ and $0.50 < z < 0.65$ redshift bins on the top and bottom rows. 

For SPT-undetected clusters (blue), $\bsel > 1$ always holds.  These clusters have substantial richness but no SZe signal, indicating that they tend to be heavily projected systems with coherently boosted richness and lensing.

For SPT-detected clusters (red), $\bsel \approx 1$ at high richness and $\bsel < 1$ at low richness. In the low-richness bins, an SPT detection requires a massive halo, so the few SPT-detected clusters there are the rare massive halos whose richness scattered well below the mean richness--mass relation. Because the richness and lensing residuals at fixed mass are positively correlated [Eq.~(\ref{eq:optical_selection_effect})], a low richness residual comes with a low large-scale lensing residual, and the stacked lensing of these clusters falls below that of a mass-matched sample, $\bsel < 1$.

\section{Inference}
\label{sec:inference}

\begin{table*}[]
  \centering
  \begin{tabular}{cccl}
  Parameter & Prior & Posterior & Description \\
  \hline
  \hline
  \multicolumn{4}{l}{$\lambda$-mass scaling relation: log-normal model from \citet{grandisSelectionFunctionClusters2025}}\\ 
  $\ln A_{\lambda}$ & $\mathcal{U}(-100, 100)$ & $3.94^{+0.03}_{-0.03}$ & intercept of the $\lambda$-mass relation \\
  $B_{\lambda}$ & $\mathcal{U}(0, 10)$ & $0.95^{+0.02}_{-0.01}$ & slope of $\lambda$-mass relation \\
  $C_{\lambda}$ & $\mathcal{U}(-5, 5)$ & $0.09^{+0.11}_{-0.11}$ & redshift evolution of intercept of $\lambda$-mass relation\\
  $\sigma_\lambda$ & $\mathcal{U}(0.05, 0.6)$ & $0.09^{+0.05}_{-0.03}$ & intrinsic scatter of the $\lambda$-mass relation at high mass \\
  $B_\sigma$ & $\mathcal{U}(-0.5, 1.5)$ & $0.16^{+0.74}_{-0.43}$ & mass dependence of the intrinsic scatter below $M_{\rm break}$ \\
  $\log_{10} M_{\rm break}$ & $\mathcal{U}(11, 16)$ & $12.66^{+2.19}_{-1.13}$ & mass below which the intrinsic scatter rises  \\
  \hline
  \multicolumn{4}{l}{$\zeta$-mass scaling relation: log-normal model from \citet{bocquetSPTClustersHST2024a}}\\ 
  $\ln A_{\mathrm{\zeta}}$ & $\mathcal{U}(-100, 100)$ & $0.65^{+0.07}_{-0.08}$ & intercept of $\zeta$-mass relation \\
  $B_{\mathrm{\zeta}}$ & $\mathcal{U}(0, 10)$ & $1.64^{+0.05}_{-0.05}$ & slope of $\zeta$-mass relation\\
  $C_{\mathrm{\zeta}}$ & $\mathcal{U}(-5, 5)$ & $0.41^{+0.30}_{-0.30}$ & redshift evolution of intercept of $\zeta$-mass relation\\
  $\sigma_\zeta$ & $\mathcal{U}(0.01, 1)$ & $0.37^{+0.08}_{-0.07}$ & scatter of $\zeta$-mass relation \\
  $\rho_{\lambda\zeta}$ & $\mathcal{U}(-0.95, 0.95)$ & $-0.33^{+0.34}_{-0.37}$ & correlation between the $\lambda$ and $\zeta$ scatter at fixed halo mass \\
  \hline
  \multicolumn{4}{l}{Optical selection effect bias: correlated scatter model from \citet{wuOpticalSelectionBias2022}}\\
  $\omega_1$ & $\mathcal{U}(-0.16, 0.16)$ & $-0.08^{+0.07}_{-0.05}$ & $\rholds \sigmaDS$, 1-halo regime ($\sigmaDS = 0.16$) \\
  $\omega_2$ & $\mathcal{U}(-0.62, 0.62)$ & $0.26^{+0.06}_{-0.06}$ & $\rholds \sigmaDS$, 2-halo regime ($\sigmaDS = 0.62$) \\
  \hline
  \multicolumn{4}{l}{Miscentering: Gaussian priors from the DES Y3 mis-centering calibration of \citet{kellyDarkEnergySurvey2024}} \\
  $\tau$ & $\mathcal{N}(0.23, 0.05)$ & $0.28^{+0.04}_{-0.04}$ & characteristic miscentering offset \\
  $\fcen$ & $\mathcal{N}(0.87, 0.04)$ & $0.89^{+0.03}_{-0.03}$ & fraction of well-centered clusters \\
  \hline
  \multicolumn{4}{l}{Baryonic correction model: correlated Gaussian prior from the X-ray analysis of \citet{grandisDeterminingBaryonImpact2024}} \\
  $\log_{10} M_c$ & $\mathcal{N}(14.53, 0.20)$ & $14.49^{+0.17}_{-0.18}$ & characteristic mass scale where gas profile slope shallows \\
  $\mu$ & $\mathcal{N}(0.54, 0.10)$ & $0.53^{+0.09}_{-0.09}$ & mass dependence of gas profile slope \\
  $\theta_{\mathrm{ej}}$ & $\mathcal{N}(3.59, 1.39)$ & $3.73^{+1.26}_{-1.05}$ & gas ejection radius in units of $r_{200c}$\\
  $\gamma$ & $\mathcal{N}(1.14, 0.13)$ & $1.16^{+0.11}_{-0.10}$ & transition sharpness of gas profile slope \\
  $\delta $ & $\mathcal{N}(7.31, 1.61)$ & $7.54^{+1.45}_{-1.22}$ & outer slope of gas profile \\
  $\eta$ & $\mathcal{N}(0.25, 0.03)$ & $0.25^{+0.03}_{-0.02}$ & mass dependence of the stellar fraction in satellite galaxies \\
  $\eta_\delta$ & $\mathcal{N}(0.23, 0.10)$ & $0.23^{+0.09}_{-0.09}$ & mass dependence of stellar fraction in central galaxy
  \\  \hline
  \end{tabular}
  \caption{Model parameters for fitting the observable density and stacked lensing. The scaling-relation parameters are described in Section~\ref{sec:scaling_relations}, the projection-effect parameters $\omega_1$ and $\omega_2$ in Section~\ref{sec:correlated_scatter}, the miscentering parameters in Section~\ref{sec:lensing_sys}, and the baryonic correction model parameters in Section~\ref{sec:bcm}. We use uninformative priors for mass-$\zeta$, mass-$\lambda$ scaling relations, and projection effect parameters $\omega_1$ and $\omega_2$. We use priors from the Dark Energy Survey Y3 miscentering analysis from \citet{kellyDarkEnergySurvey2024} for miscentering and X-ray-informed priors from \citet{grandisDeterminingBaryonImpact2024} on the baryonic correction model.
  The bounds on $\omega_1$ and $\omega_2$ follow from $|\rholds| \leq 1$ with $\sigmaDS$ measured in the Cardinal simulation by \citet{zhouForecastingConstraintsOptical2024}.
  The Posterior column lists the marginal posterior median and 68\% credible interval.
  }
  \label{tab:param}
\end{table*}

We summarize the parameters of our model and their priors in Table~\ref{tab:param}.
We use flat priors for parameters related to observable-mass relations and projection effects.
For the miscentering parameter priors, we use the DES Y3 miscentering posterior informed by XMM-Newton and Chandra observations \cite{kellyDarkEnergySurvey2024}. 
For the baryonic-effect parameters, we use the posteriors from \citet{grandisDeterminingBaryonImpact2024}, which are derived from X-ray observations, with gas and stellar mass fraction informed by cluster weak lensing. 

\subsection{Sampling and Implementation}

We use the Python implementation of the affine-invariant ensemble sampler \texttt{emcee} \cite{goodmanEnsembleSamplersAffine2010, foreman-mackeyEmceeMCMCHammer2013} to sample from the posterior. We use \texttt{cupy}, which is a GPU-accelerated drop-in replacement for \texttt{numpy} and \texttt{scipy}, to build our model to enable parallel computation for the density likelihood on the cluster observable and mass grid. For 88 walkers, each sampler step takes about 3.4 seconds on four NVIDIA A100 graphical processing units (GPUs). The fast speed makes our methodology very scalable to include more clusters and extendable to more complicated models.  We run 26,678 steps with 88 walkers, initialized from the posterior of a shorter preliminary run, and discard the first 2,000 steps (about two autocorrelation times) as burn-in. The maximum autocorrelation time is 1,086 steps, yielding an effective sample size of 23 per walker, about 2,000 in total.

\subsection{Posterior Predictive Distribution}

To examine our model fit and understand the constraining power we have at the data-vector level, we compute the posterior predictive distributions (PPDs) for both the observable distribution and stacked cluster lensing.

In general, for given observed data $y$ and model parameter $\Omega$, the PPD is 
\beq
  P(\tilde{y} \mid y)=\int P(\tilde{y} \mid \Omega) P(\Omega \mid y) \dd \Omega,
\eeq
where $\tilde{y}$ is the predicted unobserved data, $P(\Omega\mid y)$ is the posterior, and $P(\tilde{y} \mid \Omega)$ is the sampling distribution.
For cluster observables $\lambda$ and $\xi$, we use Monte-Carlo simulations to generate cluster observables, following \citet{zhouForecastingConstraintsOptical2024}. 

For stacked cluster lensing, we use the model output described in Section~\ref{sec:lensing_likelihood} to generate mean stacked lensing predictions.

\subsection{Goodness of fit tests}

To quantify the goodness-of-fit of our observable density model to the data, we calculate the Bayesian $p$-value \cite{gelmanBayesianDataAnalysis2004}, 
defined as the probability that the simulated data could be more extreme (less likely under the model) than the observed data, as measured by the test quantity $T$:
\beq
p_B=\operatorname{Pr}\left(T\left(y^{\text {sim}}, \Omega\right) \leq T(y^{\text {obs}}, \Omega) \mid y^{\text {obs}}\right),
\eeq
where $y^{\text{obs}}$ is the observed data and $y^{\text{sim}}$ is the simulated data. A very small $p$-value indicates a misfit, whereas a large $p$-value only indicates the absence of evidence for a misfit under the test statistic $T$. 

In practice, we first sample our posterior. At each sampled posterior point $\Omega$, we simulate cluster observables $O^{\mathrm{sim}} = \{O^{\mathrm{sim}}_i\mid i = 1,2,...,N_{\mathrm{obs}}\}$ with our model. Then we compare $T(O^{\mathrm{sim}}, \Omega)$ with $T(O^{\mathrm{obs}}\mid \Omega)$ with $T$ chosen to be the density log-likelihood of clusters,

\beq
  T(O, \Omega) = \sum_i^{N_{\mathrm{obs}}} \ln P(O_i \mid S_i, z_i, \Omega). 
\eeq

We then empirically estimate $p_B$ by calculating the fraction of sampled posterior with ${T(O^{\mathrm{sim}}, \Omega) \leq T(O^{\mathrm{obs}}, \Omega)}$. 

To quantify the goodness-of-fit of our stacked lensing model, we calculate the reduced $\chi^2$ value for all bins,
\beq
\chi^2 = \sum_{j,k,l} \left(g^{\mathrm{model}}_{t,jkl} - g^{\mathrm{data}}_{jkl}\right)^T \mathrm{C}_{jkl}^{-1} \left(g^{\mathrm{model}}_{t,jkl} - g^{\mathrm{data}}_{jkl}\right),
\eeq
where the sum runs over the 48 stacks in redshift, $\lambda$, and $\xi$ (SPT-detected, SPT-undetected, and DES-only) bins, each with 12 radial bins, for a total of 576 data points, and $\mathrm{C}_{jkl}$ is the covariance used in the lensing likelihood (Sec.~\ref{sec:lensing_likelihood}). We evaluate $g^{\mathrm{model}}_{t,jkl}$ at the median of the posterior predictive distribution and define the reduced $\chi^2$ as $\chi^2$ per data point.

\section{Results}
\label{sec:results}

In this section, we present the main results of the paper. We first show the posterior distribution and assess the goodness of fit of our model to the data using the Bayesian $p$-value for the density model and the reduced $\chi^2$ for the lensing model. We then examine the posterior predictive distribution for the optical selection bias, as well as the inferred scaling relations and mass distributions.

\subsection{Posterior distributions}

\begin{figure*}
  \centering
  \includegraphics[width=\textwidth]{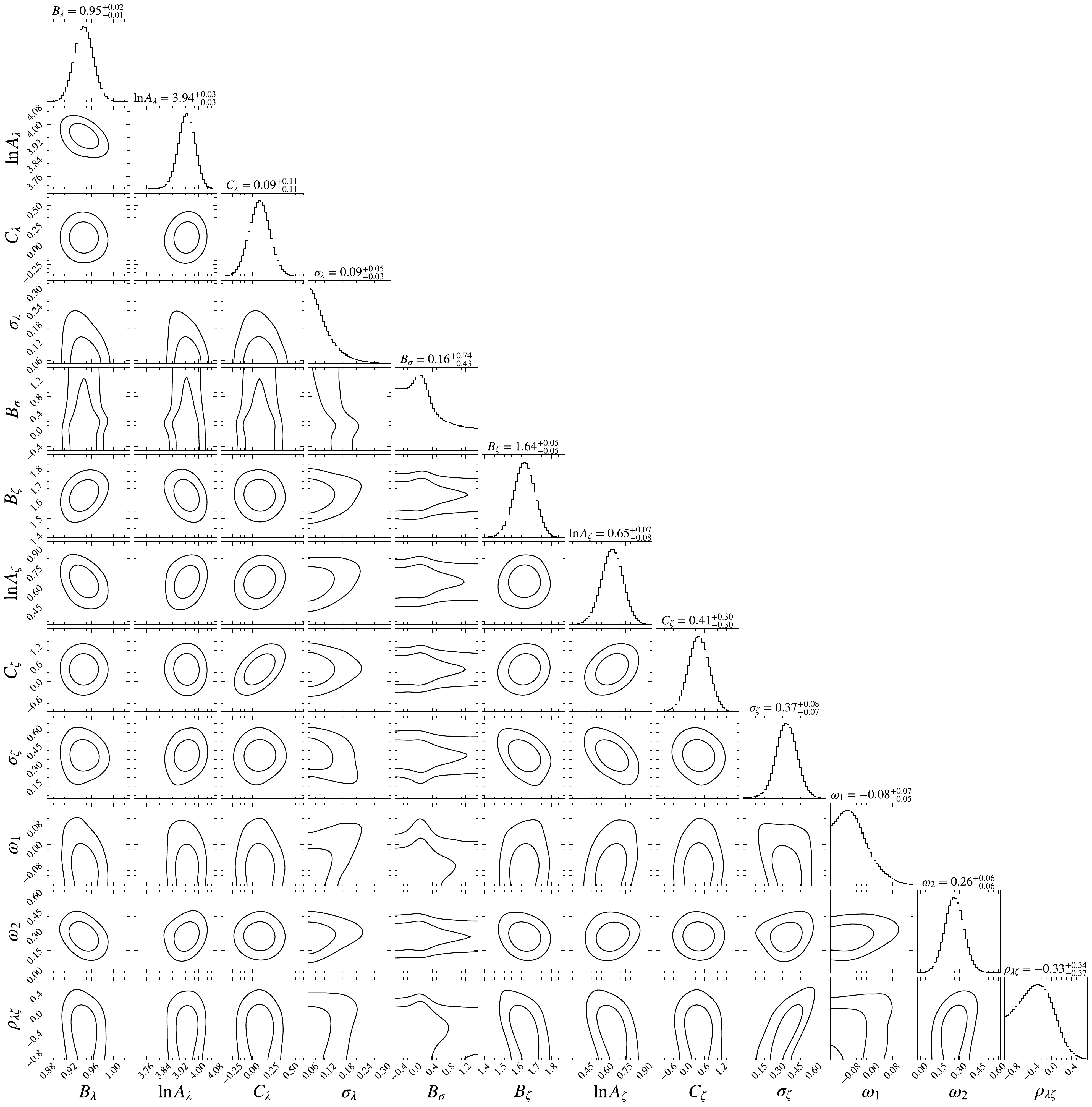}
  \caption{Posterior distribution of the scaling-relation and correlated-scatter parameters. The contours enclose 68\% and 95\% of the two-dimensional marginal posteriors, and the numbers above the diagonal panels give the marginal posterior median and 68\% credible interval of each parameter. The miscentering and baryonic correction model parameters are omitted because their posteriors follow their priors. The 2-halo correlated-scatter parameter, $\omega_2 = 0.26^{+0.06}_{-0.06}$, is four standard deviations from zero, whereas the 1-halo parameter $\omega_1$ is only weakly constrained and consistent with zero. }
  \label{fig:posterior}
\end{figure*}

Figure~\ref{fig:posterior} shows the 68\% and 95\% credible regions of the posterior distributions. We do not show the baryonic and miscentering parameters since they are prior-dominated.
The marginalized constraints are also listed in Table~\ref{tab:param}. 

The key result is the detection of a non-zero correlated-scatter parameter at large scales, $\omega_2 = 0.26^{+0.06}_{-0.06}$, four standard deviations from zero. The small-scale parameter, $\omega_1 = -0.08^{+0.07}_{-0.05}$, is only weakly constrained: its posterior concentration is only slightly better than the prior range and is consistent both with zero and with the smaller positive correlation expected in the one-halo regime from simulations \cite{sunayamaImpactProjectionEffects2020, wuOpticalSelectionBias2022}. The large-scale term is the one our data require, as expected for projection effects (Sec.~\ref{sec:discussion}).

\subsection{Goodness of fit}

In Fig.~\ref{fig:density_ppd}, we show the observed cluster densities and the PPD from our fit. 
In the left panel, we show the $\lambda$ distribution of SPT-undetected clusters and the $3\sigma$ PPD. 
In the middle panel, we show the $\xi$ distribution of the SPT-only clusters outside the DES footprint and the $3\sigma$ PPD. 
In the right panel, we show the joint distribution of $\lambda$ and $\xi$ along with contours that contain the 68\%, 95\%, and 99\% PPD densities. 
We do not show the DES-only sample because it is equivalent to the combination of SPT-detected and undetected. 
We see that the PPD agrees well with the data. The posterior predictive $p$-value is 0.53, so there is no evidence of a misfit of the observable distribution. 

In Fig.~\ref{fig:lensing}, we show the stacked cluster lensing signal (points) along with the $3\sigma$ PPD in redshift and $\lambda$ bins, split into SPT-detected (red) and SPT-undetected (blue).  
In the upper right corner of each subpanel, we show the number of clusters in each stack.
This figure shows clusters in the DES-SPT overlapping region, while Appendix~\ref{appendix:lensing_no_spt} and Fig.~\ref{fig:lensing_no_spt} present the lensing for the DES-only region. 

Our model fits the stacked lensing data well with a reduced $\chi^2$ of 1.17 for all bins. 
For the highest redshift bins, our model predictions are lower than the data points for SPT-detected clusters, which hints at an inflexibility of the redshift dependence in our scaling relation. 

\begin{figure*}
  \centering
  \includegraphics[width=\textwidth]{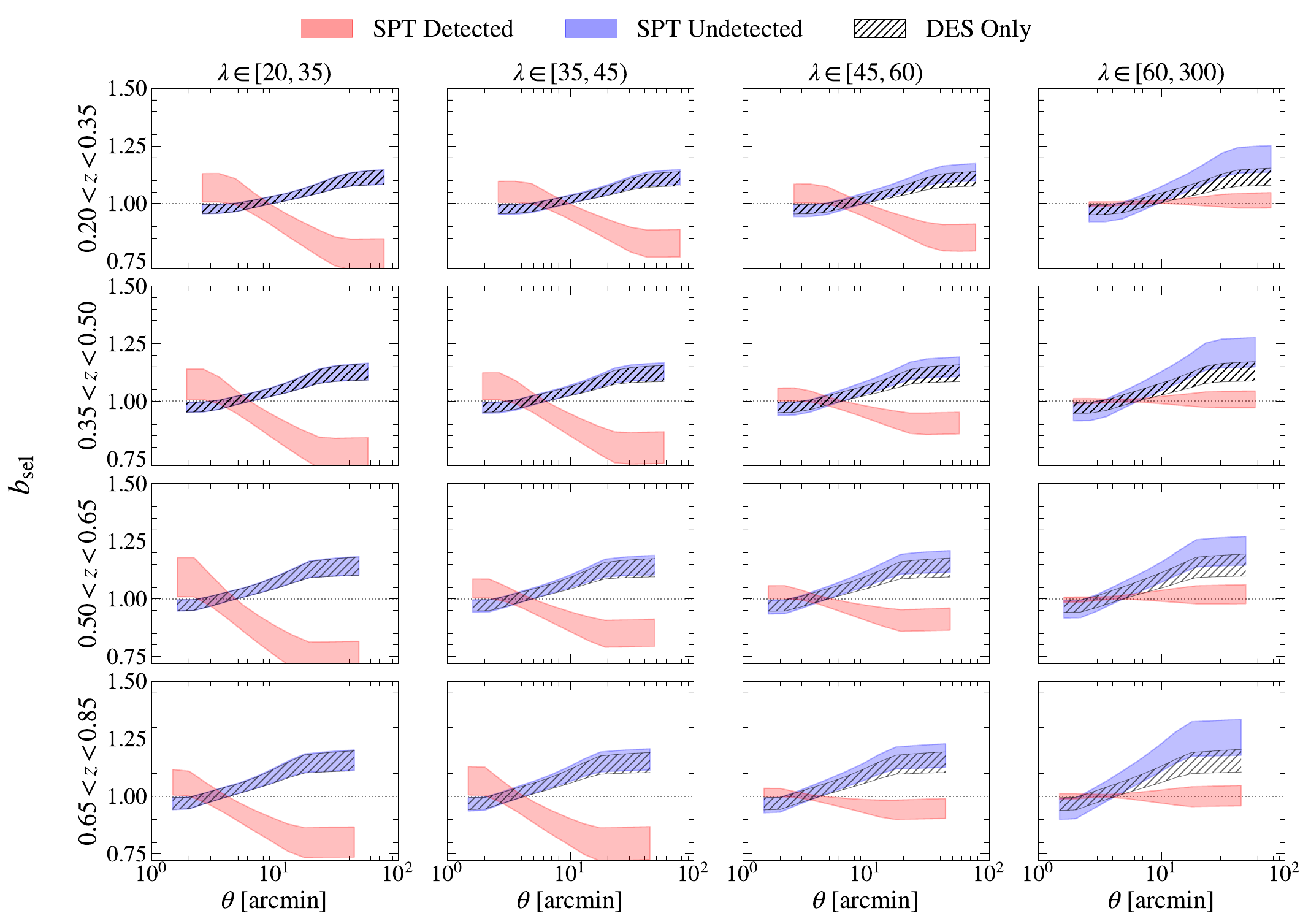}
  \caption{
  Optical selection bias of cluster lensing [Eq.~(\ref{eq:bsel})] inferred from the posterior for the three samples.  The SPT-undetected sample (blue) shows the strongest bias. These are low-mass halos with a significant boost in richness and lensing due to line-of-sight projection.  
  The DES-only sample (gray) agrees with the SPT-undetected sample at low richness and has a lower selection bias at high richness.
  The SPT-detected sample (red) has a low selection bias at high richness but is underbiased ($\bsel<1$) at low richness.  These underbiased clusters represent a small population of low-richness clusters with high $\xi$.
  }
  \label{fig:bsel_ppd}
\end{figure*}

\subsection{Selection bias}

Fig.~\ref{fig:bsel_ppd} shows the posterior predictive distribution of the lensing selection bias $\bsel$ [Eq.~(\ref{eq:bsel})], evaluated with $\omega_1$ and $\omega_2$ drawn from the posterior.
The posterior gives $\omega_2 = 0.26^{+0.06}_{-0.06}$ in the two-halo regime and $\omega_1 = -0.08^{+0.07}_{-0.05}$ in the one-halo regime (Table~\ref{tab:param}), so the bias is confined to large scales. For the $\lambda$-selected samples at $\rp \gtrsim 5\,h^{-1}$Mpc, $\bsel$ is 1.13--1.21 for SPT-undetected clusters, rising with richness, and 1.12--1.13 for DES-only clusters, whereas within $0.8\,h^{-1}$Mpc it is 0.95--0.97 for both. This is consistent with prior studies \cite{sunayamaImpactProjectionEffects2020, toCombinationClusterNumber2021, wuOpticalSelectionBias2022, sunayamaObservationalConstraintsAnisotropic2023, salcedoConsistencyDarkEnergy2024}: $\lambda$ selection boosts the large-scale lensing while leaving the small-scale lensing nearly unbiased. The SPT-undetected clusters follow the trend of $\lambda$ selection alone because they essentially are the richness-selected sample: SPT detects only 0.4\% of the clusters at $\lambda = 20$--35 and 35\% at $\lambda > 60$, so the undetected clusters are the low-mass systems with coherently upscattered richness and lensing that dominate any richness-selected sample. For SPT-detected clusters, the large-scale bias flips from $\bsel \approx 1.00$ at $\lambda > 60$ to 0.90, 0.82, and 0.77 in the $\lambda = 45$--60, 35--45, and 20--35 bins. This is the expected consequence of the positive $\lambda$--lensing correlation in our model: at a given richness, SPT detection selects the most massive halos, whose richness lies below the mean relation for their mass, and the positive correlation then implies a lensing signal below the mean for their mass as well. As richness decreases, the SPT-detected clusters correspond to an increasingly rare population with high mass and downscattered richness, causing this flip in $\bsel$ (also see Fig.~\ref{fig:mass_distribution} below).

Finally, we note that $\bsel$ for SPT-detected clusters can be significant.  
This implies that the optical selection effect not only impacts cluster mass calibration of optically selected clusters but also could potentially bias mass calibration of SPT-detected clusters when $\lambda$ is treated as another observable or selection variable \cite{costanziCosmologicalConstraintsY12021, bocquetSPTClustersHST2024a, bocquetSPTClustersHST2024b}. Even if richness is not explicitly included in an SZe analysis, most SZe cluster catalogs require optical confirmation \cite{kleinSPTSZMCMFExtension2024}, which could implicitly include a richness selection and thus a bias.

\begin{figure*}
  \centering
  \includegraphics[width=\textwidth]{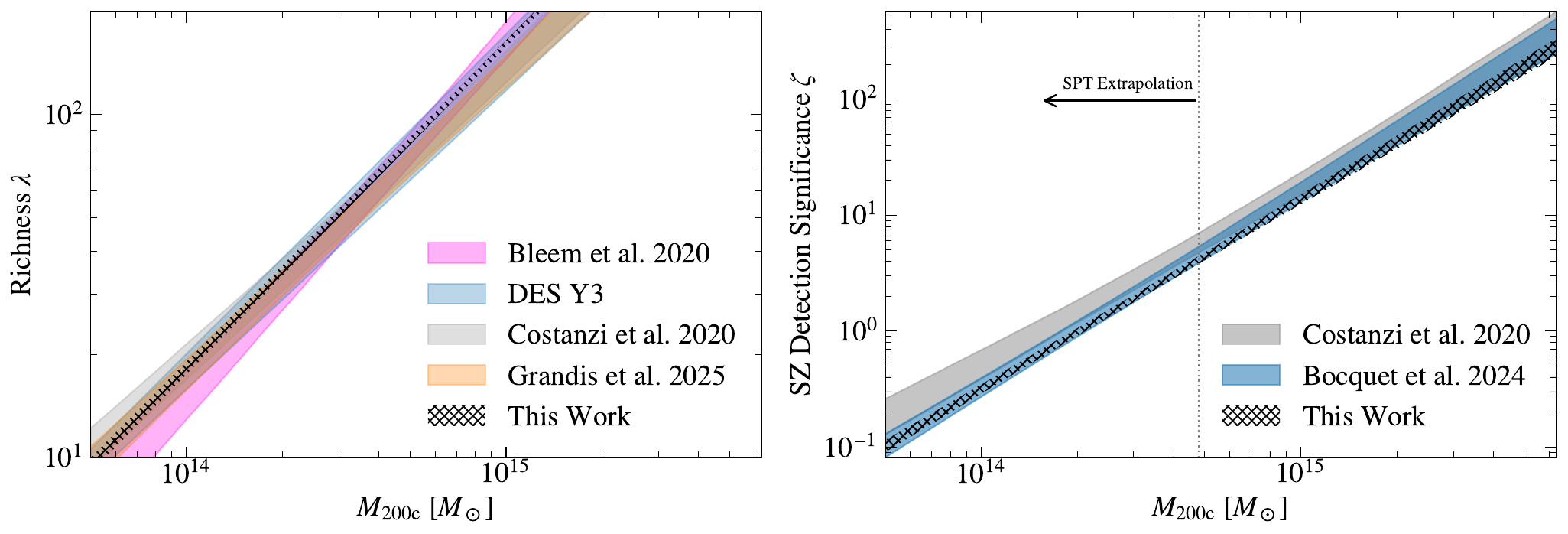}
  \caption{
  Observable--mass relations $\langle \lambda \mid M \rangle$ (left) and $\langle \zeta \mid M \rangle$ (right) at $z = 0.45$ from this work (black hatched) compared with results from the literature. Each band is the 68\% range of the relation over the posterior samples of the respective analysis. Our relations are consistent with the previous results over the mass range of our sample. The SZe-selected samples constrain the relations only above the dotted line in the right panel ($4.8\times10^{14}\,M_\odot$). Their bands at lower mass are extrapolations of the fitted power laws.
  }
  \label{fig:scaling_relation_ppd}
\end{figure*}

\subsection{Scaling relation}
In Fig.~\ref{fig:scaling_relation_ppd}, we show the posterior predictive $\lambda$-mass and $\xi$-mass scaling relation from this work along with results from previous studies. For the $\lambda$-mass relation, we compare our results with \citet{costanziCosmologicalConstraintsY12021}, who use clusters with DES Y1 and SPT-SZ detections to constrain observable-mass relations; \citet{abbottDarkEnergySurvey2025}, who utilize cluster abundances, cluster weak lensing, and cross-correlations between galaxies and clusters from DES Y3; \citet{bleemSPTpolExtendedCluster2020}, who measure optical richnesses for SZe-selected clusters from the SPTpol Extended Cluster Survey; and \citet{grandisSelectionFunctionClusters2025}, who cross-match DES Y3 \redmapper clusters with SPT detections to calibrate the optical selection function. For the $\xi$-mass relation, we compare with \citet{costanziCosmologicalConstraintsY12021} and \citet{bocquetSPTClustersHST2024b}, the latter of which combines SPT-selected clusters with DES-based $\lambda$ and weak lensing measurements from both DES and HST. 

Our posterior predictive $\lambda$-mass and $\zeta$-mass relations are consistent with all of these results over the mass range of our sample. For the SZe-selected samples, the comparison below $\sim 4\times10^{14}\,M_\odot$ relies on an extrapolation of their fitted relations, indicated by the dotted line in the right panel. Our richness scatter implies a mass scatter at fixed richness of $\sigma_{\ln M \mid \lambda} \approx 0.15$ (68\% range 0.14--0.19) at $\lambda = 70$, at the low end of the broad range $0.14$--$0.55$ inferred from archival X-ray temperatures of DES Y1 clusters \cite{farahiMassVarianceArchival2019, kellyDarkEnergySurvey2024}. The scatter depends on richness: it decreases from $\sigma_{\ln M \mid \lambda} \approx 0.34$ at $\lambda = 20$ to $0.17$ at $\lambda = 70$ and $0.12$ at $\lambda = 200$. The tighter constraint comes from jointly modeling the covariance of the richness, SZe, and lensing scatter, in particular the SZe signal and non-detections of the richness-selected clusters. 

\begin{figure*}
  \centering
  \includegraphics[width=\textwidth]{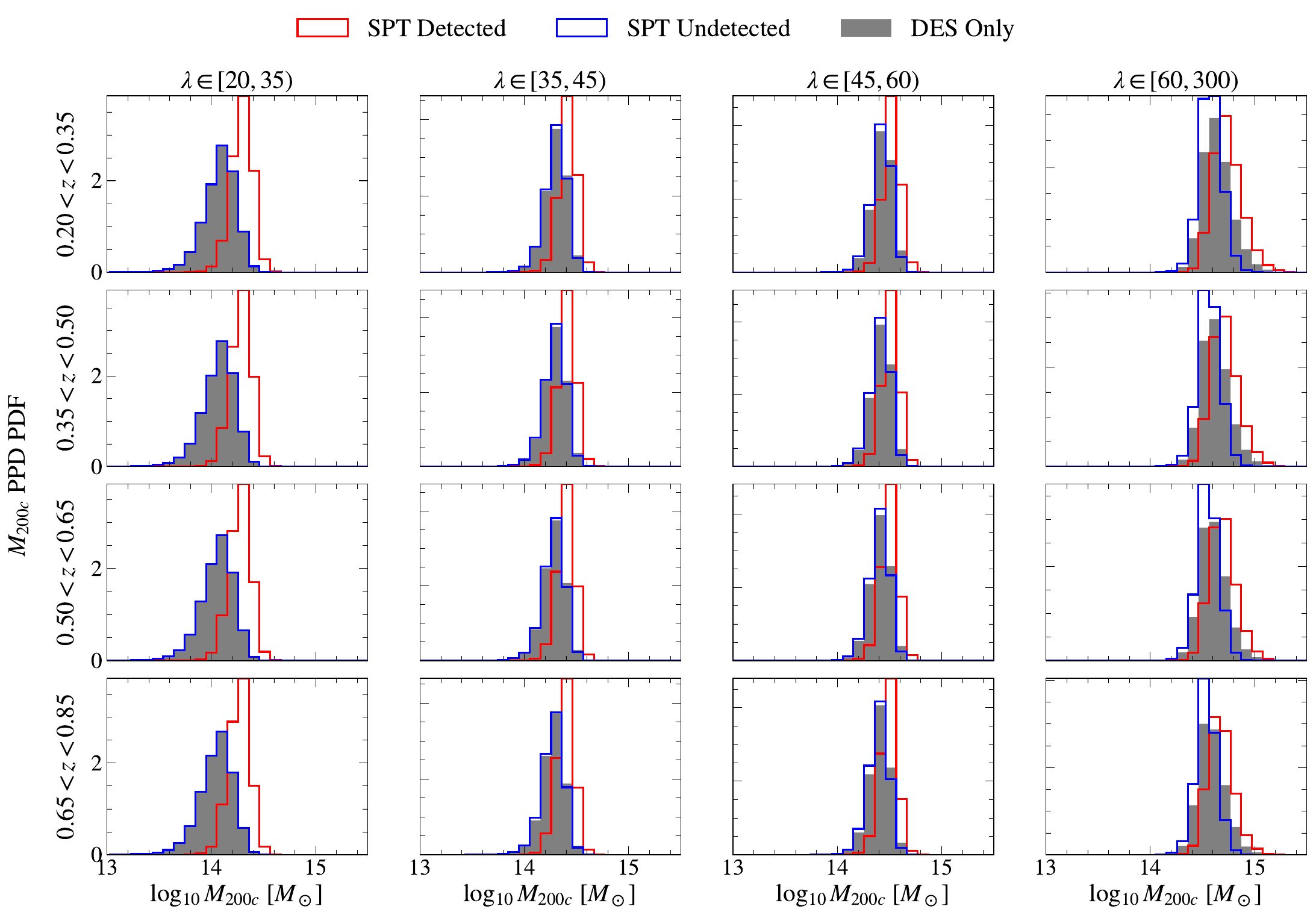}
  \caption{Posterior predictive cluster mass distribution in each redshift and $\lambda$ bin.
  The red histogram is the mass distribution of SPT-detected clusters, and the blue histogram is for SPT-undetected clusters.  The gray histogram shows the DES-only clusters (representing no SZe selection).
  SPT-detected clusters always have higher mass than SPT-undetected clusters.  For high-richness bins, the gray and blue are well separated; for low-richness bins, most clusters are SPT-undetected, and thus gray and blue overlap almost completely.
  In the lowest $\lambda$ bin, 0.5\% of the SPT-undetected and DES-only clusters have $\mtwoh < 10^{13.5}\,\hiMsun$.
  }
  \label{fig:mass_distribution}
\end{figure*}

\subsection{Mass distribution}
In Fig.~\ref{fig:mass_distribution}, we show the posterior predictive mass distribution in redshift and $\lambda$ bins, again split into SPT-detected (red), SPT-undetected (blue), and DES-only (gray).

In all bins, the SPT-detected clusters are more massive than the SPT-undetected ones.
In high-richness bins, SPT-detected and undetected clusters contribute to approximately 1/3 and 2/3 of the entire sample (see the legend of Fig.~\ref{fig:lensing}).
The fraction of SPT-detected clusters is reduced to approximately 0.5\% in the low-richness bins. 
Therefore, the SPT-undetected and DES-only PDFs are almost identical for low-richness bins.  In these bins, the SPT-detected clusters are dominated by the rare massive halos with significantly downscattered richness, leading to the underbiased lensing ($\bsel < 1$) we have seen in Fig.~\ref{fig:bsel_ppd}. The lowest $\lambda$ bin has a low-mass tail, with 0.5\% of its SPT-undetected clusters below $\mtwoh = 10^{13.5}\,\hiMsun$, the observational counterpart of the tail seen in simulated \redmapper catalogs \cite{farahiGalaxyClusterMass2016} and in previous DES analyses \cite{salcedoConsistencyDarkEnergy2024, abbottDarkEnergySurvey2025}. 

The contrast between the mass distribution of SZe-detected and undetected objects in a redshift and $\lambda$ bin highlights the power of combining observables. 
The SZe detection can pick up the high-mass clusters in the bin and thus constrain the fraction of the clusters with $\lambda$ contaminated by galaxies from line-of-sight structures \citep{grandisExploringContaminationDESY12021, grandisSelectionFunctionClusters2025}. 

\section{Discussion}
In this section we discuss analysis choices in this paper and their implications. We first examine the consequences of fixing the cosmology, then the use of stacked rather than individual cluster lensing, the extension and scaling of the framework to additional observables and larger samples, and finally its complementarity to other approaches to cluster cosmology.

\label{sec:discussion}

{\bf Impact of fixed cosmology.} In this paper, we constrain scaling relations and optical selection bias under a fixed cosmology. We leave out the model components that are most sensitive to cosmology to avoid biasing the estimation of the scaling relation and optical selection effect parameters. The first model component is the total number of halos within the survey volume appearing in the Poisson factor in Eq.~\ref{eq:full_density}, which depends on cosmology through the survey volume and the normalization of the halo mass function. We drop this factor and use only the normalized cluster density. 
The second model component sensitive to cosmology is the redshift distribution of clusters, which depends on the volume element and on the growth of structure; we condition the observable distribution of each cluster on its redshift instead of treating the redshift as an observable, so neither the survey volume nor the normalization of the halo mass function enters our likelihood. 
In a future paper, where we relax cosmology, we plan to incorporate these components to maximize information to constrain cosmology.

{\bf Individual vs.~Stacked Lensing} The statistical framework we propose in this work can accommodate both stacked cluster lensing as well as individual cluster lensing \cite{bocquetSPTClustersHST2024a, grandisSRGEROSITAAllSky2024}. In fact, we predict the mean individual cluster lensing given observables in Eq.~(\ref{eq:mean_ind_lensing}) and then stack later. With the depth of the upcoming deep lensing data \citep{spergelWideFieldInfrarRedSurvey2015,ivezicLSSTScienceDrivers2019}, it is possible that we would have a combination of stacked and individual cluster lensing profiles with high SNR \citep{zhouAccurateGalaxyCluster2026}. The statistical framework we propose in this paper would be optimal for utilizing these data. Note that we do not first fit NFW profiles to lensing profiles in bins to obtain mean cluster masses. The model predicts the lensing profiles directly from the observables. Given the scale dependence of the optical selection bias, fitting unbiased model profiles to biased data profiles might lead to mass biases that are not calibratable \citep{ndeImpactProjectioninducedOptical2026}.

{\bf Extendability and Scalability} In addition to relaxing cosmology, the statistical framework for combining cluster observables proposed in this work can be readily extended to an arbitrary number of observables, including X-ray count rate and stellar mass in the stellar halo \cite{huangOuterStellarMass2022,xuOutskirtStellarMass2025}. For instance, by further incorporating X-ray count rates, we could model any cluster detected in X-ray, millimeter, or optical bands, even if it is only observed in a single wavelength. Moreover, combining observables not only amplifies statistical power but also provides opportunities to probe physics and systematics. For example, just like a high-$\lambda$ SZe-undetected cluster is more likely to suffer from the optical selection effect, a cluster that is detected in X-ray but not by SZe is more likely to be a cool core cluster or have AGN contamination. 

Moreover, we can extend our model by using observables as a function of the radius from the cluster center to constrain radially dependent scaling relation parameters. For example, instead of using scalar $\xi$, we can use Compton-$y$ profiles. In this way, we can gain information about the correlation between different observables at different radii, which can further our understanding of the underlying physics that shaped these correlations \citep{farahiCorrelationsDarkMatter2022}. Given the high efficiency of our GPU implementation, the pipeline is highly scalable and can readily accommodate these more complex datasets without approaching computational limits. Should larger samples or more complex physics make the likelihood evaluation impractical, the same population model can serve as the forward model for simulation-based inference \cite{cranmerFrontierSimulationbasedInference2020, farahiOASISObservationAwareSimulationBased2026}, which learns the posterior from simulated catalogs and can incorporate effects that are difficult to write in closed form. 

{\bf Complementarity to Other Cluster Cosmology Approaches.} The general framework of combining observables is not limited to the models adopted in this paper but is compatible with and complementary to other modeling approaches to cluster cosmology. For optical clusters, \citet{sunayamaOpticalClusterCosmology2024} uses a parametric form of anisotropic boost to model HSC cluster abundance and lensing. \citet{salcedoCosmologicalConstraintsDark2025} uses simulation-based forward modeling of DES cluster abundance and cluster lensing. \citet{abbottDarkEnergySurvey2025} cross-correlates DES galaxies with clusters. Adding ICM observables to these analyses, as described in this work, can better constrain the scaling-relation and projection-effect parameters without sacrificing the statistical power of optically selected clusters. For ICM-selected clusters, current cosmological analyses model the ICM observables with an observable-mass relation \cite{bocquetMultiprobeCosmologyAbundance2025,bocquetSPTClustersHST2024b, ghirardiniSRGEROSITAAllSky2024}, while forward modeling with baryon pasting is an emerging alternative \cite{osatoBaryonPastingAlgorithm2022,keruzoreOptimizationQualityAssessment2023, lauBaryonPastingUchuu2025}. The framework in this paper is compatible with both. In either case, adding optically selected clusters below the ICM detection threshold can improve the constraints on the scaling relation at the low-mass end and potentially constrain the correlation between the ICM and optical observables.

\section{Summary}

In this paper, we carry out the mass calibration of DES and SPT clusters, including the optical selection effect in our model, using all clusters in the DES Y3 and SPT-SZ, SPTpol-500d, and SPTpol-ECS. The data vector consists of the richness and SZe significance of 19,986 \redmapper clusters with $\lambda \ge 20$ and $0.2 \le z < 0.85$, of which 488 are detected by SPT and 13,926 lie in the SPT footprint without a detection and enter as upper limits on $\xi$, together with 120 SPT-only clusters outside DES and the stacked DES Y3 weak-lensing profiles in bins of redshift, richness, and SPT detection. We model the richness, SZe signal, and lensing signal at fixed halo mass as a multivariate log-normal distribution whose correlated scatter encodes the selection effects, with a scale-dependent correlation $\omega(\rp)$ between the richness and lensing residuals describing line-of-sight projections, and we fit the observable density and the stacked lensing jointly at fixed cosmology, including baryonic and miscentering corrections. We place tight constraints on the correlated scatter between $\lambda$ and lensing at a given halo mass in the two-halo regime. The correlated scatter is detected at four standard deviations ($\omega_2 = 0.26^{+0.06}_{-0.06}$) and translates into an empirical calibration of the optical selection bias of the weak-lensing signal (Fig.~\ref{fig:bsel_ppd}): for richness-selected clusters the large-scale lensing signal is boosted by 12--21\%, with no significant bias within $0.8\,h^{-1}$Mpc, whereas for SPT-detected clusters the large-scale signal is lowered by up to 23\% at $\lambda = 20$--35 and is unbiased at $\lambda > 60$. Our richness--mass and SZe--mass relations agree with previous DES and SPT analyses over the mass range of our sample, while the SZe scatter at fixed mass is larger than that found for SPT-selected samples. Treating selection biases via correlations among intrinsic scatters provides a principled way of modeling them within the context of Bayesian population models. Our method also provides a conceptual avenue to model other selection biases, like assembly bias in galaxy-galaxy lensing. We develop a flexible statistical framework for combining multiwavelength cluster datasets, requiring only that a cluster be detected in at least one of the constituent surveys. The framework is highly scalable and can be readily extended to accommodate more complex physical models and scaled to use more clusters. The statistical framework can also be easily adapted to strengthen current modeling techniques.

Current and future multiwavelength cluster datasets will offer a prime opportunity to apply the inference framework developed in this study. For a cluster survey, as long as the selection function for cluster observables is well-understood, we can use the framework in this work to combine it with other surveys. For optical and infrared surveys, the Nancy Roman Space Telescope \cite{spergelWideFieldInfrarRedSurvey2015} and Vera C. Rubin Observatory \cite{ivezicLSSTScienceDrivers2019} will find tens of thousands of clusters and billions of lensing source galaxies. In the X-ray regime, the eROSITA all-sky survey will yield vast catalogs of X-ray-selected clusters \cite{predehlEROSITAXrayTelescope2021, bulbulSRGEROSITAAllSky2024}, with the forthcoming Athena observatory \cite{barconsAthenaXrayObservatory2015} providing deep, targeted follow-up to characterize their detailed thermodynamic properties. Finally, millimeter-wave surveys from facilities like the Simons Observatory \cite{adeSimonsObservatoryScience2019}, SPT-3G \cite{bensonSPT3GNextgenerationCosmic2014}, and the Atacama Cosmology Telescope \cite{swetzOverviewAtacamaCosmology2011} will deliver massive catalogs of clusters selected via the SZe, enabling a fully integrated, multi-probe view of cluster physics and cosmology. The SPT-3G main-field cluster catalog \cite{bleemGalaxyClustersSelected2026} is already available and detects a much larger fraction of the \redmapper clusters in the DES area it covers than SPT-SZ. Applying our framework to DES and SPT-3G will extend the joint mass calibration to much lower masses.

Following further validation of our pipeline against more realistic mock catalogs, a subsequent paper will incorporate cosmological parameters into the model to perform a fully blinded cosmological analysis. Given the unprecedented volume and quality of upcoming cluster datasets, the methodology proposed here will be instrumental in unlocking the full potential of cluster cosmology.

\section*{Acknowledgement}

This work was supported by the U.S. Department of Energy, Office of Science, Office of High Energy Physics, under Award Number DE-SC0010107. SG acknowledges support provided by the Austrian Research Promotion Agency (FFG) and the Federal Ministry of the Republic of Austria for Innovation, Mobility and Infrastructure (BMIMI) via the Austrian Space Applications Programme with grant numbers 899537, 900565, 911971, and 928759. This research was funded in part by
the Austrian Science Fund (FWF) 10.55776/F101300.

AF acknowledges support from the National Science Foundation under Cooperative Agreement 2421782 and the Simons Foundation grant MPS-AI-00010515 awarded to NSF-Simons AI Institute for Cosmic Origins--CosmicAI, \href{https://www.cosmicai.org/}{https://www.cosmicai.org/}. 

The South Pole Telescope program is supported by the National Science Foundation (NSF) through award OPP-2332483. Partial support is also provided by the Kavli Institute of Cosmological Physics at the University of Chicago. 
Argonne National Laboratory’s work was supported by the U.S. Department of Energy, Office of High Energy Physics, under contract DE-AC02-06CH11357. 
The Melbourne authors acknowledge support from the Australian Research Council’s Discovery Project scheme (No. DP260100705). 

Funding for the DES Projects has been provided by the U.S. Department of Energy, the U.S. National Science Foundation, the Ministry of Science and Education of Spain, 
the Science and Technology Facilities Council of the United Kingdom, the Higher Education Funding Council for England, the National Center for Supercomputing 
Applications at the University of Illinois at Urbana-Champaign, the Kavli Institute of Cosmological Physics at the University of Chicago, 
the Center for Cosmology and Astro-Particle Physics at the Ohio State University,
the Mitchell Institute for Fundamental Physics and Astronomy at Texas A\&M University, Financiadora de Estudos e Projetos, 
Funda{\c c}{\~a}o Carlos Chagas Filho de Amparo {\`a} Pesquisa do Estado do Rio de Janeiro, Conselho Nacional de Desenvolvimento Cient{\'i}fico e Tecnol{\'o}gico and 
the Minist{\'e}rio da Ci{\^e}ncia, Tecnologia e Inova{\c c}{\~a}o, the Deutsche Forschungsgemeinschaft and the Collaborating Institutions in the Dark Energy Survey. 

The Collaborating Institutions are Argonne National Laboratory, the University of California at Santa Cruz, the University of Cambridge, Centro de Investigaciones Energ{\'e}ticas, 
Medioambientales y Tecnol{\'o}gicas-Madrid, the University of Chicago, University College London, the DES-Brazil Consortium, the University of Edinburgh, 
the Eidgen{\"o}ssische Technische Hochschule (ETH) Z{\"u}rich, 
Fermi National Accelerator Laboratory, the University of Illinois at Urbana-Champaign, the Institut de Ci{\`e}ncies de l'Espai (IEEC/CSIC), 
the Institut de F{\'i}sica d'Altes Energies, Lawrence Berkeley National Laboratory, the Ludwig-Maximilians Universit{\"a}t M{\"u}nchen and the associated Excellence Cluster Universe, 
the University of Michigan, NSF NOIRLab, the University of Nottingham, The Ohio State University, the University of Pennsylvania, the University of Portsmouth, 
SLAC National Accelerator Laboratory, Stanford University, the University of Sussex, Texas A\&M University, and the OzDES Membership Consortium.

Based in part on observations at NSF Cerro Tololo Inter-American Observatory at NSF NOIRLab (NOIRLab Prop. ID 2012B-0001; PI: J. Frieman), which is managed by the Association of Universities for Research in Astronomy (AURA) under a cooperative agreement with the National Science Foundation.

The DES data management system is supported by the National Science Foundation under Grant Numbers AST-1138766 and AST-1536171.
Data access is enabled by Jetstream2 and OSN at Indiana University through allocation PHY240006: Dark Energy Survey from the Advanced Cyberinfrastructure Coordination Ecosystem: Services and Support (ACCESS) program, which is supported by U.S. National Science Foundation grants 2138259, 2138286, 2138307, 2137603, and 2138296.
The DES participants from Spanish institutions are partially supported by MICINN under grants PID2021-123012, PID2021-128989 PID2022-141079, SEV-2016-0588, CEX2020-001058-M and CEX2020-001007-S, some of which include ERDF funds from the European Union. IFAE is partially funded by the CERCA program of the Generalitat de Catalunya.

We  acknowledge support from the Brazilian Instituto Nacional de Ci\^enciae Tecnologia (INCT) do e-Universo (CNPq grant 465376/2014-2).

This document was prepared by the DES Collaboration using the resources of the Fermi National Accelerator Laboratory (Fermilab), a U.S. Department of Energy, Office of Science, Office of High Energy Physics HEP User Facility. Fermilab is managed by Fermi Forward Discovery Group, LLC, acting under Contract No. 89243024CSC000002.

\appendix

\section{Lensing Measurement}
\label{app:lensing_measurement}

Here $e_{t,i}$ and $e_{\times,i}$ are the tangential and cross components of the \textsc{metacalibration} ellipticity of source $i$ relative to the lens, $w_i^{\rm s}$ its shape weight, $\mathcal{R}_i$ its shear response, and $b$ the SOMPz tomographic bin of the source \citep{mylesDarkEnergySurvey2021}.

For each lens bin and radial bin, we compute the following statistics:
\begin{itemize}
  \item The raw tangential and cross shear 
  \beq
  g_{\alpha}=\frac{\sum_{b=2,3,4} w^b \sum_{i \in b} w_i^{\mathrm{s}} e_{\alpha, i}}{\sum_{b=2,3,4} w^b \sum_{i \in b} w_i^{\mathrm{s}} \mathcal{R}_i} \text { for } \alpha \in\{t, \times\},  
\eeq
where $i$ indicates
the source-lens pair, and $b$ indicates the tomographic redshift bin ($b=1$ is omitted). Source bin $b$ is used only for lenses with $z_{\rm l} < z_{{\rm med},b}$, the median redshift of the bin (0.475, 0.745, and 0.945 for $b = 2, 3, 4$), and the weights $w^b$ are the mean inverse critical surface density of the bin at the lens redshift.
  \item The average angular separation between source-lens pair
  \beq
  \bar{\theta} =\frac{\sum_{b=2,3,4} w^b \sum_{i \in b} w_i^{\mathrm{s}} \theta_{i} \mathcal{R}_i}{\sum_{b=2,3,4} w^b \sum_{i \in b} w_i^{\mathrm{s}} \mathcal{R}_i},
  \eeq
  where $\theta_i$ is the angular separation between the $i^{th}$ pair. 
  \item The average physical separation between source-lens pairs
  \beq
  \bar{R} =\frac{\sum_{b=2,3,4} w^b \sum_{i \in b} w_i^{\mathrm{s}} R_{i} \mathcal{R}_i}{\sum_{b=2,3,4} w^b \sum_{i \in b} w_i^{\mathrm{s}} \mathcal{R}_i},
  \eeq
  \item The local source redshift distribution
  \beq
  P(z_\text{s}) =\frac{\sum_{b=2,3,4} w^b \sum_{i \in b} w_i^{\mathrm{s}} \mathcal{R}_i P(z_\text{s}|\hat{c}_i)}{\sum_{b=2,3,4} w^b \sum_{i \in b} w_i^{\mathrm{s}} \mathcal{R}_i},
  \label{eq:Pzs}
  \eeq
  constructed using the SOM-cell based estimates $P(z_\text{s}|\hat{c}_i)$ for each source,
  \item the effective number of sources 
  \beq
    N_\text{eff} = \frac{\left(\sum_{b=2,3,4} w^b \sum_{i \in b} w_i^{\mathrm{s}} \mathcal{R}_i\right)^2}{\sum_{b=2,3,4} 
    \sum_{i \in b}  
    (w^b w_i^{\mathrm{s}} \mathcal{R}_i)^2},
  \eeq
which is used to estimate the variance in the shape noise of sources, given by
  \beq
    \text{Var}\left[g_{\alpha} \right] = \frac{\sigma_\epsilon^2}{N_\text{eff}},
    \label{eq:var_g}
  \eeq
  with the effective shape noise variance $\sigma_\epsilon^2$ computed for each lens redshift bin according to the discussion in Sec 3.3.3 of \citet{grandisSRGEROSITAAllSky2024}, and
  \item the empirical photometric redshift distributions
  \beq
  \begin{split}
  P_\beta(z_{\mathrm{s}}) &= \frac{\sum_{b=2,3,4} w^b \sum_{i \in b} w_i^{\mathrm{s}} \mathcal{R}_i \mathbb{I}(z_{\mathrm{s}}, \hat{z}_{\beta,i})}{\sum_{b=2,3,4} w^b \sum_{i \in b} w_i^{\mathrm{s}} \mathcal{R}_i}, \\
  &\quad \text{for } \beta \in \{\texttt{dnf}, \texttt{bpz}\}
  \end{split}
  \eeq
  where $\mathbb{I}(z_\text{s}, \hat{z}_{\beta,i})=(z_{\text{s},+} - z_{\text{s},-})^{-1}$ if the measured photometric redshift estimate $ \hat{z}_{\beta,i}$ falls into the source redshift bin $(z_{\text{s},-}, z_{\text{s},+})$. Effectively, this is a shear-response-weighted, normalized histogram.
\end{itemize}

\section{Observational weak lensing systematics}
\label{appendix:contamination}
Having selected the sources with weights based on the tomographic redshift bin selection, we can readily propagate the calibration of the multiplicative shear bias and the photometric redshift distribution of the tomographic source bins to our measurement following Eq. 17 in \citet{grandisSRGEROSITAAllSky2024} that takes into account the correlated uncertainty between shear calibration and photometric redshift calibration, induced mainly by blending \citep[see][for an extended discussion]{maccrannDarkEnergySurvey2021}.

Besides photo-z and shear calibration, the second major source of observational systematics in our measurement is cluster member contamination (alternatively also known as boost factor). While this effect can be significantly mitigated with scale cuts for galaxy-galaxy lensing studies, our data vector is significantly more impacted by this effect for three reasons: (1) we explicitly seek to measure weak lensing close to our source to pick up the signal from the cluster's own mass distribution (called the 1-halo term in the context of the halo model), (2) galaxy clusters are associated with more significant overdensities of galaxies than the bright galaxies used for galaxy-galaxy lensing, and (3) we are making a very lenient source background selection. Taken together, this implies that many galaxies associated with our clusters leak into our source sample. As they carry no weak lensing signal, they dilute the shear estimator. The strength of this dilution is proportional to $1-f_\text {cl}(R)$, the fraction of cluster members at a distance $R$ from a given lens sample. To complete the analogy to galaxy-galaxy lensing, consider that the 'boost factor' $\mathcal{B}(R) = (1-f_\text{cl}(R))^{-1}$. 

\begin{figure*}
  \centering
  \includegraphics[width=\textwidth]{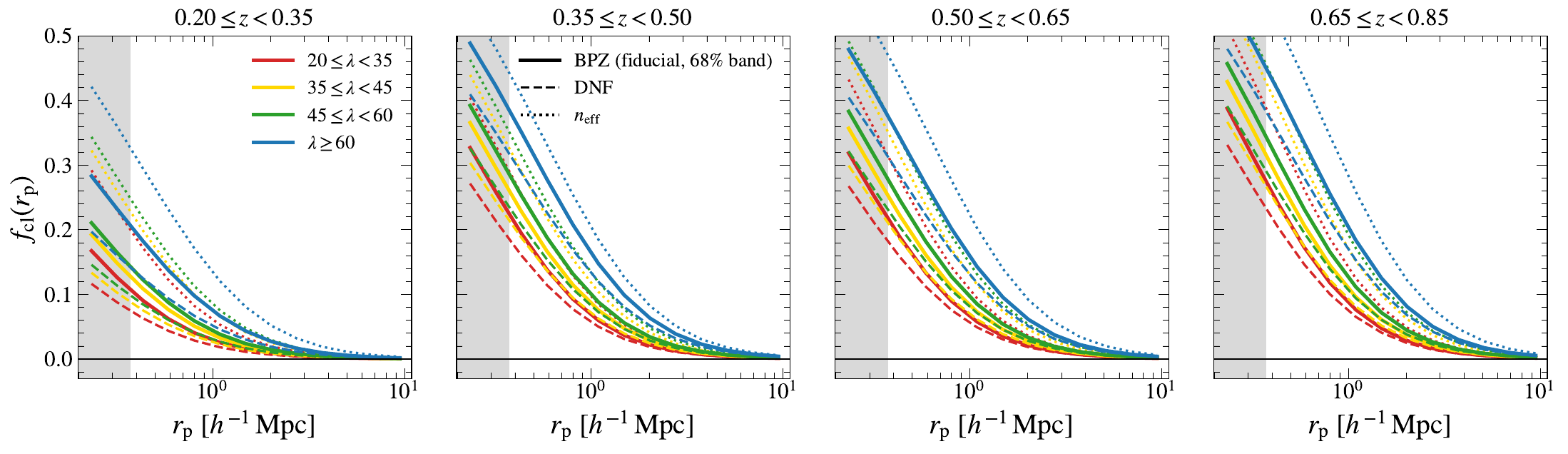}
  \caption{Cluster-member contamination fraction $f_{\rm cl}(r_{\rm p})$ of the source sample in bins of redshift (panels) and $\lambda$ (colors). Solid lines show the fiducial estimate from the decomposition of the \textsc{bpz} source redshift distributions, with the 68\% range of the Monte-Carlo realizations; dashed and dotted lines show the estimates from \textsc{dnf} and from the effective source number density, respectively. The gray band marks the scales excluded from the lensing fit. The lensing model is multiplied by $1 - f_{\rm cl}(r_{\rm p})$ using the median of the \textsc{bpz} realizations.}
  \label{fig:member_contamination}
\end{figure*}

We compute three estimates of the cluster-member contamination fraction, shown in Fig.~\ref{fig:member_contamination}, following the prescriptions in Sec. 3.3.1 of \citet{grandisSRGEROSITAAllSky2024}. One fit is based on the source number density derived from the effective number of sources. This estimate is equivalent to the typical estimate used in galaxy-galaxy lensing, which is provided directly from the correlation function between the lens and source positions. In the context of cluster weak lensing, it has been shown that this estimator can be biased by several percent because of source blending and magnification \cite{kleinebreilSRGEROSITAAllSky2025}. We therefore supplement this estimator with one based on the decomposition of source redshift distributions. \citet{bocquetSPTClustersHST2024a} noted that the estimate based on SOM-based photo-zs does not yield a decomposition signal. We therefore follow that work's recommendation to use the photo-z histograms derived from \textsc{bpz} and \textsc{dnf} estimates. We tested that using \textsc{bpz} and \textsc{dnf} makes no practical difference to our conclusions. Therefore, we only show the results computed with \textsc{bpz}. The lensing model is multiplied by $1 - f_{\rm cl}(r_{\rm p})$ using the median of the \textsc{bpz} realizations.

\section{Baryonification}
\label{sec:bcm}
Baryonic feedback modifies the mass distribution in and around halos. We model this effect using an analytic implementation of the baryonic feedback model of \cite{giriEmulationBaryonicEffects2021}. For a cluster halo of a given mass, we model a ``baryonified'' halo profile as,
\beq
\rho_\mathrm{dmb}(r) = \rho_\mathrm{clm}(r) + \rho_\mathrm{gas}(r) + \rho_\mathrm{cga}(r),
\eeq
a sum of collisionless matter (clm), gas, and central galaxy (cga) profiles. These density profiles are normalized such that the total mass is conserved,
\begin{align}
\int_0^\infty &\left( \rho_\mathrm{clm}(r) + \rho_\mathrm{gas}(r) + \rho_\mathrm{cga}(r) \right) 4 \pi r^2 \dd r \nonumber \\
&= \int_0^\infty \rho_\mathrm{nfw}(r) 4 \pi r^2 \dd r = M_{\rm tot}
\end{align}
where $\rho_\mathrm{nfw}$ is the truncated NFW \cite{navarroUniversalDensityProfile1997} profile,
\beq
\rho_\mathrm{nfw}(r \mid M, c) = \frac{\rho_0}{\frac{r}{r_s} ( 1 + \frac{r}{r_s})^2} \frac{1}{(1 + (\frac{r}{r_t})^2)^2},
\label{eq:nfw}
\eeq
for a halo of mass $M$ (we use $M_{200c}$ throughout) with radius $r_{200c}$, scale radius $r_s$, and concentration $c = r_{200c} / r_s$; the truncation radius is $r_t = 4\, r_{200c}$.

The central galaxy profile is described by an exponentially truncated power law,
\beq
\rho_\mathrm{cga} (r) = \frac{f_\mathrm{cga}(M)}{4 \pi^{3/2} R_h r^2} \mathrm{exp}\left[ - \frac{r}{2 R_h} \right],
\eeq
where $f_\mathrm{cga}(M)$ is the fraction of stars in the central galaxy, and $R_h = 0.015\, r_{200c}$ is the stellar half-light radius. The gas profile is written following \citet{schneiderNewMethodQuantify2015, schneiderQuantifyingBaryonEffects2019},
\beq
\rho_\mathrm{gas} \propto \frac{\Omega_b / \Omega_m - f_\mathrm{star}(M)}{\left[ 1 + 10 \left( \frac{r}{r_{200c}}\right) \right]^{\beta(M)} \left[ 1 + \left(\frac{r}{\theta_\mathrm{ej} r_{200c}} \right)^\gamma \right]^\frac{\delta - \beta(M)}{\gamma}},
\eeq
where $f_\mathrm{star}(M)$ is the total fraction of stars in haloes. The parameter $\beta$ introduces an additional halo-mass dependence,
\beq
\beta(M) = \frac{3(M/M_c)^\mu}{1 + (M / M_c)^\mu},
\eeq
with $M_c$ and $\mu$ free parameters (Table~\ref{tab:param}). The initial collisionless matter profile is written as
\beq
\rho_\mathrm{clm}(r) = \left[ \frac{\Omega_\mathrm{dm}}{\Omega_m} + f_\mathrm{sga}(M) \right]\rho_\mathrm{nfw}(r \mid M, c),
\eeq
where $\rho_\mathrm{nfw}(r | M, c)$ is the NFW profile with concentrations given by the mass--concentration relation of \citet{ishiyamaUchuuSimulationsData2021} and $f_\mathrm{sga}$ is the stellar fraction from satellite galaxies and intracluster light. This profile is then modified according to the ``adiabatic relaxation'' prescription of \cite{schneiderQuantifyingBaryonEffects2019}, which relates the total mass of the initial NFW profile $m_i$ enclosed within radius $r_i$ to that of the final baryonified profile,
\beq
\frac{r_f}{r_i} - 1 = a \left[ \left( \frac{m_i}{m_f}\right)^n - 1\right],
\eeq
where $a = 0.3$ and $n=2$ as in \cite{schneiderQuantifyingBaryonEffects2019}, and $m_i$ and $m_f$ are the masses enclosed within the initial and final radii, $r_i$ and $r_f$, of a mass shell, given by
\begin{align}
m_i &= m_\mathrm{nfw}(r_i), \nonumber \\
m_f &= f_\mathrm{clm} m_\mathrm{nfw}(r_i) + m_\mathrm{cga}(r_f) + m_\mathrm{gas}(r_f),
\end{align}
where $m_X(r)$ denotes the mass of component $X$ enclosed within radius $r$ and $f_\mathrm{clm} = \Omega_\mathrm{dm}/\Omega_m + f_\mathrm{sga}$. These equations are iteratively solved for,
\beq
\zeta(r_i) = r_f / r_i
\eeq
to obtain the final collisionless matter profile,
\beq
\rho_\mathrm{clm}(r) = \frac{f_\mathrm{clm}}{4\pi r^2} \frac{d}{dr} m_\mathrm{nfw}\left(r / \zeta(r) \right).
\eeq
Finally, we relate the various stellar-to-halo fractions via,
\beq
f_\mathrm{sga}(M) = f_\mathrm{star}(M) - f_\mathrm{cga}(M),
\eeq
where
\beq
f_i(M) = 0.055 \left( \frac{M}{M_s} \right)^{-\eta_i}.
\eeq
where $M_s = 2.5 \times 10^{11} h^{-1} M_\odot$ and $\eta_\mathrm{star} = \eta$ and $\eta_\mathrm{cga} = \eta + \eta_\delta$. The full set of baryonification parameters is summarized in Table~\ref{tab:param}.

To apply the baryonic effect correction, as described in Section~\ref{sec:lensing_likelihood}, for a given set of our model parameters, we infer the mass distribution for each cluster given observables. From this, we calculate an average mass $M$ for each cluster. For each of our observable bins in $\lambda$, redshift $z$, and $\xi$ bins, we compute a mean surface density profile $\langle \Sigma \rangle$, mass $\langle M \rangle$, and concentration $\langle c \rangle$ using the mass-concentration relation from \citet{ishiyamaUchuuSimulationsData2021} for the bin.

Given an assumed mass and concentration for each of our bins, we calculate a baryonic correction to the cluster surface density profile $\Sigma$,
\beq
f_b(\rp \mid M, z; \theta_{b}) = \frac{\Sigma_\mathrm{dmb} (\rp \mid M, z; \theta_b) }{ \Sigma_\mathrm{dmo} (\rp \mid M, z)}
\eeq
where the parameter vector $\theta_b$ consists of baryonification parameters, and $\Sigma$ is calculated from either a baryonified or dark matter-only halo density profile. To do this, we first model the two-halo term contribution to the density profile,
\beq
\rho^{2h}(r) = \rho_m \times (b_h \xi_{mm} + 1)
\eeq
where $\rho_m = \Omega_m \rho_c$ is the cosmological matter density, the halo bias $b_h$ is computed using \citet{tinkerLargescaleBiasDark2010} and $\xi_{mm}$ is computed using \citet{eisensteinPowerSpectraCold1999}. The halo-matter cross-correlation function is modeled then as a sum of 1- and 2-halo terms,
\beq
\xi_{hm}(r) = \frac{\rho^{1h}(r) + \rho^{2h}(r)}{\rho_m} - 1.
\eeq
where $\rho^{1h}$ is the one-halo density profile, $\rho_\mathrm{dmb}$ or $\rho_\mathrm{nfw}$. This correlation function is then further converted into a surface density profile,
\beq
\Sigma (\rp) = \rho_m \int \dd \pi \, \xi_{hm}(\rp, \pi).
\eeq
This correction multiplies the well-centered profile of Sec.~\ref{sec:lensing_sys}, $\Sigmacen(\rp) = f_b(\rp)\,\Sigmanosys(\rp)$ [Eq.~(\ref{eq:baryon_sigma})].

\begin{widetext}
\section{Estimating $\xi$ with Gauss Hermite Integral}
\label{appendix:gauss_hermite}

The probability of observing $\xi$ at a given halo mass and redshift $P(\xi \mid M,z,\Omega)$ is approximated with a Gauss-Hermite integral. We evaluate the integrand on a 500-step grid from $\log_{10} M \in [12, 16]$ and numerically integrate the probability integrals. 

To evaluate the likelihood of the biased $\xi$, we marginalize over the unbiased $\zeta$ with Gauss-Hermite estimation, 
\begin{align}
  P(\xi \mid \Omega) &= \int P(\xi \mid \zeta)\,P(\ln\zeta \mid \Omega) \; \dd \ln\zeta\\
  &= \int \frac{1}{\sqrt{2\pi}} \exp\left(-\frac{1}{2} \left(\xi - \sqrt{\zeta^2 +3}\right)^2 \right) \frac{1}{\sqrt{2\pi \sigma_{\zeta}^2}}\exp\left( -\frac{\left( \ln\zeta - \mu_\zeta \right)^2}{2 \sigma_\zeta^2} \right) \; \dd \ln\zeta ,
\end{align}
where $\mu_\zeta = \langle \ln\zeta \mid M \rangle$. Integrating over $\ln\zeta$ rather than $\zeta$ is a change of variables, $P(\zeta)\,\dd\zeta = P(\ln\zeta)\,\dd\ln\zeta$; since $\ln\zeta$ is Gaussian at fixed mass, the weight of the integral is a Gaussian, which is what the Gauss--Hermite rule requires.

Define
\beq
  x=\frac{\ln\zeta-\mu_\zeta}{\sqrt{2} \sigma_\zeta}.
\eeq
We have 
\beq
  P(\xi \mid \Omega)=\int_{-\infty}^{+\infty} e^{-x^2}\left[\frac{1}{ \sqrt{2} {\pi}} \exp \left(-\frac{1}{2}\left(\xi-\sqrt{e^{2\left(\mu_\zeta+\sqrt{2} \sigma_\zeta x\right)}+3}\right)^2\right)\right] \dd x, 
\eeq
which can be approximated with the standard Gaussian-Hermite approximation,
\beq
  P(\xi \mid \Omega) \approx \sum_{i=1}^{N_{g h}} w_i\left[\frac{1}{\sqrt{\pi} \sqrt{2 \pi}} \exp \left(-\frac{1}{2}\left(\xi-\sqrt{e^{2\left(\mu_\zeta+\sqrt{2} \sigma_\zeta x_i\right)}+3}\right)^2\right)\right].
\eeq
These expressions hold for $\zeta > \zeta_{\rm min}$. Below $\zeta_{\rm min}$ the model uses $\xi \mid \zeta \sim \mathcal{N}(\zeta, 1)$ (Sec.~\ref{sec:scaling_relations}), so $\sqrt{e^{2(\mu_\zeta + \sqrt{2}\sigma_\zeta x_i)} + 3}$ is replaced by $e^{\mu_\zeta + \sqrt{2}\sigma_\zeta x_i}$ at those nodes.
$w_i$ is the standard Gauss-Hermite weight,
\beq
  w_i=\frac{2^{n-1} n!\sqrt{\pi}}{n^2\left[H_{n-1}\left(x_i\right)\right]^2}.
\eeq
$H_n(x)$ are the physicists' version of the Hermite polynomials. $x_i$ are the roots of $H_n(x)$. We use $N_{gh} = 100$ sample points, i.e., the roots $x_i$ of $H_{100}(x)$ with their weights $w_i$. Compared to high-resolution Monte-Carlo approximations of the likelihood, the mean absolute error of the Gauss-Hermite approximation is $\sim 10^{-5}$, which is sufficient for our purposes. 

\end{widetext}

\section{Lensing of DES clusters outside of SPT footprint}
\label{appendix:lensing_no_spt}

\begin{figure*}
  \centering
  \includegraphics[width=\textwidth]{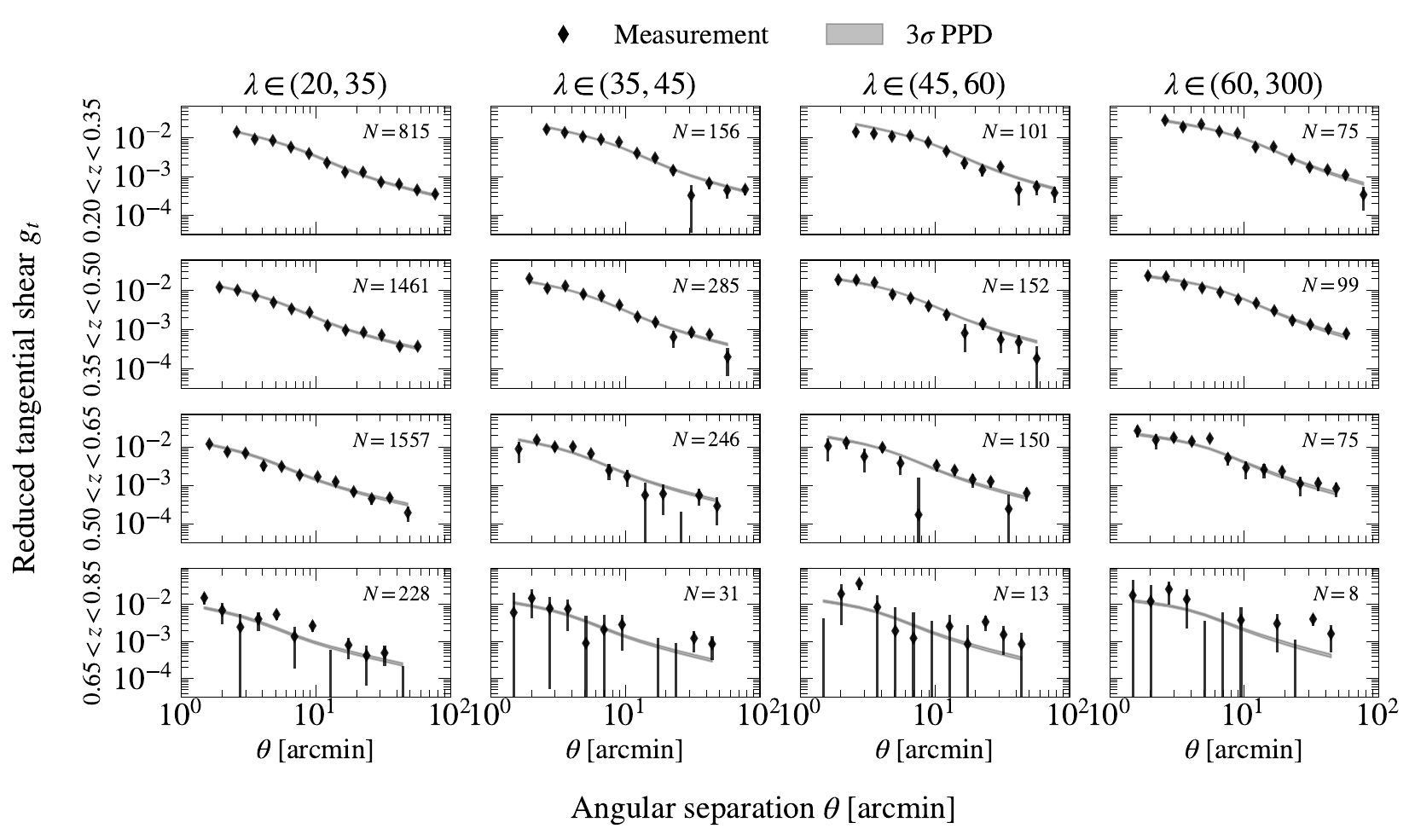}
  \caption{Stacked lensing measurement and PPD of DES clusters not in the SPT footprint. The black points with error bars are the measurements, and the gray bands are 3$\sigma$ PPD. The number of clusters in each bin is shown in the top right corner of each panel.}
  \label{fig:lensing_no_spt}
\end{figure*}

In Fig.~\ref{fig:lensing_no_spt}, we show the measurement and the PPD of stacked lensing in redshift and $\lambda$ bins of DES clusters that are not in the SPT footprint. We see that the model fits the data well. 

\section{Cluster lensing semi-analytical covariance matrix}
\label{app:cov}

We calculate the cluster lensing covariance matrices by combining the empirical shape noise based on the effective number of sources [Eq.~(\ref{eq:var_g})] and the analytical approach in \cite{wuCovarianceMatricesGalaxy2019}. 
The empirical shape noise includes source weights and responses but lacks the contribution from large-scale structure, while the opposite is true for the analytical method. 
We then calculate the full analytic covariance matrix with clustering and multiply the matrix by the scaling factor. We describe the details below. 

The analytic covariance matrix is given by
\beq
\begin{aligned}
\mathrm{Cov}\left[\gamma_t(\theta_1), \gamma_t(\theta_2)\right]
=
\frac{1}{4\pi f_{\mathrm{sky}}}
\int \frac{\ell\, d\ell}{2\pi}\,
\hat{J}_2(\ell \theta_1)\,
\hat{J}_2(\ell \theta_2)\,
\\
\left[
\left(C_\ell^{hh} + \frac{1}{n_h^{\mathrm{(2D)}}}\right)
\left(C_\ell^{\kappa\kappa} + \frac{\sigma_\gamma^2}{n_s^{\mathrm{(2D)}}}\right)
+
\left(C_\ell^{h\kappa}\right)^2
\right],
\label{eq:cov}
\end{aligned}
\eeq
where $f_{\mathrm{sky}}$ is the fractional sky coverage, and $\hat{J}_2(\ell \theta)$ is the bin-averaged Bessel function centered on $\theta$. 

The terms in the first parentheses correspond to the noise in the cluster density field, determined by the cluster angular power spectrum $C_\ell^{hh} = b^2 C_\ell^{mm}$ and 2D number density $n_h^{\mathrm{(2D)}}$.
For each cluster subsample, we use the cluster counts from the cluster catalog and the cluster bias estimated from the mean cluster mass and the halo bias-mass relation from \cite{tinkerLargescaleBiasDark2010} and the maximum-a-posteriori scaling relations from \cite{bocquetSPTClustersHST2024b}.

The terms in the second parenthesis correspond to the noise in the lensing field, determined by the convergence power spectrum $C_\ell^{\kappa\kappa}$, the shape noise per galaxy $\sigma_\gamma^2$, and the 2D source density $n_s^{\mathrm{(2D)}}$. 
Our source galaxy has 
$n_{\rm src} = 3.8~{\rm arcmin}^{-2}$
and $\sigma_\gamma = 0.2635 + 0.0300 z_{\rm cl}^2 - 0.0008 z_{\rm cl}$ \cite{grandisSRGEROSITAAllSky2024}.
We use the source distribution $P(z_{\rm s})$ measured based on Eq.~(\ref{eq:Pzs}). We fit a function $P(z_{\rm s}) \propto z^m e^{-(z/z_*)^\beta}$ to the measured $P(z_{\rm s})$ to obtain a smooth distribution.
We then use this smoothed distribution to calculate the covariance matrix. 

We assume $h=0.7$, $\Omega_{\rm DE}$=0.7, $\Omega_m$=0.3, and $\sigma_8$=0.8. The covariance matrices are insensitive to these cosmological parameters. The $C_\ell^{hk}$ term in Eq.~(\ref{eq:cov}) is subdominant and is ignored in the calculation.

We first calculate the analytic covariance matrix without clustering (ignoring $C_\ell^{hh}$ and $C_\ell^{\kappa\kappa}$), and we compare it with the empirical shape noise. We calculate the rescaling factors
\beq
f = \frac{\delta g_t}{\sigma^{\text{no-clustering}}(\gamma_t)}, 
\eeq
where $\sigma^{\text{no-clustering}}$ denotes the square root of the diagonal elements of the covariances without $C_\ell^{hh}$ and $C_\ell^{\kappa\kappa}$ terms, and 
$\delta g_t$ is from Eq.~(\ref{eq:var_g}).
We take the mean for scales above 2 arcmin. 
These factors account for source weights and responses, as well as any small discrepancies in the source distribution. They are close to 1 at low redshifts and can reach 1.8 for high-redshift, high-$\lambda$ clusters. We then rescale the full analytic covariance matrix by $f^2$. The end product is a covariance matrix that matches the empirical shape noise at small scales and includes contributions from the large-scale structure. 

\bibliographystyle{apsrev4-2-author-truncate-noarxiv}
\bibliography{betterbibtex}%

\begin{thebibliography}{141}%
\makeatletter
\providecommand \@ifxundefined [1]{%
 \@ifx{#1\undefined}
}%
\providecommand \@ifnum [1]{%
 \ifnum #1\expandafter \@firstoftwo
 \else \expandafter \@secondoftwo
 \fi
}%
\providecommand \@ifx [1]{%
 \ifx #1\expandafter \@firstoftwo
 \else \expandafter \@secondoftwo
 \fi
}%
\providecommand \natexlab [1]{#1}%
\providecommand \enquote  [1]{``#1''}%
\providecommand \bibnamefont  [1]{#1}%
\providecommand \bibfnamefont [1]{#1}%
\providecommand \citenamefont [1]{#1}%
\providecommand \href@noop [0]{\@secondoftwo}%
\providecommand \href [0]{\begingroup \@sanitize@url \@href}%
\providecommand \@href[1]{\@@startlink{#1}\@@href}%
\providecommand \@@href[1]{\endgroup#1\@@endlink}%
\providecommand \@sanitize@url [0]{\catcode `\\12\catcode `\$12\catcode
  `\&12\catcode `\#12\catcode `\^12\catcode `\_12\catcode `\%12\relax}%
\providecommand \@@startlink[1]{}%
\providecommand \@@endlink[0]{}%
\providecommand \url  [0]{\begingroup\@sanitize@url \@url }%
\providecommand \@url [1]{\endgroup\@href {#1}{\urlprefix }}%
\providecommand \urlprefix  [0]{URL }%
\providecommand \Eprint[0]{\href }%
\providecommand \doibase [0]{https://doi.org/}%
\providecommand \selectlanguage [0]{\@gobble}%
\providecommand \bibinfo  [0]{\@secondoftwo}%
\providecommand \bibfield  [0]{\@secondoftwo}%
\providecommand \translation [1]{[#1]}%
\providecommand \BibitemOpen [0]{}%
\providecommand \bibitemStop [0]{}%
\providecommand \bibitemNoStop [0]{.\EOS\space}%
\providecommand \EOS [0]{\spacefactor3000\relax}%
\providecommand \BibitemShut  [1]{\csname bibitem#1\endcsname}%
\let\auto@bib@innerbib\@empty
\bibitem [{\citenamefont {Allen}\ \emph {et~al.}(2011)\citenamefont {Allen},
  \citenamefont {Evrard},\ and\ \citenamefont
  {Mantz}}]{allenCosmologicalParametersObservations2011}%
  \BibitemOpen
  \bibfield  {author} {\bibinfo {author} {\bibfnamefont {S.~W.}\ \bibnamefont
  {Allen}}, \bibinfo {author} {\bibfnamefont {A.~E.}\ \bibnamefont {Evrard}},\
  and\ \bibinfo {author} {\bibfnamefont {A.~B.}\ \bibnamefont {Mantz}},\ }\href
  {https://doi.org/10.1146/annurev-astro-081710-102514} {\bibfield  {journal}
  {\bibinfo  {journal} {Annu. Rev. Astron. Astrophys.}\ }\textbf {\bibinfo
  {volume} {49}},\ \bibinfo {pages} {409} (\bibinfo {year} {2011})}\BibitemShut
  {NoStop}%
\bibitem [{\citenamefont {Weinberg}\ \emph {et~al.}(2013)\citenamefont
  {Weinberg}, \citenamefont {Mortonson}, \citenamefont {Eisenstein},
  \citenamefont {Hirata}, \citenamefont {Riess},\ and\ \citenamefont
  {Rozo}}]{weinbergObservationalProbesCosmic2013}%
  \BibitemOpen
  \bibfield  {author} {\bibinfo {author} {\bibfnamefont {D.~H.}\ \bibnamefont
  {Weinberg}}, \bibinfo {author} {\bibfnamefont {M.~J.}\ \bibnamefont
  {Mortonson}}, \bibinfo {author} {\bibfnamefont {D.~J.}\ \bibnamefont
  {Eisenstein}}, \emph {et~al.},\ }\href
  {https://doi.org/10.1016/j.physrep.2013.05.001} {\bibfield  {journal}
  {\bibinfo  {journal} {Phys. Rep.}\ }\textbf {\bibinfo {volume} {530}},\
  \bibinfo {pages} {87} (\bibinfo {year} {2013})}\BibitemShut {NoStop}%
\bibitem [{\citenamefont {Huterer}\ \emph {et~al.}(2015)\citenamefont
  {Huterer}, \citenamefont {Kirkby}, \citenamefont {Bean}, \citenamefont
  {Connolly}, \citenamefont {Dawson}, \citenamefont {Dodelson}, \citenamefont
  {Evrard}, \citenamefont {Jain}, \citenamefont {Jarvis}, \citenamefont
  {Linder}, \citenamefont {Mandelbaum}, \citenamefont {May}, \citenamefont
  {Raccanelli}, \citenamefont {Reid}, \citenamefont {Rozo}, \citenamefont
  {Schmidt}, \citenamefont {Sehgal}, \citenamefont {Slosar}, \citenamefont
  {{van Engelen}}, \citenamefont {Wu},\ and\ \citenamefont
  {Zhao}}]{hutererGrowthCosmicStructure2015}%
  \BibitemOpen
  \bibfield  {author} {\bibinfo {author} {\bibfnamefont {D.}~\bibnamefont
  {Huterer}}, \bibinfo {author} {\bibfnamefont {D.}~\bibnamefont {Kirkby}},
  \bibinfo {author} {\bibfnamefont {R.}~\bibnamefont {Bean}}, \emph {et~al.},\
  }\href {https://doi.org/10.1016/j.astropartphys.2014.07.004} {\bibfield
  {journal} {\bibinfo  {journal} {Astropart. Phys.}\ }\textbf {\bibinfo
  {volume} {63}},\ \bibinfo {pages} {23} (\bibinfo {year} {2015})}\BibitemShut
  {NoStop}%
\bibitem [{\citenamefont {Kellogg}\ \emph {et~al.}(1973)\citenamefont
  {Kellogg}, \citenamefont {Murray}, \citenamefont {Giacconi}, \citenamefont
  {Tananbaum},\ and\ \citenamefont {Gursky}}]{kelloggClustersGalaxiesWide1973}%
  \BibitemOpen
  \bibfield  {author} {\bibinfo {author} {\bibfnamefont {E.}~\bibnamefont
  {Kellogg}}, \bibinfo {author} {\bibfnamefont {S.}~\bibnamefont {Murray}},
  \bibinfo {author} {\bibfnamefont {R.}~\bibnamefont {Giacconi}}, \emph
  {et~al.},\ }\href {https://doi.org/10.1086/181311} {\bibfield  {journal}
  {\bibinfo  {journal} {Astrophys. J.}\ }\textbf {\bibinfo {volume} {185}},\
  \bibinfo {pages} {L13} (\bibinfo {year} {1973})}\BibitemShut {NoStop}%
\bibitem [{\citenamefont {Mitchell}\ \emph {et~al.}(1976)\citenamefont
  {Mitchell}, \citenamefont {Culhane}, \citenamefont {Davison},\ and\
  \citenamefont {Ives}}]{mitchellAriel5Observations1976}%
  \BibitemOpen
  \bibfield  {author} {\bibinfo {author} {\bibfnamefont {R.~J.}\ \bibnamefont
  {Mitchell}}, \bibinfo {author} {\bibfnamefont {J.~L.}\ \bibnamefont
  {Culhane}}, \bibinfo {author} {\bibfnamefont {P.~J.~N.}\ \bibnamefont
  {Davison}},\ and\ \bibinfo {author} {\bibfnamefont {J.~C.}\ \bibnamefont
  {Ives}},\ }\href {https://doi.org/10.1093/mnras/175.1.29P} {\bibfield
  {journal} {\bibinfo  {journal} {Mon. Not. R. Astron. Soc.}\ }\textbf
  {\bibinfo {volume} {175}},\ \bibinfo {pages} {29P} (\bibinfo {year}
  {1976})}\BibitemShut {NoStop}%
\bibitem [{\citenamefont {Sarazin}(1986)}]{sarazinXrayEmissionClusters1986}%
  \BibitemOpen
  \bibfield  {author} {\bibinfo {author} {\bibfnamefont {C.~L.}\ \bibnamefont
  {Sarazin}},\ }\href {https://doi.org/10.1103/RevModPhys.58.1} {\bibfield
  {journal} {\bibinfo  {journal} {Rev. Mod. Phys.}\ }\textbf {\bibinfo {volume}
  {58}},\ \bibinfo {pages} {1} (\bibinfo {year} {1986})}\BibitemShut {NoStop}%
\bibitem [{\citenamefont {Rosati}\ \emph {et~al.}(2002)\citenamefont {Rosati},
  \citenamefont {Borgani},\ and\ \citenamefont
  {Norman}}]{rosatiEvolutionXrayClusters2002}%
  \BibitemOpen
  \bibfield  {author} {\bibinfo {author} {\bibfnamefont {P.}~\bibnamefont
  {Rosati}}, \bibinfo {author} {\bibfnamefont {S.}~\bibnamefont {Borgani}},\
  and\ \bibinfo {author} {\bibfnamefont {C.}~\bibnamefont {Norman}},\ }\href
  {https://doi.org/10.1146/annurev.astro.40.120401.150547} {\bibfield
  {journal} {\bibinfo  {journal} {Annu. Rev. Astron. Astrophys.}\ }\textbf
  {\bibinfo {volume} {40}},\ \bibinfo {pages} {539} (\bibinfo {year}
  {2002})}\BibitemShut {NoStop}%
\bibitem [{\citenamefont {Vikhlinin}\ \emph {et~al.}(2006)\citenamefont
  {Vikhlinin}, \citenamefont {Kravtsov}, \citenamefont {Forman}, \citenamefont
  {Jones}, \citenamefont {Markevitch}, \citenamefont {Murray},\ and\
  \citenamefont {Van~Speybroeck}}]{vikhlininChandraSampleNearby2006}%
  \BibitemOpen
  \bibfield  {author} {\bibinfo {author} {\bibfnamefont {A.}~\bibnamefont
  {Vikhlinin}}, \bibinfo {author} {\bibfnamefont {A.}~\bibnamefont {Kravtsov}},
  \bibinfo {author} {\bibfnamefont {W.}~\bibnamefont {Forman}}, \emph
  {et~al.},\ }\href {https://doi.org/10.1086/500288} {\bibfield  {journal}
  {\bibinfo  {journal} {Astrophys. J.}\ }\textbf {\bibinfo {volume} {640}},\
  \bibinfo {pages} {691} (\bibinfo {year} {2006})}\BibitemShut {NoStop}%
\bibitem [{\citenamefont {Bulbul}\ \emph {et~al.}(2024)\citenamefont {Bulbul},
  \citenamefont {Liu}, \citenamefont {Kluge}, \citenamefont {Zhang},
  \citenamefont {Sanders}, \citenamefont {Bahar}, \citenamefont {Ghirardini},
  \citenamefont {Artis}, \citenamefont {Seppi}, \citenamefont {Garrel},
  \citenamefont {{Ramos-Ceja}}, \citenamefont {Comparat}, \citenamefont
  {Balzer}, \citenamefont {B{\"o}ckmann}, \citenamefont {Br{\"u}ggen},
  \citenamefont {Clerc}, \citenamefont {Dennerl}, \citenamefont {Dolag},
  \citenamefont {Freyberg}, \citenamefont {Grandis}, \citenamefont {Gruen},
  \citenamefont {Kleinebreil}, \citenamefont {Krippendorf}, \citenamefont
  {Lamer}, \citenamefont {Merloni}, \citenamefont {Migkas}, \citenamefont
  {Nandra}, \citenamefont {Pacaud}, \citenamefont {Predehl}, \citenamefont
  {Reiprich}, \citenamefont {Schrabback}, \citenamefont {Veronica},
  \citenamefont {Weller},\ and\ \citenamefont
  {Zelmer}}]{bulbulSRGEROSITAAllSky2024}%
  \BibitemOpen
  \bibfield  {author} {\bibinfo {author} {\bibfnamefont {E.}~\bibnamefont
  {Bulbul}}, \bibinfo {author} {\bibfnamefont {A.}~\bibnamefont {Liu}},
  \bibinfo {author} {\bibfnamefont {M.}~\bibnamefont {Kluge}}, \emph {et~al.},\
  }\href {https://doi.org/10.1051/0004-6361/202348264} {\bibfield  {journal}
  {\bibinfo  {journal} {Astron. Astrophys.}\ }\textbf {\bibinfo {volume}
  {685}},\ \bibinfo {pages} {A106} (\bibinfo {year} {2024})}\BibitemShut
  {NoStop}%
\bibitem [{\citenamefont {Zwicky}(1937)}]{zwickyMassesNebulaeClusters1937}%
  \BibitemOpen
  \bibfield  {author} {\bibinfo {author} {\bibfnamefont {F.}~\bibnamefont
  {Zwicky}},\ }\href {https://doi.org/10.1086/143864} {\bibfield  {journal}
  {\bibinfo  {journal} {Astrophys. J.}\ }\textbf {\bibinfo {volume} {86}},\
  \bibinfo {pages} {217} (\bibinfo {year} {1937})}\BibitemShut {NoStop}%
\bibitem [{\citenamefont {Lynds}\ and\ \citenamefont
  {Petrosian}(1989)}]{lyndsLuminousArcsClusters1989}%
  \BibitemOpen
  \bibfield  {author} {\bibinfo {author} {\bibfnamefont {R.}~\bibnamefont
  {Lynds}}\ and\ \bibinfo {author} {\bibfnamefont {V.}~\bibnamefont
  {Petrosian}},\ }\href {https://doi.org/10.1086/166989} {\bibfield  {journal}
  {\bibinfo  {journal} {Astrophys. J.}\ }\textbf {\bibinfo {volume} {336}},\
  \bibinfo {pages} {1} (\bibinfo {year} {1989})}\BibitemShut {NoStop}%
\bibitem [{\citenamefont {Gonzalez}\ \emph {et~al.}(2005)\citenamefont
  {Gonzalez}, \citenamefont {Zabludoff},\ and\ \citenamefont
  {Zaritsky}}]{gonzalezIntraclusterLightNearby2005}%
  \BibitemOpen
  \bibfield  {author} {\bibinfo {author} {\bibfnamefont {A.~H.}\ \bibnamefont
  {Gonzalez}}, \bibinfo {author} {\bibfnamefont {A.~I.}\ \bibnamefont
  {Zabludoff}},\ and\ \bibinfo {author} {\bibfnamefont {D.}~\bibnamefont
  {Zaritsky}},\ }\href {https://doi.org/10.1086/425896} {\bibfield  {journal}
  {\bibinfo  {journal} {Astrophys. J.}\ }\textbf {\bibinfo {volume} {618}},\
  \bibinfo {pages} {195} (\bibinfo {year} {2005})}\BibitemShut {NoStop}%
\bibitem [{\citenamefont {Zibetti}\ \emph {et~al.}(2005)\citenamefont
  {Zibetti}, \citenamefont {White}, \citenamefont {Schneider},\ and\
  \citenamefont {Brinkmann}}]{zibettiIntergalacticStars0252005}%
  \BibitemOpen
  \bibfield  {author} {\bibinfo {author} {\bibfnamefont {S.}~\bibnamefont
  {Zibetti}}, \bibinfo {author} {\bibfnamefont {S.~D.~M.}\ \bibnamefont
  {White}}, \bibinfo {author} {\bibfnamefont {D.~P.}\ \bibnamefont
  {Schneider}},\ and\ \bibinfo {author} {\bibfnamefont {J.}~\bibnamefont
  {Brinkmann}},\ }\href {https://doi.org/10.1111/j.1365-2966.2005.08817.x}
  {\bibfield  {journal} {\bibinfo  {journal} {Mon. Not. R. Astron. Soc.}\
  }\textbf {\bibinfo {volume} {358}},\ \bibinfo {pages} {949} (\bibinfo {year}
  {2005})}\BibitemShut {NoStop}%
\bibitem [{\citenamefont {Wittman}\ \emph {et~al.}(2006)\citenamefont
  {Wittman}, \citenamefont {Dell'Antonio}, \citenamefont {Hughes},
  \citenamefont {Margoniner}, \citenamefont {Tyson}, \citenamefont {Cohen},\
  and\ \citenamefont {Norman}}]{wittmanFirstResultsShearselected2006}%
  \BibitemOpen
  \bibfield  {author} {\bibinfo {author} {\bibfnamefont {D.}~\bibnamefont
  {Wittman}}, \bibinfo {author} {\bibfnamefont {I.~P.}\ \bibnamefont
  {Dell'Antonio}}, \bibinfo {author} {\bibfnamefont {J.~P.}\ \bibnamefont
  {Hughes}}, \emph {et~al.},\ }\href {https://doi.org/10.1086/502621}
  {\bibfield  {journal} {\bibinfo  {journal} {Astrophys. J.}\ }\textbf
  {\bibinfo {volume} {643}},\ \bibinfo {pages} {128} (\bibinfo {year}
  {2006})}\BibitemShut {NoStop}%
\bibitem [{\citenamefont {Kneib}\ and\ \citenamefont
  {Natarajan}(2011)}]{kneibClusterLenses2011}%
  \BibitemOpen
  \bibfield  {author} {\bibinfo {author} {\bibfnamefont {J.-P.}\ \bibnamefont
  {Kneib}}\ and\ \bibinfo {author} {\bibfnamefont {P.}~\bibnamefont
  {Natarajan}},\ }\href {https://doi.org/10.1007/s00159-011-0047-3} {\bibfield
  {journal} {\bibinfo  {journal} {Astron. Astrophys. Rev.}\ }\textbf {\bibinfo
  {volume} {19}},\ \bibinfo {pages} {47} (\bibinfo {year} {2011})}\BibitemShut
  {NoStop}%
\bibitem [{\citenamefont {Oguri}\ and\ \citenamefont
  {Takada}(2011)}]{oguriCombiningClusterObservables2011}%
  \BibitemOpen
  \bibfield  {author} {\bibinfo {author} {\bibfnamefont {M.}~\bibnamefont
  {Oguri}}\ and\ \bibinfo {author} {\bibfnamefont {M.}~\bibnamefont {Takada}},\
  }\href {https://doi.org/10.1103/PhysRevD.83.023008} {\bibfield  {journal}
  {\bibinfo  {journal} {Phys. Rev. D}\ }\textbf {\bibinfo {volume} {83}},\
  \bibinfo {pages} {023008} (\bibinfo {year} {2011})}\BibitemShut {NoStop}%
\bibitem [{\citenamefont {Oguri}(2014)}]{oguriClusterFindingAlgorithm2014}%
  \BibitemOpen
  \bibfield  {author} {\bibinfo {author} {\bibfnamefont {M.}~\bibnamefont
  {Oguri}},\ }\href {https://doi.org/10.1093/mnras/stu1446} {\bibfield
  {journal} {\bibinfo  {journal} {Mon. Not. R. Astron. Soc.}\ }\textbf
  {\bibinfo {volume} {444}},\ \bibinfo {pages} {147} (\bibinfo {year}
  {2014})}\BibitemShut {NoStop}%
\bibitem [{\citenamefont {Rykoff}\ \emph {et~al.}(2014)\citenamefont {Rykoff},
  \citenamefont {Rozo}, \citenamefont {Busha}, \citenamefont {Cunha},
  \citenamefont {Finoguenov}, \citenamefont {Evrard}, \citenamefont {Hao},
  \citenamefont {Koester}, \citenamefont {Leauthaud}, \citenamefont {Nord},
  \citenamefont {Pierre}, \citenamefont {Reddick}, \citenamefont {Sadibekova},
  \citenamefont {Sheldon},\ and\ \citenamefont
  {Wechsler}}]{rykoffRedMaPPerAlgorithmSDSS2014}%
  \BibitemOpen
  \bibfield  {author} {\bibinfo {author} {\bibfnamefont {E.~S.}\ \bibnamefont
  {Rykoff}}, \bibinfo {author} {\bibfnamefont {E.}~\bibnamefont {Rozo}},
  \bibinfo {author} {\bibfnamefont {M.~T.}\ \bibnamefont {Busha}}, \emph
  {et~al.},\ }\href {https://doi.org/10.1088/0004-637X/785/2/104} {\bibfield
  {journal} {\bibinfo  {journal} {Astrophys. J.}\ }\textbf {\bibinfo {volume}
  {785}},\ \bibinfo {pages} {104} (\bibinfo {year} {2014})}\BibitemShut
  {NoStop}%
\bibitem [{\citenamefont {Rykoff}\ \emph {et~al.}(2016)\citenamefont {Rykoff},
  \citenamefont {Rozo}, \citenamefont {Hollowood}, \citenamefont
  {{Bermeo-Hernandez}}, \citenamefont {Jeltema}, \citenamefont {Mayers},
  \citenamefont {Romer}, \citenamefont {Rooney}, \citenamefont {Saro},
  \citenamefont {Vergara~Cervantes}, \citenamefont {Wechsler}, \citenamefont
  {Wilcox}, \citenamefont {Abbott}, \citenamefont {Abdalla}, \citenamefont
  {Allam}, \citenamefont {Annis}, \citenamefont {{Benoit-L{\'e}vy}},
  \citenamefont {Bernstein}, \citenamefont {Bertin}, \citenamefont {Brooks},
  \citenamefont {Burke}, \citenamefont {Capozzi}, \citenamefont
  {Carnero~Rosell}, \citenamefont {Carrasco~Kind}, \citenamefont {Castander},
  \citenamefont {Childress}, \citenamefont {Collins}, \citenamefont {Cunha},
  \citenamefont {D'Andrea}, \citenamefont {{da Costa}}, \citenamefont {Davis},
  \citenamefont {Desai}, \citenamefont {Diehl}, \citenamefont {Dietrich},
  \citenamefont {Doel}, \citenamefont {Evrard}, \citenamefont {Finley},
  \citenamefont {Flaugher}, \citenamefont {Fosalba}, \citenamefont {Frieman},
  \citenamefont {Glazebrook}, \citenamefont {Goldstein}, \citenamefont {Gruen},
  \citenamefont {Gruendl}, \citenamefont {Gutierrez}, \citenamefont {Hilton},
  \citenamefont {Honscheid}, \citenamefont {Hoyle}, \citenamefont {James},
  \citenamefont {Kay}, \citenamefont {Kuehn}, \citenamefont {Kuropatkin},
  \citenamefont {Lahav}, \citenamefont {Lewis}, \citenamefont {Lidman},
  \citenamefont {Lima}, \citenamefont {Maia}, \citenamefont {Mann},
  \citenamefont {Marshall}, \citenamefont {Martini}, \citenamefont {Melchior},
  \citenamefont {Miller}, \citenamefont {Miquel}, \citenamefont {Mohr},
  \citenamefont {Nichol}, \citenamefont {Nord}, \citenamefont {Ogando},
  \citenamefont {Plazas}, \citenamefont {Reil}, \citenamefont {Sahl{\'e}n},
  \citenamefont {Sanchez}, \citenamefont {Santiago}, \citenamefont {Scarpine},
  \citenamefont {Schubnell}, \citenamefont {{Sevilla-Noarbe}}, \citenamefont
  {Smith}, \citenamefont {{Soares-Santos}}, \citenamefont {Sobreira},
  \citenamefont {Stott}, \citenamefont {Suchyta}, \citenamefont {Swanson},
  \citenamefont {Tarle}, \citenamefont {Thomas}, \citenamefont {Tucker},
  \citenamefont {Uddin}, \citenamefont {Viana}, \citenamefont {Vikram},
  \citenamefont {Walker}, \citenamefont {Zhang},\ and\ \citenamefont {{DES
  Collaboration}}}]{rykoffRedMaPPerGalaxyCluster2016}%
  \BibitemOpen
  \bibfield  {author} {\bibinfo {author} {\bibfnamefont {E.~S.}\ \bibnamefont
  {Rykoff}}, \bibinfo {author} {\bibfnamefont {E.}~\bibnamefont {Rozo}},
  \bibinfo {author} {\bibfnamefont {D.}~\bibnamefont {Hollowood}}, \emph
  {et~al.},\ }\href {https://doi.org/10.3847/0067-0049/224/1/1} {\bibfield
  {journal} {\bibinfo  {journal} {Astrophys. J. Suppl. Ser.}\ }\textbf
  {\bibinfo {volume} {224}},\ \bibinfo {pages} {1} (\bibinfo {year}
  {2016})}\BibitemShut {NoStop}%
\bibitem [{\citenamefont {Bellagamba}\ \emph {et~al.}(2018)\citenamefont
  {Bellagamba}, \citenamefont {Roncarelli}, \citenamefont {Maturi},\ and\
  \citenamefont {Moscardini}}]{bellagambaAMICOOptimizedDetection2018}%
  \BibitemOpen
  \bibfield  {author} {\bibinfo {author} {\bibfnamefont {F.}~\bibnamefont
  {Bellagamba}}, \bibinfo {author} {\bibfnamefont {M.}~\bibnamefont
  {Roncarelli}}, \bibinfo {author} {\bibfnamefont {M.}~\bibnamefont {Maturi}},\
  and\ \bibinfo {author} {\bibfnamefont {L.}~\bibnamefont {Moscardini}},\
  }\href {https://doi.org/10.1093/mnras/stx2701} {\bibfield  {journal}
  {\bibinfo  {journal} {Mon. Not. R. Astron. Soc.}\ }\textbf {\bibinfo {volume}
  {473}},\ \bibinfo {pages} {5221} (\bibinfo {year} {2018})}\BibitemShut
  {NoStop}%
\bibitem [{\citenamefont {Wen}\ \emph {et~al.}(2018)\citenamefont {Wen},
  \citenamefont {Han},\ and\ \citenamefont
  {Yang}}]{wenCatalogueClustersGalaxies2018}%
  \BibitemOpen
  \bibfield  {author} {\bibinfo {author} {\bibfnamefont {Z.~L.}\ \bibnamefont
  {Wen}}, \bibinfo {author} {\bibfnamefont {J.~L.}\ \bibnamefont {Han}},\ and\
  \bibinfo {author} {\bibfnamefont {F.}~\bibnamefont {Yang}},\ }\href
  {https://doi.org/10.1093/mnras/stx3189} {\bibfield  {journal} {\bibinfo
  {journal} {Mon. Not. R. Astron. Soc.}\ }\textbf {\bibinfo {volume} {475}},\
  \bibinfo {pages} {343} (\bibinfo {year} {2018})}\BibitemShut {NoStop}%
\bibitem [{\citenamefont {Zhang}\ \emph {et~al.}(2019)\citenamefont {Zhang},
  \citenamefont {Yanny}, \citenamefont {Palmese}, \citenamefont {Gruen},
  \citenamefont {To}, \citenamefont {Rykoff}, \citenamefont {Leung},
  \citenamefont {Collins}, \citenamefont {Hilton}, \citenamefont {Abbott},
  \citenamefont {Annis}, \citenamefont {Avila}, \citenamefont {Bertin},
  \citenamefont {Brooks}, \citenamefont {Burke}, \citenamefont {Rosell},
  \citenamefont {Kind}, \citenamefont {Carretero}, \citenamefont {Cunha},
  \citenamefont {D'Andrea}, \citenamefont {Da~Costa}, \citenamefont
  {De~Vicente}, \citenamefont {Desai}, \citenamefont {Diehl}, \citenamefont
  {Dietrich}, \citenamefont {Doel}, \citenamefont {{Drlica-Wagner}},
  \citenamefont {Eifler}, \citenamefont {Evrard}, \citenamefont {Flaugher},
  \citenamefont {Fosalba}, \citenamefont {Frieman}, \citenamefont
  {{Garc'Ia-Bellido}}, \citenamefont {Gaztanaga}, \citenamefont {Gerdes},
  \citenamefont {Gruendl}, \citenamefont {Gschwend}, \citenamefont {Gutierrez},
  \citenamefont {Hartley}, \citenamefont {Hollowood}, \citenamefont
  {Honscheid}, \citenamefont {Hoyle}, \citenamefont {James}, \citenamefont
  {Jeltema}, \citenamefont {Kuehn}, \citenamefont {Kuropatkin}, \citenamefont
  {Li}, \citenamefont {Lima}, \citenamefont {Maia}, \citenamefont {March},
  \citenamefont {Marshall}, \citenamefont {Melchior}, \citenamefont
  {Menanteau}, \citenamefont {Miller}, \citenamefont {Miquel}, \citenamefont
  {Mohr}, \citenamefont {Ogando}, \citenamefont {Plazas}, \citenamefont
  {Romer}, \citenamefont {Sanchez}, \citenamefont {Scarpine}, \citenamefont
  {Schubnell}, \citenamefont {Serrano}, \citenamefont {{Sevilla-Noarbe}},
  \citenamefont {Smith}, \citenamefont {{Soares-Santos}}, \citenamefont
  {Sobreira}, \citenamefont {Suchyta}, \citenamefont {Swanson}, \citenamefont
  {Tarle}, \citenamefont {Thomas},\ and\ \citenamefont
  {Wester}}]{zhangDarkEnergySurvey2019}%
  \BibitemOpen
  \bibfield  {author} {\bibinfo {author} {\bibfnamefont {Y.}~\bibnamefont
  {Zhang}}, \bibinfo {author} {\bibfnamefont {B.}~\bibnamefont {Yanny}},
  \bibinfo {author} {\bibfnamefont {A.}~\bibnamefont {Palmese}}, \emph
  {et~al.},\ }\href {https://doi.org/10.3847/1538-4357/ab0dfd} {\bibfield
  {journal} {\bibinfo  {journal} {Astrophys. J.}\ }\textbf {\bibinfo {volume}
  {874}},\ \bibinfo {pages} {165} (\bibinfo {year} {2019})}\BibitemShut
  {NoStop}%
\bibitem [{\citenamefont {Umetsu}(2020)}]{umetsuClustergalaxyWeakLensing2020}%
  \BibitemOpen
  \bibfield  {author} {\bibinfo {author} {\bibfnamefont {K.}~\bibnamefont
  {Umetsu}},\ }\href {https://doi.org/10.1007/s00159-020-00129-w} {\bibfield
  {journal} {\bibinfo  {journal} {Astron. Astrophys. Rev.}\ }\textbf {\bibinfo
  {volume} {28}},\ \bibinfo {pages} {7} (\bibinfo {year} {2020})}\BibitemShut
  {NoStop}%
\bibitem [{\citenamefont {Aguena}\ \emph
  {et~al.}(2021{\natexlab{a}})\citenamefont {Aguena}, \citenamefont {Benoist},
  \citenamefont {{da Costa}}, \citenamefont {Ogando}, \citenamefont {Gschwend},
  \citenamefont {{Sampaio-Santos}}, \citenamefont {Lima}, \citenamefont {Maia},
  \citenamefont {Allam}, \citenamefont {Avila}, \citenamefont {Bacon},
  \citenamefont {Bertin}, \citenamefont {Bhargava}, \citenamefont {Brooks},
  \citenamefont {Carnero~Rosell}, \citenamefont {Carrasco~Kind}, \citenamefont
  {Carretero}, \citenamefont {Costanzi}, \citenamefont {De~Vicente},
  \citenamefont {Desai}, \citenamefont {Diehl}, \citenamefont {Doel},
  \citenamefont {Everett}, \citenamefont {Evrard}, \citenamefont {Ferrero},
  \citenamefont {Fert{\'e}}, \citenamefont {Flaugher}, \citenamefont {Fosalba},
  \citenamefont {Frieman}, \citenamefont {{Garc{\'i}a-Bellido}}, \citenamefont
  {Giles}, \citenamefont {Gruendl}, \citenamefont {Gutierrez}, \citenamefont
  {Hinton}, \citenamefont {Hollowood}, \citenamefont {Honscheid}, \citenamefont
  {James}, \citenamefont {Jeltema}, \citenamefont {Kuehn}, \citenamefont
  {Kuropatkin}, \citenamefont {Lahav}, \citenamefont {Melchior}, \citenamefont
  {Miquel}, \citenamefont {Morgan}, \citenamefont {Palmese}, \citenamefont
  {{Paz-Chinch{\'o}n}}, \citenamefont {Plazas}, \citenamefont {Romer},
  \citenamefont {Sanchez}, \citenamefont {Santiago}, \citenamefont {Schubnell},
  \citenamefont {Serrano}, \citenamefont {{Sevilla-Noarbe}}, \citenamefont
  {Smith}, \citenamefont {{Soares-Santos}}, \citenamefont {Suchyta},
  \citenamefont {Tarle}, \citenamefont {To}, \citenamefont {Tucker},\ and\
  \citenamefont {Wilkinson}}]{aguenaWaZPGalaxyCluster2021}%
  \BibitemOpen
  \bibfield  {author} {\bibinfo {author} {\bibfnamefont {M.}~\bibnamefont
  {Aguena}}, \bibinfo {author} {\bibfnamefont {C.}~\bibnamefont {Benoist}},
  \bibinfo {author} {\bibfnamefont {L.~N.}\ \bibnamefont {{da Costa}}}, \emph
  {et~al.},\ }\href {https://doi.org/10.1093/mnras/stab264} {\bibfield
  {journal} {\bibinfo  {journal} {Mon. Not. R. Astron. Soc.}\ }\textbf
  {\bibinfo {volume} {502}},\ \bibinfo {pages} {4435} (\bibinfo {year}
  {2021}{\natexlab{a}})}\BibitemShut {NoStop}%
\bibitem [{\citenamefont {Kluge}\ \emph {et~al.}(2021)\citenamefont {Kluge},
  \citenamefont {Bender}, \citenamefont {Riffeser}, \citenamefont {Goessl},
  \citenamefont {Hopp}, \citenamefont {Schmidt},\ and\ \citenamefont
  {Ries}}]{klugePhotometricDissectionIntracluster2021}%
  \BibitemOpen
  \bibfield  {author} {\bibinfo {author} {\bibfnamefont {M.}~\bibnamefont
  {Kluge}}, \bibinfo {author} {\bibfnamefont {R.}~\bibnamefont {Bender}},
  \bibinfo {author} {\bibfnamefont {A.}~\bibnamefont {Riffeser}}, \emph
  {et~al.},\ }\href {https://doi.org/10.3847/1538-4365/abcda6} {\bibfield
  {journal} {\bibinfo  {journal} {Astrophys. J. Suppl. Ser.}\ }\textbf
  {\bibinfo {volume} {252}},\ \bibinfo {pages} {27} (\bibinfo {year}
  {2021})}\BibitemShut {NoStop}%
\bibitem [{\citenamefont {Li}\ \emph {et~al.}(2022)\citenamefont {Li},
  \citenamefont {Huang}, \citenamefont {Leauthaud}, \citenamefont {Moustakas},
  \citenamefont {Danieli}, \citenamefont {Greene}, \citenamefont {Abraham},
  \citenamefont {Ardila}, \citenamefont {{Kado-Fong}}, \citenamefont
  {Lokhorst}, \citenamefont {Lupton},\ and\ \citenamefont
  {Price}}]{liReachingEdgeProbing2022}%
  \BibitemOpen
  \bibfield  {author} {\bibinfo {author} {\bibfnamefont {J.}~\bibnamefont
  {Li}}, \bibinfo {author} {\bibfnamefont {S.}~\bibnamefont {Huang}}, \bibinfo
  {author} {\bibfnamefont {A.}~\bibnamefont {Leauthaud}}, \emph {et~al.},\
  }\href {https://doi.org/10.1093/mnras/stac2121} {\bibfield  {journal}
  {\bibinfo  {journal} {Mon. Not. R. Astron. Soc.}\ }\textbf {\bibinfo {volume}
  {515}},\ \bibinfo {pages} {5335} (\bibinfo {year} {2022})}\BibitemShut
  {NoStop}%
\bibitem [{\citenamefont {Montes}(2022)}]{montesFaintLightGroups2022}%
  \BibitemOpen
  \bibfield  {author} {\bibinfo {author} {\bibfnamefont {M.}~\bibnamefont
  {Montes}},\ }\href {https://doi.org/10.1038/s41550-022-01616-z} {\bibfield
  {journal} {\bibinfo  {journal} {Nat. Astron.}\ }\textbf {\bibinfo {volume}
  {6}},\ \bibinfo {pages} {308} (\bibinfo {year} {2022})}\BibitemShut {NoStop}%
\bibitem [{\citenamefont {Zhou}\ \emph {et~al.}(2023)\citenamefont {Zhou},
  \citenamefont {Tong}, \citenamefont {Troxel}, \citenamefont {Blazek},
  \citenamefont {Lin}, \citenamefont {Bacon}, \citenamefont {Bleem},
  \citenamefont {Chang}, \citenamefont {Costanzi}, \citenamefont {DeRose},
  \citenamefont {Dietrich}, \citenamefont {{Drlica-Wagner}}, \citenamefont
  {Gruen}, \citenamefont {Gruendl}, \citenamefont {Hoyle}, \citenamefont
  {Jarvis}, \citenamefont {MacCrann}, \citenamefont {Mawdsley}, \citenamefont
  {McClintock}, \citenamefont {Melchior}, \citenamefont {Prat}, \citenamefont
  {Pujol}, \citenamefont {Rozo}, \citenamefont {Rykoff}, \citenamefont
  {Samuroff}, \citenamefont {Sheldon}, \citenamefont {Shin}, \citenamefont
  {Rosell}, \citenamefont {Yanny}, \citenamefont {S{\'a}nchez}, \citenamefont
  {Tucker}, \citenamefont {{Sevilla-Noarbe}}, \citenamefont {Zuntz},
  \citenamefont {Varga}, \citenamefont {Zhang}, \citenamefont {Alves},
  \citenamefont {Amon}, \citenamefont {Bertin}, \citenamefont {Brooks},
  \citenamefont {Burke}, \citenamefont {Kind}, \citenamefont {{da Costa}},
  \citenamefont {Davis}, \citenamefont {De~Vicente}, \citenamefont {Desai},
  \citenamefont {Diehl}, \citenamefont {Doel}, \citenamefont {Everett},
  \citenamefont {Ferrero}, \citenamefont {Flaugher}, \citenamefont {Frieman},
  \citenamefont {Gerdes}, \citenamefont {Gutierrez}, \citenamefont {Hinton},
  \citenamefont {Hollowood}, \citenamefont {Honscheid}, \citenamefont {James},
  \citenamefont {Jeltema}, \citenamefont {Kuehn}, \citenamefont {Lahav},
  \citenamefont {Lima}, \citenamefont {Marshall}, \citenamefont
  {{Mena-Fern{\'a}ndez}}, \citenamefont {Menanteau}, \citenamefont {Miquel},
  \citenamefont {Palmese}, \citenamefont {{Paz-Chinch{\'o}n}}, \citenamefont
  {Pieres}, \citenamefont {Malag{\'o}n}, \citenamefont {Porredon},
  \citenamefont {Raveri}, \citenamefont {Romer}, \citenamefont {Sanchez},
  \citenamefont {Smith}, \citenamefont {{Soares-Santos}}, \citenamefont
  {Suchyta}, \citenamefont {Swanson}, \citenamefont {Tarle}, \citenamefont
  {To}, \citenamefont {Weaverdyck}, \citenamefont {Weller},\ and\ \citenamefont
  {Wiseman}}]{zhouIntrinsicAlignmentRed2023}%
  \BibitemOpen
  \bibfield  {author} {\bibinfo {author} {\bibfnamefont {C.}~\bibnamefont
  {Zhou}}, \bibinfo {author} {\bibfnamefont {A.}~\bibnamefont {Tong}}, \bibinfo
  {author} {\bibfnamefont {M.~A.}\ \bibnamefont {Troxel}}, \emph {et~al.},\
  }\href {https://doi.org/10.1093/mnras/stad2712} {\bibfield  {journal}
  {\bibinfo  {journal} {Mon. Not. R. Astron. Soc.}\ }\textbf {\bibinfo {volume}
  {526}},\ \bibinfo {pages} {323} (\bibinfo {year} {2023})}\BibitemShut
  {NoStop}%
\bibitem [{\citenamefont {Fu}\ \emph {et~al.}(2024)\citenamefont {Fu},
  \citenamefont {Dell'Antonio}, \citenamefont {Escalante}, \citenamefont
  {Nelson}, \citenamefont {Englert}, \citenamefont {Helhoski}, \citenamefont
  {Shinde}, \citenamefont {Brockland}, \citenamefont {LaDuca}, \citenamefont
  {Larkin}, \citenamefont {Paris}, \citenamefont {Weiner}, \citenamefont
  {Black}, \citenamefont {Chary}, \citenamefont {Clowe}, \citenamefont
  {Cooper}, \citenamefont {Donahue}, \citenamefont {Evrard}, \citenamefont
  {Lacy}, \citenamefont {Lauer}, \citenamefont {Liu}, \citenamefont {McCleary},
  \citenamefont {Meneghetti}, \citenamefont {Miyatake}, \citenamefont {Montes},
  \citenamefont {Natarajan}, \citenamefont {Ntampaka}, \citenamefont
  {Pierpaoli}, \citenamefont {Postman}, \citenamefont {Sohn}, \citenamefont
  {Turner}, \citenamefont {Umetsu}, \citenamefont {Utsumi},\ and\ \citenamefont
  {Wilson}}]{fuLoVoCCSIIWeak2024}%
  \BibitemOpen
  \bibfield  {author} {\bibinfo {author} {\bibfnamefont {S.}~\bibnamefont
  {Fu}}, \bibinfo {author} {\bibfnamefont {I.}~\bibnamefont {Dell'Antonio}},
  \bibinfo {author} {\bibfnamefont {Z.}~\bibnamefont {Escalante}}, \emph
  {et~al.},\ }\href {https://doi.org/10.3847/1538-4357/ad67c6} {\bibfield
  {journal} {\bibinfo  {journal} {Astrophys. J.}\ }\textbf {\bibinfo {volume}
  {974}},\ \bibinfo {pages} {69} (\bibinfo {year} {2024})}\BibitemShut
  {NoStop}%
\bibitem [{\citenamefont {Zhang}\ \emph
  {et~al.}(2024{\natexlab{a}})\citenamefont {Zhang}, \citenamefont
  {{Golden-Marx}}, \citenamefont {Ogando}, \citenamefont {Yanny}, \citenamefont
  {Rykoff}, \citenamefont {Allam}, \citenamefont {Aguena}, \citenamefont
  {Bacon}, \citenamefont {Bocquet}, \citenamefont {Brooks}, \citenamefont
  {Carnero~Rosell}, \citenamefont {Carretero}, \citenamefont {Cheng},
  \citenamefont {Conselice}, \citenamefont {Costanzi}, \citenamefont {{da
  Costa}}, \citenamefont {Pereira}, \citenamefont {Davis}, \citenamefont
  {Desai}, \citenamefont {Diehl}, \citenamefont {Doel}, \citenamefont
  {Ferrero}, \citenamefont {Flaugher}, \citenamefont {Frieman}, \citenamefont
  {Gruen}, \citenamefont {Gruendl}, \citenamefont {Hinton}, \citenamefont
  {Hollowood}, \citenamefont {Honscheid}, \citenamefont {James}, \citenamefont
  {Jeltema}, \citenamefont {Kuehn}, \citenamefont {Kuropatkin}, \citenamefont
  {Lahav}, \citenamefont {Lee}, \citenamefont {Lima}, \citenamefont
  {{Mena-Fern{\'a}ndez}}, \citenamefont {Miquel}, \citenamefont {Palmese},
  \citenamefont {Pieres}, \citenamefont {Plazas~Malag{\'o}n}, \citenamefont
  {Romer}, \citenamefont {Sanchez}, \citenamefont {Smith}, \citenamefont
  {Suchyta}, \citenamefont {Tarle}, \citenamefont {To}, \citenamefont {Tucker},
  \citenamefont {Weaverdyck},\ and\ \citenamefont {{DES
  Collaboration}}}]{zhangDarkEnergySurvey2024}%
  \BibitemOpen
  \bibfield  {author} {\bibinfo {author} {\bibfnamefont {Y.}~\bibnamefont
  {Zhang}}, \bibinfo {author} {\bibfnamefont {J.~B.}\ \bibnamefont
  {{Golden-Marx}}}, \bibinfo {author} {\bibfnamefont {R.~L.~C.}\ \bibnamefont
  {Ogando}}, \emph {et~al.},\ }\href {https://doi.org/10.1093/mnras/stae1165}
  {\bibfield  {journal} {\bibinfo  {journal} {Mon. Not. R. Astron. Soc.}\
  }\textbf {\bibinfo {volume} {531}},\ \bibinfo {pages} {510} (\bibinfo {year}
  {2024}{\natexlab{a}})}\BibitemShut {NoStop}%
\bibitem [{\citenamefont {Benoist}\ \emph {et~al.}(2025)\citenamefont
  {Benoist}, \citenamefont {Aguena}, \citenamefont {{da Costa}}, \citenamefont
  {Gschwend}, \citenamefont {Allam}, \citenamefont {Alves}, \citenamefont
  {{Andrade-Oliveira}}, \citenamefont {Bacon}, \citenamefont {Bleem},
  \citenamefont {Brooks}, \citenamefont {Carnero~Rosell}, \citenamefont
  {Carretero}, \citenamefont {Castander}, \citenamefont {Costanzi},
  \citenamefont {De~Vicente}, \citenamefont {Desai}, \citenamefont {Dodelson},
  \citenamefont {Doel}, \citenamefont {Everett}, \citenamefont {Flaugher},
  \citenamefont {Frieman}, \citenamefont {{Garcia-Bellido}}, \citenamefont
  {Giannini}, \citenamefont {Giles}, \citenamefont {Gruendl}, \citenamefont
  {Gutierrez}, \citenamefont {Hinton}, \citenamefont {Hollowood}, \citenamefont
  {Honscheid}, \citenamefont {James}, \citenamefont {Kuehn}, \citenamefont
  {Lee}, \citenamefont {Marshall}, \citenamefont {{Mena-Fern{\'a}ndez}},
  \citenamefont {Miquel}, \citenamefont {Plazas~Malag{\'o}n}, \citenamefont
  {Romer}, \citenamefont {Sanchez}, \citenamefont {Santiago}, \citenamefont
  {Sevilla}, \citenamefont {Smith}, \citenamefont {Suchyta}, \citenamefont
  {Tarle}, \citenamefont {Weaverdyck}, \citenamefont {Weller},\ and\
  \citenamefont {{da Silva Pereira}}}]{benoistFinalWaZPGalaxy2025}%
  \BibitemOpen
  \bibfield  {author} {\bibinfo {author} {\bibfnamefont {C.}~\bibnamefont
  {Benoist}}, \bibinfo {author} {\bibfnamefont {M.}~\bibnamefont {Aguena}},
  \bibinfo {author} {\bibfnamefont {L.}~\bibnamefont {{da Costa}}}, \emph
  {et~al.},\ }\href {https://doi.org/10.48550/arXiv.2507.05360} {\enquote
  {\bibinfo {title} {The final {{WaZP}} galaxy cluster catalog of the {{Dark
  Energy Survey}} and comparison with {{SZE}} data},}\ } (\bibinfo {year}
  {2025})\BibitemShut {NoStop}%
\bibitem [{\citenamefont {Chen}\ \emph {et~al.}(2025)\citenamefont {Chen},
  \citenamefont {Chiu}, \citenamefont {Oguri}, \citenamefont {Lin},
  \citenamefont {Miyatake}, \citenamefont {Miyazaki}, \citenamefont {More},
  \citenamefont {Hamana}, \citenamefont {Rau}, \citenamefont {Sunayama},
  \citenamefont {Sugiyama},\ and\ \citenamefont
  {Takada}}]{chenWeakLensingShearSelectedGalaxy2025}%
  \BibitemOpen
  \bibfield  {author} {\bibinfo {author} {\bibfnamefont {K.-F.}\ \bibnamefont
  {Chen}}, \bibinfo {author} {\bibfnamefont {I.-N.}\ \bibnamefont {Chiu}},
  \bibinfo {author} {\bibfnamefont {M.}~\bibnamefont {Oguri}}, \emph {et~al.},\
  }\href {https://doi.org/10.33232/001c.128019} {\bibfield  {journal} {\bibinfo
   {journal} {Open J. Astrophys.}\ }\textbf {\bibinfo {volume} {8}},\ \bibinfo
  {pages} {2} (\bibinfo {year} {2025})}\BibitemShut {NoStop}%
\bibitem [{\citenamefont {Englert}\ \emph {et~al.}(2025)\citenamefont
  {Englert}, \citenamefont {Dell'Antonio},\ and\ \citenamefont
  {Montes}}]{englertIntraclusterLightAbell2025}%
  \BibitemOpen
  \bibfield  {author} {\bibinfo {author} {\bibfnamefont {A.~M.}\ \bibnamefont
  {Englert}}, \bibinfo {author} {\bibfnamefont {I.}~\bibnamefont
  {Dell'Antonio}},\ and\ \bibinfo {author} {\bibfnamefont {M.}~\bibnamefont
  {Montes}},\ }\href {https://doi.org/10.3847/2041-8213/ade8f1} {\bibfield
  {journal} {\bibinfo  {journal} {Astrophys. J. Lett.}\ }\textbf {\bibinfo
  {volume} {989}},\ \bibinfo {pages} {L2} (\bibinfo {year} {2025})}\BibitemShut
  {NoStop}%
\bibitem [{\citenamefont {Sunyaev}\ and\ \citenamefont
  {Zeldovich}(1970)}]{sunyaevSmallscaleFluctuationsRelic1970}%
  \BibitemOpen
  \bibfield  {author} {\bibinfo {author} {\bibfnamefont {R.~A.}\ \bibnamefont
  {Sunyaev}}\ and\ \bibinfo {author} {\bibfnamefont {{\relax Ya}.~B.}\
  \bibnamefont {Zeldovich}},\ }\href {https://doi.org/10.1007/BF00653471}
  {\bibfield  {journal} {\bibinfo  {journal} {Astrophys. Space Sci.}\ }\textbf
  {\bibinfo {volume} {7}},\ \bibinfo {pages} {3} (\bibinfo {year}
  {1970})}\BibitemShut {NoStop}%
\bibitem [{\citenamefont {Sunyaev}\ and\ \citenamefont
  {Zeldovich}(1972)}]{sunyaevObservationsRelicRadiation1972}%
  \BibitemOpen
  \bibfield  {author} {\bibinfo {author} {\bibfnamefont {R.~A.}\ \bibnamefont
  {Sunyaev}}\ and\ \bibinfo {author} {\bibfnamefont {{\relax Ya}.~B.}\
  \bibnamefont {Zeldovich}},\ }\href@noop {} {\bibfield  {journal} {\bibinfo
  {journal} {Comments Astrophys. Space Phys.}\ }\textbf {\bibinfo {volume}
  {4}},\ \bibinfo {pages} {173} (\bibinfo {year} {1972})}\BibitemShut {NoStop}%
\bibitem [{\citenamefont {Zaldarriaga}\ and\ \citenamefont
  {Seljak}(1999)}]{zaldarriagaReconstructingProjectedMatter1999}%
  \BibitemOpen
  \bibfield  {author} {\bibinfo {author} {\bibfnamefont {M.}~\bibnamefont
  {Zaldarriaga}}\ and\ \bibinfo {author} {\bibfnamefont {U.}~\bibnamefont
  {Seljak}},\ }\href {https://doi.org/10.1103/PhysRevD.59.123507} {\bibfield
  {journal} {\bibinfo  {journal} {Phys. Rev. D}\ }\textbf {\bibinfo {volume}
  {59}},\ \bibinfo {pages} {123507} (\bibinfo {year} {1999})}\BibitemShut
  {NoStop}%
\bibitem [{\citenamefont {Carlstrom}\ \emph {et~al.}(2002)\citenamefont
  {Carlstrom}, \citenamefont {Holder},\ and\ \citenamefont
  {Reese}}]{carlstromCosmologySunyaevZeldovichEffect2002}%
  \BibitemOpen
  \bibfield  {author} {\bibinfo {author} {\bibfnamefont {J.~E.}\ \bibnamefont
  {Carlstrom}}, \bibinfo {author} {\bibfnamefont {G.~P.}\ \bibnamefont
  {Holder}},\ and\ \bibinfo {author} {\bibfnamefont {E.~D.}\ \bibnamefont
  {Reese}},\ }\href {https://doi.org/10.1146/annurev.astro.40.060401.093803}
  {\bibfield  {journal} {\bibinfo  {journal} {Annu. Rev. Astron. Astrophys.}\
  }\textbf {\bibinfo {volume} {40}},\ \bibinfo {pages} {643} (\bibinfo {year}
  {2002})}\BibitemShut {NoStop}%
\bibitem [{\citenamefont {Lewis}\ and\ \citenamefont
  {Challinor}(2006)}]{lewisWeakGravitationalLensing2006}%
  \BibitemOpen
  \bibfield  {author} {\bibinfo {author} {\bibfnamefont {A.}~\bibnamefont
  {Lewis}}\ and\ \bibinfo {author} {\bibfnamefont {A.}~\bibnamefont
  {Challinor}},\ }\href {https://doi.org/10.1016/j.physrep.2006.03.002}
  {\bibfield  {journal} {\bibinfo  {journal} {Phys. Rep.}\ }\textbf {\bibinfo
  {volume} {429}},\ \bibinfo {pages} {1} (\bibinfo {year} {2006})}\BibitemShut
  {NoStop}%
\bibitem [{\citenamefont {Baxter}\ \emph {et~al.}(2015)\citenamefont {Baxter},
  \citenamefont {Keisler}, \citenamefont {Dodelson}, \citenamefont {Aird},
  \citenamefont {Allen}, \citenamefont {Ashby}, \citenamefont {Bautz},
  \citenamefont {Bayliss}, \citenamefont {Benson}, \citenamefont {Bleem},
  \citenamefont {Bocquet}, \citenamefont {Brodwin}, \citenamefont {Carlstrom},
  \citenamefont {Chang}, \citenamefont {Chiu}, \citenamefont {Cho},
  \citenamefont {Clocchiatti}, \citenamefont {Crawford}, \citenamefont
  {Crites}, \citenamefont {Desai}, \citenamefont {Dietrich}, \citenamefont {{de
  Haan}}, \citenamefont {Dobbs}, \citenamefont {Foley}, \citenamefont {Forman},
  \citenamefont {George}, \citenamefont {Gladders}, \citenamefont {Gonzalez},
  \citenamefont {Halverson}, \citenamefont {Harrington}, \citenamefont
  {Hennig}, \citenamefont {Hoekstra}, \citenamefont {Holder}, \citenamefont
  {Holzapfel}, \citenamefont {Hou}, \citenamefont {Hrubes}, \citenamefont
  {Jones}, \citenamefont {Knox}, \citenamefont {Lee}, \citenamefont {Leitch},
  \citenamefont {Liu}, \citenamefont {Lueker}, \citenamefont {{Luong-Van}},
  \citenamefont {Mantz}, \citenamefont {Marrone}, \citenamefont {McDonald},
  \citenamefont {McMahon}, \citenamefont {Meyer}, \citenamefont {Millea},
  \citenamefont {Mocanu}, \citenamefont {Murray}, \citenamefont {Padin},
  \citenamefont {Pryke}, \citenamefont {Reichardt}, \citenamefont {Rest},
  \citenamefont {Ruhl}, \citenamefont {Saliwanchik}, \citenamefont {Saro},
  \citenamefont {Sayre}, \citenamefont {Schaffer}, \citenamefont {Shirokoff},
  \citenamefont {Song}, \citenamefont {Spieler}, \citenamefont {Stalder},
  \citenamefont {Stanford}, \citenamefont {Staniszewski}, \citenamefont
  {Stark}, \citenamefont {Story}, \citenamefont {{van Engelen}}, \citenamefont
  {Vanderlinde}, \citenamefont {Vieira}, \citenamefont {Vikhlinin},
  \citenamefont {Williamson}, \citenamefont {Zahn},\ and\ \citenamefont
  {Zenteno}}]{baxterMeasurementGravitationalLensing2015}%
  \BibitemOpen
  \bibfield  {author} {\bibinfo {author} {\bibfnamefont {E.~J.}\ \bibnamefont
  {Baxter}}, \bibinfo {author} {\bibfnamefont {R.}~\bibnamefont {Keisler}},
  \bibinfo {author} {\bibfnamefont {S.}~\bibnamefont {Dodelson}}, \emph
  {et~al.},\ }\href {https://doi.org/10.1088/0004-637X/806/2/247} {\bibfield
  {journal} {\bibinfo  {journal} {Astrophys. J.}\ }\textbf {\bibinfo {volume}
  {806}},\ \bibinfo {pages} {247} (\bibinfo {year} {2015})}\BibitemShut
  {NoStop}%
\bibitem [{\citenamefont {Bleem}\ \emph {et~al.}(2015)\citenamefont {Bleem},
  \citenamefont {Stalder}, \citenamefont {{de Haan}}, \citenamefont {Aird},
  \citenamefont {Allen}, \citenamefont {Applegate}, \citenamefont {Ashby},
  \citenamefont {Bautz}, \citenamefont {Bayliss}, \citenamefont {Benson},
  \citenamefont {Bocquet}, \citenamefont {Brodwin}, \citenamefont {Carlstrom},
  \citenamefont {Chang}, \citenamefont {Chiu}, \citenamefont {Cho},
  \citenamefont {Clocchiatti}, \citenamefont {Crawford}, \citenamefont
  {Crites}, \citenamefont {Desai}, \citenamefont {Dietrich}, \citenamefont
  {Dobbs}, \citenamefont {Foley}, \citenamefont {Forman}, \citenamefont
  {George}, \citenamefont {Gladders}, \citenamefont {Gonzalez}, \citenamefont
  {Halverson}, \citenamefont {Hennig}, \citenamefont {Hoekstra}, \citenamefont
  {Holder}, \citenamefont {Holzapfel}, \citenamefont {Hrubes}, \citenamefont
  {Jones}, \citenamefont {Keisler}, \citenamefont {Knox}, \citenamefont {Lee},
  \citenamefont {Leitch}, \citenamefont {Liu}, \citenamefont {Lueker},
  \citenamefont {{Luong-Van}}, \citenamefont {Mantz}, \citenamefont {Marrone},
  \citenamefont {McDonald}, \citenamefont {McMahon}, \citenamefont {Meyer},
  \citenamefont {Mocanu}, \citenamefont {Mohr}, \citenamefont {Murray},
  \citenamefont {Padin}, \citenamefont {Pryke}, \citenamefont {Reichardt},
  \citenamefont {Rest}, \citenamefont {Ruel}, \citenamefont {Ruhl},
  \citenamefont {Saliwanchik}, \citenamefont {Saro}, \citenamefont {Sayre},
  \citenamefont {Schaffer}, \citenamefont {Schrabback}, \citenamefont
  {Shirokoff}, \citenamefont {Song}, \citenamefont {Spieler}, \citenamefont
  {Stanford}, \citenamefont {Staniszewski}, \citenamefont {Stark},
  \citenamefont {Story}, \citenamefont {Stubbs}, \citenamefont {Vanderlinde},
  \citenamefont {Vieira}, \citenamefont {Vikhlinin}, \citenamefont
  {Williamson}, \citenamefont {Zahn},\ and\ \citenamefont
  {Zenteno}}]{bleemGalaxyClustersDiscovered2015}%
  \BibitemOpen
  \bibfield  {author} {\bibinfo {author} {\bibfnamefont {L.~E.}\ \bibnamefont
  {Bleem}}, \bibinfo {author} {\bibfnamefont {B.}~\bibnamefont {Stalder}},
  \bibinfo {author} {\bibfnamefont {T.}~\bibnamefont {{de Haan}}}, \emph
  {et~al.},\ }\href {https://doi.org/10.1088/0067-0049/216/2/27} {\bibfield
  {journal} {\bibinfo  {journal} {Astrophys. J. Suppl. Ser.}\ }\textbf
  {\bibinfo {volume} {216}},\ \bibinfo {pages} {27} (\bibinfo {year}
  {2015})}\BibitemShut {NoStop}%
\bibitem [{\citenamefont {Madhavacheril}\ \emph {et~al.}(2015)\citenamefont
  {Madhavacheril}, \citenamefont {Sehgal}, \citenamefont {Allison},
  \citenamefont {Battaglia}, \citenamefont {Bond}, \citenamefont {Calabrese},
  \citenamefont {Caligiuri}, \citenamefont {Coughlin}, \citenamefont
  {Crichton}, \citenamefont {Datta}, \citenamefont {Devlin}, \citenamefont
  {Dunkley}, \citenamefont {D{\"u}nner}, \citenamefont {Fogarty}, \citenamefont
  {Grace}, \citenamefont {Hajian}, \citenamefont {Hasselfield}, \citenamefont
  {Hill}, \citenamefont {Hilton}, \citenamefont {Hincks}, \citenamefont
  {Hlozek}, \citenamefont {Hughes}, \citenamefont {Kosowsky}, \citenamefont
  {Louis}, \citenamefont {Lungu}, \citenamefont {McMahon}, \citenamefont
  {Moodley}, \citenamefont {Munson}, \citenamefont {Naess}, \citenamefont
  {Nati}, \citenamefont {Newburgh}, \citenamefont {Niemack}, \citenamefont
  {Page}, \citenamefont {Partridge}, \citenamefont {Schmitt}, \citenamefont
  {Sherwin}, \citenamefont {Sievers}, \citenamefont {Spergel}, \citenamefont
  {Staggs}, \citenamefont {Thornton}, \citenamefont {Van~Engelen},
  \citenamefont {Ward}, \citenamefont {Wollack},\ and\ \citenamefont {{Atacama
  Cosmology Telescope
  Collaboration}}}]{madhavacherilEvidenceLensingCosmic2015}%
  \BibitemOpen
  \bibfield  {author} {\bibinfo {author} {\bibfnamefont {M.}~\bibnamefont
  {Madhavacheril}}, \bibinfo {author} {\bibfnamefont {N.}~\bibnamefont
  {Sehgal}}, \bibinfo {author} {\bibfnamefont {R.}~\bibnamefont {Allison}},
  \emph {et~al.},\ }\href {https://doi.org/10.1103/PhysRevLett.114.151302}
  {\bibfield  {journal} {\bibinfo  {journal} {Phys. Rev. Lett.}\ }\textbf
  {\bibinfo {volume} {114}},\ \bibinfo {pages} {151302} (\bibinfo {year}
  {2015})}\BibitemShut {NoStop}%
\bibitem [{\citenamefont {{Planck Collaboration}}\ \emph
  {et~al.}(2016)\citenamefont {{Planck Collaboration}}, \citenamefont {Ade},
  \citenamefont {Aghanim}, \citenamefont {Arnaud}, \citenamefont {Ashdown},
  \citenamefont {Aumont}, \citenamefont {Baccigalupi}, \citenamefont {Banday},
  \citenamefont {Barreiro}, \citenamefont {Barrena}, \citenamefont {Bartlett},
  \citenamefont {Bartolo}, \citenamefont {Battaner}, \citenamefont {Battye},
  \citenamefont {Benabed}, \citenamefont {Beno{\^i}t}, \citenamefont
  {{Benoit-L{\'e}vy}}, \citenamefont {Bernard}, \citenamefont {Bersanelli},
  \citenamefont {Bielewicz}, \citenamefont {Bikmaev}, \citenamefont
  {B{\"o}hringer}, \citenamefont {Bonaldi}, \citenamefont {Bonavera},
  \citenamefont {Bond}, \citenamefont {Borrill}, \citenamefont {Bouchet},
  \citenamefont {Bucher}, \citenamefont {Burenin}, \citenamefont {Burigana},
  \citenamefont {Butler}, \citenamefont {Calabrese}, \citenamefont {Cardoso},
  \citenamefont {Carvalho}, \citenamefont {Catalano}, \citenamefont
  {Challinor}, \citenamefont {Chamballu}, \citenamefont {Chary}, \citenamefont
  {Chiang}, \citenamefont {Chon}, \citenamefont {Christensen}, \citenamefont
  {Clements}, \citenamefont {Colombi}, \citenamefont {Colombo}, \citenamefont
  {Combet}, \citenamefont {Comis}, \citenamefont {Couchot}, \citenamefont
  {Coulais}, \citenamefont {Crill}, \citenamefont {Curto}, \citenamefont
  {Cuttaia}, \citenamefont {Dahle}, \citenamefont {Danese}, \citenamefont
  {Davies}, \citenamefont {Davis}, \citenamefont {{de Bernardis}},
  \citenamefont {{de Rosa}}, \citenamefont {{de Zotti}}, \citenamefont
  {Delabrouille}, \citenamefont {D{\'e}sert}, \citenamefont {Dickinson},
  \citenamefont {Diego}, \citenamefont {Dolag}, \citenamefont {Dole},
  \citenamefont {Donzelli}, \citenamefont {Dor{\'e}}, \citenamefont {Douspis},
  \citenamefont {Ducout}, \citenamefont {Dupac}, \citenamefont {Efstathiou},
  \citenamefont {Eisenhardt}, \citenamefont {Elsner}, \citenamefont
  {En{\ss}lin}, \citenamefont {Eriksen}, \citenamefont {Falgarone},
  \citenamefont {Fergusson}, \citenamefont {Feroz}, \citenamefont {Ferragamo},
  \citenamefont {Finelli}, \citenamefont {Forni}, \citenamefont {Frailis},
  \citenamefont {Fraisse}, \citenamefont {Franceschi}, \citenamefont {Frejsel},
  \citenamefont {Galeotta}, \citenamefont {Galli}, \citenamefont {Ganga},
  \citenamefont {{G{\'e}nova-Santos}}, \citenamefont {Giard}, \citenamefont
  {{Giraud-H{\'e}raud}}, \citenamefont {Gjerl{\o}w}, \citenamefont
  {{Gonz{\'a}lez-Nuevo}}, \citenamefont {G{\'o}rski}, \citenamefont {Grainge},
  \citenamefont {Gratton}, \citenamefont {Gregorio}, \citenamefont {Gruppuso},
  \citenamefont {Gudmundsson}, \citenamefont {Hansen}, \citenamefont {Hanson},
  \citenamefont {Harrison}, \citenamefont {Hempel}, \citenamefont
  {{Henrot-Versill{\'e}}}, \citenamefont {{Hern{\'a}ndez-Monteagudo}},
  \citenamefont {Herranz}, \citenamefont {Hildebrandt}, \citenamefont {Hivon},
  \citenamefont {Hobson}, \citenamefont {Holmes}, \citenamefont {Hornstrup},
  \citenamefont {Hovest}, \citenamefont {Huffenberger}, \citenamefont {Hurier},
  \citenamefont {Jaffe}, \citenamefont {Jaffe}, \citenamefont {Jin},
  \citenamefont {Jones}, \citenamefont {Juvela}, \citenamefont {Keih{\"a}nen},
  \citenamefont {Keskitalo}, \citenamefont {Khamitov}, \citenamefont {Kisner},
  \citenamefont {Kneissl}, \citenamefont {Knoche}, \citenamefont {Kunz},
  \citenamefont {{Kurki-Suonio}}, \citenamefont {Lagache}, \citenamefont
  {Lamarre}, \citenamefont {Lasenby}, \citenamefont {Lattanzi}, \citenamefont
  {Lawrence}, \citenamefont {Leonardi}, \citenamefont {Lesgourgues},
  \citenamefont {Levrier}, \citenamefont {Liguori}, \citenamefont {Lilje},
  \citenamefont {{Linden-V{\o}rnle}}, \citenamefont {{L{\'o}pez-Caniego}},
  \citenamefont {Lubin}, \citenamefont {{Mac{\'i}as-P{\'e}rez}}, \citenamefont
  {Maggio}, \citenamefont {Maino}, \citenamefont {Mak}, \citenamefont
  {Mandolesi}, \citenamefont {Mangilli}, \citenamefont {Martin}, \citenamefont
  {{Mart{\'i}nez-Gonz{\'a}lez}}, \citenamefont {Masi}, \citenamefont
  {Matarrese}, \citenamefont {Mazzotta}, \citenamefont {McGehee}, \citenamefont
  {Mei}, \citenamefont {Melchiorri}, \citenamefont {Melin}, \citenamefont
  {Mendes}, \citenamefont {Mennella}, \citenamefont {Migliaccio}, \citenamefont
  {Mitra}, \citenamefont {{Miville-Desch{\^e}nes}}, \citenamefont {Moneti},
  \citenamefont {Montier}, \citenamefont {Morgante}, \citenamefont {Mortlock},
  \citenamefont {Moss}, \citenamefont {Munshi}, \citenamefont {Murphy},
  \citenamefont {Naselsky}, \citenamefont {Nastasi}, \citenamefont {Nati},
  \citenamefont {Natoli}, \citenamefont {Netterfield}, \citenamefont
  {{N{\o}rgaard-Nielsen}}, \citenamefont {Noviello}, \citenamefont {Novikov},
  \citenamefont {Novikov}, \citenamefont {Olamaie}, \citenamefont {Oxborrow},
  \citenamefont {Paci}, \citenamefont {Pagano}, \citenamefont {Pajot},
  \citenamefont {Paoletti}, \citenamefont {Pasian}, \citenamefont {Patanchon},
  \citenamefont {Pearson}, \citenamefont {Perdereau}, \citenamefont {Perotto},
  \citenamefont {Perrott}, \citenamefont {Perrotta}, \citenamefont {Pettorino},
  \citenamefont {Piacentini}, \citenamefont {Piat}, \citenamefont {Pierpaoli},
  \citenamefont {Pietrobon}, \citenamefont {Plaszczynski}, \citenamefont
  {Pointecouteau}, \citenamefont {Polenta}, \citenamefont {Pratt},
  \citenamefont {Pr{\'e}zeau}, \citenamefont {Prunet},\ and\ \citenamefont
  {Puget}}]{planckcollaborationPlanck2015Results2016}%
  \BibitemOpen
  \bibfield  {author} {\bibinfo {author} {\bibnamefont {{Planck
  Collaboration}}}, \bibinfo {author} {\bibfnamefont {P.~A.~R.}\ \bibnamefont
  {Ade}}, \bibinfo {author} {\bibfnamefont {N.}~\bibnamefont {Aghanim}}, \emph
  {et~al.},\ }\href {https://doi.org/10.1051/0004-6361/201525823} {\bibfield
  {journal} {\bibinfo  {journal} {Astron. Astrophys.}\ }\textbf {\bibinfo
  {volume} {594}},\ \bibinfo {pages} {A27} (\bibinfo {year}
  {2016})}\BibitemShut {NoStop}%
\bibitem [{\citenamefont {Bleem}\ \emph {et~al.}(2020)\citenamefont {Bleem},
  \citenamefont {Bocquet}, \citenamefont {Stalder}, \citenamefont {Gladders},
  \citenamefont {Ade}, \citenamefont {Allen}, \citenamefont {Anderson},
  \citenamefont {Annis}, \citenamefont {Ashby}, \citenamefont {Austermann},
  \citenamefont {Avila}, \citenamefont {Avva}, \citenamefont {Bayliss},
  \citenamefont {Beall}, \citenamefont {Bechtol}, \citenamefont {Bender},
  \citenamefont {Benson}, \citenamefont {Bertin}, \citenamefont {Bianchini},
  \citenamefont {Blake}, \citenamefont {Brodwin}, \citenamefont {Brooks},
  \citenamefont {{Buckley-Geer}}, \citenamefont {Burke}, \citenamefont
  {Carlstrom}, \citenamefont {Rosell}, \citenamefont {Carrasco~Kind},
  \citenamefont {Carretero}, \citenamefont {Chang}, \citenamefont {Chiang},
  \citenamefont {Citron}, \citenamefont {Moran}, \citenamefont {Costanzi},
  \citenamefont {Crawford}, \citenamefont {Crites}, \citenamefont {{da Costa}},
  \citenamefont {{de Haan}}, \citenamefont {De~Vicente}, \citenamefont {Desai},
  \citenamefont {Diehl}, \citenamefont {Dietrich}, \citenamefont {Dobbs},
  \citenamefont {Eifler}, \citenamefont {Everett}, \citenamefont {Flaugher},
  \citenamefont {Floyd}, \citenamefont {Frieman}, \citenamefont {Gallicchio},
  \citenamefont {{Garc{\'i}a-Bellido}}, \citenamefont {George}, \citenamefont
  {Gerdes}, \citenamefont {Gilbert}, \citenamefont {Gruen}, \citenamefont
  {Gruendl}, \citenamefont {Gschwend}, \citenamefont {Gupta}, \citenamefont
  {Gutierrez}, \citenamefont {Halverson}, \citenamefont {Harrington},
  \citenamefont {Henning}, \citenamefont {Heymans}, \citenamefont {Holder},
  \citenamefont {Hollowood}, \citenamefont {Holzapfel}, \citenamefont
  {Honscheid}, \citenamefont {Hrubes}, \citenamefont {Huang}, \citenamefont
  {Hubmayr}, \citenamefont {Irwin}, \citenamefont {James}, \citenamefont
  {Jeltema}, \citenamefont {Joudaki}, \citenamefont {Khullar}, \citenamefont
  {Klein}, \citenamefont {Knox}, \citenamefont {Kuropatkin}, \citenamefont
  {Lee}, \citenamefont {Li}, \citenamefont {Lidman}, \citenamefont {Lowitz},
  \citenamefont {MacCrann}, \citenamefont {Mahler}, \citenamefont {Maia},
  \citenamefont {Marshall}, \citenamefont {McDonald}, \citenamefont {McMahon},
  \citenamefont {Melchior}, \citenamefont {Menanteau}, \citenamefont {Meyer},
  \citenamefont {Miquel}, \citenamefont {Mocanu}, \citenamefont {Mohr},
  \citenamefont {Montgomery}, \citenamefont {Nadolski}, \citenamefont {Natoli},
  \citenamefont {Nibarger}, \citenamefont {Noble}, \citenamefont {Novosad},
  \citenamefont {Padin}, \citenamefont {Palmese}, \citenamefont {Parkinson},
  \citenamefont {Patil}, \citenamefont {{Paz-Chinch{\'o}n}}, \citenamefont
  {Plazas}, \citenamefont {Pryke}, \citenamefont {Ramachandra}, \citenamefont
  {Reichardt}, \citenamefont {Remolina~Gonz{\'a}lez}, \citenamefont {Romer},
  \citenamefont {Roodman}, \citenamefont {Ruhl}, \citenamefont {Rykoff},
  \citenamefont {Saliwanchik}, \citenamefont {Sanchez}, \citenamefont {Saro},
  \citenamefont {Sayre}, \citenamefont {Schaffer}, \citenamefont {Schrabback},
  \citenamefont {Serrano}, \citenamefont {Sharon}, \citenamefont {Sievers},
  \citenamefont {Smecher}, \citenamefont {Smith}, \citenamefont
  {{Soares-Santos}}, \citenamefont {Stark}, \citenamefont {Story},
  \citenamefont {Suchyta}, \citenamefont {Tarle}, \citenamefont {Tucker},
  \citenamefont {Vanderlinde}, \citenamefont {Veach}, \citenamefont {Vieira},
  \citenamefont {Wang}, \citenamefont {Weller}, \citenamefont {Whitehorn},
  \citenamefont {Wu}, \citenamefont {Yefremenko},\ and\ \citenamefont
  {Zhang}}]{bleemSPTpolExtendedCluster2020}%
  \BibitemOpen
  \bibfield  {author} {\bibinfo {author} {\bibfnamefont {L.~E.}\ \bibnamefont
  {Bleem}}, \bibinfo {author} {\bibfnamefont {S.}~\bibnamefont {Bocquet}},
  \bibinfo {author} {\bibfnamefont {B.}~\bibnamefont {Stalder}}, \emph
  {et~al.},\ }\href {https://doi.org/10.3847/1538-4365/ab6993} {\bibfield
  {journal} {\bibinfo  {journal} {Astrophys. J. Suppl. Ser.}\ }\textbf
  {\bibinfo {volume} {247}},\ \bibinfo {pages} {25} (\bibinfo {year}
  {2020})}\BibitemShut {NoStop}%
\bibitem [{\citenamefont {Hilton}\ \emph {et~al.}(2021)\citenamefont {Hilton},
  \citenamefont {Sif{\'o}n}, \citenamefont {Naess}, \citenamefont
  {Madhavacheril}, \citenamefont {Oguri}, \citenamefont {Rozo}, \citenamefont
  {Rykoff}, \citenamefont {Abbott}, \citenamefont {Adhikari}, \citenamefont
  {Aguena}, \citenamefont {Aiola}, \citenamefont {Allam}, \citenamefont
  {Amodeo}, \citenamefont {Amon}, \citenamefont {Annis}, \citenamefont
  {Ansarinejad}, \citenamefont {{Aros-Bunster}}, \citenamefont {Austermann},
  \citenamefont {Avila}, \citenamefont {Bacon}, \citenamefont {Battaglia},
  \citenamefont {Beall}, \citenamefont {Becker}, \citenamefont {Bernstein},
  \citenamefont {Bertin}, \citenamefont {Bhandarkar}, \citenamefont {Bhargava},
  \citenamefont {Bond}, \citenamefont {Brooks}, \citenamefont {Burke},
  \citenamefont {Calabrese}, \citenamefont {Carrasco~Kind}, \citenamefont
  {Carretero}, \citenamefont {Choi}, \citenamefont {Choi}, \citenamefont
  {Conselice}, \citenamefont {{da Costa}}, \citenamefont {Costanzi},
  \citenamefont {Crichton}, \citenamefont {Crowley}, \citenamefont
  {D{\"u}nner}, \citenamefont {Denison}, \citenamefont {Devlin}, \citenamefont
  {Dicker}, \citenamefont {Diehl}, \citenamefont {Dietrich}, \citenamefont
  {Doel}, \citenamefont {Duff}, \citenamefont {Duivenvoorden}, \citenamefont
  {Dunkley}, \citenamefont {Everett}, \citenamefont {Ferraro}, \citenamefont
  {Ferrero}, \citenamefont {Fert{\'e}}, \citenamefont {Flaugher}, \citenamefont
  {Frieman}, \citenamefont {Gallardo}, \citenamefont {{Garc{\'i}a-Bellido}},
  \citenamefont {Gaztanaga}, \citenamefont {Gerdes}, \citenamefont {Giles},
  \citenamefont {Golec}, \citenamefont {Gralla}, \citenamefont {Grandis},
  \citenamefont {Gruen}, \citenamefont {Gruendl}, \citenamefont {Gschwend},
  \citenamefont {Gutierrez}, \citenamefont {Han}, \citenamefont {Hartley},
  \citenamefont {Hasselfield}, \citenamefont {Hill}, \citenamefont {Hilton},
  \citenamefont {Hincks}, \citenamefont {Hinton}, \citenamefont {Ho},
  \citenamefont {Honscheid}, \citenamefont {Hoyle}, \citenamefont {Hubmayr},
  \citenamefont {Huffenberger}, \citenamefont {Hughes}, \citenamefont
  {Jaelani}, \citenamefont {Jain}, \citenamefont {James}, \citenamefont
  {Jeltema}, \citenamefont {Kent}, \citenamefont {Knowles}, \citenamefont
  {Koopman}, \citenamefont {Kuehn}, \citenamefont {Lahav}, \citenamefont
  {Lima}, \citenamefont {Lin}, \citenamefont {Lokken}, \citenamefont {Loubser},
  \citenamefont {MacCrann}, \citenamefont {Maia}, \citenamefont {Marriage},
  \citenamefont {Martin}, \citenamefont {McMahon}, \citenamefont {Melchior},
  \citenamefont {Menanteau}, \citenamefont {Miquel}, \citenamefont {Miyatake},
  \citenamefont {Moodley}, \citenamefont {Morgan}, \citenamefont {Mroczkowski},
  \citenamefont {Nati}, \citenamefont {Newburgh}, \citenamefont {Niemack},
  \citenamefont {Nishizawa}, \citenamefont {Ogando}, \citenamefont
  {{Orlowski-Scherer}}, \citenamefont {Page}, \citenamefont {Palmese},
  \citenamefont {Partridge}, \citenamefont {{Paz-Chinch{\'o}n}}, \citenamefont
  {Phakathi}, \citenamefont {Plazas}, \citenamefont {Robertson}, \citenamefont
  {Romer}, \citenamefont {Carnero~Rosell}, \citenamefont {Salatino},
  \citenamefont {Sanchez}, \citenamefont {Schaan}, \citenamefont {Schillaci},
  \citenamefont {Sehgal}, \citenamefont {Serrano}, \citenamefont {Shin},
  \citenamefont {Simon}, \citenamefont {Smith}, \citenamefont
  {{Soares-Santos}}, \citenamefont {Spergel}, \citenamefont {Staggs},
  \citenamefont {Storer}, \citenamefont {Suchyta}, \citenamefont {Swanson},
  \citenamefont {Tarle}, \citenamefont {Thomas}, \citenamefont {To},
  \citenamefont {Trac}, \citenamefont {Ullom}, \citenamefont {Vale},
  \citenamefont {Van~Lanen}, \citenamefont {Vavagiakis}, \citenamefont
  {De~Vicente}, \citenamefont {Wilkinson}, \citenamefont {Wollack},
  \citenamefont {Xu},\ and\ \citenamefont
  {Zhang}}]{hiltonAtacamaCosmologyTelescope2021}%
  \BibitemOpen
  \bibfield  {author} {\bibinfo {author} {\bibfnamefont {M.}~\bibnamefont
  {Hilton}}, \bibinfo {author} {\bibfnamefont {C.}~\bibnamefont {Sif{\'o}n}},
  \bibinfo {author} {\bibfnamefont {S.}~\bibnamefont {Naess}}, \emph {et~al.},\
  }\href {https://doi.org/10.3847/1538-4365/abd023} {\bibfield  {journal}
  {\bibinfo  {journal} {Astrophys. J. Suppl. Ser.}\ }\textbf {\bibinfo {volume}
  {253}},\ \bibinfo {pages} {3} (\bibinfo {year} {2021})}\BibitemShut {NoStop}%
\bibitem [{\citenamefont {Bleem}\ \emph {et~al.}(2024)\citenamefont {Bleem},
  \citenamefont {Klein}, \citenamefont {Abbot}, \citenamefont {Ade},
  \citenamefont {Aguena}, \citenamefont {Alves}, \citenamefont {Anderson},
  \citenamefont {{Andrade-Oliveira}}, \citenamefont {Ansarinejad},
  \citenamefont {Archipley}, \citenamefont {Ashby}, \citenamefont {Austermann},
  \citenamefont {Bacon}, \citenamefont {Beall}, \citenamefont {Bender},
  \citenamefont {Benson}, \citenamefont {Bianchini}, \citenamefont {Bocquet},
  \citenamefont {Brooks}, \citenamefont {Burke}, \citenamefont {Calzadilla},
  \citenamefont {Carlstrom}, \citenamefont {Carnero~Rosell}, \citenamefont
  {Carretero}, \citenamefont {Chang}, \citenamefont {Chaubal}, \citenamefont
  {Chiang}, \citenamefont {Chou}, \citenamefont {Citron}, \citenamefont
  {Corbett~Moran}, \citenamefont {Costanzi}, \citenamefont {Constanzi},
  \citenamefont {Crawford}, \citenamefont {Crites}, \citenamefont {{da Costa}},
  \citenamefont {{de Haan}}, \citenamefont {De~Vicente}, \citenamefont {Desai},
  \citenamefont {Dobbs}, \citenamefont {Doel}, \citenamefont {Everett},
  \citenamefont {Ferrero}, \citenamefont {Flaugher}, \citenamefont {Floyd},
  \citenamefont {Friedel}, \citenamefont {Frieman}, \citenamefont {Gallicchio},
  \citenamefont {{Garc'ia-Bellido}}, \citenamefont {Gatti}, \citenamefont
  {George}, \citenamefont {Giannini}, \citenamefont {Grandis}, \citenamefont
  {Gruen}, \citenamefont {Gruendl}, \citenamefont {Gupta}, \citenamefont
  {Gutierrez}, \citenamefont {Halverson}, \citenamefont {Hinton}, \citenamefont
  {Hinton}, \citenamefont {Holder}, \citenamefont {Hollowood}, \citenamefont
  {Holzapfel}, \citenamefont {Honscheid}, \citenamefont {Hrubes}, \citenamefont
  {Huang}, \citenamefont {Hubmayr}, \citenamefont {Irwin}, \citenamefont
  {{Mena-Fern{\'a}ndez}}, \citenamefont {James}, \citenamefont
  {K{\'e}ruzor{\'e}}, \citenamefont {Knox}, \citenamefont {Kuehn},
  \citenamefont {Lahav}, \citenamefont {Lee}, \citenamefont {Lee},
  \citenamefont {Li}, \citenamefont {Lowitz}, \citenamefont {Marshal},
  \citenamefont {McDonald}, \citenamefont {McMahon}, \citenamefont {Menanteau},
  \citenamefont {Meyer}, \citenamefont {Miquel}, \citenamefont {Mohr},
  \citenamefont {Montgomery}, \citenamefont {Myles}, \citenamefont {Natoli},
  \citenamefont {Nibarger}, \citenamefont {Noble}, \citenamefont {Novosad},
  \citenamefont {Ogando}, \citenamefont {Padin}, \citenamefont {Patil},
  \citenamefont {Pereira}, \citenamefont {Pieres}, \citenamefont
  {Plazas~Malag'on}, \citenamefont {Pryke}, \citenamefont {Reichardt},
  \citenamefont {{Rodr'iguez-Monroy}}, \citenamefont {Romer}, \citenamefont
  {Ruhl}, \citenamefont {Saliwanchik}, \citenamefont {Salvati}, \citenamefont
  {Sanchez}, \citenamefont {Saro}, \citenamefont {Schaffer}, \citenamefont
  {Schrabback}, \citenamefont {{Sevilla-Noarbe}}, \citenamefont {Sievers},
  \citenamefont {Smecher}, \citenamefont {Smith}, \citenamefont
  {Somboonpanyakul}, \citenamefont {Stalder}, \citenamefont {Stark},
  \citenamefont {Suchyta}, \citenamefont {Swanson}, \citenamefont {Tarle},
  \citenamefont {To}, \citenamefont {Tucker}, \citenamefont {Veach},
  \citenamefont {Vieira}, \citenamefont {Vincenzi}, \citenamefont {Wang},
  \citenamefont {Weller}, \citenamefont {Whitehorn}, \citenamefont {Wiseman},
  \citenamefont {Wu}, \citenamefont {Yefremenko}, \citenamefont {Zebrowski},\
  and\ \citenamefont {Zhang}}]{bleemGalaxyClustersDiscovered2024}%
  \BibitemOpen
  \bibfield  {author} {\bibinfo {author} {\bibfnamefont {L.~E.}\ \bibnamefont
  {Bleem}}, \bibinfo {author} {\bibfnamefont {M.}~\bibnamefont {Klein}},
  \bibinfo {author} {\bibfnamefont {T.~M.~C.}\ \bibnamefont {Abbot}}, \emph
  {et~al.},\ }\href {https://doi.org/10.21105/astro.2311.07512} {\bibfield
  {journal} {\bibinfo  {journal} {Open J. Astrophys.}\ }\textbf {\bibinfo
  {volume} {7}},\ \bibinfo {pages} {13} (\bibinfo {year} {2024})}\BibitemShut
  {NoStop}%
\bibitem [{\citenamefont {Madhavacheril}\ \emph {et~al.}(2024)\citenamefont
  {Madhavacheril}, \citenamefont {Qu}, \citenamefont {Sherwin}, \citenamefont
  {MacCrann}, \citenamefont {Li}, \citenamefont {{Abril-Cabezas}},
  \citenamefont {Ade}, \citenamefont {Aiola}, \citenamefont {Alford},
  \citenamefont {Amiri}, \citenamefont {Amodeo}, \citenamefont {An},
  \citenamefont {Atkins}, \citenamefont {Austermann}, \citenamefont
  {Battaglia}, \citenamefont {Battistelli}, \citenamefont {Beall},
  \citenamefont {Bean}, \citenamefont {Beringue}, \citenamefont {Bhandarkar},
  \citenamefont {Biermann}, \citenamefont {Bolliet}, \citenamefont {Bond},
  \citenamefont {Cai}, \citenamefont {Calabrese}, \citenamefont {Calafut},
  \citenamefont {Capalbo}, \citenamefont {Carrero}, \citenamefont {Challinor},
  \citenamefont {Chesmore}, \citenamefont {Cho}, \citenamefont {Choi},
  \citenamefont {Clark}, \citenamefont {C{\'o}rdova~Rosado}, \citenamefont
  {Cothard}, \citenamefont {Coughlin}, \citenamefont {Coulton}, \citenamefont
  {Crowley}, \citenamefont {Dalal}, \citenamefont {Darwish}, \citenamefont
  {Devlin}, \citenamefont {Dicker}, \citenamefont {Doze}, \citenamefont
  {Duell}, \citenamefont {Duff}, \citenamefont {Duivenvoorden}, \citenamefont
  {Dunkley}, \citenamefont {D{\"u}nner}, \citenamefont {Fanfani}, \citenamefont
  {Fankhanel}, \citenamefont {Farren}, \citenamefont {Ferraro}, \citenamefont
  {Freundt}, \citenamefont {Fuzia}, \citenamefont {Gallardo}, \citenamefont
  {Garrido}, \citenamefont {Givans}, \citenamefont {Gluscevic}, \citenamefont
  {Golec}, \citenamefont {Guan}, \citenamefont {Hall}, \citenamefont {Halpern},
  \citenamefont {Han}, \citenamefont {Harrison}, \citenamefont {Hasselfield},
  \citenamefont {Healy}, \citenamefont {Henderson}, \citenamefont {Hensley},
  \citenamefont {{Herv{\'i}as-Caimapo}}, \citenamefont {Hill}, \citenamefont
  {Hilton}, \citenamefont {Hilton}, \citenamefont {Hincks}, \citenamefont
  {Hlo{\v z}ek}, \citenamefont {Ho}, \citenamefont {Huber}, \citenamefont
  {Hubmayr}, \citenamefont {Huffenberger}, \citenamefont {Hughes},
  \citenamefont {Irwin}, \citenamefont {Isopi}, \citenamefont {Jense},
  \citenamefont {Keller}, \citenamefont {Kim}, \citenamefont {Knowles},
  \citenamefont {Koopman}, \citenamefont {Kosowsky}, \citenamefont {Kramer},
  \citenamefont {Kusiak}, \citenamefont {La~Posta}, \citenamefont {Lague},
  \citenamefont {Lakey}, \citenamefont {Lee}, \citenamefont {Li}, \citenamefont
  {Limon}, \citenamefont {Lokken}, \citenamefont {Louis}, \citenamefont
  {Lungu}, \citenamefont {MacInnis}, \citenamefont {Maldonado}, \citenamefont
  {Maldonado}, \citenamefont {{Mallaby-Kay}}, \citenamefont {Marques},
  \citenamefont {McMahon}, \citenamefont {Mehta}, \citenamefont {Menanteau},
  \citenamefont {Moodley}, \citenamefont {Morris}, \citenamefont {Mroczkowski},
  \citenamefont {Naess}, \citenamefont {Namikawa}, \citenamefont {Nati},
  \citenamefont {Newburgh}, \citenamefont {Nicola}, \citenamefont {Niemack},
  \citenamefont {Nolta}, \citenamefont {{Orlowski-Scherer}}, \citenamefont
  {Page}, \citenamefont {Pandey}, \citenamefont {Partridge}, \citenamefont
  {Prince}, \citenamefont {Puddu}, \citenamefont {Radiconi}, \citenamefont
  {Robertson}, \citenamefont {Rojas}, \citenamefont {Sakuma}, \citenamefont
  {Salatino}, \citenamefont {Schaan}, \citenamefont {Schmitt}, \citenamefont
  {Sehgal}, \citenamefont {Shaikh}, \citenamefont {Sierra}, \citenamefont
  {Sievers}, \citenamefont {Sif{\'o}n}, \citenamefont {Simon}, \citenamefont
  {Sonka}, \citenamefont {Spergel}, \citenamefont {Staggs}, \citenamefont
  {Storer}, \citenamefont {Switzer}, \citenamefont {Tampier}, \citenamefont
  {Thornton}, \citenamefont {Trac}, \citenamefont {Treu}, \citenamefont
  {Tucker}, \citenamefont {Ullom}, \citenamefont {Vale}, \citenamefont
  {Van~Engelen}, \citenamefont {Van~Lanen}, \citenamefont {{van Marrewijk}},
  \citenamefont {Vargas}, \citenamefont {Vavagiakis}, \citenamefont {Wagoner},
  \citenamefont {Wang}, \citenamefont {Wenzl}, \citenamefont {Wollack},
  \citenamefont {Xu}, \citenamefont {Zago},\ and\ \citenamefont
  {Zheng}}]{madhavacherilAtacamaCosmologyTelescope2024}%
  \BibitemOpen
  \bibfield  {author} {\bibinfo {author} {\bibfnamefont {M.~S.}\ \bibnamefont
  {Madhavacheril}}, \bibinfo {author} {\bibfnamefont {F.~J.}\ \bibnamefont
  {Qu}}, \bibinfo {author} {\bibfnamefont {B.~D.}\ \bibnamefont {Sherwin}},
  \emph {et~al.},\ }\href {https://doi.org/10.3847/1538-4357/acff5f} {\bibfield
   {journal} {\bibinfo  {journal} {Astrophys. J.}\ }\textbf {\bibinfo {volume}
  {962}},\ \bibinfo {pages} {113} (\bibinfo {year} {2024})}\BibitemShut
  {NoStop}%
\bibitem [{\citenamefont {Qu}\ \emph {et~al.}(2024)\citenamefont {Qu},
  \citenamefont {Sherwin}, \citenamefont {Madhavacheril}, \citenamefont {Han},
  \citenamefont {Crowley}, \citenamefont {{Abril-Cabezas}}, \citenamefont
  {Ade}, \citenamefont {Aiola}, \citenamefont {Alford}, \citenamefont {Amiri},
  \citenamefont {Amodeo}, \citenamefont {An}, \citenamefont {Atkins},
  \citenamefont {Austermann}, \citenamefont {Battaglia}, \citenamefont
  {Battistelli}, \citenamefont {Beall}, \citenamefont {Bean}, \citenamefont
  {Beringue}, \citenamefont {Bhandarkar}, \citenamefont {Biermann},
  \citenamefont {Bolliet}, \citenamefont {Bond}, \citenamefont {Cai},
  \citenamefont {Calabrese}, \citenamefont {Calafut}, \citenamefont {Capalbo},
  \citenamefont {Carrero}, \citenamefont {Carron}, \citenamefont {Challinor},
  \citenamefont {Chesmore}, \citenamefont {Cho}, \citenamefont {Choi},
  \citenamefont {Clark}, \citenamefont {C{\'o}rdova~Rosado}, \citenamefont
  {Cothard}, \citenamefont {Coughlin}, \citenamefont {Coulton}, \citenamefont
  {Dalal}, \citenamefont {Darwish}, \citenamefont {Devlin}, \citenamefont
  {Dicker}, \citenamefont {Doze}, \citenamefont {Duell}, \citenamefont {Duff},
  \citenamefont {Duivenvoorden}, \citenamefont {Dunkley}, \citenamefont
  {D{\"u}nner}, \citenamefont {Fanfani}, \citenamefont {Fankhanel},
  \citenamefont {Farren}, \citenamefont {Ferraro}, \citenamefont {Freundt},
  \citenamefont {Fuzia}, \citenamefont {Gallardo}, \citenamefont {Garrido},
  \citenamefont {Gluscevic}, \citenamefont {Golec}, \citenamefont {Guan},
  \citenamefont {Halpern}, \citenamefont {Harrison}, \citenamefont
  {Hasselfield}, \citenamefont {Healy}, \citenamefont {Henderson},
  \citenamefont {Hensley}, \citenamefont {{Herv{\'i}as-Caimapo}}, \citenamefont
  {Hill}, \citenamefont {Hilton}, \citenamefont {Hilton}, \citenamefont
  {Hincks}, \citenamefont {Hlo{\v z}ek}, \citenamefont {Ho}, \citenamefont
  {Huber}, \citenamefont {Hubmayr}, \citenamefont {Huffenberger}, \citenamefont
  {Hughes}, \citenamefont {Irwin}, \citenamefont {Isopi}, \citenamefont
  {Jense}, \citenamefont {Keller}, \citenamefont {Kim}, \citenamefont
  {Knowles}, \citenamefont {Koopman}, \citenamefont {Kosowsky}, \citenamefont
  {Kramer}, \citenamefont {Kusiak}, \citenamefont {La~Posta}, \citenamefont
  {Lague}, \citenamefont {Lakey}, \citenamefont {Lee}, \citenamefont {Li},
  \citenamefont {Li}, \citenamefont {Limon}, \citenamefont {Lokken},
  \citenamefont {Louis}, \citenamefont {Lungu}, \citenamefont {MacCrann},
  \citenamefont {MacInnis}, \citenamefont {Maldonado}, \citenamefont
  {Maldonado}, \citenamefont {{Mallaby-Kay}}, \citenamefont {Marques},
  \citenamefont {McMahon}, \citenamefont {Mehta}, \citenamefont {Menanteau},
  \citenamefont {Moodley}, \citenamefont {Morris}, \citenamefont {Mroczkowski},
  \citenamefont {Naess}, \citenamefont {Namikawa}, \citenamefont {Nati},
  \citenamefont {Newburgh}, \citenamefont {Nicola}, \citenamefont {Niemack},
  \citenamefont {Nolta}, \citenamefont {{Orlowski-Scherer}}, \citenamefont
  {Page}, \citenamefont {Pandey}, \citenamefont {Partridge}, \citenamefont
  {Prince}, \citenamefont {Puddu}, \citenamefont {Radiconi}, \citenamefont
  {Robertson}, \citenamefont {Rojas}, \citenamefont {Sakuma}, \citenamefont
  {Salatino}, \citenamefont {Schaan}, \citenamefont {Schmitt}, \citenamefont
  {Sehgal}, \citenamefont {Shaikh}, \citenamefont {Sierra}, \citenamefont
  {Sievers}, \citenamefont {Sif{\'o}n}, \citenamefont {Simon}, \citenamefont
  {Sonka}, \citenamefont {Spergel}, \citenamefont {Staggs}, \citenamefont
  {Storer}, \citenamefont {Switzer}, \citenamefont {Tampier}, \citenamefont
  {Thornton}, \citenamefont {Trac}, \citenamefont {Treu}, \citenamefont
  {Tucker}, \citenamefont {Ullom}, \citenamefont {Vale}, \citenamefont
  {Van~Engelen}, \citenamefont {Van~Lanen}, \citenamefont {{van Marrewijk}},
  \citenamefont {Vargas}, \citenamefont {Vavagiakis}, \citenamefont {Wagoner},
  \citenamefont {Wang}, \citenamefont {Wenzl}, \citenamefont {Wollack},
  \citenamefont {Xu}, \citenamefont {Zago},\ and\ \citenamefont
  {Zheng}}]{quAtacamaCosmologyTelescope2024}%
  \BibitemOpen
  \bibfield  {author} {\bibinfo {author} {\bibfnamefont {F.~J.}\ \bibnamefont
  {Qu}}, \bibinfo {author} {\bibfnamefont {B.~D.}\ \bibnamefont {Sherwin}},
  \bibinfo {author} {\bibfnamefont {M.~S.}\ \bibnamefont {Madhavacheril}},
  \emph {et~al.},\ }\href {https://doi.org/10.3847/1538-4357/acfe06} {\bibfield
   {journal} {\bibinfo  {journal} {Astrophys. J.}\ }\textbf {\bibinfo {volume}
  {962}},\ \bibinfo {pages} {112} (\bibinfo {year} {2024})}\BibitemShut
  {NoStop}%
\bibitem [{\citenamefont {Govoni}\ and\ \citenamefont
  {Feretti}(2004)}]{govoniMagneticFieldsClusters2004}%
  \BibitemOpen
  \bibfield  {author} {\bibinfo {author} {\bibfnamefont {F.}~\bibnamefont
  {Govoni}}\ and\ \bibinfo {author} {\bibfnamefont {L.}~\bibnamefont
  {Feretti}},\ }\href {https://doi.org/10.1142/S0218271804005080} {\bibfield
  {journal} {\bibinfo  {journal} {Int. J. Mod. Phys. D}\ }\textbf {\bibinfo
  {volume} {13}},\ \bibinfo {pages} {1549} (\bibinfo {year}
  {2004})}\BibitemShut {NoStop}%
\bibitem [{\citenamefont {En{\ss}lin}\ and\ \citenamefont
  {Vogt}(2006)}]{ensslinMagneticTurbulenceCool2006}%
  \BibitemOpen
  \bibfield  {author} {\bibinfo {author} {\bibfnamefont {T.~A.}\ \bibnamefont
  {En{\ss}lin}}\ and\ \bibinfo {author} {\bibfnamefont {C.}~\bibnamefont
  {Vogt}},\ }\href {https://doi.org/10.1051/0004-6361:20053518} {\bibfield
  {journal} {\bibinfo  {journal} {Astron. Astrophys.}\ }\textbf {\bibinfo
  {volume} {453}},\ \bibinfo {pages} {447} (\bibinfo {year}
  {2006})}\BibitemShut {NoStop}%
\bibitem [{\citenamefont {Feretti}\ \emph {et~al.}(2012)\citenamefont
  {Feretti}, \citenamefont {Giovannini}, \citenamefont {Govoni},\ and\
  \citenamefont {Murgia}}]{ferettiClustersGalaxiesObservational2012}%
  \BibitemOpen
  \bibfield  {author} {\bibinfo {author} {\bibfnamefont {L.}~\bibnamefont
  {Feretti}}, \bibinfo {author} {\bibfnamefont {G.}~\bibnamefont {Giovannini}},
  \bibinfo {author} {\bibfnamefont {F.}~\bibnamefont {Govoni}},\ and\ \bibinfo
  {author} {\bibfnamefont {M.}~\bibnamefont {Murgia}},\ }\href
  {https://doi.org/10.1007/s00159-012-0054-z} {\bibfield  {journal} {\bibinfo
  {journal} {Astron. Astrophys. Rev.}\ }\textbf {\bibinfo {volume} {20}},\
  \bibinfo {pages} {54} (\bibinfo {year} {2012})}\BibitemShut {NoStop}%
\bibitem [{\citenamefont {{van Weeren}}\ \emph {et~al.}(2019)\citenamefont
  {{van Weeren}}, \citenamefont {{de Gasperin}}, \citenamefont {Akamatsu},
  \citenamefont {Br{\"u}ggen}, \citenamefont {Feretti}, \citenamefont {Kang},
  \citenamefont {Stroe},\ and\ \citenamefont
  {Zandanel}}]{vanweerenDiffuseRadioEmission2019}%
  \BibitemOpen
  \bibfield  {author} {\bibinfo {author} {\bibfnamefont {R.~J.}\ \bibnamefont
  {{van Weeren}}}, \bibinfo {author} {\bibfnamefont {F.}~\bibnamefont {{de
  Gasperin}}}, \bibinfo {author} {\bibfnamefont {H.}~\bibnamefont {Akamatsu}},
  \emph {et~al.},\ }\href {https://doi.org/10.1007/s11214-019-0584-z}
  {\bibfield  {journal} {\bibinfo  {journal} {Space Sci. Rev.}\ }\textbf
  {\bibinfo {volume} {215}},\ \bibinfo {pages} {16} (\bibinfo {year}
  {2019})}\BibitemShut {NoStop}%
\bibitem [{\citenamefont {{Pandey-Pommier}}\ \emph {et~al.}(2026)\citenamefont
  {{Pandey-Pommier}}, \citenamefont {McKean}, \citenamefont {Harrison},\ and\
  \citenamefont {Cress}}]{pandey-pommierComprehensiveSKASurvey2026}%
  \BibitemOpen
  \bibfield  {author} {\bibinfo {author} {\bibfnamefont {M.}~\bibnamefont
  {{Pandey-Pommier}}}, \bibinfo {author} {\bibfnamefont {J.}~\bibnamefont
  {McKean}}, \bibinfo {author} {\bibfnamefont {I.}~\bibnamefont {Harrison}},\
  and\ \bibinfo {author} {\bibfnamefont {C.~M.}\ \bibnamefont {Cress}},\ }\href
  {https://doi.org/10.48550/arXiv.2608.19159} {\enquote {\bibinfo {title} {A
  comprehensive {{SKA Survey}} of lensing clusters and lensed galaxies},}\ }
  (\bibinfo {year} {2026})\BibitemShut {NoStop}%
\bibitem [{\citenamefont {Cohn}\ and\ \citenamefont
  {Battaglia}(2020)}]{cohnMultiwavelengthClusterMass2020}%
  \BibitemOpen
  \bibfield  {author} {\bibinfo {author} {\bibfnamefont {J.~D.}\ \bibnamefont
  {Cohn}}\ and\ \bibinfo {author} {\bibfnamefont {N.}~\bibnamefont
  {Battaglia}},\ }\href {https://doi.org/10.1093/mnras/stz3087} {\bibfield
  {journal} {\bibinfo  {journal} {Mon. Not. R. Astron. Soc.}\ }\textbf
  {\bibinfo {volume} {491}},\ \bibinfo {pages} {1575} (\bibinfo {year}
  {2020})}\BibitemShut {NoStop}%
\bibitem [{\citenamefont {Bocquet}\ \emph
  {et~al.}(2024{\natexlab{a}})\citenamefont {Bocquet}, \citenamefont {Grandis},
  \citenamefont {Bleem}, \citenamefont {Klein}, \citenamefont {Mohr},
  \citenamefont {Aguena}, \citenamefont {Alarcon}, \citenamefont {Allam},
  \citenamefont {Allen}, \citenamefont {Alves}, \citenamefont {Amon},
  \citenamefont {Ansarinejad}, \citenamefont {Bacon}, \citenamefont {Bayliss},
  \citenamefont {Bechtol}, \citenamefont {Becker}, \citenamefont {Benson},
  \citenamefont {Bernstein}, \citenamefont {Brodwin}, \citenamefont {Brooks},
  \citenamefont {Campos}, \citenamefont {Canning}, \citenamefont {Carlstrom},
  \citenamefont {Carnero~Rosell}, \citenamefont {Carrasco~Kind}, \citenamefont
  {Carretero}, \citenamefont {Cawthon}, \citenamefont {Chang}, \citenamefont
  {Chen}, \citenamefont {Choi}, \citenamefont {Cordero}, \citenamefont
  {Costanzi}, \citenamefont {{da Costa}}, \citenamefont {Pereira},
  \citenamefont {Davis}, \citenamefont {DeRose}, \citenamefont {Desai},
  \citenamefont {{de Haan}}, \citenamefont {De~Vicente}, \citenamefont {Diehl},
  \citenamefont {Dodelson}, \citenamefont {Doel}, \citenamefont {Doux},
  \citenamefont {{Drlica-Wagner}}, \citenamefont {Eckert}, \citenamefont
  {{Elvin-Poole}}, \citenamefont {Everett}, \citenamefont {Ferrero},
  \citenamefont {Fert{\'e}}, \citenamefont {Flores}, \citenamefont {Frieman},
  \citenamefont {{Garc{\'i}a-Bellido}}, \citenamefont {Gatti}, \citenamefont
  {Giannini}, \citenamefont {Gladders}, \citenamefont {Gruen}, \citenamefont
  {Gruendl}, \citenamefont {Harrison}, \citenamefont {Hartley}, \citenamefont
  {Herner}, \citenamefont {Hinton}, \citenamefont {Hollowood}, \citenamefont
  {Holzapfel}, \citenamefont {Honscheid}, \citenamefont {Huang}, \citenamefont
  {Huff}, \citenamefont {James}, \citenamefont {Jarvis}, \citenamefont
  {Khullar}, \citenamefont {Kim}, \citenamefont {Kraft}, \citenamefont {Kuehn},
  \citenamefont {Kuropatkin}, \citenamefont {K{\'e}ruzor{\'e}}, \citenamefont
  {Lee}, \citenamefont {Leget}, \citenamefont {MacCrann}, \citenamefont
  {Mahler}, \citenamefont {Mantz}, \citenamefont {Marshall}, \citenamefont
  {McCullough}, \citenamefont {McDonald}, \citenamefont {{Mena-Fern{\'a}ndez}},
  \citenamefont {Miquel}, \citenamefont {Myles}, \citenamefont
  {{Navarro-Alsina}}, \citenamefont {Ogando}, \citenamefont {Palmese},
  \citenamefont {Pandey}, \citenamefont {Pieres}, \citenamefont
  {Plazas~Malag{\'o}n}, \citenamefont {Prat}, \citenamefont {Raveri},
  \citenamefont {Reichardt}, \citenamefont {Roberson}, \citenamefont {Rollins},
  \citenamefont {Romer}, \citenamefont {Romero}, \citenamefont {Roodman},
  \citenamefont {Ross}, \citenamefont {Rykoff}, \citenamefont {Salvati},
  \citenamefont {S{\'a}nchez}, \citenamefont {Sanchez}, \citenamefont
  {Sanchez~Cid}, \citenamefont {Saro}, \citenamefont {Schrabback},
  \citenamefont {Schubnell}, \citenamefont {Secco}, \citenamefont
  {{Sevilla-Noarbe}}, \citenamefont {Sharon}, \citenamefont {Sheldon},
  \citenamefont {Shin}, \citenamefont {Smith}, \citenamefont {Somboonpanyakul},
  \citenamefont {Stalder}, \citenamefont {Stark}, \citenamefont {Strazzullo},
  \citenamefont {Suchyta}, \citenamefont {Swanson}, \citenamefont {Tarle},
  \citenamefont {To}, \citenamefont {Troxel}, \citenamefont {Tutusaus},
  \citenamefont {Varga}, \citenamefont {{von der Linden}}, \citenamefont
  {Weaverdyck}, \citenamefont {Weller}, \citenamefont {Wiseman}, \citenamefont
  {Yanny}, \citenamefont {Yin}, \citenamefont {Young}, \citenamefont {Zhang},
  \citenamefont {Zuntz}, \citenamefont {{(The DES}},\ and\ \citenamefont {{SPT
  Collaborations)}}}]{bocquetSPTClustersHST2024a}%
  \BibitemOpen
  \bibfield  {author} {\bibinfo {author} {\bibfnamefont {S.}~\bibnamefont
  {Bocquet}}, \bibinfo {author} {\bibfnamefont {S.}~\bibnamefont {Grandis}},
  \bibinfo {author} {\bibfnamefont {L.~E.}\ \bibnamefont {Bleem}}, \emph
  {et~al.},\ }\href {https://doi.org/10.1103/PhysRevD.110.083509} {\bibfield
  {journal} {\bibinfo  {journal} {Phys. Rev. D}\ }\textbf {\bibinfo {volume}
  {110}},\ \bibinfo {pages} {083509} (\bibinfo {year}
  {2024}{\natexlab{a}})}\BibitemShut {NoStop}%
\bibitem [{\citenamefont {Bocquet}\ \emph
  {et~al.}(2024{\natexlab{b}})\citenamefont {Bocquet}, \citenamefont {Grandis},
  \citenamefont {Bleem}, \citenamefont {Klein}, \citenamefont {Mohr},
  \citenamefont {Schrabback}, \citenamefont {Abbott}, \citenamefont {Ade},
  \citenamefont {Aguena}, \citenamefont {Alarcon}, \citenamefont {Allam},
  \citenamefont {Allen}, \citenamefont {Alves}, \citenamefont {Amon},
  \citenamefont {Anderson}, \citenamefont {Annis}, \citenamefont {Ansarinejad},
  \citenamefont {Austermann}, \citenamefont {Avila}, \citenamefont {Bacon},
  \citenamefont {Bayliss}, \citenamefont {Beall}, \citenamefont {Bechtol},
  \citenamefont {Becker}, \citenamefont {Bender}, \citenamefont {Benson},
  \citenamefont {Bernstein}, \citenamefont {Bhargava}, \citenamefont
  {Bianchini}, \citenamefont {Brodwin}, \citenamefont {Brooks}, \citenamefont
  {Bryant}, \citenamefont {Campos}, \citenamefont {Canning}, \citenamefont
  {Carlstrom}, \citenamefont {Carnero~Rosell}, \citenamefont {Carrasco~Kind},
  \citenamefont {Carretero}, \citenamefont {Castander}, \citenamefont
  {Cawthon}, \citenamefont {Chang}, \citenamefont {Chang}, \citenamefont
  {Chaubal}, \citenamefont {Chen}, \citenamefont {Chiang}, \citenamefont
  {Choi}, \citenamefont {Chou}, \citenamefont {Citron}, \citenamefont
  {Corbett~Moran}, \citenamefont {Cordero}, \citenamefont {Costanzi},
  \citenamefont {Crawford}, \citenamefont {Crites}, \citenamefont {{da Costa}},
  \citenamefont {Pereira}, \citenamefont {Davis}, \citenamefont {Davis},
  \citenamefont {DeRose}, \citenamefont {Desai}, \citenamefont {{de Haan}},
  \citenamefont {Diehl}, \citenamefont {Dobbs}, \citenamefont {Dodelson},
  \citenamefont {Doux}, \citenamefont {{Drlica-Wagner}}, \citenamefont
  {Eckert}, \citenamefont {{Elvin-Poole}}, \citenamefont {Everett},
  \citenamefont {Everett}, \citenamefont {Ferrero}, \citenamefont {Fert{\'e}},
  \citenamefont {Flores}, \citenamefont {Frieman}, \citenamefont {Gallicchio},
  \citenamefont {{Garc{\'i}a-Bellido}}, \citenamefont {Gatti}, \citenamefont
  {George}, \citenamefont {Giannini}, \citenamefont {Gladders}, \citenamefont
  {Gruen}, \citenamefont {Gruendl}, \citenamefont {Gupta}, \citenamefont
  {Gutierrez}, \citenamefont {Halverson}, \citenamefont {Harrison},
  \citenamefont {Hartley}, \citenamefont {Herner}, \citenamefont {Hinton},
  \citenamefont {Holder}, \citenamefont {Hollowood}, \citenamefont {Holzapfel},
  \citenamefont {Honscheid}, \citenamefont {Hrubes}, \citenamefont {Huang},
  \citenamefont {Hubmayr}, \citenamefont {Huff}, \citenamefont {Huterer},
  \citenamefont {Irwin}, \citenamefont {James}, \citenamefont {Jarvis},
  \citenamefont {Khullar}, \citenamefont {Kim}, \citenamefont {Knox},
  \citenamefont {Kraft}, \citenamefont {Krause}, \citenamefont {Kuehn},
  \citenamefont {Kuropatkin}, \citenamefont {K{\'e}ruzor{\'e}}, \citenamefont
  {Lahav}, \citenamefont {Lee}, \citenamefont {Leget}, \citenamefont {Li},
  \citenamefont {Lin}, \citenamefont {Lowitz}, \citenamefont {MacCrann},
  \citenamefont {Mahler}, \citenamefont {Mantz}, \citenamefont {Marshall},
  \citenamefont {McCullough}, \citenamefont {McDonald}, \citenamefont
  {McMahon}, \citenamefont {{Mena-Fern{\'a}ndez}}, \citenamefont {Menanteau},
  \citenamefont {Meyer}, \citenamefont {Miquel}, \citenamefont {Montgomery},
  \citenamefont {Myles}, \citenamefont {Natoli}, \citenamefont
  {{Navarro-Alsina}}, \citenamefont {Nibarger}, \citenamefont {Noble},
  \citenamefont {Novosad}, \citenamefont {Ogando}, \citenamefont {Omori},
  \citenamefont {Padin}, \citenamefont {Pandey}, \citenamefont {Paschos},
  \citenamefont {Patil}, \citenamefont {Pieres}, \citenamefont
  {Plazas~Malag{\'o}n}, \citenamefont {Porredon}, \citenamefont {Prat},
  \citenamefont {Pryke}, \citenamefont {Raveri}, \citenamefont {Reichardt},
  \citenamefont {Roberson}, \citenamefont {Rollins}, \citenamefont {Romero},
  \citenamefont {Roodman}, \citenamefont {Ruhl}, \citenamefont {Rykoff},
  \citenamefont {Saliwanchik}, \citenamefont {Salvati}, \citenamefont
  {S{\'a}nchez}, \citenamefont {Sanchez}, \citenamefont {Sanchez~Cid},
  \citenamefont {Saro}, \citenamefont {Schaffer}, \citenamefont {Secco},
  \citenamefont {{Sevilla-Noarbe}}, \citenamefont {Sharon}, \citenamefont
  {Sheldon}, \citenamefont {Shin}, \citenamefont {Sievers}, \citenamefont
  {Smecher}, \citenamefont {Smith}, \citenamefont {Somboonpanyakul},
  \citenamefont {Sommer}, \citenamefont {Stalder}, \citenamefont {Stark},
  \citenamefont {Stephen}, \citenamefont {Strazzullo}, \citenamefont {Suchyta},
  \citenamefont {Tarle}, \citenamefont {To}, \citenamefont {Troxel},
  \citenamefont {Tucker}, \citenamefont {Tutusaus}, \citenamefont {Varga},
  \citenamefont {Veach}, \citenamefont {Vieira}, \citenamefont {Vikhlinin},
  \citenamefont {{von der Linden}}, \citenamefont {Wang}, \citenamefont
  {Weaverdyck}, \citenamefont {Weller}, \citenamefont {Whitehorn},
  \citenamefont {Wu}, \citenamefont {Yanny}, \citenamefont {Yefremenko},
  \citenamefont {Yin}, \citenamefont {Young}, \citenamefont {Zebrowski},
  \citenamefont {Zhang}, \citenamefont {Zohren}, \citenamefont {Zuntz},
  \citenamefont {{(SPT}},\ and\ \citenamefont {{DES
  Collaborations)}}}]{bocquetSPTClustersHST2024b}%
  \BibitemOpen
  \bibfield  {author} {\bibinfo {author} {\bibfnamefont {S.}~\bibnamefont
  {Bocquet}}, \bibinfo {author} {\bibfnamefont {S.}~\bibnamefont {Grandis}},
  \bibinfo {author} {\bibfnamefont {L.~E.}\ \bibnamefont {Bleem}}, \emph
  {et~al.},\ }\href {https://doi.org/10.1103/PhysRevD.110.083510} {\bibfield
  {journal} {\bibinfo  {journal} {Phys. Rev. D}\ }\textbf {\bibinfo {volume}
  {110}},\ \bibinfo {pages} {083510} (\bibinfo {year}
  {2024}{\natexlab{b}})}\BibitemShut {NoStop}%
\bibitem [{\citenamefont {Bocquet}\ \emph {et~al.}(2025)\citenamefont
  {Bocquet}, \citenamefont {Grandis}, \citenamefont {Krause}, \citenamefont
  {To}, \citenamefont {Bleem}, \citenamefont {Klein}, \citenamefont {Mohr},
  \citenamefont {Schrabback}, \citenamefont {Alarcon}, \citenamefont {Alves},
  \citenamefont {Amon}, \citenamefont {{Andrade-Oliveira}}, \citenamefont
  {Baxter}, \citenamefont {Bechtol}, \citenamefont {Becker}, \citenamefont
  {Bernstein}, \citenamefont {Blazek}, \citenamefont {Camacho}, \citenamefont
  {Campos}, \citenamefont {Carnero~Rosell}, \citenamefont {Carrasco~Kind},
  \citenamefont {Cawthon}, \citenamefont {Chang}, \citenamefont {Chen},
  \citenamefont {Choi}, \citenamefont {Cordero}, \citenamefont {Crocce},
  \citenamefont {Davis}, \citenamefont {DeRose}, \citenamefont {Diehl},
  \citenamefont {Dodelson}, \citenamefont {Doux}, \citenamefont
  {{Drlica-Wagner}}, \citenamefont {Eckert}, \citenamefont {Eifler},
  \citenamefont {Elsner}, \citenamefont {{Elvin-Poole}}, \citenamefont
  {Everett}, \citenamefont {Fang}, \citenamefont {Fert{\'e}}, \citenamefont
  {Fosalba}, \citenamefont {Friedrich}, \citenamefont {Frieman}, \citenamefont
  {Gatti}, \citenamefont {Giannini}, \citenamefont {Gruen}, \citenamefont
  {Gruendl}, \citenamefont {Harrison}, \citenamefont {Hartley}, \citenamefont
  {Herner}, \citenamefont {Huang}, \citenamefont {Huff}, \citenamefont
  {Huterer}, \citenamefont {Jarvis}, \citenamefont {Kuropatkin}, \citenamefont
  {Leget}, \citenamefont {Lemos}, \citenamefont {Liddle}, \citenamefont
  {MacCrann}, \citenamefont {McCullough}, \citenamefont {Muir}, \citenamefont
  {Myles}, \citenamefont {{Navarro-Alsina}}, \citenamefont {Pandey},
  \citenamefont {Park}, \citenamefont {Porredon}, \citenamefont {Prat},
  \citenamefont {Raveri}, \citenamefont {Rollins}, \citenamefont {Roodman},
  \citenamefont {Rosenfeld}, \citenamefont {Rykoff}, \citenamefont
  {S{\'a}nchez}, \citenamefont {Sanchez}, \citenamefont {Secco}, \citenamefont
  {{Sevilla-Noarbe}}, \citenamefont {Sheldon}, \citenamefont {Shin},
  \citenamefont {Troxel}, \citenamefont {Tutusaus}, \citenamefont {Varga},
  \citenamefont {Weaverdyck}, \citenamefont {Wechsler}, \citenamefont {Wu},
  \citenamefont {Yanny}, \citenamefont {Yin}, \citenamefont {Zhang},
  \citenamefont {Zuntz}, \citenamefont {Abbott}, \citenamefont {Ade},
  \citenamefont {Aguena}, \citenamefont {Allam}, \citenamefont {Allen},
  \citenamefont {Anderson}, \citenamefont {Ansarinejad}, \citenamefont
  {Austermann}, \citenamefont {Bayliss}, \citenamefont {Beall}, \citenamefont
  {Bender}, \citenamefont {Benson}, \citenamefont {Bianchini}, \citenamefont
  {Brodwin}, \citenamefont {Brooks}, \citenamefont {Bryant}, \citenamefont
  {Burke}, \citenamefont {Canning}, \citenamefont {Carlstrom}, \citenamefont
  {Carretero}, \citenamefont {Castander}, \citenamefont {Chang}, \citenamefont
  {Chaubal}, \citenamefont {Chiang}, \citenamefont {Chou}, \citenamefont
  {Citron}, \citenamefont {Corbett~Moran}, \citenamefont {Costanzi},
  \citenamefont {Crawford}, \citenamefont {Crites}, \citenamefont {{da Costa}},
  \citenamefont {Pereira}, \citenamefont {Davis}, \citenamefont {{de Haan}},
  \citenamefont {Dobbs}, \citenamefont {Doel}, \citenamefont {Everett},
  \citenamefont {Farahi}, \citenamefont {Flaugher}, \citenamefont {Flores},
  \citenamefont {Floyd}, \citenamefont {Gallicchio}, \citenamefont {Gaztanaga},
  \citenamefont {George}, \citenamefont {Gladders}, \citenamefont {Gupta},
  \citenamefont {Gutierrez}, \citenamefont {Halverson}, \citenamefont {Hinton},
  \citenamefont {{Hlavacek-Larrondo}}, \citenamefont {Holder}, \citenamefont
  {Hollowood}, \citenamefont {Holzapfel}, \citenamefont {Hrubes}, \citenamefont
  {Huang}, \citenamefont {Hubmayr}, \citenamefont {Irwin}, \citenamefont
  {James}, \citenamefont {K{\'e}ruzor{\'e}}, \citenamefont {Khullar},
  \citenamefont {Kim}, \citenamefont {Knox}, \citenamefont {Kraft},
  \citenamefont {Kuehn}, \citenamefont {Lahav}, \citenamefont {Lee},
  \citenamefont {Lee}, \citenamefont {Li}, \citenamefont {Lidman},
  \citenamefont {Lima}, \citenamefont {Lowitz}, \citenamefont {Mahler},
  \citenamefont {Mantz}, \citenamefont {Marshall}, \citenamefont {McDonald},
  \citenamefont {McMahon}, \citenamefont {{Mena-Fern{\'a}ndez}}, \citenamefont
  {Meyer}, \citenamefont {Miquel}, \citenamefont {Montgomery}, \citenamefont
  {Natoli}, \citenamefont {Nibarger}, \citenamefont {Noble}, \citenamefont
  {Novosad}, \citenamefont {Ogando}, \citenamefont {Padin}, \citenamefont
  {Paschos}, \citenamefont {Patil}, \citenamefont {Plazas~Malag{\'o}n},
  \citenamefont {Pryke}, \citenamefont {Reichardt}, \citenamefont {Roberson},
  \citenamefont {Romer}, \citenamefont {Romero}, \citenamefont {Ruhl},
  \citenamefont {Saliwanchik}, \citenamefont {Salvati}, \citenamefont
  {Samuroff}, \citenamefont {Sanchez}, \citenamefont {Santiago}, \citenamefont
  {Sarkar}, \citenamefont {Saro}, \citenamefont {Schaffer}, \citenamefont
  {Sharon}, \citenamefont {Sievers}, \citenamefont {Smecher}, \citenamefont
  {Smith}, \citenamefont {Somboonpanyakul}, \citenamefont {Sommer},
  \citenamefont {Stalder}, \citenamefont {Stark},\ and\ \citenamefont
  {Stephen}}]{bocquetMultiprobeCosmologyAbundance2025}%
  \BibitemOpen
  \bibfield  {author} {\bibinfo {author} {\bibfnamefont {S.}~\bibnamefont
  {Bocquet}}, \bibinfo {author} {\bibfnamefont {S.}~\bibnamefont {Grandis}},
  \bibinfo {author} {\bibfnamefont {E.}~\bibnamefont {Krause}}, \emph
  {et~al.},\ }\href {https://doi.org/10.1103/PhysRevD.111.063533} {\bibfield
  {journal} {\bibinfo  {journal} {Phys. Rev. D}\ }\textbf {\bibinfo {volume}
  {111}},\ \bibinfo {pages} {063533} (\bibinfo {year} {2025})}\BibitemShut
  {NoStop}%
\bibitem [{\citenamefont {Grandis}\ \emph
  {et~al.}(2024{\natexlab{a}})\citenamefont {Grandis}, \citenamefont
  {Ghirardini}, \citenamefont {Bocquet}, \citenamefont {Garrel}, \citenamefont
  {Mohr}, \citenamefont {Liu}, \citenamefont {Kluge}, \citenamefont {Kimmig},
  \citenamefont {Reiprich}, \citenamefont {Alarcon}, \citenamefont {Amon},
  \citenamefont {Artis}, \citenamefont {Bahar}, \citenamefont {Balzer},
  \citenamefont {Bechtol}, \citenamefont {Becker}, \citenamefont {Bernstein},
  \citenamefont {Bulbul}, \citenamefont {Campos}, \citenamefont
  {Carnero~Rosell}, \citenamefont {Carrasco~Kind}, \citenamefont {Cawthon},
  \citenamefont {Chang}, \citenamefont {Chen}, \citenamefont {Chiu},
  \citenamefont {Choi}, \citenamefont {Clerc}, \citenamefont {Comparat},
  \citenamefont {Cordero}, \citenamefont {Davis}, \citenamefont {Derose},
  \citenamefont {Diehl}, \citenamefont {Dodelson}, \citenamefont {Doux},
  \citenamefont {{Drlica-Wagner}}, \citenamefont {Eckert}, \citenamefont
  {{Elvin-Poole}}, \citenamefont {Everett}, \citenamefont {Ferte},
  \citenamefont {Gatt}, \citenamefont {Giannini}, \citenamefont {Giles},
  \citenamefont {Gruen}, \citenamefont {Gruendl}, \citenamefont {Harrison},
  \citenamefont {Hartley}, \citenamefont {Herner}, \citenamefont {Huf},
  \citenamefont {Kleinebreil}, \citenamefont {Kuropatkin}, \citenamefont
  {Leget}, \citenamefont {Maccrann}, \citenamefont {Mccullough}, \citenamefont
  {Merloni}, \citenamefont {Myles}, \citenamefont {Nandra}, \citenamefont
  {{Navarro-Alsina}}, \citenamefont {Okabe}, \citenamefont {Pacaud},
  \citenamefont {Pandey}, \citenamefont {Prat}, \citenamefont {Predehl},
  \citenamefont {Ramos}, \citenamefont {Raveri}, \citenamefont {Rollins},
  \citenamefont {Roodman}, \citenamefont {Ross}, \citenamefont {Rykoff},
  \citenamefont {Sanchez}, \citenamefont {Sanders}, \citenamefont {Schrabback},
  \citenamefont {Secco}, \citenamefont {Seppi}, \citenamefont
  {{Sevilla-Noarbe}}, \citenamefont {Sheldon}, \citenamefont {Shin},
  \citenamefont {Troxel}, \citenamefont {Tutusaus}, \citenamefont {Varga},
  \citenamefont {Wu}, \citenamefont {Yanny}, \citenamefont {Yin}, \citenamefont
  {Zhang}, \citenamefont {Zhang}, \citenamefont {Alves}, \citenamefont
  {Bhargava}, \citenamefont {Brooks}, \citenamefont {Burke}, \citenamefont
  {Carretero}, \citenamefont {Costanzi}, \citenamefont {{da Costa}},
  \citenamefont {Pereira}, \citenamefont {De~Vicente}, \citenamefont {Desai},
  \citenamefont {Doel}, \citenamefont {Ferrero}, \citenamefont {Flaugher},
  \citenamefont {Friedel}, \citenamefont {Frieman}, \citenamefont
  {{Garc{\'i}a-Bellido}}, \citenamefont {Gutierrez}, \citenamefont {Hinton},
  \citenamefont {Hollowood}, \citenamefont {Honscheid}, \citenamefont {James},
  \citenamefont {Jeffrey}, \citenamefont {Lahav}, \citenamefont {Lee},
  \citenamefont {Marshall}, \citenamefont {Menanteau}, \citenamefont {Ogando},
  \citenamefont {Pieres}, \citenamefont {Plazas~Malag{\'o}n}, \citenamefont
  {Romer}, \citenamefont {Sanchez}, \citenamefont {Schubnell}, \citenamefont
  {Smith}, \citenamefont {Suchyta}, \citenamefont {Swanson}, \citenamefont
  {Tarle}, \citenamefont {Weaverdyck},\ and\ \citenamefont
  {Weller}}]{grandisSRGEROSITAAllSky2024}%
  \BibitemOpen
  \bibfield  {author} {\bibinfo {author} {\bibfnamefont {S.}~\bibnamefont
  {Grandis}}, \bibinfo {author} {\bibfnamefont {V.}~\bibnamefont {Ghirardini}},
  \bibinfo {author} {\bibfnamefont {S.}~\bibnamefont {Bocquet}}, \emph
  {et~al.},\ }\href {https://doi.org/10.48550/arXiv.2402.08455} {\enquote
  {\bibinfo {title} {The {{SRG}}/{{eROSITA All-Sky Survey}}: {{Dark Energy
  Survey Year}} 3 {{Weak Gravitational Lensing}} by {{eRASS1}} selected
  {{Galaxy Clusters}}},}\ } (\bibinfo {year} {2024}{\natexlab{a}})\BibitemShut
  {NoStop}%
\bibitem [{\citenamefont {Ghirardini}\ \emph {et~al.}(2024)\citenamefont
  {Ghirardini}, \citenamefont {Bulbul}, \citenamefont {Artis}, \citenamefont
  {Clerc}, \citenamefont {Garrel}, \citenamefont {Grandis}, \citenamefont
  {Kluge}, \citenamefont {Liu}, \citenamefont {Bahar}, \citenamefont {Balzer},
  \citenamefont {Chiu}, \citenamefont {Comparat}, \citenamefont {Gruen},
  \citenamefont {Kleinebreil}, \citenamefont {Krippendorf}, \citenamefont
  {Merloni}, \citenamefont {Nandra}, \citenamefont {Okabe}, \citenamefont
  {Pacaud}, \citenamefont {Predehl}, \citenamefont {{Ramos-Ceja}},
  \citenamefont {Reiprich}, \citenamefont {Sanders}, \citenamefont
  {Schrabback}, \citenamefont {Seppi}, \citenamefont {Zelmer}, \citenamefont
  {Zhang}, \citenamefont {Bornemann}, \citenamefont {Brunner}, \citenamefont
  {Burwitz}, \citenamefont {Coutinho}, \citenamefont {Dennerl}, \citenamefont
  {Freyberg}, \citenamefont {Friedrich}, \citenamefont {Gaida}, \citenamefont
  {Gueguen}, \citenamefont {Haberl}, \citenamefont {Kink}, \citenamefont
  {Lamer}, \citenamefont {Li}, \citenamefont {Liu}, \citenamefont {Maitra},
  \citenamefont {Meidinger}, \citenamefont {Mueller}, \citenamefont {Miyatake},
  \citenamefont {Miyazaki}, \citenamefont {Robrade}, \citenamefont {Schwope},\
  and\ \citenamefont {Stewart}}]{ghirardiniSRGEROSITAAllSky2024}%
  \BibitemOpen
  \bibfield  {author} {\bibinfo {author} {\bibfnamefont {V.}~\bibnamefont
  {Ghirardini}}, \bibinfo {author} {\bibfnamefont {E.}~\bibnamefont {Bulbul}},
  \bibinfo {author} {\bibfnamefont {E.}~\bibnamefont {Artis}}, \emph {et~al.},\
  }\href@noop {} {\enquote {\bibinfo {title} {The {{SRG}}/{{eROSITA All-Sky
  Survey}}: {{Cosmology Constraints}} from {{Cluster Abundances}} in the
  {{Western Galactic Hemisphere}}},}\ } (\bibinfo {year} {2024})\BibitemShut
  {NoStop}%
\bibitem [{\citenamefont {Costanzi}\ \emph {et~al.}(2021)\citenamefont
  {Costanzi}, \citenamefont {Saro}, \citenamefont {Bocquet}, \citenamefont
  {Abbott}, \citenamefont {Aguena}, \citenamefont {Allam}, \citenamefont
  {Amara}, \citenamefont {Annis}, \citenamefont {Avila}, \citenamefont {Bacon},
  \citenamefont {Benson}, \citenamefont {Bhargava}, \citenamefont {Brooks},
  \citenamefont {{Buckley-Geer}}, \citenamefont {Burke}, \citenamefont
  {Rosell}, \citenamefont {Kind}, \citenamefont {Carretero}, \citenamefont
  {Choi}, \citenamefont {{da Costa}}, \citenamefont {Pereira}, \citenamefont
  {De~Vicente}, \citenamefont {Desai}, \citenamefont {Diehl}, \citenamefont
  {Dietrich}, \citenamefont {Doel}, \citenamefont {Eifler}, \citenamefont
  {Everett}, \citenamefont {Ferrero}, \citenamefont {Fert{\'e}}, \citenamefont
  {Flaugher}, \citenamefont {Fosalba}, \citenamefont {Frieman}, \citenamefont
  {{Garc{\'i}a-Bellido}}, \citenamefont {Gaztanaga}, \citenamefont {Gerdes},
  \citenamefont {Giannantonio}, \citenamefont {Giles}, \citenamefont {Grandis},
  \citenamefont {Gruen}, \citenamefont {Gruendl}, \citenamefont {Gupta},
  \citenamefont {Gutierrez}, \citenamefont {Hartley}, \citenamefont {Hinton},
  \citenamefont {Hollowood}, \citenamefont {Honscheid}, \citenamefont {James},
  \citenamefont {Jeltema}, \citenamefont {Krause}, \citenamefont {Kuehn},
  \citenamefont {Kuropatkin}, \citenamefont {Lahav}, \citenamefont {Lima},
  \citenamefont {MacCrann}, \citenamefont {Maia}, \citenamefont {Marshall},
  \citenamefont {Menanteau}, \citenamefont {Miquel}, \citenamefont {Mohr},
  \citenamefont {Morgan}, \citenamefont {Myles}, \citenamefont {Ogando},
  \citenamefont {Palmese}, \citenamefont {{Paz-Chinch{\'o}n}}, \citenamefont
  {Plazas}, \citenamefont {Rapetti}, \citenamefont {Reichardt}, \citenamefont
  {Romer}, \citenamefont {Roodman}, \citenamefont {Ruppin}, \citenamefont
  {Salvati}, \citenamefont {Samuroff}, \citenamefont {Sanchez}, \citenamefont
  {Scarpine}, \citenamefont {Serrano}, \citenamefont {{Sevilla-Noarbe}},
  \citenamefont {Singh}, \citenamefont {Smith}, \citenamefont
  {{Soares-Santos}}, \citenamefont {Stark}, \citenamefont {Suchyta},
  \citenamefont {Swanson}, \citenamefont {Tarle}, \citenamefont {Thomas},
  \citenamefont {To}, \citenamefont {Tucker}, \citenamefont {Varga},
  \citenamefont {Wechsler},\ and\ \citenamefont
  {Zhang}}]{costanziCosmologicalConstraintsY12021}%
  \BibitemOpen
  \bibfield  {author} {\bibinfo {author} {\bibfnamefont {M.}~\bibnamefont
  {Costanzi}}, \bibinfo {author} {\bibfnamefont {A.}~\bibnamefont {Saro}},
  \bibinfo {author} {\bibfnamefont {S.}~\bibnamefont {Bocquet}}, \emph
  {et~al.},\ }\href {https://doi.org/10.1103/PhysRevD.103.043522} {\bibfield
  {journal} {\bibinfo  {journal} {Phys. Rev. D}\ }\textbf {\bibinfo {volume}
  {103}},\ \bibinfo {pages} {043522} (\bibinfo {year} {2021})}\BibitemShut
  {NoStop}%
\bibitem [{\citenamefont {Grandis}\ \emph {et~al.}(2021)\citenamefont
  {Grandis}, \citenamefont {Mohr}, \citenamefont {Costanzi}, \citenamefont
  {Saro}, \citenamefont {Bocquet}, \citenamefont {Klein}, \citenamefont
  {Aguena}, \citenamefont {Allam}, \citenamefont {Annis}, \citenamefont
  {Ansarinejad}, \citenamefont {Bacon}, \citenamefont {Bertin}, \citenamefont
  {Bleem}, \citenamefont {Brooks}, \citenamefont {Burke}, \citenamefont
  {Rosel}, \citenamefont {Kind}, \citenamefont {Carretero}, \citenamefont
  {Castander}, \citenamefont {Choi}, \citenamefont {{da Costa}}, \citenamefont
  {De~Vincente}, \citenamefont {Desai}, \citenamefont {Diehl}, \citenamefont
  {Dietrich}, \citenamefont {Doel}, \citenamefont {Eifler}, \citenamefont
  {Everett}, \citenamefont {Ferrero}, \citenamefont {Floyd}, \citenamefont
  {Fosalba}, \citenamefont {Frieman}, \citenamefont {{Garc{\'i}a-Bellido}},
  \citenamefont {Gaztanaga}, \citenamefont {Gruen}, \citenamefont {Gruendl},
  \citenamefont {Gschwend}, \citenamefont {Gupta}, \citenamefont {Gutierrez},
  \citenamefont {Hinton}, \citenamefont {Hollowood}, \citenamefont {Honscheid},
  \citenamefont {James}, \citenamefont {Jeltema}, \citenamefont {Kuehn},
  \citenamefont {Lahav}, \citenamefont {Lidman}, \citenamefont {Lima},
  \citenamefont {Maia}, \citenamefont {March}, \citenamefont {Marshall},
  \citenamefont {Melchior}, \citenamefont {Menanteau}, \citenamefont {Miquel},
  \citenamefont {Morgan}, \citenamefont {Myles}, \citenamefont {Ogando},
  \citenamefont {Palmese}, \citenamefont {{Paz-Chinch{\'o}n}}, \citenamefont
  {Plazas}, \citenamefont {Reichardt}, \citenamefont {Romer}, \citenamefont
  {Sanchez}, \citenamefont {Scarpine}, \citenamefont {Serrano}, \citenamefont
  {{Sevilla-Noarbe}}, \citenamefont {Singh}, \citenamefont {Smith},
  \citenamefont {Suchyta}, \citenamefont {Swanson}, \citenamefont {Tarle},
  \citenamefont {Thomas}, \citenamefont {To}, \citenamefont {Weller},
  \citenamefont {Wilkinson},\ and\ \citenamefont
  {Wu}}]{grandisExploringContaminationDESY12021}%
  \BibitemOpen
  \bibfield  {author} {\bibinfo {author} {\bibfnamefont {S.}~\bibnamefont
  {Grandis}}, \bibinfo {author} {\bibfnamefont {J.~J.}\ \bibnamefont {Mohr}},
  \bibinfo {author} {\bibfnamefont {M.}~\bibnamefont {Costanzi}}, \emph
  {et~al.},\ }\href {https://doi.org/10.1093/mnras/stab869} {\bibfield
  {journal} {\bibinfo  {journal} {Mon. Not. R. Astron. Soc.}\ }\textbf
  {\bibinfo {volume} {504}},\ \bibinfo {pages} {1253} (\bibinfo {year}
  {2021})}\BibitemShut {NoStop}%
\bibitem [{\citenamefont {Grandis}\ \emph {et~al.}(2025)\citenamefont
  {Grandis}, \citenamefont {Costanzi}, \citenamefont {Mohr}, \citenamefont
  {Bleem}, \citenamefont {Wu}, \citenamefont {Aguena}, \citenamefont {Allam},
  \citenamefont {{Andrade-Oliveira}}, \citenamefont {Bocquet}, \citenamefont
  {Brooks}, \citenamefont {Carnero~Rosell}, \citenamefont {Carretero},
  \citenamefont {{da Costa}}, \citenamefont {Pereira}, \citenamefont {Davis},
  \citenamefont {Desai}, \citenamefont {Diehl}, \citenamefont {Doel},
  \citenamefont {Everett}, \citenamefont {Flaugher}, \citenamefont {Frieman},
  \citenamefont {{Garc{\'i}a-Bellido}}, \citenamefont {Gaztanaga},
  \citenamefont {Gruen}, \citenamefont {Gruendl}, \citenamefont {Gutierrez},
  \citenamefont {Hinton}, \citenamefont {{Hlacacek-Larrondo}}, \citenamefont
  {Hollowood}, \citenamefont {Honscheid}, \citenamefont {James}, \citenamefont
  {Klein}, \citenamefont {Marshall}, \citenamefont {{Mena-Fern{\'a}ndez}},
  \citenamefont {Miquel}, \citenamefont {Palmese}, \citenamefont
  {Plazas~Malag{\'o}n}, \citenamefont {Reichardt}, \citenamefont {Romer},
  \citenamefont {Samuroff}, \citenamefont {Sanchez~Cid}, \citenamefont
  {Sanchez}, \citenamefont {Santiago}, \citenamefont {Saro}, \citenamefont
  {{Sevilla-Noarbe}}, \citenamefont {Smith}, \citenamefont {{Soares-Santos}},
  \citenamefont {Sommer}, \citenamefont {Suchyta}, \citenamefont {Tarle},
  \citenamefont {To}, \citenamefont {Tucker}, \citenamefont {Weaverdyck},
  \citenamefont {Weller},\ and\ \citenamefont
  {Wiseman}}]{grandisSelectionFunctionClusters2025}%
  \BibitemOpen
  \bibfield  {author} {\bibinfo {author} {\bibfnamefont {S.}~\bibnamefont
  {Grandis}}, \bibinfo {author} {\bibfnamefont {M.}~\bibnamefont {Costanzi}},
  \bibinfo {author} {\bibfnamefont {J.~J.}\ \bibnamefont {Mohr}}, \emph
  {et~al.},\ }\href {https://doi.org/10.1051/0004-6361/202554177} {\bibfield
  {journal} {\bibinfo  {journal} {Astron. Astrophys.}\ }\textbf {\bibinfo
  {volume} {700}},\ \bibinfo {pages} {A15} (\bibinfo {year}
  {2025})}\BibitemShut {NoStop}%
\bibitem [{\citenamefont {Zhou}\ \emph {et~al.}(2024)\citenamefont {Zhou},
  \citenamefont {Wu}, \citenamefont {Salcedo}, \citenamefont {Grandis},
  \citenamefont {Jeltema}, \citenamefont {Leauthaud}, \citenamefont {Costanzi},
  \citenamefont {Sunayama}, \citenamefont {Weinberg}, \citenamefont {Zhang},
  \citenamefont {Rozo}, \citenamefont {To}, \citenamefont {Bocquet},
  \citenamefont {Varga},\ and\ \citenamefont
  {Kwiecien}}]{zhouForecastingConstraintsOptical2024}%
  \BibitemOpen
  \bibfield  {author} {\bibinfo {author} {\bibfnamefont {C.}~\bibnamefont
  {Zhou}}, \bibinfo {author} {\bibfnamefont {H.-Y.}\ \bibnamefont {Wu}},
  \bibinfo {author} {\bibfnamefont {A.~N.}\ \bibnamefont {Salcedo}}, \emph
  {et~al.},\ }\href {https://doi.org/10.1103/PhysRevD.110.103508} {\bibfield
  {journal} {\bibinfo  {journal} {Phys. Rev. D}\ }\textbf {\bibinfo {volume}
  {110}},\ \bibinfo {pages} {103508} (\bibinfo {year} {2024})}\BibitemShut
  {NoStop}%
\bibitem [{\citenamefont {Bartelmann}\ and\ \citenamefont
  {Schneider}(2001)}]{bartelmannWeakGravitationalLensing2001}%
  \BibitemOpen
  \bibfield  {author} {\bibinfo {author} {\bibfnamefont {M.}~\bibnamefont
  {Bartelmann}}\ and\ \bibinfo {author} {\bibfnamefont {P.}~\bibnamefont
  {Schneider}},\ }\href {https://doi.org/10.1016/S0370-1573(00)00082-X}
  {\bibfield  {journal} {\bibinfo  {journal} {Phys. Rep.}\ }\textbf {\bibinfo
  {volume} {340}},\ \bibinfo {pages} {291} (\bibinfo {year}
  {2001})}\BibitemShut {NoStop}%
\bibitem [{\citenamefont {Johnston}\ \emph {et~al.}(2007)\citenamefont
  {Johnston}, \citenamefont {Sheldon}, \citenamefont {Wechsler}, \citenamefont
  {Rozo}, \citenamefont {Koester}, \citenamefont {Frieman}, \citenamefont
  {McKay}, \citenamefont {Evrard}, \citenamefont {Becker},\ and\ \citenamefont
  {Annis}}]{johnstonCrosscorrelationWeakLensing2007}%
  \BibitemOpen
  \bibfield  {author} {\bibinfo {author} {\bibfnamefont {D.~E.}\ \bibnamefont
  {Johnston}}, \bibinfo {author} {\bibfnamefont {E.~S.}\ \bibnamefont
  {Sheldon}}, \bibinfo {author} {\bibfnamefont {R.~H.}\ \bibnamefont
  {Wechsler}}, \emph {et~al.},\ }\href
  {https://doi.org/10.48550/arXiv.0709.1159} {\enquote {\bibinfo {title}
  {Cross-correlation {{Weak Lensing}} of {{SDSS}} galaxy {{Clusters II}}:
  {{Cluster Density Profiles}} and the {{Mass}}--{{Richness Relation}}},}\ }
  (\bibinfo {year} {2007})\BibitemShut {NoStop}%
\bibitem [{\citenamefont {Mandelbaum}\ \emph {et~al.}(2010)\citenamefont
  {Mandelbaum}, \citenamefont {Seljak}, \citenamefont {Baldauf},\ and\
  \citenamefont {Smith}}]{mandelbaumPrecisionClusterMass2010}%
  \BibitemOpen
  \bibfield  {author} {\bibinfo {author} {\bibfnamefont {R.}~\bibnamefont
  {Mandelbaum}}, \bibinfo {author} {\bibfnamefont {U.}~\bibnamefont {Seljak}},
  \bibinfo {author} {\bibfnamefont {T.}~\bibnamefont {Baldauf}},\ and\ \bibinfo
  {author} {\bibfnamefont {R.~E.}\ \bibnamefont {Smith}},\ }\href
  {https://doi.org/10.1111/j.1365-2966.2010.16619.x} {\bibfield  {journal}
  {\bibinfo  {journal} {Mon. Not. R. Astron. Soc.}\ }\textbf {\bibinfo {volume}
  {405}},\ \bibinfo {pages} {2078} (\bibinfo {year} {2010})}\BibitemShut
  {NoStop}%
\bibitem [{\citenamefont {Rozo}\ \emph {et~al.}(2011)\citenamefont {Rozo},
  \citenamefont {Wu},\ and\ \citenamefont
  {Schmidt}}]{rozoStackedWeakLensing2011}%
  \BibitemOpen
  \bibfield  {author} {\bibinfo {author} {\bibfnamefont {E.}~\bibnamefont
  {Rozo}}, \bibinfo {author} {\bibfnamefont {H.-Y.}\ \bibnamefont {Wu}},\ and\
  \bibinfo {author} {\bibfnamefont {F.}~\bibnamefont {Schmidt}},\ }\href
  {https://doi.org/10.1088/0004-637X/735/2/118} {\bibfield  {journal} {\bibinfo
   {journal} {Astrophys. J.}\ }\textbf {\bibinfo {volume} {735}},\ \bibinfo
  {pages} {118} (\bibinfo {year} {2011})}\BibitemShut {NoStop}%
\bibitem [{\citenamefont {Aguena}\ \emph
  {et~al.}(2021{\natexlab{b}})\citenamefont {Aguena}, \citenamefont {Avestruz},
  \citenamefont {Combet}, \citenamefont {Fu}, \citenamefont {Herbonnet},
  \citenamefont {Malz}, \citenamefont {{Penna-Lima}}, \citenamefont {Ricci},
  \citenamefont {Vitenti}, \citenamefont {Baumont}, \citenamefont {Fan},
  \citenamefont {Fong}, \citenamefont {Ho}, \citenamefont {Kirby},
  \citenamefont {Payerne}, \citenamefont {Boutigny}, \citenamefont {Lee},
  \citenamefont {Liu}, \citenamefont {McClintock}, \citenamefont {Miyatake},
  \citenamefont {Sif{\'o}n}, \citenamefont {{von der Linden}}, \citenamefont
  {Wu}, \citenamefont {Yoon},\ and\ \citenamefont
  {Collaboration}}]{aguenaCLMMLSSTDESCCluster2021}%
  \BibitemOpen
  \bibfield  {author} {\bibinfo {author} {\bibfnamefont {M.}~\bibnamefont
  {Aguena}}, \bibinfo {author} {\bibfnamefont {C.}~\bibnamefont {Avestruz}},
  \bibinfo {author} {\bibfnamefont {C.}~\bibnamefont {Combet}}, \emph
  {et~al.},\ }\href {https://doi.org/10.1093/mnras/stab2764} {\bibfield
  {journal} {\bibinfo  {journal} {Mon. Not. R. Astron. Soc.}\ }\textbf
  {\bibinfo {volume} {508}},\ \bibinfo {pages} {6092} (\bibinfo {year}
  {2021}{\natexlab{b}})}\BibitemShut {NoStop}%
\bibitem [{\citenamefont {Wu}\ \emph {et~al.}(2021)\citenamefont {Wu},
  \citenamefont {Weinberg}, \citenamefont {Salcedo},\ and\ \citenamefont
  {Wibking}}]{wuCosmologyGalaxyCluster2021}%
  \BibitemOpen
  \bibfield  {author} {\bibinfo {author} {\bibfnamefont {H.-Y.}\ \bibnamefont
  {Wu}}, \bibinfo {author} {\bibfnamefont {D.~H.}\ \bibnamefont {Weinberg}},
  \bibinfo {author} {\bibfnamefont {A.~N.}\ \bibnamefont {Salcedo}},\ and\
  \bibinfo {author} {\bibfnamefont {B.~D.}\ \bibnamefont {Wibking}},\ }\href
  {https://doi.org/10.3847/1538-4357/abdc23} {\bibfield  {journal} {\bibinfo
  {journal} {Astrophys. J.}\ }\textbf {\bibinfo {volume} {910}},\ \bibinfo
  {pages} {28} (\bibinfo {year} {2021})}\BibitemShut {NoStop}%
\bibitem [{\citenamefont {Mantz}\ \emph {et~al.}(2016)\citenamefont {Mantz},
  \citenamefont {Allen}, \citenamefont {Morris}, \citenamefont {{von der
  Linden}}, \citenamefont {Applegate}, \citenamefont {Kelly}, \citenamefont
  {Burke}, \citenamefont {Donovan},\ and\ \citenamefont
  {Ebeling}}]{mantzWeighingGiantsGalaxy2016}%
  \BibitemOpen
  \bibfield  {author} {\bibinfo {author} {\bibfnamefont {A.~B.}\ \bibnamefont
  {Mantz}}, \bibinfo {author} {\bibfnamefont {S.~W.}\ \bibnamefont {Allen}},
  \bibinfo {author} {\bibfnamefont {R.~G.}\ \bibnamefont {Morris}}, \emph
  {et~al.},\ }\href {https://doi.org/10.1093/mnras/stw2250} {\bibfield
  {journal} {\bibinfo  {journal} {Mon. Not. R. Astron. Soc.}\ }\textbf
  {\bibinfo {volume} {463}},\ \bibinfo {pages} {3582} (\bibinfo {year}
  {2016})}\BibitemShut {NoStop}%
\bibitem [{\citenamefont {Melchior}\ \emph {et~al.}(2017)\citenamefont
  {Melchior}, \citenamefont {Gruen}, \citenamefont {McClintock}, \citenamefont
  {Varga}, \citenamefont {Sheldon}, \citenamefont {Rozo}, \citenamefont
  {Amara}, \citenamefont {Becker}, \citenamefont {Benson}, \citenamefont
  {Bermeo}, \citenamefont {Bridle}, \citenamefont {Clampitt}, \citenamefont
  {Dietrich}, \citenamefont {Hartley}, \citenamefont {Hollowood}, \citenamefont
  {Jain}, \citenamefont {Jarvis}, \citenamefont {Jeltema}, \citenamefont
  {Kacprzak}, \citenamefont {MacCrann}, \citenamefont {Rykoff}, \citenamefont
  {Saro}, \citenamefont {Suchyta}, \citenamefont {Troxel}, \citenamefont
  {Zuntz}, \citenamefont {Bonnett}, \citenamefont {Plazas}, \citenamefont
  {Abbott}, \citenamefont {Abdalla}, \citenamefont {Annis}, \citenamefont
  {{Benoit-L{\'e}vy}}, \citenamefont {Bernstein}, \citenamefont {Bertin},
  \citenamefont {Brooks}, \citenamefont {{Buckley-Geer}}, \citenamefont
  {Carnero~Rosell}, \citenamefont {Carrasco~Kind}, \citenamefont {Carretero},
  \citenamefont {Cunha}, \citenamefont {D'Andrea}, \citenamefont {{da Costa}},
  \citenamefont {Desai}, \citenamefont {Eifler}, \citenamefont {Flaugher},
  \citenamefont {Fosalba}, \citenamefont {{Garc{\'i}a-Bellido}}, \citenamefont
  {Gaztanaga}, \citenamefont {Gerdes}, \citenamefont {Gruendl}, \citenamefont
  {Gschwend}, \citenamefont {Gutierrez}, \citenamefont {Honscheid},
  \citenamefont {James}, \citenamefont {Kirk}, \citenamefont {Krause},
  \citenamefont {Kuehn}, \citenamefont {Kuropatkin}, \citenamefont {Lahav},
  \citenamefont {Lima}, \citenamefont {Maia}, \citenamefont {March},
  \citenamefont {Martini}, \citenamefont {Menanteau}, \citenamefont {Miller},
  \citenamefont {Miquel}, \citenamefont {Mohr}, \citenamefont {Nichol},
  \citenamefont {Ogando}, \citenamefont {Romer}, \citenamefont {Sanchez},
  \citenamefont {Scarpine}, \citenamefont {{Sevilla-Noarbe}}, \citenamefont
  {Smith}, \citenamefont {{Soares-Santos}}, \citenamefont {Sobreira},
  \citenamefont {Swanson}, \citenamefont {Tarle}, \citenamefont {Thomas},
  \citenamefont {Walker}, \citenamefont {Weller},\ and\ \citenamefont
  {Zhang}}]{melchiorWeaklensingMassCalibration2017}%
  \BibitemOpen
  \bibfield  {author} {\bibinfo {author} {\bibfnamefont {P.}~\bibnamefont
  {Melchior}}, \bibinfo {author} {\bibfnamefont {D.}~\bibnamefont {Gruen}},
  \bibinfo {author} {\bibfnamefont {T.}~\bibnamefont {McClintock}}, \emph
  {et~al.},\ }\href {https://doi.org/10.1093/mnras/stx1053} {\bibfield
  {journal} {\bibinfo  {journal} {Mon. Not. R. Astron. Soc.}\ }\textbf
  {\bibinfo {volume} {469}},\ \bibinfo {pages} {4899} (\bibinfo {year}
  {2017})}\BibitemShut {NoStop}%
\bibitem [{\citenamefont {Simet}\ \emph {et~al.}(2017)\citenamefont {Simet},
  \citenamefont {McClintock}, \citenamefont {Mandelbaum}, \citenamefont {Rozo},
  \citenamefont {Rykoff}, \citenamefont {Sheldon},\ and\ \citenamefont
  {Wechsler}}]{simetWeakLensingMeasurement2017}%
  \BibitemOpen
  \bibfield  {author} {\bibinfo {author} {\bibfnamefont {M.}~\bibnamefont
  {Simet}}, \bibinfo {author} {\bibfnamefont {T.}~\bibnamefont {McClintock}},
  \bibinfo {author} {\bibfnamefont {R.}~\bibnamefont {Mandelbaum}}, \emph
  {et~al.},\ }\href {https://doi.org/10.1093/mnras/stw3250} {\bibfield
  {journal} {\bibinfo  {journal} {Mon. Not. R. Astron. Soc.}\ }\textbf
  {\bibinfo {volume} {466}},\ \bibinfo {pages} {3103} (\bibinfo {year}
  {2017})}\BibitemShut {NoStop}%
\bibitem [{\citenamefont {McClintock}\ \emph {et~al.}(2019)\citenamefont
  {McClintock}, \citenamefont {Varga}, \citenamefont {Gruen}, \citenamefont
  {Rozo}, \citenamefont {Rykoff}, \citenamefont {Shin}, \citenamefont
  {Melchior}, \citenamefont {DeRose}, \citenamefont {Seitz}, \citenamefont
  {Dietrich}, \citenamefont {Sheldon}, \citenamefont {Zhang}, \citenamefont
  {{von der Linden}}, \citenamefont {Jeltema}, \citenamefont {Mantz},
  \citenamefont {Romer}, \citenamefont {Allen}, \citenamefont {Becker},
  \citenamefont {Bermeo}, \citenamefont {Bhargava}, \citenamefont {Costanzi},
  \citenamefont {Everett}, \citenamefont {Farahi}, \citenamefont {Hamaus},
  \citenamefont {Hartley}, \citenamefont {Hollowood}, \citenamefont {Hoyle},
  \citenamefont {Israel}, \citenamefont {Li}, \citenamefont {MacCrann},
  \citenamefont {Morris}, \citenamefont {Palmese}, \citenamefont {Plazas},
  \citenamefont {Pollina}, \citenamefont {Rau}, \citenamefont {Simet},
  \citenamefont {{Soares-Santos}}, \citenamefont {Troxel}, \citenamefont
  {Cervantes}, \citenamefont {Wechsler}, \citenamefont {Zuntz}, \citenamefont
  {Abbott}, \citenamefont {Abdalla}, \citenamefont {Allam}, \citenamefont
  {Annis}, \citenamefont {Avila}, \citenamefont {Bridle}, \citenamefont
  {Brooks}, \citenamefont {Burke}, \citenamefont {Rosell}, \citenamefont
  {Kind}, \citenamefont {Carretero}, \citenamefont {Castander}, \citenamefont
  {Crocce}, \citenamefont {Cunha}, \citenamefont {D'Andrea}, \citenamefont {{da
  Costa}}, \citenamefont {Davis}, \citenamefont {De~Vicente}, \citenamefont
  {Diehl}, \citenamefont {Doel}, \citenamefont {{Drlica-Wagner}}, \citenamefont
  {Evrard}, \citenamefont {Flaugher}, \citenamefont {Fosalba}, \citenamefont
  {Frieman}, \citenamefont {{Garc{\'i}a-Bellido}}, \citenamefont {Gaztanaga},
  \citenamefont {Gerdes}, \citenamefont {Giannantonio}, \citenamefont
  {Gruendl}, \citenamefont {Gutierrez}, \citenamefont {Honscheid},
  \citenamefont {James}, \citenamefont {Kirk}, \citenamefont {Krause},
  \citenamefont {Kuehn}, \citenamefont {Lahav}, \citenamefont {Li},
  \citenamefont {Lima}, \citenamefont {March}, \citenamefont {Marshall},
  \citenamefont {Menanteau}, \citenamefont {Miquel}, \citenamefont {Mohr},
  \citenamefont {Nord}, \citenamefont {Ogando}, \citenamefont {Roodman},
  \citenamefont {Sanchez}, \citenamefont {Scarpine}, \citenamefont {Schindler},
  \citenamefont {{Sevilla-Noarbe}}, \citenamefont {Smith}, \citenamefont
  {Smith}, \citenamefont {Sobreira}, \citenamefont {Suchyta}, \citenamefont
  {Swanson}, \citenamefont {Tarle}, \citenamefont {Tucker}, \citenamefont
  {Vikram}, \citenamefont {Walker},\ and\ \citenamefont
  {Weller}}]{mcclintockDarkEnergySurvey2019}%
  \BibitemOpen
  \bibfield  {author} {\bibinfo {author} {\bibfnamefont {T.}~\bibnamefont
  {McClintock}}, \bibinfo {author} {\bibfnamefont {T.~N.}\ \bibnamefont
  {Varga}}, \bibinfo {author} {\bibfnamefont {D.}~\bibnamefont {Gruen}}, \emph
  {et~al.},\ }\href {https://doi.org/10.1093/mnras/sty2711} {\bibfield
  {journal} {\bibinfo  {journal} {Mon. Not. R. Astron. Soc.}\ }\textbf
  {\bibinfo {volume} {482}},\ \bibinfo {pages} {1352} (\bibinfo {year}
  {2019})}\BibitemShut {NoStop}%
\bibitem [{\citenamefont {Miyatake}\ \emph {et~al.}(2019)\citenamefont
  {Miyatake}, \citenamefont {Battaglia}, \citenamefont {Hilton}, \citenamefont
  {Medezinski}, \citenamefont {Nishizawa}, \citenamefont {More}, \citenamefont
  {Aiola}, \citenamefont {Bahcall}, \citenamefont {Bond}, \citenamefont
  {Calabrese}, \citenamefont {Choi}, \citenamefont {Devlin}, \citenamefont
  {Dunkley}, \citenamefont {Dunner}, \citenamefont {Fuzia}, \citenamefont
  {Gallardo}, \citenamefont {Gralla}, \citenamefont {Hasselfield},
  \citenamefont {Halpern}, \citenamefont {Hikage}, \citenamefont {Hill},
  \citenamefont {Hincks}, \citenamefont {Hlo{\v z}ek}, \citenamefont
  {Huffenberger}, \citenamefont {Hughes}, \citenamefont {Koopman},
  \citenamefont {Kosowsky}, \citenamefont {Louis}, \citenamefont
  {Madhavacheril}, \citenamefont {McMahon}, \citenamefont {Mandelbaum},
  \citenamefont {Marriage}, \citenamefont {Maurin}, \citenamefont {Miyazaki},
  \citenamefont {Moodley}, \citenamefont {Murata}, \citenamefont {Naess},
  \citenamefont {Newburgh}, \citenamefont {Niemack}, \citenamefont
  {Nishimichi}, \citenamefont {Okabe}, \citenamefont {Oguri}, \citenamefont
  {Osato}, \citenamefont {Page}, \citenamefont {Partridge}, \citenamefont
  {Robertson}, \citenamefont {Sehgal}, \citenamefont {Sherwin}, \citenamefont
  {Shirasaki}, \citenamefont {Sievers}, \citenamefont {Sif{\'o}n},
  \citenamefont {Simon}, \citenamefont {Spergel}, \citenamefont {Staggs},
  \citenamefont {Stein}, \citenamefont {Takada}, \citenamefont {Trac},
  \citenamefont {Umetsu}, \citenamefont {van Engelen},\ and\ \citenamefont
  {Wollack}}]{miyatakeWeaklensingMassCalibration2019}%
  \BibitemOpen
  \bibfield  {author} {\bibinfo {author} {\bibfnamefont {H.}~\bibnamefont
  {Miyatake}}, \bibinfo {author} {\bibfnamefont {N.}~\bibnamefont {Battaglia}},
  \bibinfo {author} {\bibfnamefont {M.}~\bibnamefont {Hilton}}, \emph
  {et~al.},\ }\href {https://doi.org/10.3847/1538-4357/ab0af0} {\bibfield
  {journal} {\bibinfo  {journal} {Astrophys. J.}\ }\textbf {\bibinfo {volume}
  {875}},\ \bibinfo {pages} {63} (\bibinfo {year} {2019})}\BibitemShut
  {NoStop}%
\bibitem [{\citenamefont {Mulroy}\ \emph {et~al.}(2019)\citenamefont {Mulroy},
  \citenamefont {Farahi}, \citenamefont {Evrard}, \citenamefont {Smith},
  \citenamefont {Finoguenov}, \citenamefont {O'Donnell}, \citenamefont
  {Marrone}, \citenamefont {Abdulla}, \citenamefont {Bourdin}, \citenamefont
  {Carlstrom}, \citenamefont {D{\'e}mocl{\`e}s}, \citenamefont {Haines},
  \citenamefont {Martino}, \citenamefont {Mazzotta}, \citenamefont {McGee},\
  and\ \citenamefont {Okabe}}]{mulroyLoCuSSScalingRelations2019}%
  \BibitemOpen
  \bibfield  {author} {\bibinfo {author} {\bibfnamefont {S.~L.}\ \bibnamefont
  {Mulroy}}, \bibinfo {author} {\bibfnamefont {A.}~\bibnamefont {Farahi}},
  \bibinfo {author} {\bibfnamefont {A.~E.}\ \bibnamefont {Evrard}}, \emph
  {et~al.},\ }\href {https://doi.org/10.1093/mnras/sty3484} {\bibfield
  {journal} {\bibinfo  {journal} {Mon. Not. R. Astron. Soc.}\ }\textbf
  {\bibinfo {volume} {484}},\ \bibinfo {pages} {60} (\bibinfo {year}
  {2019})}\BibitemShut {NoStop}%
\bibitem [{\citenamefont {Huang}\ \emph {et~al.}(2022)\citenamefont {Huang},
  \citenamefont {Leauthaud}, \citenamefont {Bradshaw}, \citenamefont {Hearin},
  \citenamefont {Behroozi}, \citenamefont {Lange}, \citenamefont {Greene},
  \citenamefont {DeRose}, \citenamefont {Speagle},\ and\ \citenamefont
  {Xhakaj}}]{huangOuterStellarMass2022}%
  \BibitemOpen
  \bibfield  {author} {\bibinfo {author} {\bibfnamefont {S.}~\bibnamefont
  {Huang}}, \bibinfo {author} {\bibfnamefont {A.}~\bibnamefont {Leauthaud}},
  \bibinfo {author} {\bibfnamefont {C.}~\bibnamefont {Bradshaw}}, \emph
  {et~al.},\ }\href {https://doi.org/10.1093/mnras/stac1680} {\bibfield
  {journal} {\bibinfo  {journal} {Mon. Not. R. Astron. Soc.}\ }\textbf
  {\bibinfo {volume} {515}},\ \bibinfo {pages} {4722} (\bibinfo {year}
  {2022})}\BibitemShut {NoStop}%
\bibitem [{\citenamefont {Chiu}\ \emph {et~al.}(2022)\citenamefont {Chiu},
  \citenamefont {Ghirardini}, \citenamefont {Liu}, \citenamefont {Grandis},
  \citenamefont {Bulbul}, \citenamefont {Bahar}, \citenamefont {Comparat},
  \citenamefont {Bocquet}, \citenamefont {Clerc}, \citenamefont {Klein},
  \citenamefont {Liu}, \citenamefont {Li}, \citenamefont {Miyatake},
  \citenamefont {Mohr}, \citenamefont {More}, \citenamefont {Oguri},
  \citenamefont {Okabe}, \citenamefont {Pacaud}, \citenamefont {{Ramos-Ceja}},
  \citenamefont {Reiprich}, \citenamefont {Schrabback},\ and\ \citenamefont
  {Umetsu}}]{chiuEROSITAFinalEquatorialDepth2022}%
  \BibitemOpen
  \bibfield  {author} {\bibinfo {author} {\bibfnamefont {I.~N.}\ \bibnamefont
  {Chiu}}, \bibinfo {author} {\bibfnamefont {V.}~\bibnamefont {Ghirardini}},
  \bibinfo {author} {\bibfnamefont {A.}~\bibnamefont {Liu}}, \emph {et~al.},\
  }\href {https://doi.org/10.1051/0004-6361/202141755} {\bibfield  {journal}
  {\bibinfo  {journal} {Astron. Astrophys.}\ }\textbf {\bibinfo {volume}
  {661}},\ \bibinfo {pages} {A11} (\bibinfo {year} {2022})}\BibitemShut
  {NoStop}%
\bibitem [{\citenamefont {Ansarinejad}\ \emph {et~al.}(2024)\citenamefont
  {Ansarinejad}, \citenamefont {Raghunathan}, \citenamefont {Abbott},
  \citenamefont {Ade}, \citenamefont {Aguena}, \citenamefont {Alves},
  \citenamefont {Anderson}, \citenamefont {{Andrade-Oliveira}}, \citenamefont
  {Archipley}, \citenamefont {Balkenhol}, \citenamefont {Benabed},
  \citenamefont {Bender}, \citenamefont {Benson}, \citenamefont {Bertin},
  \citenamefont {Bianchini}, \citenamefont {Bleem}, \citenamefont {Bocquet},
  \citenamefont {Bouchet}, \citenamefont {Brooks}, \citenamefont {Bryant},
  \citenamefont {Burke}, \citenamefont {Camphuis}, \citenamefont {Carlstrom},
  \citenamefont {Carnero~Rosell}, \citenamefont {Carretero}, \citenamefont
  {Castander}, \citenamefont {Cecil}, \citenamefont {Chang}, \citenamefont
  {Chaubal}, \citenamefont {Chichura}, \citenamefont {Chou}, \citenamefont
  {Coerver}, \citenamefont {Costanzi}, \citenamefont {Crawford}, \citenamefont
  {Cukierman}, \citenamefont {{da Costa}}, \citenamefont {Daley}, \citenamefont
  {Davis}, \citenamefont {{de Haan}}, \citenamefont {Desai}, \citenamefont
  {De~Vicente}, \citenamefont {Dibert}, \citenamefont {Dobbs}, \citenamefont
  {Doel}, \citenamefont {Doussot}, \citenamefont {Doux}, \citenamefont
  {Dutcher}, \citenamefont {Everett}, \citenamefont {Feng}, \citenamefont
  {Ferguson}, \citenamefont {Ferrero}, \citenamefont {Fichman}, \citenamefont
  {Foster}, \citenamefont {Frieman}, \citenamefont {Galli}, \citenamefont
  {Gambrel}, \citenamefont {{Garc{\'i}a-Bellido}}, \citenamefont {Gardner},
  \citenamefont {Gaztanaga}, \citenamefont {Ge}, \citenamefont {Giannini},
  \citenamefont {{Goeckner-Wald}}, \citenamefont {Grandis}, \citenamefont
  {Gruendl}, \citenamefont {Gualtieri}, \citenamefont {Guidi}, \citenamefont
  {Guns}, \citenamefont {Gutierrez}, \citenamefont {Halverson}, \citenamefont
  {Hinton}, \citenamefont {Hivon}, \citenamefont {Holder}, \citenamefont
  {Hollowood}, \citenamefont {Holzapfel}, \citenamefont {Honscheid},
  \citenamefont {Hood}, \citenamefont {Huang}, \citenamefont {James},
  \citenamefont {K{\'e}ruzor{\'e}}, \citenamefont {Knox}, \citenamefont
  {Korman}, \citenamefont {Kuo}, \citenamefont {Lee}, \citenamefont {Lee},
  \citenamefont {Levy}, \citenamefont {Lowitz}, \citenamefont {Lu},
  \citenamefont {Maniyar}, \citenamefont {Marshall}, \citenamefont
  {{Mena-Fern{\'a}ndez}}, \citenamefont {Menanteau}, \citenamefont {Miquel},
  \citenamefont {Millea}, \citenamefont {Mohr}, \citenamefont {Montgomery},
  \citenamefont {Nakato}, \citenamefont {Natoli}, \citenamefont {Noble},
  \citenamefont {Novosad}, \citenamefont {Ogando}, \citenamefont {Omori},
  \citenamefont {Padin}, \citenamefont {Palmese}, \citenamefont {Pan},
  \citenamefont {Paschos}, \citenamefont {Pereira}, \citenamefont {Pieres},
  \citenamefont {Plazas~Malag{\'o}n}, \citenamefont {Prabhu}, \citenamefont
  {Quan}, \citenamefont {Rahlin}, \citenamefont {Rahimi}, \citenamefont
  {Reichardt}, \citenamefont {Reil}, \citenamefont {Romer}, \citenamefont
  {Rouble}, \citenamefont {Ruhl}, \citenamefont {Sanchez}, \citenamefont
  {Sanchez~Cid}, \citenamefont {Schiappucci}, \citenamefont {{Sevilla-Noarbe}},
  \citenamefont {Smecher}, \citenamefont {Smith}, \citenamefont {Sobrin},
  \citenamefont {Stark}, \citenamefont {Stephen}, \citenamefont {Suchyta},
  \citenamefont {Suzuki}, \citenamefont {Swanson}, \citenamefont {Tandoi},
  \citenamefont {Tarle}, \citenamefont {Thompson}, \citenamefont {Thorne},
  \citenamefont {Trendafilova}, \citenamefont {Tucker}, \citenamefont {Umilta},
  \citenamefont {Vieira}, \citenamefont {Wang}, \citenamefont {Weaverdyck},
  \citenamefont {Whitehorn}, \citenamefont {Wiseman}, \citenamefont {Wu},
  \citenamefont {Yefremenko}, \citenamefont {Young}, \citenamefont {Zebrowski},
  \citenamefont {{The SPT-3G collaboration}},\ and\ \citenamefont {{The DES
  collaboration}}}]{ansarinejadMassCalibrationYear32024}%
  \BibitemOpen
  \bibfield  {author} {\bibinfo {author} {\bibfnamefont {B.}~\bibnamefont
  {Ansarinejad}}, \bibinfo {author} {\bibfnamefont {S.}~\bibnamefont
  {Raghunathan}}, \bibinfo {author} {\bibfnamefont {T.~M.~C.}\ \bibnamefont
  {Abbott}}, \emph {et~al.},\ }\href
  {https://doi.org/10.1088/1475-7516/2024/07/024} {\bibfield  {journal}
  {\bibinfo  {journal} {J. Cosmol. Astropart. Phys.}\ }\textbf {\bibinfo
  {volume} {2024}},\ \bibinfo {pages} {024}}\BibitemShut {NoStop}%
\bibitem [{\citenamefont {Kwiecien}\ \emph {et~al.}(2025)\citenamefont
  {Kwiecien}, \citenamefont {Jeltema}, \citenamefont {Leauthaud}, \citenamefont
  {Huang}, \citenamefont {Rykoff}, \citenamefont {Heydenreich}, \citenamefont
  {Lange}, \citenamefont {Everett}, \citenamefont {Zhou}, \citenamefont
  {Kelly}, \citenamefont {Zhang}, \citenamefont {Shin}, \citenamefont
  {{Golden-Marx}}, \citenamefont {Marshall}, \citenamefont {Aguena},
  \citenamefont {Allam}, \citenamefont {Bocquet}, \citenamefont {Brooks},
  \citenamefont {Rosell}, \citenamefont {Carretero}, \citenamefont {{da
  Costa}}, \citenamefont {Pereira}, \citenamefont {Davis}, \citenamefont
  {De~Vicente}, \citenamefont {Doel}, \citenamefont {Ferrero}, \citenamefont
  {Flaugher}, \citenamefont {Frieman}, \citenamefont {{Garc{\'i}a-Bellido}},
  \citenamefont {Gatti}, \citenamefont {Gaztanaga}, \citenamefont {Giannini},
  \citenamefont {Gruen}, \citenamefont {Gruendl}, \citenamefont {Gutierrez},
  \citenamefont {Hinton}, \citenamefont {Hollowood}, \citenamefont {Honscheid},
  \citenamefont {James}, \citenamefont {Lee}, \citenamefont {Miquel},
  \citenamefont {Pieres}, \citenamefont {Malag{\'o}n}, \citenamefont {Romer},
  \citenamefont {Samuroff}, \citenamefont {Sanchez}, \citenamefont {Santiago},
  \citenamefont {{Sevilla-Noarbe}}, \citenamefont {Smith}, \citenamefont
  {Suchyta}, \citenamefont {Swanson}, \citenamefont {Tarle}, \citenamefont
  {Tucker}, \citenamefont {Vikram}, \citenamefont {Weaverdyck},\ and\
  \citenamefont {Wiseman}}]{kwiecienImprovingGalaxyCluster2025}%
  \BibitemOpen
  \bibfield  {author} {\bibinfo {author} {\bibfnamefont {M.}~\bibnamefont
  {Kwiecien}}, \bibinfo {author} {\bibfnamefont {T.}~\bibnamefont {Jeltema}},
  \bibinfo {author} {\bibfnamefont {A.}~\bibnamefont {Leauthaud}}, \emph
  {et~al.},\ }\href {https://doi.org/10.1103/1j5f-cmkg} {\bibfield  {journal}
  {\bibinfo  {journal} {Phys. Rev. D}\ }\textbf {\bibinfo {volume} {111}},\
  \bibinfo {pages} {123524} (\bibinfo {year} {2025})}\BibitemShut {NoStop}%
\bibitem [{\citenamefont {Chiu}\ \emph {et~al.}(2025)\citenamefont {Chiu},
  \citenamefont {Ghirardini}, \citenamefont {Grandis}, \citenamefont {Okabe},
  \citenamefont {Artis}, \citenamefont {Bulbul}, \citenamefont {Emre~Bahar},
  \citenamefont {Balzer}, \citenamefont {Clerc}, \citenamefont {Comparat},
  \citenamefont {Hsieh}, \citenamefont {Kleinebreil}, \citenamefont {Kluge},
  \citenamefont {Liu}, \citenamefont {{Monteiro-Oliveira}}, \citenamefont
  {Oguri}, \citenamefont {Pacaud}, \citenamefont {Ramos~Ceja}, \citenamefont
  {Reiprich}, \citenamefont {Sanders}, \citenamefont {Schrabback},
  \citenamefont {Seppi}, \citenamefont {Sommer}, \citenamefont {Tam},
  \citenamefont {Umetsu},\ and\ \citenamefont
  {Zhang}}]{chiuSRGEROSITAAllSky2025}%
  \BibitemOpen
  \bibfield  {author} {\bibinfo {author} {\bibfnamefont {I.-N.}\ \bibnamefont
  {Chiu}}, \bibinfo {author} {\bibfnamefont {V.}~\bibnamefont {Ghirardini}},
  \bibinfo {author} {\bibfnamefont {S.}~\bibnamefont {Grandis}}, \emph
  {et~al.},\ }\href {https://doi.org/10.1051/0004-6361/202554942} {\bibfield
  {journal} {\bibinfo  {journal} {Astron. Astrophys.}\ }\textbf {\bibinfo
  {volume} {704}},\ \bibinfo {pages} {A110} (\bibinfo {year}
  {2025})}\BibitemShut {NoStop}%
\bibitem [{\citenamefont {Sunayama}\ \emph {et~al.}(2020)\citenamefont
  {Sunayama}, \citenamefont {Park}, \citenamefont {Takada}, \citenamefont
  {Kobayashi}, \citenamefont {Nishimichi}, \citenamefont {Kurita},
  \citenamefont {More}, \citenamefont {Oguri},\ and\ \citenamefont
  {Osato}}]{sunayamaImpactProjectionEffects2020}%
  \BibitemOpen
  \bibfield  {author} {\bibinfo {author} {\bibfnamefont {T.}~\bibnamefont
  {Sunayama}}, \bibinfo {author} {\bibfnamefont {Y.}~\bibnamefont {Park}},
  \bibinfo {author} {\bibfnamefont {M.}~\bibnamefont {Takada}}, \emph
  {et~al.},\ }\href {https://doi.org/10.1093/mnras/staa1646} {\bibfield
  {journal} {\bibinfo  {journal} {Mon. Not. R. Astron. Soc.}\ }\textbf
  {\bibinfo {volume} {496}},\ \bibinfo {pages} {4468} (\bibinfo {year}
  {2020})}\BibitemShut {NoStop}%
\bibitem [{\citenamefont {Wu}\ \emph {et~al.}(2022)\citenamefont {Wu},
  \citenamefont {Costanzi}, \citenamefont {To}, \citenamefont {Salcedo},
  \citenamefont {Weinberg}, \citenamefont {Annis}, \citenamefont {Bocquet},
  \citenamefont {{da~Silva~Pereira}}, \citenamefont {DeRose}, \citenamefont
  {Esteves}, \citenamefont {Farahi}, \citenamefont {Grandis}, \citenamefont
  {Rozo}, \citenamefont {Rykoff}, \citenamefont {Varga}, \citenamefont
  {Wechsler}, \citenamefont {Zeng}, \citenamefont {Zhang}, \citenamefont
  {Zhang},\ and\ \citenamefont {{DES
  Collaboration}}}]{wuOpticalSelectionBias2022}%
  \BibitemOpen
  \bibfield  {author} {\bibinfo {author} {\bibfnamefont {H.-Y.}\ \bibnamefont
  {Wu}}, \bibinfo {author} {\bibfnamefont {M.}~\bibnamefont {Costanzi}},
  \bibinfo {author} {\bibfnamefont {C.-H.}\ \bibnamefont {To}}, \emph
  {et~al.},\ }\href {https://doi.org/10.1093/mnras/stac2048} {\bibfield
  {journal} {\bibinfo  {journal} {Mon. Not. R. Astron. Soc.}\ }\textbf
  {\bibinfo {volume} {515}},\ \bibinfo {pages} {4471} (\bibinfo {year}
  {2022})}\BibitemShut {NoStop}%
\bibitem [{\citenamefont {Lee}\ \emph {et~al.}(2024)\citenamefont {Lee},
  \citenamefont {Wu}, \citenamefont {Salcedo}, \citenamefont {Sunayama},
  \citenamefont {Costanzi}, \citenamefont {Myles}, \citenamefont {Cao},
  \citenamefont {Rozo}, \citenamefont {To}, \citenamefont {Weinberg},
  \citenamefont {Yang},\ and\ \citenamefont
  {Zhou}}]{leeOpticalGalaxyCluster2024}%
  \BibitemOpen
  \bibfield  {author} {\bibinfo {author} {\bibfnamefont {A.}~\bibnamefont
  {Lee}}, \bibinfo {author} {\bibfnamefont {H.-Y.}\ \bibnamefont {Wu}},
  \bibinfo {author} {\bibfnamefont {A.~N.}\ \bibnamefont {Salcedo}}, \emph
  {et~al.},\ }\href {https://doi.org/10.48550/arXiv.2410.02497} {\enquote
  {\bibinfo {title} {Optical galaxy cluster mock catalogs with realistic
  projection effects: Validations with the {{SDSS redMaPPer}} clusters},}\ }
  (\bibinfo {year} {2024})\BibitemShut {NoStop}%
\bibitem [{\citenamefont {Nde}\ \emph {et~al.}(2026)\citenamefont {Nde},
  \citenamefont {Wu}, \citenamefont {Cao}, \citenamefont {Kamau}, \citenamefont
  {Tamosiunas}, \citenamefont {To},\ and\ \citenamefont
  {Zhou}}]{ndeImpactProjectioninducedOptical2026}%
  \BibitemOpen
  \bibfield  {author} {\bibinfo {author} {\bibfnamefont {T.~N.}\ \bibnamefont
  {Nde}}, \bibinfo {author} {\bibfnamefont {H.-Y.}\ \bibnamefont {Wu}},
  \bibinfo {author} {\bibfnamefont {S.}~\bibnamefont {Cao}}, \emph {et~al.},\
  }\href {https://doi.org/10.1103/h9xq-k9gh} {\bibfield  {journal} {\bibinfo
  {journal} {Phys. Rev. D}\ }\textbf {\bibinfo {volume} {113}},\ \bibinfo
  {pages} {063559} (\bibinfo {year} {2026})}\BibitemShut {NoStop}%
\bibitem [{\citenamefont {Abbott}\ \emph {et~al.}(2020)\citenamefont {Abbott},
  \citenamefont {Aguena}, \citenamefont {Alarcon}, \citenamefont {Allam},
  \citenamefont {Allen}, \citenamefont {Annis}, \citenamefont {Avila},
  \citenamefont {Bacon}, \citenamefont {Bechtol}, \citenamefont {Bermeo},
  \citenamefont {Bernstein}, \citenamefont {Bertin}, \citenamefont {Bhargava},
  \citenamefont {Bocquet}, \citenamefont {Brooks}, \citenamefont {Brout},
  \citenamefont {{Buckley-Geer}}, \citenamefont {Burke}, \citenamefont
  {Carnero~Rosell}, \citenamefont {Carrasco~Kind}, \citenamefont {Carretero},
  \citenamefont {Castander}, \citenamefont {Cawthon}, \citenamefont {Chang},
  \citenamefont {Chen}, \citenamefont {Choi}, \citenamefont {Costanzi},
  \citenamefont {Crocce}, \citenamefont {{da Costa}}, \citenamefont {Davis},
  \citenamefont {De~Vicente}, \citenamefont {DeRose}, \citenamefont {Desai},
  \citenamefont {Diehl}, \citenamefont {Dietrich}, \citenamefont {Dodelson},
  \citenamefont {Doel}, \citenamefont {{Drlica-Wagner}}, \citenamefont
  {Eckert}, \citenamefont {Eifler}, \citenamefont {{Elvin-Poole}},
  \citenamefont {Estrada}, \citenamefont {Everett}, \citenamefont {Evrard},
  \citenamefont {Farahi}, \citenamefont {Ferrero}, \citenamefont {Flaugher},
  \citenamefont {Fosalba}, \citenamefont {Frieman}, \citenamefont
  {{Garc{\'i}a-Bellido}}, \citenamefont {Gatti}, \citenamefont {Gaztanaga},
  \citenamefont {Gerdes}, \citenamefont {Giannantonio}, \citenamefont {Giles},
  \citenamefont {Grandis}, \citenamefont {Gruen}, \citenamefont {Gruendl},
  \citenamefont {Gschwend}, \citenamefont {Gutierrez}, \citenamefont {Hartley},
  \citenamefont {Hinton}, \citenamefont {Hollowood}, \citenamefont {Honscheid},
  \citenamefont {Hoyle}, \citenamefont {Huterer}, \citenamefont {James},
  \citenamefont {Jarvis}, \citenamefont {Jeltema}, \citenamefont {Johnson},
  \citenamefont {Johnson}, \citenamefont {Kent}, \citenamefont {Krause},
  \citenamefont {Kron}, \citenamefont {Kuehn}, \citenamefont {Kuropatkin},
  \citenamefont {Lahav}, \citenamefont {Li}, \citenamefont {Lidman},
  \citenamefont {Lima}, \citenamefont {Lin}, \citenamefont {MacCrann},
  \citenamefont {Maia}, \citenamefont {Mantz}, \citenamefont {Marshall},
  \citenamefont {Martini}, \citenamefont {Mayers}, \citenamefont {Melchior},
  \citenamefont {{Mena-Fern{\'a}ndez}}, \citenamefont {Menanteau},
  \citenamefont {Miquel}, \citenamefont {Mohr}, \citenamefont {Nichol},
  \citenamefont {Nord}, \citenamefont {Ogando}, \citenamefont {Palmese},
  \citenamefont {{Paz-Chinch{\'o}n}}, \citenamefont {Plazas}, \citenamefont
  {Prat}, \citenamefont {Rau}, \citenamefont {Romer}, \citenamefont {Roodman},
  \citenamefont {Rooney}, \citenamefont {Rozo}, \citenamefont {Rykoff},
  \citenamefont {Sako}, \citenamefont {Samuroff}, \citenamefont {S{\'a}nchez},
  \citenamefont {Sanchez}, \citenamefont {Saro}, \citenamefont {Scarpine},
  \citenamefont {Schubnell}, \citenamefont {Scolnic}, \citenamefont {Serrano},
  \citenamefont {{Sevilla-Noarbe}}, \citenamefont {Sheldon}, \citenamefont
  {Smith}, \citenamefont {Smith}, \citenamefont {Suchyta}, \citenamefont
  {Swanson}, \citenamefont {Tarle}, \citenamefont {Thomas}, \citenamefont {To},
  \citenamefont {Troxel}, \citenamefont {Tucker}, \citenamefont {Varga},
  \citenamefont {{von der Linden}}, \citenamefont {Walker}, \citenamefont
  {Wechsler}, \citenamefont {Weller}, \citenamefont {Wilkinson}, \citenamefont
  {Wu}, \citenamefont {Yanny}, \citenamefont {Zhang}, \citenamefont {Zhang},
  \citenamefont {Zuntz},\ and\ \citenamefont {{DES
  Collaboration}}}]{abbottDarkEnergySurvey2020}%
  \BibitemOpen
  \bibfield  {author} {\bibinfo {author} {\bibfnamefont {T.~M.~C.}\
  \bibnamefont {Abbott}}, \bibinfo {author} {\bibfnamefont {M.}~\bibnamefont
  {Aguena}}, \bibinfo {author} {\bibfnamefont {A.}~\bibnamefont {Alarcon}},
  \emph {et~al.},\ }\href {https://doi.org/10.1103/PhysRevD.102.023509}
  {\bibfield  {journal} {\bibinfo  {journal} {Phys. Rev. D}\ }\textbf {\bibinfo
  {volume} {102}},\ \bibinfo {pages} {023509} (\bibinfo {year}
  {2020})}\BibitemShut {NoStop}%
\bibitem [{\citenamefont {Vanderlinde}\ \emph {et~al.}(2010)\citenamefont
  {Vanderlinde}, \citenamefont {Crawford}, \citenamefont {{de Haan}},
  \citenamefont {Dudley}, \citenamefont {Shaw}, \citenamefont {Ade},
  \citenamefont {Aird}, \citenamefont {Benson}, \citenamefont {Bleem},
  \citenamefont {Brodwin}, \citenamefont {Carlstrom}, \citenamefont {Chang},
  \citenamefont {Crites}, \citenamefont {Desai}, \citenamefont {Dobbs},
  \citenamefont {Foley}, \citenamefont {George}, \citenamefont {Gladders},
  \citenamefont {Hall}, \citenamefont {Halverson}, \citenamefont {High},
  \citenamefont {Holder}, \citenamefont {Holzapfel}, \citenamefont {Hrubes},
  \citenamefont {Joy}, \citenamefont {Keisler}, \citenamefont {Knox},
  \citenamefont {Lee}, \citenamefont {Leitch}, \citenamefont {Loehr},
  \citenamefont {Lueker}, \citenamefont {Marrone}, \citenamefont {McMahon},
  \citenamefont {Mehl}, \citenamefont {Meyer}, \citenamefont {Mohr},
  \citenamefont {Montroy}, \citenamefont {Ngeow}, \citenamefont {Padin},
  \citenamefont {Plagge}, \citenamefont {Pryke}, \citenamefont {Reichardt},
  \citenamefont {Rest}, \citenamefont {Ruel}, \citenamefont {Ruhl},
  \citenamefont {Schaffer}, \citenamefont {Shirokoff}, \citenamefont {Song},
  \citenamefont {Spieler}, \citenamefont {Stalder}, \citenamefont
  {Staniszewski}, \citenamefont {Stark}, \citenamefont {Stubbs}, \citenamefont
  {{van Engelen}}, \citenamefont {Vieira}, \citenamefont {Williamson},
  \citenamefont {Yang}, \citenamefont {Zahn},\ and\ \citenamefont
  {Zenteno}}]{vanderlindeGalaxyClustersSelected2010}%
  \BibitemOpen
  \bibfield  {author} {\bibinfo {author} {\bibfnamefont {K.}~\bibnamefont
  {Vanderlinde}}, \bibinfo {author} {\bibfnamefont {T.~M.}\ \bibnamefont
  {Crawford}}, \bibinfo {author} {\bibfnamefont {T.}~\bibnamefont {{de Haan}}},
  \emph {et~al.},\ }\href {https://doi.org/10.1088/0004-637X/722/2/1180}
  {\bibfield  {journal} {\bibinfo  {journal} {Astrophys. J.}\ }\textbf
  {\bibinfo {volume} {722}},\ \bibinfo {pages} {1180} (\bibinfo {year}
  {2010})}\BibitemShut {NoStop}%
\bibitem [{\citenamefont {Bocquet}\ \emph {et~al.}(2019)\citenamefont
  {Bocquet}, \citenamefont {Dietrich}, \citenamefont {Schrabback},
  \citenamefont {Bleem}, \citenamefont {Klein}, \citenamefont {Allen},
  \citenamefont {Applegate}, \citenamefont {Ashby}, \citenamefont {Bautz},
  \citenamefont {Bayliss}, \citenamefont {Benson}, \citenamefont {Brodwin},
  \citenamefont {Bulbul}, \citenamefont {Canning}, \citenamefont {Capasso},
  \citenamefont {Carlstrom}, \citenamefont {Chang}, \citenamefont {Chiu},
  \citenamefont {Cho}, \citenamefont {Clocchiatti}, \citenamefont {Crawford},
  \citenamefont {Crites}, \citenamefont {{de Haan}}, \citenamefont {Desai},
  \citenamefont {Dobbs}, \citenamefont {Foley}, \citenamefont {Forman},
  \citenamefont {Garmire}, \citenamefont {George}, \citenamefont {Gladders},
  \citenamefont {Gonzalez}, \citenamefont {Grandis}, \citenamefont {Gupta},
  \citenamefont {Halverson}, \citenamefont {{Hlavacek-Larrondo}}, \citenamefont
  {Hoekstra}, \citenamefont {Holder}, \citenamefont {Holzapfel}, \citenamefont
  {Hou}, \citenamefont {Hrubes}, \citenamefont {Huang}, \citenamefont {Jones},
  \citenamefont {Khullar}, \citenamefont {Knox}, \citenamefont {Kraft},
  \citenamefont {Lee}, \citenamefont {{von der Linden}}, \citenamefont
  {{Luong-Van}}, \citenamefont {Mantz}, \citenamefont {Marrone}, \citenamefont
  {McDonald}, \citenamefont {McMahon}, \citenamefont {Meyer}, \citenamefont
  {Mocanu}, \citenamefont {Mohr}, \citenamefont {Morris}, \citenamefont
  {Padin}, \citenamefont {Patil}, \citenamefont {Pryke}, \citenamefont
  {Rapetti}, \citenamefont {Reichardt}, \citenamefont {Rest}, \citenamefont
  {Ruhl}, \citenamefont {Saliwanchik}, \citenamefont {Saro}, \citenamefont
  {Sayre}, \citenamefont {Schaffer}, \citenamefont {Shirokoff}, \citenamefont
  {Stalder}, \citenamefont {Stanford}, \citenamefont {Staniszewski},
  \citenamefont {Stark}, \citenamefont {Story}, \citenamefont {Strazzullo},
  \citenamefont {Stubbs}, \citenamefont {Vanderlinde}, \citenamefont {Vieira},
  \citenamefont {Vikhlinin}, \citenamefont {Williamson},\ and\ \citenamefont
  {Zenteno}}]{bocquetClusterCosmologyConstraints2019}%
  \BibitemOpen
  \bibfield  {author} {\bibinfo {author} {\bibfnamefont {S.}~\bibnamefont
  {Bocquet}}, \bibinfo {author} {\bibfnamefont {J.~P.}\ \bibnamefont
  {Dietrich}}, \bibinfo {author} {\bibfnamefont {T.}~\bibnamefont
  {Schrabback}}, \emph {et~al.},\ }\href
  {https://doi.org/10.3847/1538-4357/ab1f10} {\bibfield  {journal} {\bibinfo
  {journal} {Astrophys. J.}\ }\textbf {\bibinfo {volume} {878}},\ \bibinfo
  {pages} {55} (\bibinfo {year} {2019})}\BibitemShut {NoStop}%
\bibitem [{\citenamefont {Costanzi}\ \emph {et~al.}(2019)\citenamefont
  {Costanzi}, \citenamefont {Rozo}, \citenamefont {Simet}, \citenamefont
  {Zhang}, \citenamefont {Evrard}, \citenamefont {Mantz}, \citenamefont
  {Rykoff}, \citenamefont {Jeltema}, \citenamefont {Gruen}, \citenamefont
  {Allen}, \citenamefont {McClintock}, \citenamefont {Romer}, \citenamefont
  {{von der Linden}}, \citenamefont {Farahi}, \citenamefont {DeRose},
  \citenamefont {Varga}, \citenamefont {Weller}, \citenamefont {Giles},
  \citenamefont {Hollowood}, \citenamefont {Bhargava}, \citenamefont
  {{Bermeo-Hernandez}}, \citenamefont {Chen}, \citenamefont {Abbott},
  \citenamefont {Abdalla}, \citenamefont {Avila}, \citenamefont {Bechtol},
  \citenamefont {Brooks}, \citenamefont {{Buckley-Geer}}, \citenamefont
  {Burke}, \citenamefont {Rosell}, \citenamefont {Kind}, \citenamefont
  {Carretero}, \citenamefont {Crocce}, \citenamefont {Cunha}, \citenamefont
  {{da Costa}}, \citenamefont {Davis}, \citenamefont {De~Vicente},
  \citenamefont {Diehl}, \citenamefont {Dietrich}, \citenamefont {Doel},
  \citenamefont {Eifler}, \citenamefont {Estrada}, \citenamefont {Flaugher},
  \citenamefont {Fosalba}, \citenamefont {Frieman}, \citenamefont
  {{Garc{\'i}a-Bellido}}, \citenamefont {Gaztanaga}, \citenamefont {Gerdes},
  \citenamefont {Giannantonio}, \citenamefont {Gruendl}, \citenamefont
  {Gschwend}, \citenamefont {Gutierrez}, \citenamefont {Hartley}, \citenamefont
  {Honscheid}, \citenamefont {Hoyle}, \citenamefont {James}, \citenamefont
  {Krause}, \citenamefont {Kuehn}, \citenamefont {Kuropatkin}, \citenamefont
  {Lima}, \citenamefont {Lin}, \citenamefont {Maia}, \citenamefont {March},
  \citenamefont {Marshall}, \citenamefont {Martini}, \citenamefont {Menanteau},
  \citenamefont {Miller}, \citenamefont {Miquel}, \citenamefont {Mohr},
  \citenamefont {Ogando}, \citenamefont {Plazas}, \citenamefont {Roodman},
  \citenamefont {Sanchez}, \citenamefont {Scarpine}, \citenamefont {Schindler},
  \citenamefont {Schubnell}, \citenamefont {Serrano}, \citenamefont
  {{Sevilla-Noarbe}}, \citenamefont {Sheldon}, \citenamefont {Smith},
  \citenamefont {{Soares-Santos}}, \citenamefont {Sobreira}, \citenamefont
  {Suchyta}, \citenamefont {Swanson}, \citenamefont {Tarle}, \citenamefont
  {Thomas},\ and\ \citenamefont
  {Wechsler}}]{costanziMethodsClusterCosmology2019}%
  \BibitemOpen
  \bibfield  {author} {\bibinfo {author} {\bibfnamefont {M.}~\bibnamefont
  {Costanzi}}, \bibinfo {author} {\bibfnamefont {E.}~\bibnamefont {Rozo}},
  \bibinfo {author} {\bibfnamefont {M.}~\bibnamefont {Simet}}, \emph {et~al.},\
  }\href {https://doi.org/10.1093/mnras/stz1949} {\bibfield  {journal}
  {\bibinfo  {journal} {Mon. Not. R. Astron. Soc.}\ }\textbf {\bibinfo {volume}
  {488}},\ \bibinfo {pages} {4779} (\bibinfo {year} {2019})}\BibitemShut
  {NoStop}%
\bibitem [{\citenamefont {Abbott}\ \emph {et~al.}(2018)\citenamefont {Abbott},
  \citenamefont {Abdalla}, \citenamefont {Allam}, \citenamefont {Amara},
  \citenamefont {Annis}, \citenamefont {Asorey}, \citenamefont {Avila},
  \citenamefont {Ballester}, \citenamefont {Banerji}, \citenamefont
  {Barkhouse}, \citenamefont {Baruah}, \citenamefont {Baumer}, \citenamefont
  {Bechtol}, \citenamefont {Becker}, \citenamefont {{Benoit-L{\'e}vy}},
  \citenamefont {Bernstein}, \citenamefont {Bertin}, \citenamefont {Blazek},
  \citenamefont {Bocquet}, \citenamefont {Brooks}, \citenamefont {Brout},
  \citenamefont {{Buckley-Geer}}, \citenamefont {Burke}, \citenamefont {Busti},
  \citenamefont {Campisano}, \citenamefont {{Cardiel-Sas}}, \citenamefont
  {Carnero~Rosell}, \citenamefont {Carrasco~Kind}, \citenamefont {Carretero},
  \citenamefont {Castander}, \citenamefont {Cawthon}, \citenamefont {Chang},
  \citenamefont {Chen}, \citenamefont {Conselice}, \citenamefont {Costa},
  \citenamefont {Crocce}, \citenamefont {Cunha}, \citenamefont {D'Andrea},
  \citenamefont {{da Costa}}, \citenamefont {Das}, \citenamefont {Daues},
  \citenamefont {Davis}, \citenamefont {Davis}, \citenamefont {De~Vicente},
  \citenamefont {DePoy}, \citenamefont {DeRose}, \citenamefont {Desai},
  \citenamefont {Diehl}, \citenamefont {Dietrich}, \citenamefont {Dodelson},
  \citenamefont {Doel}, \citenamefont {{Drlica-Wagner}}, \citenamefont
  {Eifler}, \citenamefont {Elliott}, \citenamefont {Evrard}, \citenamefont
  {Farahi}, \citenamefont {Fausti~Neto}, \citenamefont {Fernandez},
  \citenamefont {Finley}, \citenamefont {Flaugher}, \citenamefont {Foley},
  \citenamefont {Fosalba}, \citenamefont {Friedel}, \citenamefont {Frieman},
  \citenamefont {{Garc{\'i}a-Bellido}}, \citenamefont {Gaztanaga},
  \citenamefont {Gerdes}, \citenamefont {Giannantonio}, \citenamefont {Gill},
  \citenamefont {Glazebrook}, \citenamefont {Goldstein}, \citenamefont {Gower},
  \citenamefont {Gruen}, \citenamefont {Gruendl}, \citenamefont {Gschwend},
  \citenamefont {Gupta}, \citenamefont {Gutierrez}, \citenamefont {Hamilton},
  \citenamefont {Hartley}, \citenamefont {Hinton}, \citenamefont {Hislop},
  \citenamefont {Hollowood}, \citenamefont {Honscheid}, \citenamefont {Hoyle},
  \citenamefont {Huterer}, \citenamefont {Jain}, \citenamefont {James},
  \citenamefont {Jeltema}, \citenamefont {Johnson}, \citenamefont {Johnson},
  \citenamefont {Kacprzak}, \citenamefont {Kent}, \citenamefont {Khullar},
  \citenamefont {Klein}, \citenamefont {Kovacs}, \citenamefont {Koziol},
  \citenamefont {Krause}, \citenamefont {Kremin}, \citenamefont {Kron},
  \citenamefont {Kuehn}, \citenamefont {Kuhlmann}, \citenamefont {Kuropatkin},
  \citenamefont {Lahav}, \citenamefont {Lasker}, \citenamefont {Li},
  \citenamefont {Li}, \citenamefont {Liddle}, \citenamefont {Lima},
  \citenamefont {Lin}, \citenamefont {{L{\'o}pez-Reyes}}, \citenamefont
  {MacCrann}, \citenamefont {Maia}, \citenamefont {Maloney}, \citenamefont
  {Manera}, \citenamefont {March}, \citenamefont {Marriner}, \citenamefont
  {Marshall}, \citenamefont {Martini}, \citenamefont {McClintock},
  \citenamefont {McKay}, \citenamefont {McMahon}, \citenamefont {Melchior},
  \citenamefont {Menanteau}, \citenamefont {Miller}, \citenamefont {Miquel},
  \citenamefont {Mohr}, \citenamefont {Morganson}, \citenamefont {Mould},
  \citenamefont {Neilsen}, \citenamefont {Nichol}, \citenamefont {Nogueira},
  \citenamefont {Nord}, \citenamefont {Nugent}, \citenamefont {Nunes},
  \citenamefont {Ogando}, \citenamefont {Old}, \citenamefont {Pace},
  \citenamefont {Palmese}, \citenamefont {{Paz-Chinch{\'o}n}}, \citenamefont
  {Peiris}, \citenamefont {Percival}, \citenamefont {Petravick}, \citenamefont
  {Plazas}, \citenamefont {Poh}, \citenamefont {Pond}, \citenamefont
  {Porredon}, \citenamefont {Pujol}, \citenamefont {Refregier}, \citenamefont
  {Reil}, \citenamefont {Ricker}, \citenamefont {Rollins}, \citenamefont
  {Romer}, \citenamefont {Roodman}, \citenamefont {Rooney}, \citenamefont
  {Ross}, \citenamefont {Rykoff}, \citenamefont {Sako}, \citenamefont
  {Sanchez}, \citenamefont {Sanchez}, \citenamefont {Santiago}, \citenamefont
  {Saro}, \citenamefont {Scarpine}, \citenamefont {Scolnic}, \citenamefont
  {Serrano}, \citenamefont {{Sevilla-Noarbe}}, \citenamefont {Sheldon},
  \citenamefont {Shipp}, \citenamefont {Silveira}, \citenamefont {Smith},
  \citenamefont {Smith}, \citenamefont {Smith}, \citenamefont
  {{Soares-Santos}}, \citenamefont {Sobreira}, \citenamefont {Song},
  \citenamefont {Stebbins}, \citenamefont {Suchyta}, \citenamefont {Sullivan},
  \citenamefont {Swanson}, \citenamefont {Tarle}, \citenamefont {Thaler},
  \citenamefont {Thomas}, \citenamefont {Thomas}, \citenamefont {Troxel},
  \citenamefont {Tucker}, \citenamefont {Vikram}, \citenamefont {Vivas},
  \citenamefont {Walker}, \citenamefont {Wechsler}, \citenamefont {Weller},
  \citenamefont {Wester}, \citenamefont {Wolf}, \citenamefont {Wu},
  \citenamefont {Yanny}, \citenamefont {Zenteno}, \citenamefont {Zhang},
  \citenamefont {Zuntz}, \citenamefont {{DES Collaboration}}, \citenamefont
  {Juneau}, \citenamefont {Fitzpatrick},\ and\ \citenamefont
  {Nikutta}}]{abbottDarkEnergySurvey2018}%
  \BibitemOpen
  \bibfield  {author} {\bibinfo {author} {\bibfnamefont {T.~M.~C.}\
  \bibnamefont {Abbott}}, \bibinfo {author} {\bibfnamefont {F.~B.}\
  \bibnamefont {Abdalla}}, \bibinfo {author} {\bibfnamefont {S.}~\bibnamefont
  {Allam}}, \emph {et~al.},\ }\href {https://doi.org/10.3847/1538-4365/aae9f0}
  {\bibfield  {journal} {\bibinfo  {journal} {Astrophys. J. Suppl. Ser.}\
  }\textbf {\bibinfo {volume} {239}},\ \bibinfo {pages} {18} (\bibinfo {year}
  {2018})}\BibitemShut {NoStop}%
\bibitem [{\citenamefont {Abbott}\ \emph {et~al.}(2025)\citenamefont {Abbott},
  \citenamefont {Aguena}, \citenamefont {Alarcon}, \citenamefont {Amon},
  \citenamefont {Anbajagane}, \citenamefont {{Andrade-Oliveira}}, \citenamefont
  {Avila}, \citenamefont {Allam}, \citenamefont {Bacon}, \citenamefont
  {Becker}, \citenamefont {Bhargava}, \citenamefont {Blazek}, \citenamefont
  {Bocquet}, \citenamefont {Brooks}, \citenamefont {Rosell}, \citenamefont
  {Carretero}, \citenamefont {Castander}, \citenamefont {Chang}, \citenamefont
  {Choi}, \citenamefont {Conselice}, \citenamefont {Costanzi}, \citenamefont
  {Crocce}, \citenamefont {{da Costa}}, \citenamefont {Pereira}, \citenamefont
  {Doux}, \citenamefont {Davis}, \citenamefont {Desai}, \citenamefont {Diehl},
  \citenamefont {Dodelson}, \citenamefont {Doel}, \citenamefont
  {{Elvin-Poole}}, \citenamefont {Esteves}, \citenamefont {Everett},
  \citenamefont {Farahi}, \citenamefont {Fert{\'e}}, \citenamefont {Flaugher},
  \citenamefont {{Garc{\'i}a-Bellido}}, \citenamefont {Gatti}, \citenamefont
  {Giannini}, \citenamefont {Giles}, \citenamefont {Grandis}, \citenamefont
  {Gruen}, \citenamefont {Gruendl}, \citenamefont {Gutierrez}, \citenamefont
  {Harrison}, \citenamefont {Hinton}, \citenamefont {Hollowood}, \citenamefont
  {Honscheid}, \citenamefont {Jeffrey}, \citenamefont {Jeltema}, \citenamefont
  {Krause}, \citenamefont {Lahav}, \citenamefont {Lee}, \citenamefont {Lidman},
  \citenamefont {Lima}, \citenamefont {Lin}, \citenamefont {Mohr},
  \citenamefont {Marshall}, \citenamefont {McCullough}, \citenamefont
  {{Mena-Fern'andez}}, \citenamefont {Miquel}, \citenamefont {Muir},
  \citenamefont {Myles}, \citenamefont {Ogando}, \citenamefont {Palmese},
  \citenamefont {Paterno}, \citenamefont {Malag{\'o}n}, \citenamefont
  {Porredon}, \citenamefont {Prat}, \citenamefont {Romer}, \citenamefont
  {Roodman}, \citenamefont {Rozo}, \citenamefont {Rykoff}, \citenamefont
  {Rosenfeld}, \citenamefont {Sanchez}, \citenamefont {Cid}, \citenamefont
  {{Sevilla-Noarbe}}, \citenamefont {Smith}, \citenamefont {Suchyta},
  \citenamefont {Shin}, \citenamefont {Tarle}, \citenamefont {Thomas},
  \citenamefont {To}, \citenamefont {Troxel}, \citenamefont {Vikram},
  \citenamefont {Walker}, \citenamefont {Weinberg}, \citenamefont {Weaverdyck},
  \citenamefont {Wechsler}, \citenamefont {Weller}, \citenamefont {Wu},
  \citenamefont {Yamamoto}, \citenamefont {Yanny}, \citenamefont {Zhang},
  \citenamefont {Zhou},\ and\ \citenamefont {{DES
  Collaboration}}}]{abbottDarkEnergySurvey2025}%
  \BibitemOpen
  \bibfield  {author} {\bibinfo {author} {\bibfnamefont {T.~M.~C.}\
  \bibnamefont {Abbott}}, \bibinfo {author} {\bibfnamefont {M.}~\bibnamefont
  {Aguena}}, \bibinfo {author} {\bibfnamefont {A.}~\bibnamefont {Alarcon}},
  \emph {et~al.},\ }\href {https://doi.org/10.1103/3dzh-d8f5} {\bibfield
  {journal} {\bibinfo  {journal} {Phys. Rev. D}\ }\textbf {\bibinfo {volume}
  {112}},\ \bibinfo {pages} {083535} (\bibinfo {year} {2025})}\BibitemShut
  {NoStop}%
\bibitem [{\citenamefont {Gatti}\ \emph {et~al.}(2021)\citenamefont {Gatti},
  \citenamefont {Sheldon}, \citenamefont {Amon}, \citenamefont {Becker},
  \citenamefont {Troxel}, \citenamefont {Choi}, \citenamefont {Doux},
  \citenamefont {MacCrann}, \citenamefont {Alsina}, \citenamefont {Harrison},
  \citenamefont {Gruen}, \citenamefont {Bernstein}, \citenamefont {Jarvis},
  \citenamefont {Secco}, \citenamefont {Fert{\'e}}, \citenamefont {Shin},
  \citenamefont {McCullough}, \citenamefont {Rollins}, \citenamefont {Chen},
  \citenamefont {Chang}, \citenamefont {Pandey}, \citenamefont {Tutusaus},
  \citenamefont {Prat}, \citenamefont {{Elvin-Poole}}, \citenamefont {Sanchez},
  \citenamefont {Plazas}, \citenamefont {Roodman}, \citenamefont {Zuntz},
  \citenamefont {Abbott}, \citenamefont {Aguena}, \citenamefont {Allam},
  \citenamefont {Annis}, \citenamefont {Avila}, \citenamefont {Bacon},
  \citenamefont {Bertin}, \citenamefont {Bhargava}, \citenamefont {Brooks},
  \citenamefont {Burke}, \citenamefont {Rosell}, \citenamefont {Kind},
  \citenamefont {Carretero}, \citenamefont {Castander}, \citenamefont
  {Conselice}, \citenamefont {Costanzi}, \citenamefont {da~Costa},
  \citenamefont {Davis}, \citenamefont {Vicente}, \citenamefont {Desai},
  \citenamefont {Diehl}, \citenamefont {Dietrich}, \citenamefont {Doel},
  \citenamefont {{Drlica-Wagner}}, \citenamefont {Eckert}, \citenamefont
  {Everett}, \citenamefont {Ferrero}, \citenamefont {Frieman}, \citenamefont
  {{Garc{\'i}a-Bellido}}, \citenamefont {Gerdes}, \citenamefont {Giannantonio},
  \citenamefont {Gruendl}, \citenamefont {Gschwend}, \citenamefont {Gutierrez},
  \citenamefont {Hartley}, \citenamefont {Hinton}, \citenamefont {Hollowood},
  \citenamefont {Honscheid}, \citenamefont {Hoyle}, \citenamefont {Huff},
  \citenamefont {Huterer}, \citenamefont {Jain}, \citenamefont {James},
  \citenamefont {Jeltema}, \citenamefont {Krause}, \citenamefont {Kron},
  \citenamefont {Kuropatkin}, \citenamefont {Lahav}, \citenamefont {Lima},
  \citenamefont {Maia}, \citenamefont {Marshall}, \citenamefont {Miquel},
  \citenamefont {Morgan}, \citenamefont {Myles}, \citenamefont {Palmese},
  \citenamefont {{Paz-Chinch{\'o}n}}, \citenamefont {Rykoff}, \citenamefont
  {Samuroff}, \citenamefont {Sanchez}, \citenamefont {Scarpine}, \citenamefont
  {Schubnell}, \citenamefont {Serrano}, \citenamefont {{Sevilla-Noarbe}},
  \citenamefont {Smith}, \citenamefont {Suchyta}, \citenamefont {Swanson},
  \citenamefont {Tarle}, \citenamefont {Thomas}, \citenamefont {To},
  \citenamefont {Tucker}, \citenamefont {Varga}, \citenamefont {Wechsler},
  \citenamefont {Weller}, \citenamefont {Wester},\ and\ \citenamefont
  {Wilkinson}}]{gattiDarkEnergySurvey2021}%
  \BibitemOpen
  \bibfield  {author} {\bibinfo {author} {\bibfnamefont {M.}~\bibnamefont
  {Gatti}}, \bibinfo {author} {\bibfnamefont {E.}~\bibnamefont {Sheldon}},
  \bibinfo {author} {\bibfnamefont {A.}~\bibnamefont {Amon}}, \emph {et~al.},\
  }\href {https://doi.org/10.1093/mnras/stab918} {\bibfield  {journal}
  {\bibinfo  {journal} {Mon. Not. R. Astron. Soc.}\ }\textbf {\bibinfo {volume}
  {504}},\ \bibinfo {pages} {4312} (\bibinfo {year} {2021})}\BibitemShut
  {NoStop}%
\bibitem [{\citenamefont {Huff}\ and\ \citenamefont
  {Mandelbaum}(2017)}]{huffMetacalibrationDirectSelfCalibration2017}%
  \BibitemOpen
  \bibfield  {author} {\bibinfo {author} {\bibfnamefont {E.}~\bibnamefont
  {Huff}}\ and\ \bibinfo {author} {\bibfnamefont {R.}~\bibnamefont
  {Mandelbaum}},\ }\href {https://doi.org/10.48550/arXiv.1702.02600} {\enquote
  {\bibinfo {title} {Metacalibration: {{Direct Self-Calibration}} of {{Biases}}
  in {{Shear Measurement}}},}\ } (\bibinfo {year} {2017})\BibitemShut {NoStop}%
\bibitem [{\citenamefont {Sheldon}\ and\ \citenamefont
  {Huff}(2017)}]{sheldonPracticalWeaklensingShear2017}%
  \BibitemOpen
  \bibfield  {author} {\bibinfo {author} {\bibfnamefont {E.~S.}\ \bibnamefont
  {Sheldon}}\ and\ \bibinfo {author} {\bibfnamefont {E.~M.}\ \bibnamefont
  {Huff}},\ }\href {https://doi.org/10.3847/1538-4357/aa704b} {\bibfield
  {journal} {\bibinfo  {journal} {Astrophys. J.}\ }\textbf {\bibinfo {volume}
  {841}},\ \bibinfo {pages} {24} (\bibinfo {year} {2017})}\BibitemShut
  {NoStop}%
\bibitem [{\citenamefont {Myles}\ \emph {et~al.}(2021)\citenamefont {Myles},
  \citenamefont {Alarcon}, \citenamefont {Amon}, \citenamefont {S{\'a}nchez},
  \citenamefont {Everett}, \citenamefont {DeRose}, \citenamefont {McCullough},
  \citenamefont {Gruen}, \citenamefont {Bernstein}, \citenamefont {Troxel},
  \citenamefont {Dodelson}, \citenamefont {Campos}, \citenamefont {MacCrann},
  \citenamefont {Yin}, \citenamefont {Raveri}, \citenamefont {Amara},
  \citenamefont {Becker}, \citenamefont {Choi}, \citenamefont {Cordero},
  \citenamefont {Eckert}, \citenamefont {Gatti}, \citenamefont {Giannini},
  \citenamefont {Gschwend}, \citenamefont {Gruendl}, \citenamefont {Harrison},
  \citenamefont {Hartley}, \citenamefont {Huff}, \citenamefont {Kuropatkin},
  \citenamefont {Lin}, \citenamefont {Masters}, \citenamefont {Miquel},
  \citenamefont {Prat}, \citenamefont {Roodman}, \citenamefont {Rykoff},
  \citenamefont {{Sevilla-Noarbe}}, \citenamefont {Sheldon}, \citenamefont
  {Wechsler}, \citenamefont {Yanny}, \citenamefont {Abbott}, \citenamefont
  {Aguena}, \citenamefont {Allam}, \citenamefont {Annis}, \citenamefont
  {Bacon}, \citenamefont {Bertin}, \citenamefont {Bhargava}, \citenamefont
  {Bridle}, \citenamefont {Brooks}, \citenamefont {Burke}, \citenamefont
  {Carnero~Rosell}, \citenamefont {Carrasco~Kind}, \citenamefont {Carretero},
  \citenamefont {Castander}, \citenamefont {Conselice}, \citenamefont
  {Costanzi}, \citenamefont {Crocce}, \citenamefont {{da Costa}}, \citenamefont
  {Pereira}, \citenamefont {Desai}, \citenamefont {Diehl}, \citenamefont
  {Eifler}, \citenamefont {{Elvin-Poole}}, \citenamefont {Evrard},
  \citenamefont {Ferrero}, \citenamefont {Fert{\'e}}, \citenamefont {Flaugher},
  \citenamefont {Fosalba}, \citenamefont {Frieman}, \citenamefont
  {{Garc{\'i}a-Bellido}}, \citenamefont {Gaztanaga}, \citenamefont
  {Giannantonio}, \citenamefont {Hinton}, \citenamefont {Hollowood},
  \citenamefont {Honscheid}, \citenamefont {Hoyle}, \citenamefont {Huterer},
  \citenamefont {James}, \citenamefont {Krause}, \citenamefont {Kuehn},
  \citenamefont {Lahav}, \citenamefont {Lima}, \citenamefont {Maia},
  \citenamefont {Marshall}, \citenamefont {Martini}, \citenamefont {Melchior},
  \citenamefont {Menanteau}, \citenamefont {Mohr}, \citenamefont {Morgan},
  \citenamefont {Muir}, \citenamefont {Ogando}, \citenamefont {Palmese},
  \citenamefont {{Paz-Chinch{\'o}n}}, \citenamefont {Plazas}, \citenamefont
  {{Rodriguez-Monroy}}, \citenamefont {Samuroff}, \citenamefont {Sanchez},
  \citenamefont {Scarpine}, \citenamefont {Secco}, \citenamefont {Serrano},
  \citenamefont {Smith}, \citenamefont {{Soares-Santos}}, \citenamefont
  {Suchyta}, \citenamefont {Swanson}, \citenamefont {Tarle}, \citenamefont
  {Thomas}, \citenamefont {To}, \citenamefont {Varga}, \citenamefont {Weller},\
  and\ \citenamefont {Wester}}]{mylesDarkEnergySurvey2021}%
  \BibitemOpen
  \bibfield  {author} {\bibinfo {author} {\bibfnamefont {J.}~\bibnamefont
  {Myles}}, \bibinfo {author} {\bibfnamefont {A.}~\bibnamefont {Alarcon}},
  \bibinfo {author} {\bibfnamefont {A.}~\bibnamefont {Amon}}, \emph {et~al.},\
  }\href {https://doi.org/10.1093/mnras/stab1515} {\bibfield  {journal}
  {\bibinfo  {journal} {Mon. Not. R. Astron. Soc.}\ }\textbf {\bibinfo {volume}
  {505}},\ \bibinfo {pages} {4249} (\bibinfo {year} {2021})}\BibitemShut
  {NoStop}%
\bibitem [{\citenamefont {De~Vicente}\ \emph {et~al.}(2016)\citenamefont
  {De~Vicente}, \citenamefont {S{\'a}nchez},\ and\ \citenamefont
  {{Sevilla-Noarbe}}}]{devicenteDNFGalaxyPhotometric2016}%
  \BibitemOpen
  \bibfield  {author} {\bibinfo {author} {\bibfnamefont {J.}~\bibnamefont
  {De~Vicente}}, \bibinfo {author} {\bibfnamefont {E.}~\bibnamefont
  {S{\'a}nchez}},\ and\ \bibinfo {author} {\bibfnamefont {I.}~\bibnamefont
  {{Sevilla-Noarbe}}},\ }\href {https://doi.org/10.1093/mnras/stw857}
  {\bibfield  {journal} {\bibinfo  {journal} {Mon. Not. R. Astron. Soc.}\
  }\textbf {\bibinfo {volume} {459}},\ \bibinfo {pages} {3078} (\bibinfo {year}
  {2016})}\BibitemShut {NoStop}%
\bibitem [{\citenamefont
  {Ben{\'i}tez}(2000)}]{benitezBayesianPhotometricRedshift2000}%
  \BibitemOpen
  \bibfield  {author} {\bibinfo {author} {\bibfnamefont {N.}~\bibnamefont
  {Ben{\'i}tez}},\ }\href {https://doi.org/10.1086/308947} {\bibfield
  {journal} {\bibinfo  {journal} {Astrophys. J.}\ }\textbf {\bibinfo {volume}
  {536}},\ \bibinfo {pages} {571} (\bibinfo {year} {2000})}\BibitemShut
  {NoStop}%
\bibitem [{\citenamefont {Barlow}(1990)}]{barlowExtendedMaximumLikelihood1990}%
  \BibitemOpen
  \bibfield  {author} {\bibinfo {author} {\bibfnamefont {R.}~\bibnamefont
  {Barlow}},\ }\href {https://doi.org/10.1016/0168-9002(90)91334-8} {\bibfield
  {journal} {\bibinfo  {journal} {Nucl. Instrum. Methods Phys. Res. Sect.
  Accel. Spectrometers Detect. Assoc. Equip.}\ }\textbf {\bibinfo {volume}
  {297}},\ \bibinfo {pages} {496} (\bibinfo {year} {1990})}\BibitemShut
  {NoStop}%
\bibitem [{\citenamefont {Mantz}\ \emph {et~al.}(2010)\citenamefont {Mantz},
  \citenamefont {Allen}, \citenamefont {Rapetti},\ and\ \citenamefont
  {Ebeling}}]{mantzObservedGrowthMassive2010}%
  \BibitemOpen
  \bibfield  {author} {\bibinfo {author} {\bibfnamefont {A.}~\bibnamefont
  {Mantz}}, \bibinfo {author} {\bibfnamefont {S.~W.}\ \bibnamefont {Allen}},
  \bibinfo {author} {\bibfnamefont {D.}~\bibnamefont {Rapetti}},\ and\ \bibinfo
  {author} {\bibfnamefont {H.}~\bibnamefont {Ebeling}},\ }\href
  {https://doi.org/10.1111/j.1365-2966.2010.16992.x} {\bibfield  {journal}
  {\bibinfo  {journal} {Mon. Not. R. Astron. Soc.}\ }\textbf {\bibinfo {volume}
  {406}},\ \bibinfo {pages} {1759} (\bibinfo {year} {2010})}\BibitemShut
  {NoStop}%
\bibitem [{\citenamefont
  {Tobin}(1958)}]{tobinEstimationRelationshipsLimited1958}%
  \BibitemOpen
  \bibfield  {author} {\bibinfo {author} {\bibfnamefont {J.}~\bibnamefont
  {Tobin}},\ }\href {https://doi.org/10.2307/1907382} {\bibfield  {journal}
  {\bibinfo  {journal} {Econometrica}\ }\textbf {\bibinfo {volume} {26}},\
  \bibinfo {pages} {24} (\bibinfo {year} {1958})}\BibitemShut {NoStop}%
\bibitem [{\citenamefont {Amemiya}(1984)}]{amemiyaTobitModelsSurvey1984}%
  \BibitemOpen
  \bibfield  {author} {\bibinfo {author} {\bibfnamefont {T.}~\bibnamefont
  {Amemiya}},\ }\href {https://doi.org/10.1016/0304-4076(84)90074-5} {\bibfield
   {journal} {\bibinfo  {journal} {J. Econom.}\ }\textbf {\bibinfo {volume}
  {24}},\ \bibinfo {pages} {3} (\bibinfo {year} {1984})}\BibitemShut {NoStop}%
\bibitem [{\citenamefont {Bocquet}\ \emph {et~al.}(2016)\citenamefont
  {Bocquet}, \citenamefont {Saro}, \citenamefont {Dolag},\ and\ \citenamefont
  {Mohr}}]{bocquetHaloMassFunction2016}%
  \BibitemOpen
  \bibfield  {author} {\bibinfo {author} {\bibfnamefont {S.}~\bibnamefont
  {Bocquet}}, \bibinfo {author} {\bibfnamefont {A.}~\bibnamefont {Saro}},
  \bibinfo {author} {\bibfnamefont {K.}~\bibnamefont {Dolag}},\ and\ \bibinfo
  {author} {\bibfnamefont {J.~J.}\ \bibnamefont {Mohr}},\ }\href
  {https://doi.org/10.1093/mnras/stv2657} {\bibfield  {journal} {\bibinfo
  {journal} {Mon. Not. R. Astron. Soc.}\ }\textbf {\bibinfo {volume} {456}},\
  \bibinfo {pages} {2361} (\bibinfo {year} {2016})}\BibitemShut {NoStop}%
\bibitem [{\citenamefont {Nishimichi}\ \emph {et~al.}(2019)\citenamefont
  {Nishimichi}, \citenamefont {Takada}, \citenamefont {Takahashi},
  \citenamefont {Osato}, \citenamefont {Shirasaki}, \citenamefont {Oogi},
  \citenamefont {Miyatake}, \citenamefont {Oguri}, \citenamefont {Murata},
  \citenamefont {Kobayashi},\ and\ \citenamefont
  {Yoshida}}]{nishimichiDarkQuestFast2019}%
  \BibitemOpen
  \bibfield  {author} {\bibinfo {author} {\bibfnamefont {T.}~\bibnamefont
  {Nishimichi}}, \bibinfo {author} {\bibfnamefont {M.}~\bibnamefont {Takada}},
  \bibinfo {author} {\bibfnamefont {R.}~\bibnamefont {Takahashi}}, \emph
  {et~al.},\ }\href {https://doi.org/10.3847/1538-4357/ab3719} {\bibfield
  {journal} {\bibinfo  {journal} {Astrophys. J.}\ }\textbf {\bibinfo {volume}
  {884}},\ \bibinfo {pages} {29} (\bibinfo {year} {2019})}\BibitemShut
  {NoStop}%
\bibitem [{\citenamefont {Ishiyama}\ \emph {et~al.}(2021)\citenamefont
  {Ishiyama}, \citenamefont {Prada}, \citenamefont {Klypin}, \citenamefont
  {Sinha}, \citenamefont {Metcalf}, \citenamefont {Jullo}, \citenamefont
  {Altieri}, \citenamefont {Cora}, \citenamefont {Croton}, \citenamefont {{de
  la Torre}}, \citenamefont {{Mill{\'a}n-Calero}}, \citenamefont {Oogi},
  \citenamefont {Ruedas},\ and\ \citenamefont
  {{Vega-Mart{\'i}nez}}}]{ishiyamaUchuuSimulationsData2021}%
  \BibitemOpen
  \bibfield  {author} {\bibinfo {author} {\bibfnamefont {T.}~\bibnamefont
  {Ishiyama}}, \bibinfo {author} {\bibfnamefont {F.}~\bibnamefont {Prada}},
  \bibinfo {author} {\bibfnamefont {A.~A.}\ \bibnamefont {Klypin}}, \emph
  {et~al.},\ }\href {https://doi.org/10.1093/mnras/stab1755} {\bibfield
  {journal} {\bibinfo  {journal} {Mon. Not. R. Astron. Soc.}\ }\textbf
  {\bibinfo {volume} {506}},\ \bibinfo {pages} {4210} (\bibinfo {year}
  {2021})}\BibitemShut {NoStop}%
\bibitem [{\citenamefont {Farahi}\ \emph {et~al.}(2022)\citenamefont {Farahi},
  \citenamefont {Nagai},\ and\ \citenamefont
  {Anbajagane}}]{farahiCorrelationsDarkMatter2022}%
  \BibitemOpen
  \bibfield  {author} {\bibinfo {author} {\bibfnamefont {A.}~\bibnamefont
  {Farahi}}, \bibinfo {author} {\bibfnamefont {D.}~\bibnamefont {Nagai}},\ and\
  \bibinfo {author} {\bibfnamefont {D.}~\bibnamefont {Anbajagane}},\ }\href
  {https://doi.org/10.3847/1538-4357/ac721e} {\bibfield  {journal} {\bibinfo
  {journal} {Astrophys. J.}\ }\textbf {\bibinfo {volume} {933}},\ \bibinfo
  {pages} {48} (\bibinfo {year} {2022})}\BibitemShut {NoStop}%
\bibitem [{\citenamefont {Zhang}\ \emph
  {et~al.}(2024{\natexlab{b}})\citenamefont {Zhang}, \citenamefont {Farahi},
  \citenamefont {Nagai}, \citenamefont {Lau}, \citenamefont {Frieman},
  \citenamefont {Ricci}, \citenamefont {{von der Linden}}, \citenamefont {Wu},\
  and\ \citenamefont {{LSST Dark Energy Science
  Collaboration}}}]{zhangImpactPropertyCovariance2024}%
  \BibitemOpen
  \bibfield  {author} {\bibinfo {author} {\bibfnamefont {Z.}~\bibnamefont
  {Zhang}}, \bibinfo {author} {\bibfnamefont {A.}~\bibnamefont {Farahi}},
  \bibinfo {author} {\bibfnamefont {D.}~\bibnamefont {Nagai}}, \emph {et~al.},\
  }\href {https://doi.org/10.1093/mnras/stae999} {\bibfield  {journal}
  {\bibinfo  {journal} {Mon. Not. R. Astron. Soc.}\ }\textbf {\bibinfo {volume}
  {530}},\ \bibinfo {pages} {3127} (\bibinfo {year}
  {2024}{\natexlab{b}})}\BibitemShut {NoStop}%
\bibitem [{\citenamefont {Giri}\ and\ \citenamefont
  {Schneider}(2021)}]{giriEmulationBaryonicEffects2021}%
  \BibitemOpen
  \bibfield  {author} {\bibinfo {author} {\bibfnamefont {S.~K.}\ \bibnamefont
  {Giri}}\ and\ \bibinfo {author} {\bibfnamefont {A.}~\bibnamefont
  {Schneider}},\ }\href {https://doi.org/10.1088/1475-7516/2021/12/046}
  {\bibfield  {journal} {\bibinfo  {journal} {J. Cosmol. Astropart. Phys.}\
  }\textbf {\bibinfo {volume} {2021}}\bibinfo  {number} { (12)},\ \bibinfo
  {pages} {046}}\BibitemShut {NoStop}%
\bibitem [{\citenamefont {Grandis}\ \emph
  {et~al.}(2024{\natexlab{b}})\citenamefont {Grandis}, \citenamefont
  {Aric{\`o}}, \citenamefont {Schneider},\ and\ \citenamefont
  {Linke}}]{grandisDeterminingBaryonImpact2024}%
  \BibitemOpen
\bibfield  {number} {  }\bibfield  {author} {\bibinfo {author} {\bibfnamefont
  {S.}~\bibnamefont {Grandis}}, \bibinfo {author} {\bibfnamefont
  {G.}~\bibnamefont {Aric{\`o}}}, \bibinfo {author} {\bibfnamefont
  {A.}~\bibnamefont {Schneider}},\ and\ \bibinfo {author} {\bibfnamefont
  {L.}~\bibnamefont {Linke}},\ }\href {https://doi.org/10.1093/mnras/stae259}
  {\bibfield  {journal} {\bibinfo  {journal} {Mon. Not. R. Astron. Soc.}\
  }\textbf {\bibinfo {volume} {528}},\ \bibinfo {pages} {4379} (\bibinfo {year}
  {2024}{\natexlab{b}})}\BibitemShut {NoStop}%
\bibitem [{\citenamefont {Eisenstein}\ and\ \citenamefont
  {Hu}(1999)}]{eisensteinPowerSpectraCold1999}%
  \BibitemOpen
  \bibfield  {author} {\bibinfo {author} {\bibfnamefont {D.~J.}\ \bibnamefont
  {Eisenstein}}\ and\ \bibinfo {author} {\bibfnamefont {W.}~\bibnamefont
  {Hu}},\ }\href {https://doi.org/10.1086/306640} {\bibfield  {journal}
  {\bibinfo  {journal} {Astrophys. J.}\ }\textbf {\bibinfo {volume} {511}},\
  \bibinfo {pages} {5} (\bibinfo {year} {1999})}\BibitemShut {NoStop}%
\bibitem [{\citenamefont {Tinker}\ \emph {et~al.}(2010)\citenamefont {Tinker},
  \citenamefont {Robertson}, \citenamefont {Kravtsov}, \citenamefont {Klypin},
  \citenamefont {Warren}, \citenamefont {Yepes},\ and\ \citenamefont
  {Gottl{\"o}ber}}]{tinkerLargescaleBiasDark2010}%
  \BibitemOpen
  \bibfield  {author} {\bibinfo {author} {\bibfnamefont {J.~L.}\ \bibnamefont
  {Tinker}}, \bibinfo {author} {\bibfnamefont {B.~E.}\ \bibnamefont
  {Robertson}}, \bibinfo {author} {\bibfnamefont {A.~V.}\ \bibnamefont
  {Kravtsov}}, \emph {et~al.},\ }\href
  {https://doi.org/10.1088/0004-637X/724/2/878} {\bibfield  {journal} {\bibinfo
   {journal} {Astrophys. J.}\ }\textbf {\bibinfo {volume} {724}},\ \bibinfo
  {pages} {878} (\bibinfo {year} {2010})}\BibitemShut {NoStop}%
\bibitem [{\citenamefont {Kelly}\ \emph {et~al.}(2024)\citenamefont {Kelly},
  \citenamefont {Jobel}, \citenamefont {Eiger}, \citenamefont {Abd},
  \citenamefont {Jeltema}, \citenamefont {Giles}, \citenamefont {Hollowood},
  \citenamefont {Wilkinson}, \citenamefont {Turner}, \citenamefont {Bhargava},
  \citenamefont {Everett}, \citenamefont {Farahi}, \citenamefont {Romer},
  \citenamefont {Rykoff}, \citenamefont {Wang}, \citenamefont {Bocquet},
  \citenamefont {Cross}, \citenamefont {Faridjoo}, \citenamefont {Franco},
  \citenamefont {Gardner}, \citenamefont {Kwiecien}, \citenamefont {Laubner},
  \citenamefont {McDaniel}, \citenamefont {O'Donnell}, \citenamefont {Sanchez},
  \citenamefont {Schmidt}, \citenamefont {Sripada}, \citenamefont {Swart},
  \citenamefont {Upsdell}, \citenamefont {Webber}, \citenamefont {Aguena},
  \citenamefont {Allam}, \citenamefont {Alves}, \citenamefont {Bacon},
  \citenamefont {Brooks}, \citenamefont {Burke}, \citenamefont
  {Carnero~Rosell}, \citenamefont {Carretero}, \citenamefont {Collins},
  \citenamefont {Costanzi}, \citenamefont {{da Costa}}, \citenamefont
  {Pereira}, \citenamefont {Davis}, \citenamefont {Doel}, \citenamefont
  {Ferrero}, \citenamefont {Frieman}, \citenamefont {{Garc{\'i}a-Bellido}},
  \citenamefont {Giannini}, \citenamefont {Gruen}, \citenamefont {Gruendl},
  \citenamefont {Hilton}, \citenamefont {Hinton}, \citenamefont {Honscheid},
  \citenamefont {James}, \citenamefont {Kuehn}, \citenamefont {Mann},
  \citenamefont {Marshall}, \citenamefont {{Mena-Fern{\'a}ndez}}, \citenamefont
  {Miller}, \citenamefont {Miquel}, \citenamefont {Myles}, \citenamefont
  {Palmese}, \citenamefont {Pieres}, \citenamefont {Plazas~Malag{\'o}n},
  \citenamefont {Rooney}, \citenamefont {Sahlen}, \citenamefont {Sanchez},
  \citenamefont {Sanchez~Cid}, \citenamefont {Schubnell}, \citenamefont
  {{Sevilla-Noarbe}}, \citenamefont {Smith}, \citenamefont {Stott},
  \citenamefont {Suchyta}, \citenamefont {Swanson}, \citenamefont {Tarle},
  \citenamefont {To}, \citenamefont {Viana}, \citenamefont {Weaverdyck},
  \citenamefont {Wiseman},\ and\ \citenamefont {{DES
  Collaboration}}}]{kellyDarkEnergySurvey2024}%
  \BibitemOpen
  \bibfield  {author} {\bibinfo {author} {\bibfnamefont {P.~M.}\ \bibnamefont
  {Kelly}}, \bibinfo {author} {\bibfnamefont {J.}~\bibnamefont {Jobel}},
  \bibinfo {author} {\bibfnamefont {O.}~\bibnamefont {Eiger}}, \emph {et~al.},\
  }\href {https://doi.org/10.1093/mnras/stae1786} {\bibfield  {journal}
  {\bibinfo  {journal} {Mon. Not. R. Astron. Soc.}\ }\textbf {\bibinfo {volume}
  {533}},\ \bibinfo {pages} {572} (\bibinfo {year} {2024})}\BibitemShut
  {NoStop}%
\bibitem [{\citenamefont {Goodman}\ and\ \citenamefont
  {Weare}(2010)}]{goodmanEnsembleSamplersAffine2010}%
  \BibitemOpen
  \bibfield  {author} {\bibinfo {author} {\bibfnamefont {J.}~\bibnamefont
  {Goodman}}\ and\ \bibinfo {author} {\bibfnamefont {J.}~\bibnamefont
  {Weare}},\ }\href {https://doi.org/10.2140/camcos.2010.5.65} {\bibfield
  {journal} {\bibinfo  {journal} {Commun. Appl. Math. Comput. Sci.}\ }\textbf
  {\bibinfo {volume} {5}},\ \bibinfo {pages} {65} (\bibinfo {year}
  {2010})}\BibitemShut {NoStop}%
\bibitem [{\citenamefont {{Foreman-Mackey}}\ \emph {et~al.}(2013)\citenamefont
  {{Foreman-Mackey}}, \citenamefont {Hogg}, \citenamefont {Lang},\ and\
  \citenamefont {Goodman}}]{foreman-mackeyEmceeMCMCHammer2013}%
  \BibitemOpen
  \bibfield  {author} {\bibinfo {author} {\bibfnamefont {D.}~\bibnamefont
  {{Foreman-Mackey}}}, \bibinfo {author} {\bibfnamefont {D.~W.}\ \bibnamefont
  {Hogg}}, \bibinfo {author} {\bibfnamefont {D.}~\bibnamefont {Lang}},\ and\
  \bibinfo {author} {\bibfnamefont {J.}~\bibnamefont {Goodman}},\ }\href
  {https://doi.org/10.1086/670067} {\bibfield  {journal} {\bibinfo  {journal}
  {Publ. Astron. Soc. Pac.}\ }\textbf {\bibinfo {volume} {125}},\ \bibinfo
  {pages} {306} (\bibinfo {year} {2013})}\BibitemShut {NoStop}%
\bibitem [{\citenamefont {Gelman}(2004)}]{gelmanBayesianDataAnalysis2004}%
  \BibitemOpen
  \bibfield  {author} {\bibinfo {author} {\bibfnamefont {A.}~\bibnamefont
  {Gelman}},\ }\href@noop {} {\emph {\bibinfo {title} {Bayesian Data
  Analysis}}},\ \bibinfo {edition} {2nd}\ ed.,\ Texts in Statistical Science\
  (\bibinfo  {publisher} {Chapman \& Hall/CRC},\ \bibinfo {address} {Boca
  Raton, Fla.},\ \bibinfo {year} {2004})\BibitemShut {NoStop}%
\bibitem [{\citenamefont {To}\ \emph {et~al.}(2021)\citenamefont {To},
  \citenamefont {Krause}, \citenamefont {Rozo}, \citenamefont {Wu},
  \citenamefont {Gruen}, \citenamefont {DeRose}, \citenamefont {Rykoff},
  \citenamefont {Wechsler}, \citenamefont {Becker}, \citenamefont {Costanzi},
  \citenamefont {Eifler}, \citenamefont {Pereira},\ and\ \citenamefont
  {Kokron}}]{toCombinationClusterNumber2021}%
  \BibitemOpen
  \bibfield  {author} {\bibinfo {author} {\bibfnamefont {C.}~\bibnamefont
  {To}}, \bibinfo {author} {\bibfnamefont {E.}~\bibnamefont {Krause}}, \bibinfo
  {author} {\bibfnamefont {E.}~\bibnamefont {Rozo}}, \emph {et~al.},\ }\href
  {https://doi.org/10.1093/mnras/stab239} {\bibfield  {journal} {\bibinfo
  {journal} {Mon. Not. R. Astron. Soc.}\ }\textbf {\bibinfo {volume} {502}},\
  \bibinfo {pages} {4093} (\bibinfo {year} {2021})}\BibitemShut {NoStop}%
\bibitem [{\citenamefont
  {Sunayama}(2023)}]{sunayamaObservationalConstraintsAnisotropic2023}%
  \BibitemOpen
  \bibfield  {author} {\bibinfo {author} {\bibfnamefont {T.}~\bibnamefont
  {Sunayama}},\ }\href {https://doi.org/10.1093/mnras/stad786} {\bibfield
  {journal} {\bibinfo  {journal} {Mon. Not. R. Astron. Soc.}\ }\textbf
  {\bibinfo {volume} {521}},\ \bibinfo {pages} {5064} (\bibinfo {year}
  {2023})}\BibitemShut {NoStop}%
\bibitem [{\citenamefont {Salcedo}\ \emph {et~al.}(2024)\citenamefont
  {Salcedo}, \citenamefont {Wu}, \citenamefont {Rozo}, \citenamefont
  {Weinberg}, \citenamefont {To}, \citenamefont {Sunayama},\ and\ \citenamefont
  {Lee}}]{salcedoConsistencyDarkEnergy2024}%
  \BibitemOpen
  \bibfield  {author} {\bibinfo {author} {\bibfnamefont {A.~N.}\ \bibnamefont
  {Salcedo}}, \bibinfo {author} {\bibfnamefont {H.-Y.}\ \bibnamefont {Wu}},
  \bibinfo {author} {\bibfnamefont {E.}~\bibnamefont {Rozo}}, \emph {et~al.},\
  }\href {https://doi.org/10.1103/PhysRevLett.133.221002} {\bibfield  {journal}
  {\bibinfo  {journal} {Phys. Rev. Lett.}\ }\textbf {\bibinfo {volume} {133}},\
  \bibinfo {pages} {221002} (\bibinfo {year} {2024})}\BibitemShut {NoStop}%
\bibitem [{\citenamefont {Klein}\ \emph {et~al.}(2024)\citenamefont {Klein},
  \citenamefont {Mohr}, \citenamefont {Bocquet}, \citenamefont {Aguena},
  \citenamefont {Allen}, \citenamefont {Alves}, \citenamefont {Ansarinejad},
  \citenamefont {Ashby}, \citenamefont {Bacon}, \citenamefont {Bayliss},
  \citenamefont {Benson}, \citenamefont {Bleem}, \citenamefont {Brodwin},
  \citenamefont {Brooks}, \citenamefont {Bulbul}, \citenamefont {Burke},
  \citenamefont {Canning}, \citenamefont {Carlstrom}, \citenamefont {Rosell},
  \citenamefont {Carretero}, \citenamefont {Chang}, \citenamefont {Conselice},
  \citenamefont {Costanzi}, \citenamefont {Crites}, \citenamefont {{da Costa}},
  \citenamefont {Pereira}, \citenamefont {Davis}, \citenamefont {De~Vicente},
  \citenamefont {Desai}, \citenamefont {{de Haan}}, \citenamefont {Dobbs},
  \citenamefont {Doel}, \citenamefont {Ferrero}, \citenamefont {Flores},
  \citenamefont {Frieman}, \citenamefont {George}, \citenamefont {Giannini},
  \citenamefont {Gladders}, \citenamefont {Gonzalez}, \citenamefont {Grandis},
  \citenamefont {Gruen}, \citenamefont {Gruendl}, \citenamefont {Gutierrez},
  \citenamefont {Halverson}, \citenamefont {Hinton}, \citenamefont {Holder},
  \citenamefont {Hollowood}, \citenamefont {Holzapfel}, \citenamefont
  {Honscheid}, \citenamefont {Hrubes}, \citenamefont {Huang}, \citenamefont
  {James}, \citenamefont {Khullar}, \citenamefont {Kim}, \citenamefont {Knox},
  \citenamefont {Kraft}, \citenamefont {K{\'e}ruzor{\'e}}, \citenamefont {Lee},
  \citenamefont {{Luong-Van}}, \citenamefont {Mahler}, \citenamefont {Mantz},
  \citenamefont {Marrone}, \citenamefont {Marshall}, \citenamefont {McDonald},
  \citenamefont {McMahon}, \citenamefont {{Mena-Fern{\'a}ndez}}, \citenamefont
  {Menanteau}, \citenamefont {Meyer}, \citenamefont {Miquel}, \citenamefont
  {Myles}, \citenamefont {Padin}, \citenamefont {Pieres}, \citenamefont
  {Plazas~Malag{\'o}n}, \citenamefont {Pryke}, \citenamefont {Reichardt},
  \citenamefont {Reil}, \citenamefont {Roberson}, \citenamefont {Romer},
  \citenamefont {Romero}, \citenamefont {Ruhl}, \citenamefont {Saliwanchik},
  \citenamefont {Salvati}, \citenamefont {Sanchez}, \citenamefont {Saro},
  \citenamefont {Schaffer}, \citenamefont {Schrabback}, \citenamefont
  {Schubnell}, \citenamefont {{Sevilla-Noarbe}}, \citenamefont {Sharon},
  \citenamefont {Shirokoff}, \citenamefont {Smith}, \citenamefont
  {Somboonpanyakul}, \citenamefont {Stalder}, \citenamefont {Stanford},
  \citenamefont {Stark}, \citenamefont {Strazzullo}, \citenamefont {Suchyta},
  \citenamefont {Swanson}, \citenamefont {Tarle}, \citenamefont {To},
  \citenamefont {Vanderlinde}, \citenamefont {Vieira}, \citenamefont {{von der
  Linden}}, \citenamefont {Weaverdyck}, \citenamefont {Williamson},
  \citenamefont {Wiseman},\ and\ \citenamefont
  {Young}}]{kleinSPTSZMCMFExtension2024}%
  \BibitemOpen
  \bibfield  {author} {\bibinfo {author} {\bibfnamefont {M.}~\bibnamefont
  {Klein}}, \bibinfo {author} {\bibfnamefont {J.~J.}\ \bibnamefont {Mohr}},
  \bibinfo {author} {\bibfnamefont {S.}~\bibnamefont {Bocquet}}, \emph
  {et~al.},\ }\href {https://doi.org/10.1093/mnras/stae1359} {\bibfield
  {journal} {\bibinfo  {journal} {Mon. Not. R. Astron. Soc.}\ }\textbf
  {\bibinfo {volume} {531}},\ \bibinfo {pages} {3973} (\bibinfo {year}
  {2024})}\BibitemShut {NoStop}%
\bibitem [{\citenamefont {Farahi}\ \emph {et~al.}(2019)\citenamefont {Farahi},
  \citenamefont {Chen}, \citenamefont {Evrard}, \citenamefont {Hollowood},
  \citenamefont {Wilkinson}, \citenamefont {Bhargava}, \citenamefont {Giles},
  \citenamefont {Romer}, \citenamefont {Jeltema}, \citenamefont {Hilton},
  \citenamefont {Bermeo}, \citenamefont {Mayers}, \citenamefont
  {Vergara~Cervantes}, \citenamefont {Rozo}, \citenamefont {Rykoff},
  \citenamefont {Collins}, \citenamefont {Costanzi}, \citenamefont {Everett},
  \citenamefont {Liddle}, \citenamefont {Mann}, \citenamefont {Mantz},
  \citenamefont {Rooney}, \citenamefont {Sahlen}, \citenamefont {Stott},
  \citenamefont {Viana}, \citenamefont {Zhang}, \citenamefont {Annis},
  \citenamefont {Avila}, \citenamefont {Brooks}, \citenamefont
  {{Buckley-Geer}}, \citenamefont {Burke}, \citenamefont {Carnero~Rosell},
  \citenamefont {Carrasco~Kind}, \citenamefont {Carretero}, \citenamefont
  {Castander}, \citenamefont {{da Costa}}, \citenamefont {De~Vicente},
  \citenamefont {Desai}, \citenamefont {Diehl}, \citenamefont {Dietrich},
  \citenamefont {Doel}, \citenamefont {Flaugher}, \citenamefont {Fosalba},
  \citenamefont {Frieman}, \citenamefont {{Garc{\'i}a-Bellido}}, \citenamefont
  {Gaztanaga}, \citenamefont {Gerdes}, \citenamefont {Gruen}, \citenamefont
  {Gruendl}, \citenamefont {Gschwend}, \citenamefont {Gutierrez}, \citenamefont
  {Honscheid}, \citenamefont {James}, \citenamefont {Krause}, \citenamefont
  {Kuehn}, \citenamefont {Kuropatkin}, \citenamefont {Lima}, \citenamefont
  {Maia}, \citenamefont {Marshall}, \citenamefont {Melchior}, \citenamefont
  {Menanteau}, \citenamefont {Miquel}, \citenamefont {Ogando}, \citenamefont
  {Plazas}, \citenamefont {Sanchez}, \citenamefont {Scarpine}, \citenamefont
  {Schubnell}, \citenamefont {Serrano}, \citenamefont {{Sevilla-Noarbe}},
  \citenamefont {Smith}, \citenamefont {Sobreira}, \citenamefont {Suchyta},
  \citenamefont {Swanson}, \citenamefont {Tarle}, \citenamefont {Thomas},
  \citenamefont {Tucker}, \citenamefont {Vikram}, \citenamefont {Walker},
  \citenamefont {Weller},\ and\ \citenamefont {{DES
  Collaboration}}}]{farahiMassVarianceArchival2019}%
  \BibitemOpen
  \bibfield  {author} {\bibinfo {author} {\bibfnamefont {A.}~\bibnamefont
  {Farahi}}, \bibinfo {author} {\bibfnamefont {X.}~\bibnamefont {Chen}},
  \bibinfo {author} {\bibfnamefont {A.~E.}\ \bibnamefont {Evrard}}, \emph
  {et~al.},\ }\href {https://doi.org/10.1093/mnras/stz2689} {\bibfield
  {journal} {\bibinfo  {journal} {Mon. Not. R. Astron. Soc.}\ }\textbf
  {\bibinfo {volume} {490}},\ \bibinfo {pages} {3341} (\bibinfo {year}
  {2019})}\BibitemShut {NoStop}%
\bibitem [{\citenamefont {Farahi}\ \emph {et~al.}(2016)\citenamefont {Farahi},
  \citenamefont {Evrard}, \citenamefont {Rozo}, \citenamefont {Rykoff},\ and\
  \citenamefont {Wechsler}}]{farahiGalaxyClusterMass2016}%
  \BibitemOpen
  \bibfield  {author} {\bibinfo {author} {\bibfnamefont {A.}~\bibnamefont
  {Farahi}}, \bibinfo {author} {\bibfnamefont {A.~E.}\ \bibnamefont {Evrard}},
  \bibinfo {author} {\bibfnamefont {E.}~\bibnamefont {Rozo}}, \emph {et~al.},\
  }\href {https://doi.org/10.1093/mnras/stw1143} {\bibfield  {journal}
  {\bibinfo  {journal} {Mon. Not. R. Astron. Soc.}\ }\textbf {\bibinfo {volume}
  {460}},\ \bibinfo {pages} {3900} (\bibinfo {year} {2016})}\BibitemShut
  {NoStop}%
\bibitem [{\citenamefont {Spergel}\ \emph {et~al.}(2015)\citenamefont
  {Spergel}, \citenamefont {Gehrels}, \citenamefont {Baltay}, \citenamefont
  {Bennett}, \citenamefont {Breckinridge}, \citenamefont {Donahue},
  \citenamefont {Dressler}, \citenamefont {Gaudi}, \citenamefont {Greene},
  \citenamefont {Guyon}, \citenamefont {Hirata}, \citenamefont {Kalirai},
  \citenamefont {Kasdin}, \citenamefont {Macintosh}, \citenamefont {Moos},
  \citenamefont {Perlmutter}, \citenamefont {Postman}, \citenamefont
  {Rauscher}, \citenamefont {Rhodes}, \citenamefont {Wang}, \citenamefont
  {Weinberg}, \citenamefont {Benford}, \citenamefont {Hudson}, \citenamefont
  {Jeong}, \citenamefont {Mellier}, \citenamefont {Traub}, \citenamefont
  {Yamada}, \citenamefont {Capak}, \citenamefont {Colbert}, \citenamefont
  {Masters}, \citenamefont {Penny}, \citenamefont {Savransky}, \citenamefont
  {Stern}, \citenamefont {Zimmerman}, \citenamefont {Barry}, \citenamefont
  {Bartusek}, \citenamefont {Carpenter}, \citenamefont {Cheng}, \citenamefont
  {Content}, \citenamefont {Dekens}, \citenamefont {Demers}, \citenamefont
  {Grady}, \citenamefont {Jackson}, \citenamefont {Kuan}, \citenamefont {Kruk},
  \citenamefont {Melton}, \citenamefont {Nemati}, \citenamefont {Parvin},
  \citenamefont {Poberezhskiy}, \citenamefont {Peddie}, \citenamefont {Ruffa},
  \citenamefont {Wallace}, \citenamefont {Whipple}, \citenamefont {Wollack},\
  and\ \citenamefont {Zhao}}]{spergelWideFieldInfrarRedSurvey2015}%
  \BibitemOpen
  \bibfield  {author} {\bibinfo {author} {\bibfnamefont {D.}~\bibnamefont
  {Spergel}}, \bibinfo {author} {\bibfnamefont {N.}~\bibnamefont {Gehrels}},
  \bibinfo {author} {\bibfnamefont {C.}~\bibnamefont {Baltay}}, \emph
  {et~al.},\ }\href {https://doi.org/10.48550/arXiv.1503.03757} {\enquote
  {\bibinfo {title} {Wide-{{Field InfrarRed Survey Telescope-Astrophysics
  Focused Telescope Assets WFIRST-AFTA}} 2015 {{Report}}},}\ } (\bibinfo {year}
  {2015})\BibitemShut {NoStop}%
\bibitem [{\citenamefont {Ivezi{\'c}}\ \emph {et~al.}(2019)\citenamefont
  {Ivezi{\'c}}, \citenamefont {Kahn}, \citenamefont {Tyson}, \citenamefont
  {Abel}, \citenamefont {Acosta}, \citenamefont {Allsman}, \citenamefont
  {Alonso}, \citenamefont {AlSayyad}, \citenamefont {Anderson}, \citenamefont
  {Andrew}, \citenamefont {Angel}, \citenamefont {Angeli}, \citenamefont
  {Ansari}, \citenamefont {Antilogus}, \citenamefont {Araujo}, \citenamefont
  {Armstrong}, \citenamefont {Arndt}, \citenamefont {Astier}, \citenamefont
  {Aubourg}, \citenamefont {Auza}, \citenamefont {Axelrod}, \citenamefont
  {Bard}, \citenamefont {Barr}, \citenamefont {Barrau}, \citenamefont
  {Bartlett}, \citenamefont {Bauer}, \citenamefont {Bauman}, \citenamefont
  {Baumont}, \citenamefont {Bechtol}, \citenamefont {Bechtol}, \citenamefont
  {Becker}, \citenamefont {Becla}, \citenamefont {Beldica}, \citenamefont
  {Bellavia}, \citenamefont {Bianco}, \citenamefont {Biswas}, \citenamefont
  {Blanc}, \citenamefont {Blazek}, \citenamefont {Blandford}, \citenamefont
  {Bloom}, \citenamefont {Bogart}, \citenamefont {Bond}, \citenamefont {Booth},
  \citenamefont {Borgland}, \citenamefont {Borne}, \citenamefont {Bosch},
  \citenamefont {Boutigny}, \citenamefont {Brackett}, \citenamefont {Bradshaw},
  \citenamefont {Brandt}, \citenamefont {Brown}, \citenamefont {Bullock},
  \citenamefont {Burchat}, \citenamefont {Burke}, \citenamefont {Cagnoli},
  \citenamefont {Calabrese}, \citenamefont {Callahan}, \citenamefont {Callen},
  \citenamefont {Carlin}, \citenamefont {Carlson}, \citenamefont
  {Chandrasekharan}, \citenamefont {{Charles-Emerson}}, \citenamefont
  {Chesley}, \citenamefont {Cheu}, \citenamefont {Chiang}, \citenamefont
  {Chiang}, \citenamefont {Chirino}, \citenamefont {Chow}, \citenamefont
  {Ciardi}, \citenamefont {Claver}, \citenamefont {{Cohen-Tanugi}},
  \citenamefont {Cockrum}, \citenamefont {Coles}, \citenamefont {Connolly},
  \citenamefont {Cook}, \citenamefont {Cooray}, \citenamefont {Covey},
  \citenamefont {Cribbs}, \citenamefont {Cui}, \citenamefont {Cutri},
  \citenamefont {Daly}, \citenamefont {Daniel}, \citenamefont {Daruich},
  \citenamefont {Daubard}, \citenamefont {Daues}, \citenamefont {Dawson},
  \citenamefont {Delgado}, \citenamefont {Dellapenna}, \citenamefont {{de
  Peyster}}, \citenamefont {{de Val-Borro}}, \citenamefont {Digel},
  \citenamefont {Doherty}, \citenamefont {Dubois}, \citenamefont
  {{Dubois-Felsmann}}, \citenamefont {Durech}, \citenamefont {Economou},
  \citenamefont {Eifler}, \citenamefont {Eracleous}, \citenamefont {Emmons},
  \citenamefont {Fausti~Neto}, \citenamefont {Ferguson}, \citenamefont
  {Figueroa}, \citenamefont {{Fisher-Levine}}, \citenamefont {Focke},
  \citenamefont {Foss}, \citenamefont {Frank}, \citenamefont {Freemon},
  \citenamefont {Gangler}, \citenamefont {Gawiser}, \citenamefont {Geary},
  \citenamefont {Gee}, \citenamefont {Geha}, \citenamefont {Gessner},
  \citenamefont {Gibson}, \citenamefont {Gilmore}, \citenamefont {Glanzman},
  \citenamefont {Glick}, \citenamefont {Goldina}, \citenamefont {Goldstein},
  \citenamefont {Goodenow}, \citenamefont {Graham}, \citenamefont {Gressler},
  \citenamefont {Gris}, \citenamefont {Guy}, \citenamefont {Guyonnet},
  \citenamefont {Haller}, \citenamefont {Harris}, \citenamefont {Hascall},
  \citenamefont {Haupt}, \citenamefont {Hernandez}, \citenamefont {Herrmann},
  \citenamefont {Hileman}, \citenamefont {Hoblitt}, \citenamefont {Hodgson},
  \citenamefont {Hogan}, \citenamefont {Howard}, \citenamefont {Huang},
  \citenamefont {Huffer}, \citenamefont {Ingraham}, \citenamefont {Innes},
  \citenamefont {Jacoby}, \citenamefont {Jain}, \citenamefont {Jammes},
  \citenamefont {Jee}, \citenamefont {Jenness}, \citenamefont {Jernigan},
  \citenamefont {Jevremovi{\'c}}, \citenamefont {Johns}, \citenamefont
  {Johnson}, \citenamefont {Johnson}, \citenamefont {Jones}, \citenamefont
  {{Juramy-Gilles}}, \citenamefont {Juri{\'c}}, \citenamefont {Kalirai},
  \citenamefont {Kallivayalil}, \citenamefont {Kalmbach}, \citenamefont
  {Kantor}, \citenamefont {Karst}, \citenamefont {Kasliwal}, \citenamefont
  {Kelly}, \citenamefont {Kessler}, \citenamefont {Kinnison}, \citenamefont
  {Kirkby}, \citenamefont {Knox}, \citenamefont {Kotov}, \citenamefont
  {Krabbendam}, \citenamefont {Krughoff}, \citenamefont {Kub{\'a}nek},
  \citenamefont {Kuczewski}, \citenamefont {Kulkarni}, \citenamefont {Ku},
  \citenamefont {Kurita}, \citenamefont {Lage}, \citenamefont {Lambert},
  \citenamefont {Lange}, \citenamefont {Langton}, \citenamefont {Le~Guillou},
  \citenamefont {Levine}, \citenamefont {Liang}, \citenamefont {Lim},
  \citenamefont {Lintott}, \citenamefont {Long}, \citenamefont {Lopez},
  \citenamefont {Lotz}, \citenamefont {Lupton}, \citenamefont {Lust},
  \citenamefont {MacArthur}, \citenamefont {Mahabal}, \citenamefont
  {Mandelbaum}, \citenamefont {Markiewicz}, \citenamefont {Marsh},
  \citenamefont {Marshall}, \citenamefont {Marshall}, \citenamefont {May},
  \citenamefont {McKercher}, \citenamefont {McQueen}, \citenamefont {Meyers},
  \citenamefont {Migliore}, \citenamefont {Miller},\ and\ \citenamefont
  {Mills}}]{ivezicLSSTScienceDrivers2019}%
  \BibitemOpen
  \bibfield  {author} {\bibinfo {author} {\bibfnamefont {{\v Z}.}~\bibnamefont
  {Ivezi{\'c}}}, \bibinfo {author} {\bibfnamefont {S.~M.}\ \bibnamefont
  {Kahn}}, \bibinfo {author} {\bibfnamefont {J.~A.}\ \bibnamefont {Tyson}},
  \emph {et~al.},\ }\href {https://doi.org/10.3847/1538-4357/ab042c} {\bibfield
   {journal} {\bibinfo  {journal} {Astrophys. J.}\ }\textbf {\bibinfo {volume}
  {873}},\ \bibinfo {pages} {111} (\bibinfo {year} {2019})}\BibitemShut
  {NoStop}%
\bibitem [{\citenamefont {Zhou}\ \emph {et~al.}(2026)\citenamefont {Zhou},
  \citenamefont {Li}, \citenamefont {Wu}, \citenamefont {{von der Linden}},
  \citenamefont {Jeltema}, \citenamefont {Shin}, \citenamefont {Birrer},
  \citenamefont {Sunayama}, \citenamefont {Fu}, \citenamefont {Adari},
  \citenamefont {Baumont}, \citenamefont {More}, \citenamefont {Englert},
  \citenamefont {Gorsuch},\ and\ \citenamefont
  {Plazas~Malag{\'o}n}}]{zhouAccurateGalaxyCluster2026}%
  \BibitemOpen
  \bibfield  {author} {\bibinfo {author} {\bibfnamefont {C.}~\bibnamefont
  {Zhou}}, \bibinfo {author} {\bibfnamefont {X.}~\bibnamefont {Li}}, \bibinfo
  {author} {\bibfnamefont {H.-Y.}\ \bibnamefont {Wu}}, \emph {et~al.},\ }\href
  {https://doi.org/10.48550/arXiv.2606.21141} {\enquote {\bibinfo {title}
  {Accurate {{Galaxy Cluster Shear}} and {{Mass Calibration}} for {{LSST}} with
  {{AnaCal}}},}\ } (\bibinfo {year} {2026})\BibitemShut {NoStop}%
\bibitem [{\citenamefont {Xu}\ \emph {et~al.}(2025)\citenamefont {Xu},
  \citenamefont {Huang}, \citenamefont {Leauthaud}, \citenamefont {Diemer},
  \citenamefont {Leidig}, \citenamefont {Cannarozzo},\ and\ \citenamefont
  {Zhou}}]{xuOutskirtStellarMass2025}%
  \BibitemOpen
  \bibfield  {author} {\bibinfo {author} {\bibfnamefont {S.}~\bibnamefont
  {Xu}}, \bibinfo {author} {\bibfnamefont {S.}~\bibnamefont {Huang}}, \bibinfo
  {author} {\bibfnamefont {A.}~\bibnamefont {Leauthaud}}, \emph {et~al.},\
  }\href {https://doi.org/10.3847/1538-4357/adeca0} {\bibfield  {journal}
  {\bibinfo  {journal} {Astrophys. J.}\ }\textbf {\bibinfo {volume} {990}},\
  \bibinfo {pages} {127} (\bibinfo {year} {2025})}\BibitemShut {NoStop}%
\bibitem [{\citenamefont {Cranmer}\ \emph {et~al.}(2020)\citenamefont
  {Cranmer}, \citenamefont {Brehmer},\ and\ \citenamefont
  {Louppe}}]{cranmerFrontierSimulationbasedInference2020}%
  \BibitemOpen
  \bibfield  {author} {\bibinfo {author} {\bibfnamefont {K.}~\bibnamefont
  {Cranmer}}, \bibinfo {author} {\bibfnamefont {J.}~\bibnamefont {Brehmer}},\
  and\ \bibinfo {author} {\bibfnamefont {G.}~\bibnamefont {Louppe}},\ }\href
  {https://doi.org/10.1073/pnas.1912789117} {\bibfield  {journal} {\bibinfo
  {journal} {Proc. Natl. Acad. Sci.}\ }\textbf {\bibinfo {volume} {117}},\
  \bibinfo {pages} {30055} (\bibinfo {year} {2020})}\BibitemShut {NoStop}%
\bibitem [{\citenamefont {Farahi}\ \emph {et~al.}(2026)\citenamefont {Farahi},
  \citenamefont {Zhou},\ and\ \citenamefont
  {Vashistha}}]{farahiOASISObservationAwareSimulationBased2026}%
  \BibitemOpen
  \bibfield  {author} {\bibinfo {author} {\bibfnamefont {A.}~\bibnamefont
  {Farahi}}, \bibinfo {author} {\bibfnamefont {C.}~\bibnamefont {Zhou}},\ and\
  \bibinfo {author} {\bibfnamefont {R.}~\bibnamefont {Vashistha}},\ }\href
  {https://doi.org/10.48550/arXiv.2606.22572} {\enquote {\bibinfo {title}
  {{{OASIS}}: {{Observation-Aware Simulation-Based Inference}} via
  {{Distributional Matching}}},}\ } (\bibinfo {year} {2026})\BibitemShut
  {NoStop}%
\bibitem [{\citenamefont {Sunayama}\ \emph {et~al.}(2024)\citenamefont
  {Sunayama}, \citenamefont {Miyatake}, \citenamefont {Sugiyama}, \citenamefont
  {More}, \citenamefont {Li}, \citenamefont {Dalal}, \citenamefont {Rau},
  \citenamefont {Shi}, \citenamefont {Chiu}, \citenamefont {Shirasaki},
  \citenamefont {Zhang},\ and\ \citenamefont
  {Nishizawa}}]{sunayamaOpticalClusterCosmology2024}%
  \BibitemOpen
  \bibfield  {author} {\bibinfo {author} {\bibfnamefont {T.}~\bibnamefont
  {Sunayama}}, \bibinfo {author} {\bibfnamefont {H.}~\bibnamefont {Miyatake}},
  \bibinfo {author} {\bibfnamefont {S.}~\bibnamefont {Sugiyama}}, \emph
  {et~al.},\ }\href {https://doi.org/10.1103/PhysRevD.110.083511} {\bibfield
  {journal} {\bibinfo  {journal} {Phys. Rev. D}\ }\textbf {\bibinfo {volume}
  {110}},\ \bibinfo {pages} {083511} (\bibinfo {year} {2024})}\BibitemShut
  {NoStop}%
\bibitem [{\citenamefont {Salcedo}\ \emph {et~al.}(2025)\citenamefont
  {Salcedo}, \citenamefont {Rozo}, \citenamefont {Wu}, \citenamefont
  {Weinberg}, \citenamefont {Chiploonkar}, \citenamefont {To}, \citenamefont
  {Cao}, \citenamefont {Rykoff}, \citenamefont {Marcelina~Gountanis},\ and\
  \citenamefont {Zhou}}]{salcedoCosmologicalConstraintsDark2025}%
  \BibitemOpen
  \bibfield  {author} {\bibinfo {author} {\bibfnamefont {A.~N.}\ \bibnamefont
  {Salcedo}}, \bibinfo {author} {\bibfnamefont {E.}~\bibnamefont {Rozo}},
  \bibinfo {author} {\bibfnamefont {H.-Y.}\ \bibnamefont {Wu}}, \emph
  {et~al.},\ }\href {https://doi.org/10.48550/arXiv.2510.25706} {\enquote
  {\bibinfo {title} {Cosmological {{Constraints}} from {{Dark Energy Survey
  Year}} 1 {{Cluster Lensing}} and {{Abundances}} with {{Simulation}}-based
  {{Forward-Modeling}}},}\ } (\bibinfo {year} {2025})\BibitemShut {NoStop}%
\bibitem [{\citenamefont {Osato}\ and\ \citenamefont
  {Nagai}(2022)}]{osatoBaryonPastingAlgorithm2022}%
  \BibitemOpen
  \bibfield  {author} {\bibinfo {author} {\bibfnamefont {K.}~\bibnamefont
  {Osato}}\ and\ \bibinfo {author} {\bibfnamefont {D.}~\bibnamefont {Nagai}},\
  }\href {https://doi.org/10.1093/mnras/stac3669} {\bibfield  {journal}
  {\bibinfo  {journal} {Mon. Not. R. Astron. Soc.}\ }\textbf {\bibinfo {volume}
  {519}},\ \bibinfo {pages} {2069} (\bibinfo {year} {2022})}\BibitemShut
  {NoStop}%
\bibitem [{\citenamefont {K{\'e}ruzor{\'e}}\ \emph {et~al.}(2023)\citenamefont
  {K{\'e}ruzor{\'e}}, \citenamefont {Bleem}, \citenamefont {Buehlmann},
  \citenamefont {Emberson}, \citenamefont {Frontiere}, \citenamefont {Habib},
  \citenamefont {Heitmann},\ and\ \citenamefont
  {Larsen}}]{keruzoreOptimizationQualityAssessment2023}%
  \BibitemOpen
  \bibfield  {author} {\bibinfo {author} {\bibfnamefont {F.}~\bibnamefont
  {K{\'e}ruzor{\'e}}}, \bibinfo {author} {\bibfnamefont {L.~E.}\ \bibnamefont
  {Bleem}}, \bibinfo {author} {\bibfnamefont {M.}~\bibnamefont {Buehlmann}},
  \emph {et~al.},\ }\href {https://doi.org/10.21105/astro.2306.13807}
  {\bibfield  {journal} {\bibinfo  {journal} {Open J. Astrophys.}\ }\textbf
  {\bibinfo {volume} {6}},\ \bibinfo {pages} {43} (\bibinfo {year}
  {2023})}\BibitemShut {NoStop}%
\bibitem [{\citenamefont {Lau}\ \emph {et~al.}(2025)\citenamefont {Lau},
  \citenamefont {Nagai}, \citenamefont {Farahi}, \citenamefont {Ishiyama},
  \citenamefont {Miyatake}, \citenamefont {Osato},\ and\ \citenamefont
  {Shirasaki}}]{lauBaryonPastingUchuu2025}%
  \BibitemOpen
  \bibfield  {author} {\bibinfo {author} {\bibfnamefont {E.~T.}\ \bibnamefont
  {Lau}}, \bibinfo {author} {\bibfnamefont {D.}~\bibnamefont {Nagai}}, \bibinfo
  {author} {\bibfnamefont {A.}~\bibnamefont {Farahi}}, \emph {et~al.},\ }\href
  {https://doi.org/10.3847/1538-4357/ada940} {\bibfield  {journal} {\bibinfo
  {journal} {Astrophys. J.}\ }\textbf {\bibinfo {volume} {980}},\ \bibinfo
  {pages} {122} (\bibinfo {year} {2025})}\BibitemShut {NoStop}%
\bibitem [{\citenamefont {Predehl}\ \emph {et~al.}(2021)\citenamefont
  {Predehl}, \citenamefont {Andritschke}, \citenamefont {Arefiev},
  \citenamefont {Babyshkin}, \citenamefont {Batanov}, \citenamefont {Becker},
  \citenamefont {B{\"o}hringer}, \citenamefont {Bogomolov}, \citenamefont
  {Boller}, \citenamefont {Borm}, \citenamefont {Bornemann}, \citenamefont
  {Br{\"a}uninger}, \citenamefont {Br{\"u}ggen}, \citenamefont {Brunner},
  \citenamefont {Brusa}, \citenamefont {Bulbul}, \citenamefont {Buntov},
  \citenamefont {Burwitz}, \citenamefont {Burkert}, \citenamefont {Clerc},
  \citenamefont {Churazov}, \citenamefont {Coutinho}, \citenamefont {Dauser},
  \citenamefont {Dennerl}, \citenamefont {Doroshenko}, \citenamefont {Eder},
  \citenamefont {Emberger}, \citenamefont {Eraerds}, \citenamefont
  {Finoguenov}, \citenamefont {Freyberg}, \citenamefont {Friedrich},
  \citenamefont {Friedrich}, \citenamefont {F{\"u}rmetz}, \citenamefont
  {Georgakakis}, \citenamefont {Gilfanov}, \citenamefont {Granato},
  \citenamefont {Grossberger}, \citenamefont {Gueguen}, \citenamefont {Gureev},
  \citenamefont {Haberl}, \citenamefont {H{\"a}lker}, \citenamefont {Hartner},
  \citenamefont {Hasinger}, \citenamefont {Huber}, \citenamefont {Ji},
  \citenamefont {v.~Kienlin}, \citenamefont {Kink}, \citenamefont {Korotkov},
  \citenamefont {Kreykenbohm}, \citenamefont {Lamer}, \citenamefont {Lomakin},
  \citenamefont {Lapshov}, \citenamefont {Liu}, \citenamefont {Maitra},
  \citenamefont {Meidinger}, \citenamefont {Menz}, \citenamefont {Merloni},
  \citenamefont {Mernik}, \citenamefont {Mican}, \citenamefont {Mohr},
  \citenamefont {M{\"u}ller}, \citenamefont {Nandra}, \citenamefont {Nazarov},
  \citenamefont {Pacaud}, \citenamefont {Pavlinsky}, \citenamefont {Perinati},
  \citenamefont {Pfeffermann}, \citenamefont {Pietschner}, \citenamefont
  {{Ramos-Ceja}}, \citenamefont {Rau}, \citenamefont {Reiffers}, \citenamefont
  {Reiprich}, \citenamefont {Robrade}, \citenamefont {Salvato}, \citenamefont
  {Sanders}, \citenamefont {Santangelo}, \citenamefont {Sasaki}, \citenamefont
  {Scheuerle}, \citenamefont {Schmid}, \citenamefont {Schmitt}, \citenamefont
  {Schwope}, \citenamefont {Shirshakov}, \citenamefont {Steinmetz},
  \citenamefont {Stewart}, \citenamefont {Str{\"u}der}, \citenamefont
  {Sunyaev}, \citenamefont {Tenzer}, \citenamefont {Tiedemann}, \citenamefont
  {Tr{\"u}mper}, \citenamefont {Voron}, \citenamefont {Weber}, \citenamefont
  {Wilms},\ and\ \citenamefont {Yaroshenko}}]{predehlEROSITAXrayTelescope2021}%
  \BibitemOpen
  \bibfield  {author} {\bibinfo {author} {\bibfnamefont {P.}~\bibnamefont
  {Predehl}}, \bibinfo {author} {\bibfnamefont {R.}~\bibnamefont
  {Andritschke}}, \bibinfo {author} {\bibfnamefont {V.}~\bibnamefont
  {Arefiev}}, \emph {et~al.},\ }\href
  {https://doi.org/10.1051/0004-6361/202039313} {\bibfield  {journal} {\bibinfo
   {journal} {Astron. Astrophys.}\ }\textbf {\bibinfo {volume} {647}},\
  \bibinfo {pages} {A1} (\bibinfo {year} {2021})}\BibitemShut {NoStop}%
\bibitem [{\citenamefont {Barcons}\ \emph {et~al.}(2015)\citenamefont
  {Barcons}, \citenamefont {Nandra}, \citenamefont {Barret}, \citenamefont
  {{den Herder}}, \citenamefont {Fabian}, \citenamefont {Piro}, \citenamefont
  {Watson},\ and\ \citenamefont {{the Athena
  Team}}}]{barconsAthenaXrayObservatory2015}%
  \BibitemOpen
  \bibfield  {author} {\bibinfo {author} {\bibfnamefont {X.}~\bibnamefont
  {Barcons}}, \bibinfo {author} {\bibfnamefont {K.}~\bibnamefont {Nandra}},
  \bibinfo {author} {\bibfnamefont {D.}~\bibnamefont {Barret}}, \emph
  {et~al.},\ }in\ \href {https://doi.org/10.1088/1742-6596/610/1/012008} {\emph
  {\bibinfo {booktitle} {Journal of {{Physics Conference Series}}}}},\ Vol.\
  \bibinfo {volume} {610}\ (\bibinfo  {publisher} {IOP},\ \bibinfo {year}
  {2015})\ p.\ \bibinfo {pages} {012008}\BibitemShut {NoStop}%
\bibitem [{\citenamefont {Ade}\ \emph {et~al.}(2019)\citenamefont {Ade},
  \citenamefont {Aguirre}, \citenamefont {Ahmed}, \citenamefont {Aiola},
  \citenamefont {Ali}, \citenamefont {Alonso}, \citenamefont {Alvarez},
  \citenamefont {Arnold}, \citenamefont {Ashton}, \citenamefont {Austermann},
  \citenamefont {Awan}, \citenamefont {Baccigalupi}, \citenamefont {Baildon},
  \citenamefont {Barron}, \citenamefont {Battaglia}, \citenamefont {Battye},
  \citenamefont {Baxter}, \citenamefont {Bazarko}, \citenamefont {Beall},
  \citenamefont {Bean}, \citenamefont {Beck}, \citenamefont {Beckman},
  \citenamefont {Beringue}, \citenamefont {Bianchini}, \citenamefont {Boada},
  \citenamefont {Boettger}, \citenamefont {Bond}, \citenamefont {Borrill},
  \citenamefont {Brown}, \citenamefont {Bruno}, \citenamefont {Bryan},
  \citenamefont {Calabrese}, \citenamefont {Calafut}, \citenamefont {Calisse},
  \citenamefont {Carron}, \citenamefont {Challinor}, \citenamefont {Chesmore},
  \citenamefont {Chinone}, \citenamefont {Chluba}, \citenamefont {Cho},
  \citenamefont {Choi}, \citenamefont {Coppi}, \citenamefont {Cothard},
  \citenamefont {Coughlin}, \citenamefont {Crichton}, \citenamefont {Crowley},
  \citenamefont {Crowley}, \citenamefont {Cukierman}, \citenamefont {D'Ewart},
  \citenamefont {D{\"u}nner}, \citenamefont {{de Haan}}, \citenamefont
  {Devlin}, \citenamefont {Dicker}, \citenamefont {Didier}, \citenamefont
  {Dobbs}, \citenamefont {Dober}, \citenamefont {Duell}, \citenamefont {Duff},
  \citenamefont {Duivenvoorden}, \citenamefont {Dunkley}, \citenamefont
  {Dusatko}, \citenamefont {Errard}, \citenamefont {Fabbian}, \citenamefont
  {Feeney}, \citenamefont {Ferraro}, \citenamefont {Flux{\`a}}, \citenamefont
  {Freese}, \citenamefont {Frisch}, \citenamefont {Frolov}, \citenamefont
  {Fuller}, \citenamefont {Fuzia}, \citenamefont {Galitzki}, \citenamefont
  {Gallardo}, \citenamefont {Tomas Galvez~Ghersi}, \citenamefont {Gao},
  \citenamefont {Gawiser}, \citenamefont {Gerbino}, \citenamefont {Gluscevic},
  \citenamefont {{Goeckner-Wald}}, \citenamefont {Golec}, \citenamefont
  {Gordon}, \citenamefont {Gralla}, \citenamefont {Green}, \citenamefont
  {Grigorian}, \citenamefont {Groh}, \citenamefont {Groppi}, \citenamefont
  {Guan}, \citenamefont {Gudmundsson}, \citenamefont {Han}, \citenamefont
  {Hargrave}, \citenamefont {Hasegawa}, \citenamefont {Hasselfield},
  \citenamefont {Hattori}, \citenamefont {Haynes}, \citenamefont {Hazumi},
  \citenamefont {He}, \citenamefont {Healy}, \citenamefont {Henderson},
  \citenamefont {{Hervias-Caimapo}}, \citenamefont {Hill}, \citenamefont
  {Hill}, \citenamefont {Hilton}, \citenamefont {Hilton}, \citenamefont
  {Hincks}, \citenamefont {Hinshaw}, \citenamefont {Hlo{\v z}ek}, \citenamefont
  {Ho}, \citenamefont {Ho}, \citenamefont {Howe}, \citenamefont {Huang},
  \citenamefont {Hubmayr}, \citenamefont {Huffenberger}, \citenamefont
  {Hughes}, \citenamefont {Ijjas}, \citenamefont {Ikape}, \citenamefont
  {Irwin}, \citenamefont {Jaffe}, \citenamefont {Jain}, \citenamefont {Jeong},
  \citenamefont {Kaneko}, \citenamefont {Karpel}, \citenamefont {Katayama},
  \citenamefont {Keating}, \citenamefont {Kernasovskiy}, \citenamefont
  {Keskitalo}, \citenamefont {Kisner}, \citenamefont {Kiuchi}, \citenamefont
  {Klein}, \citenamefont {Knowles}, \citenamefont {Koopman}, \citenamefont
  {Kosowsky}, \citenamefont {Krachmalnicoff}, \citenamefont {Kuenstner},
  \citenamefont {Kuo}, \citenamefont {Kusaka}, \citenamefont {Lashner},
  \citenamefont {Lee}, \citenamefont {Lee}, \citenamefont {Leon}, \citenamefont
  {Leung}, \citenamefont {Lewis}, \citenamefont {Li}, \citenamefont {Li},
  \citenamefont {Limon}, \citenamefont {Linder}, \citenamefont
  {{Lopez-Caraballo}}, \citenamefont {Louis}, \citenamefont {Lowry},
  \citenamefont {Lungu}, \citenamefont {Madhavacheril}, \citenamefont {Mak},
  \citenamefont {Maldonado}, \citenamefont {Mani}, \citenamefont {Mates},
  \citenamefont {Matsuda}, \citenamefont {Maurin}, \citenamefont {Mauskopf},
  \citenamefont {May}, \citenamefont {McCallum}, \citenamefont {McKenney},
  \citenamefont {McMahon}, \citenamefont {Meerburg}, \citenamefont {Meyers},
  \citenamefont {Miller}, \citenamefont {Mirmelstein}, \citenamefont {Moodley},
  \citenamefont {Munchmeyer}, \citenamefont {Munson}, \citenamefont {Naess},
  \citenamefont {Nati}, \citenamefont {Navaroli}, \citenamefont {Newburgh},
  \citenamefont {Nguyen}, \citenamefont {Niemack}, \citenamefont {Nishino},
  \citenamefont {{Orlowski-Scherer}}, \citenamefont {Page}, \citenamefont
  {Partridge}, \citenamefont {Peloton}, \citenamefont {Perrotta}, \citenamefont
  {Piccirillo}, \citenamefont {Pisano}, \citenamefont {Poletti}, \citenamefont
  {Puddu}, \citenamefont {Puglisi}, \citenamefont {Raum}, \citenamefont
  {Reichardt}, \citenamefont {Remazeilles}, \citenamefont {Rephaeli},
  \citenamefont {Riechers}, \citenamefont {Rojas}, \citenamefont {Roy},
  \citenamefont {Sadeh}, \citenamefont {Sakurai}, \citenamefont {Salatino},
  \citenamefont {Sathyanarayana~Rao}, \citenamefont {Schaan}, \citenamefont
  {Schmittfull}, \citenamefont {Sehgal},\ and\ \citenamefont
  {Seibert}}]{adeSimonsObservatoryScience2019}%
  \BibitemOpen
  \bibfield  {author} {\bibinfo {author} {\bibfnamefont {P.}~\bibnamefont
  {Ade}}, \bibinfo {author} {\bibfnamefont {J.}~\bibnamefont {Aguirre}},
  \bibinfo {author} {\bibfnamefont {Z.}~\bibnamefont {Ahmed}}, \emph {et~al.},\
  }\href {https://doi.org/10.1088/1475-7516/2019/02/056} {\bibfield  {journal}
  {\bibinfo  {journal} {J. Cosmol. Astropart. Phys.}\ }\textbf {\bibinfo
  {volume} {2019}},\ \bibinfo {pages} {056}}\BibitemShut {NoStop}%
\bibitem [{\citenamefont {Benson}\ \emph {et~al.}(2014)\citenamefont {Benson},
  \citenamefont {Ade}, \citenamefont {Ahmed}, \citenamefont {Allen},
  \citenamefont {Arnold}, \citenamefont {Austermann}, \citenamefont {Bender},
  \citenamefont {Bleem}, \citenamefont {Carlstrom}, \citenamefont {Chang},
  \citenamefont {Cho}, \citenamefont {Cliche}, \citenamefont {Crawford},
  \citenamefont {Cukierman}, \citenamefont {{de Haan}}, \citenamefont {Dobbs},
  \citenamefont {Dutcher}, \citenamefont {Everett}, \citenamefont {Gilbert},
  \citenamefont {Halverson}, \citenamefont {Hanson}, \citenamefont
  {Harrington}, \citenamefont {Hattori}, \citenamefont {Henning}, \citenamefont
  {Hilton}, \citenamefont {Holder}, \citenamefont {Holzapfel}, \citenamefont
  {Irwin}, \citenamefont {Keisler}, \citenamefont {Knox}, \citenamefont
  {Kubik}, \citenamefont {Kuo}, \citenamefont {Lee}, \citenamefont {Leitch},
  \citenamefont {Li}, \citenamefont {McDonald}, \citenamefont {Meyer},
  \citenamefont {Montgomery}, \citenamefont {Myers}, \citenamefont {Natoli},
  \citenamefont {Nguyen}, \citenamefont {Novosad}, \citenamefont {Padin},
  \citenamefont {Pan}, \citenamefont {Pearson}, \citenamefont {Reichardt},
  \citenamefont {Ruhl}, \citenamefont {Saliwanchik}, \citenamefont {Simard},
  \citenamefont {Smecher}, \citenamefont {Sayre}, \citenamefont {Shirokoff},
  \citenamefont {Stark}, \citenamefont {Story}, \citenamefont {Suzuki},
  \citenamefont {Thompson}, \citenamefont {Tucker}, \citenamefont
  {Vanderlinde}, \citenamefont {Vieira}, \citenamefont {Vikhlinin},
  \citenamefont {Wang}, \citenamefont {Yefremenko},\ and\ \citenamefont
  {Yoon}}]{bensonSPT3GNextgenerationCosmic2014}%
  \BibitemOpen
  \bibfield  {author} {\bibinfo {author} {\bibfnamefont {B.~A.}\ \bibnamefont
  {Benson}}, \bibinfo {author} {\bibfnamefont {P.~A.~R.}\ \bibnamefont {Ade}},
  \bibinfo {author} {\bibfnamefont {Z.}~\bibnamefont {Ahmed}}, \emph {et~al.},\
  }in\ \href {https://doi.org/10.1117/12.2057305} {\emph {\bibinfo {booktitle}
  {Millimeter, {{Submillimeter}}, and {{Far-Infrared Detectors}} and
  {{Instrumentation}} for {{Astronomy VII}}}}},\ Vol.\ \bibinfo {volume}
  {9153}\ (\bibinfo {address} {eprint: arXiv:1407.2973},\ \bibinfo {year}
  {2014})\ p.\ \bibinfo {pages} {91531P}\BibitemShut {NoStop}%
\bibitem [{\citenamefont {Swetz}\ \emph {et~al.}(2011)\citenamefont {Swetz},
  \citenamefont {Ade}, \citenamefont {Amiri}, \citenamefont {Appel},
  \citenamefont {Battistelli}, \citenamefont {Burger}, \citenamefont
  {Chervenak}, \citenamefont {Devlin}, \citenamefont {Dicker}, \citenamefont
  {Doriese}, \citenamefont {D{\"u}nner}, \citenamefont {{Essinger-Hileman}},
  \citenamefont {Fisher}, \citenamefont {Fowler}, \citenamefont {Halpern},
  \citenamefont {Hasselfield}, \citenamefont {Hilton}, \citenamefont {Hincks},
  \citenamefont {Irwin}, \citenamefont {Jarosik}, \citenamefont {Kaul},
  \citenamefont {Klein}, \citenamefont {Lau}, \citenamefont {Limon},
  \citenamefont {Marriage}, \citenamefont {Marsden}, \citenamefont {Martocci},
  \citenamefont {Mauskopf}, \citenamefont {Moseley}, \citenamefont
  {Netterfield}, \citenamefont {Niemack}, \citenamefont {Nolta}, \citenamefont
  {Page}, \citenamefont {Parker}, \citenamefont {Staggs}, \citenamefont
  {Stryzak}, \citenamefont {Switzer}, \citenamefont {Thornton}, \citenamefont
  {Tucker}, \citenamefont {Wollack},\ and\ \citenamefont
  {Zhao}}]{swetzOverviewAtacamaCosmology2011}%
  \BibitemOpen
  \bibfield  {author} {\bibinfo {author} {\bibfnamefont {D.~S.}\ \bibnamefont
  {Swetz}}, \bibinfo {author} {\bibfnamefont {P.~A.~R.}\ \bibnamefont {Ade}},
  \bibinfo {author} {\bibfnamefont {M.}~\bibnamefont {Amiri}}, \emph {et~al.},\
  }\href {https://doi.org/10.1088/0067-0049/194/2/41} {\bibfield  {journal}
  {\bibinfo  {journal} {Astrophys. J. Suppl. Ser.}\ }\textbf {\bibinfo {volume}
  {194}},\ \bibinfo {pages} {41} (\bibinfo {year} {2011})}\BibitemShut
  {NoStop}%
\bibitem [{\citenamefont {Bleem}\ \emph {et~al.}(2026)\citenamefont {Bleem},
  \citenamefont {Klein}, \citenamefont {Kornoelje}, \citenamefont {Bocquet},
  \citenamefont {Sobrin}, \citenamefont {Aguena}, \citenamefont {Anderes},
  \citenamefont {Anderson}, \citenamefont {{Andrade-Oliveira}}, \citenamefont
  {Ansarinejad}, \citenamefont {Archipley}, \citenamefont {Balkenhol},
  \citenamefont {Barron}, \citenamefont {Barry}, \citenamefont {Bayliss},
  \citenamefont {Benabed}, \citenamefont {Bender}, \citenamefont {Benson},
  \citenamefont {Bianchini}, \citenamefont {Bouchet}, \citenamefont {Brooks},
  \citenamefont {Burke}, \citenamefont {Calzadilla}, \citenamefont {Camilleri},
  \citenamefont {Camphuis}, \citenamefont {Campitiello}, \citenamefont
  {Carlstrom}, \citenamefont {Carnero~Rosell}, \citenamefont {Carretero},
  \citenamefont {Carron}, \citenamefont {Chang}, \citenamefont {Chichura},
  \citenamefont {Chokshi}, \citenamefont {Chou}, \citenamefont {Coerver},
  \citenamefont {Costanzi}, \citenamefont {Crawford}, \citenamefont {{da
  Costa}}, \citenamefont {Daley}, \citenamefont {Davis}, \citenamefont {{de
  Haan}}, \citenamefont {De~Vicente}, \citenamefont {Desai}, \citenamefont
  {Dibert}, \citenamefont {Diehl}, \citenamefont {Dobbs}, \citenamefont
  {Doohan}, \citenamefont {Dutcher}, \citenamefont {Everett}, \citenamefont
  {Evrard}, \citenamefont {Feng}, \citenamefont {Ferguson}, \citenamefont
  {Ferree}, \citenamefont {Fichman}, \citenamefont {Flaugher}, \citenamefont
  {Floyd}, \citenamefont {Fosdick}, \citenamefont {Foster}, \citenamefont
  {Galli}, \citenamefont {Gambrel}, \citenamefont {Gao}, \citenamefont
  {{Garc{\'i}a-Bellido}}, \citenamefont {Ge}, \citenamefont {Gladders},
  \citenamefont {Grandis}, \citenamefont {Guidi}, \citenamefont {Guns},
  \citenamefont {Gutierrez}, \citenamefont {Halverson}, \citenamefont {Hinton},
  \citenamefont {Hivon}, \citenamefont {Holder}, \citenamefont {Hollowood},
  \citenamefont {Holzapfel}, \citenamefont {Hood}, \citenamefont {Hryciuk},
  \citenamefont {Huang}, \citenamefont {Jhaveri}, \citenamefont {Kent},
  \citenamefont {K{\'e}ruzor{\'e}}, \citenamefont {Khalife}, \citenamefont
  {Khullar}, \citenamefont {Knox}, \citenamefont {Kuehn}, \citenamefont {Kuo},
  \citenamefont {Lahav}, \citenamefont {Levy}, \citenamefont {Li},
  \citenamefont {Lowitz}, \citenamefont {Lu}, \citenamefont {Lynch},
  \citenamefont {Maccarone}, \citenamefont {Mahler}, \citenamefont {Maniyar},
  \citenamefont {Marshall}, \citenamefont {Martsen}, \citenamefont {McDonald},
  \citenamefont {{Mena-Fern{\'a}ndez}}, \citenamefont {Menanteau},
  \citenamefont {Millea}, \citenamefont {Miquel}, \citenamefont {Mohr},
  \citenamefont {Montgomery}, \citenamefont {Myles}, \citenamefont {Nakato},
  \citenamefont {Natoli}, \citenamefont {Nichol}, \citenamefont {Novosad},
  \citenamefont {Ogando}, \citenamefont {Omori}, \citenamefont {Ouellette},
  \citenamefont {Pan}, \citenamefont {Phadke}, \citenamefont
  {Plazas~Malag{\'o}n}, \citenamefont {Pollak}, \citenamefont {Porredon},
  \citenamefont {Prabhu}, \citenamefont {Prat}, \citenamefont {Quan},
  \citenamefont {Raghunathan}, \citenamefont {Rahimi}, \citenamefont {Rahlin},
  \citenamefont {Reichardt}, \citenamefont {Romer}, \citenamefont {Rouble},
  \citenamefont {Ruhl}, \citenamefont {Sanchez}, \citenamefont {Sanchez~Cid},
  \citenamefont {Schrabback}, \citenamefont {{Sevilla-Noarbe}}, \citenamefont
  {Silva~Oliveira}, \citenamefont {Simpson}, \citenamefont {Somboonpanyakul},
  \citenamefont {Stark}, \citenamefont {Suchyta}, \citenamefont {Swanson},
  \citenamefont {Tandoi}, \citenamefont {To}, \citenamefont {Trendafilova},
  \citenamefont {Vieira}, \citenamefont {Vieregg}, \citenamefont {Vikram},
  \citenamefont {Vitrier}, \citenamefont {Wan}, \citenamefont {Weaverdyck},
  \citenamefont {Weller}, \citenamefont {Whitehorn}, \citenamefont {Wu},
  \citenamefont {Young}, \citenamefont {Zebrowski},\ and\ \citenamefont
  {Zhan}}]{bleemGalaxyClustersSelected2026}%
  \BibitemOpen
  \bibfield  {author} {\bibinfo {author} {\bibfnamefont {L.~E.}\ \bibnamefont
  {Bleem}}, \bibinfo {author} {\bibfnamefont {M.}~\bibnamefont {Klein}},
  \bibinfo {author} {\bibfnamefont {K.}~\bibnamefont {Kornoelje}}, \emph
  {et~al.},\ }\href {https://doi.org/10.48550/arXiv.2607.01175} {\enquote
  {\bibinfo {title} {Galaxy {{Clusters Selected}} via the
  {{Sunyaev-Zel}}'dovich {{Effect}} in 5 year data from the {{SPT-3G Main
  Survey}}},}\ } (\bibinfo {year} {2026})\BibitemShut {NoStop}%
\bibitem [{\citenamefont {MacCrann}\ \emph {et~al.}(2021)\citenamefont
  {MacCrann}, \citenamefont {Becker}, \citenamefont {McCullough}, \citenamefont
  {Amon}, \citenamefont {Gruen}, \citenamefont {Jarvis}, \citenamefont {Choi},
  \citenamefont {Troxel}, \citenamefont {Sheldon}, \citenamefont {Yanny},
  \citenamefont {Herner}, \citenamefont {Dodelson}, \citenamefont {Zuntz},
  \citenamefont {Eckert}, \citenamefont {Rollins}, \citenamefont {Varga},
  \citenamefont {Bernstein}, \citenamefont {Gruendl}, \citenamefont {Harrison},
  \citenamefont {Hartley}, \citenamefont {{Sevilla-Noarbe}}, \citenamefont
  {Pieres}, \citenamefont {Bridle}, \citenamefont {Myles}, \citenamefont
  {Alarcon}, \citenamefont {Everett}, \citenamefont {S{\'a}nchez},
  \citenamefont {Huff}, \citenamefont {Tarsitano}, \citenamefont {Gatti},
  \citenamefont {Secco}, \citenamefont {Abbott}, \citenamefont {Aguena},
  \citenamefont {Allam}, \citenamefont {Annis}, \citenamefont {Bacon},
  \citenamefont {Bertin}, \citenamefont {Brooks}, \citenamefont {Burke},
  \citenamefont {Carnero~Rosell}, \citenamefont {Carrasco~Kind}, \citenamefont
  {Carretero}, \citenamefont {Costanzi}, \citenamefont {Crocce}, \citenamefont
  {Pereira}, \citenamefont {De~Vicente}, \citenamefont {Desai}, \citenamefont
  {Diehl}, \citenamefont {Dietrich}, \citenamefont {Doel}, \citenamefont
  {Eifler}, \citenamefont {Ferrero}, \citenamefont {Fert{\'e}}, \citenamefont
  {Flaugher}, \citenamefont {Fosalba}, \citenamefont {Frieman}, \citenamefont
  {{Garc{\'i}a-Bellido}}, \citenamefont {Gaztanaga}, \citenamefont {Gerdes},
  \citenamefont {Giannantonio}, \citenamefont {Gschwend}, \citenamefont
  {Gutierrez}, \citenamefont {Hinton}, \citenamefont {Hollowood}, \citenamefont
  {Honscheid}, \citenamefont {James}, \citenamefont {Lahav}, \citenamefont
  {Lima}, \citenamefont {Maia}, \citenamefont {March}, \citenamefont
  {Marshall}, \citenamefont {Martini}, \citenamefont {Melchior}, \citenamefont
  {Menanteau}, \citenamefont {Miquel}, \citenamefont {Mohr}, \citenamefont
  {Morgan}, \citenamefont {Muir}, \citenamefont {Ogando}, \citenamefont
  {Palmese}, \citenamefont {{Paz-Chinch{\'o}n}}, \citenamefont {Plazas},
  \citenamefont {{Rodriguez-Monroy}}, \citenamefont {Roodman}, \citenamefont
  {Samuroff}, \citenamefont {Sanchez}, \citenamefont {Scarpine}, \citenamefont
  {Serrano}, \citenamefont {Smith}, \citenamefont {{Soares-Santos}},
  \citenamefont {Suchyta}, \citenamefont {Swanson}, \citenamefont {Tarle},
  \citenamefont {Thomas}, \citenamefont {To}, \citenamefont {Wilkinson},\ and\
  \citenamefont {{(DES Collaboration)}}}]{maccrannDarkEnergySurvey2021}%
  \BibitemOpen
  \bibfield  {author} {\bibinfo {author} {\bibfnamefont {N.}~\bibnamefont
  {MacCrann}}, \bibinfo {author} {\bibfnamefont {M.~R.}\ \bibnamefont
  {Becker}}, \bibinfo {author} {\bibfnamefont {J.}~\bibnamefont {McCullough}},
  \emph {et~al.},\ }\href {https://doi.org/10.1093/mnras/stab2870} {\bibfield
  {journal} {\bibinfo  {journal} {Mon. Not. R. Astron. Soc.}\ }\textbf
  {\bibinfo {volume} {509}},\ \bibinfo {pages} {3371} (\bibinfo {year}
  {2021})}\BibitemShut {NoStop}%
\bibitem [{\citenamefont {Kleinebreil}\ \emph {et~al.}(2025)\citenamefont
  {Kleinebreil}, \citenamefont {Grandis}, \citenamefont {Schrabback},
  \citenamefont {Ghirardini}, \citenamefont {Chiu}, \citenamefont {Liu},
  \citenamefont {Kluge}, \citenamefont {Reiprich}, \citenamefont {Artis},
  \citenamefont {Bahar}, \citenamefont {Balzer}, \citenamefont {Bulbul},
  \citenamefont {Clerc}, \citenamefont {Comparat}, \citenamefont {Garrel},
  \citenamefont {Gruen}, \citenamefont {Li}, \citenamefont {Miyatake},
  \citenamefont {Miyazaki}, \citenamefont {{Ramos-Ceja}}, \citenamefont
  {Sanders}, \citenamefont {Seppi}, \citenamefont {Okabe},\ and\ \citenamefont
  {Zhang}}]{kleinebreilSRGEROSITAAllSky2025}%
  \BibitemOpen
  \bibfield  {author} {\bibinfo {author} {\bibfnamefont {F.}~\bibnamefont
  {Kleinebreil}}, \bibinfo {author} {\bibfnamefont {S.}~\bibnamefont
  {Grandis}}, \bibinfo {author} {\bibfnamefont {T.}~\bibnamefont {Schrabback}},
  \emph {et~al.},\ }\href {https://doi.org/10.1051/0004-6361/202449599}
  {\bibfield  {journal} {\bibinfo  {journal} {Astron. Astrophys.}\ }\textbf
  {\bibinfo {volume} {695}},\ \bibinfo {pages} {A216} (\bibinfo {year}
  {2025})}\BibitemShut {NoStop}%
\bibitem [{\citenamefont {Navarro}\ \emph {et~al.}(1997)\citenamefont
  {Navarro}, \citenamefont {Frenk},\ and\ \citenamefont
  {White}}]{navarroUniversalDensityProfile1997}%
  \BibitemOpen
  \bibfield  {author} {\bibinfo {author} {\bibfnamefont {J.~F.}\ \bibnamefont
  {Navarro}}, \bibinfo {author} {\bibfnamefont {C.~S.}\ \bibnamefont {Frenk}},\
  and\ \bibinfo {author} {\bibfnamefont {S.~D.~M.}\ \bibnamefont {White}},\
  }\href {https://doi.org/10.1086/304888} {\bibfield  {journal} {\bibinfo
  {journal} {Astrophys. J.}\ }\textbf {\bibinfo {volume} {490}},\ \bibinfo
  {pages} {493} (\bibinfo {year} {1997})}\BibitemShut {NoStop}%
\bibitem [{\citenamefont {Schneider}\ and\ \citenamefont
  {Teyssier}(2015)}]{schneiderNewMethodQuantify2015}%
  \BibitemOpen
  \bibfield  {author} {\bibinfo {author} {\bibfnamefont {A.}~\bibnamefont
  {Schneider}}\ and\ \bibinfo {author} {\bibfnamefont {R.}~\bibnamefont
  {Teyssier}},\ }\href {https://doi.org/10.1088/1475-7516/2015/12/049}
  {\bibfield  {journal} {\bibinfo  {journal} {J. Cosmol. Astropart. Phys.}\
  }\textbf {\bibinfo {volume} {2015}},\ \bibinfo {pages} {049}}\BibitemShut
  {NoStop}%
\bibitem [{\citenamefont {Schneider}\ \emph {et~al.}(2019)\citenamefont
  {Schneider}, \citenamefont {Teyssier}, \citenamefont {Stadel}, \citenamefont
  {Chisari}, \citenamefont {Le~Brun}, \citenamefont {Amara},\ and\
  \citenamefont {Refregier}}]{schneiderQuantifyingBaryonEffects2019}%
  \BibitemOpen
  \bibfield  {author} {\bibinfo {author} {\bibfnamefont {A.}~\bibnamefont
  {Schneider}}, \bibinfo {author} {\bibfnamefont {R.}~\bibnamefont {Teyssier}},
  \bibinfo {author} {\bibfnamefont {J.}~\bibnamefont {Stadel}}, \emph
  {et~al.},\ }\href {https://doi.org/10.1088/1475-7516/2019/03/020} {\bibfield
  {journal} {\bibinfo  {journal} {J. Cosmol. Astropart. Phys.}\ }\textbf
  {\bibinfo {volume} {2019}},\ \bibinfo {pages} {020}}\BibitemShut {NoStop}%
\bibitem [{\citenamefont {Wu}\ \emph {et~al.}(2019)\citenamefont {Wu},
  \citenamefont {Weinberg}, \citenamefont {Salcedo}, \citenamefont {Wibking},\
  and\ \citenamefont {Zu}}]{wuCovarianceMatricesGalaxy2019}%
  \BibitemOpen
  \bibfield  {author} {\bibinfo {author} {\bibfnamefont {H.-Y.}\ \bibnamefont
  {Wu}}, \bibinfo {author} {\bibfnamefont {D.~H.}\ \bibnamefont {Weinberg}},
  \bibinfo {author} {\bibfnamefont {A.~N.}\ \bibnamefont {Salcedo}}, \emph
  {et~al.},\ }\href {https://doi.org/10.1093/mnras/stz2617} {\bibfield
  {journal} {\bibinfo  {journal} {Mon. Not. R. Astron. Soc.}\ }\textbf
  {\bibinfo {volume} {490}},\ \bibinfo {pages} {2606} (\bibinfo {year}
  {2019})}\BibitemShut {NoStop}%
\end{thebibliography}%

\end{document}